%Paper: hep-th/9402023
%From: "LUCA LUSANNA. INFN SEZ. DI FIRENZE. PHONE:(055) 2298141."
%%<LUSANNA@fi.infn.it>
%Date: Thu, 3 Feb 1994 18:36:08 +0100 (WET)


\magnification=1200
%**********************************************
%FONTS and FONT dimensions
%**********************************************
%\font\footfont=amr8
%\font\draftfont=amti7
%\fontdimen16\tensy=2.7pt
%\fontdimen17\tensy=2.7pt
%\fontdimen14\tensy=5pt
 %**********************************************
%FOOT AND HEADLINES
%**********************************************
\def\pagenumbers{\pageno=1\footline={\hss\tenrm\folio\hss}}
\def\today{\ifcase\month
\or January\or February\or March\or April\or May
\or June\or July\or August\or September\or October
\or November \or December\fi \space\number\day, \number\year}
\def\draft{\headline={ \the\pageno
\hfill File:\jobname, Draft Version:\today}}
%**********************************************
%COUNTERS
%**********************************************
\countdef\refno=30
\refno=0
\countdef\sectno=31
\sectno=0
\countdef\chapno=32
\chapno=0
%**********************************************
%REFS, SECTIONS and ALL THAT
%**********************************************
\def\ref{\advance \refno by 1 \ifnum\refno<10 \item{ [\the\refno ]} \else
\item{[\the\refno ]} \fi}
\outer\def\section#1#2\par
           {
           \vskip0pt plus .3\vsize\penalty-250\vskip 0pt plus-.3\vsize
           \bigskip\vskip\parskip
           \noindent\leftline{\rlap{\bf #1}
           \hskip 17pt{\bf #2}}
            \nobreak\smallskip}
\def\abstract{\vfill\eject{\bf Abstract}\smallskip}

\def\hangit#1#2\par{\setbox1=\hbox{#1\enspace}
\hangindent\wd1\hangafter=0\noindent\hskip-\wd1
\hbox{#1\enspace}\ignorespaces#2\par}
%*********************************************
%FORMATTING MACROS
%*********************************************

%**********************************************
%MISC
%**********************************************
%like \cases
%**********************************************
%SYMBOLS
%**********************************************
\def\Zint{{Z \kern -.45 em Z}}
\def\complex{{\kern .1em {\raise .47ex \hbox {$\scriptscriptstyle |$}}
\kern -.4em {\rm C}}}
\def\real{{\vrule height 1.6ex width 0.05em depth 0ex
\kern -0.06em {\rm R}}}
\def\rational{{\kern .1em {\raise .47ex \hbox{$\scripscriptstyle |$}}
\kern -.35em {\rm Q}}}
\def\natural{{\vrule height 1.6ex width .05em depth 0ex \kern -.35em {\rm N}}}
\def\vide{{{\rm O} \kern -0.7em /}}

\parskip 0.5truecm
\baselineskip=12pt
\catcode`\^^?=9
%*****************************
%MY FOOTNOTE
%*****************************
\newskip\zmineskip \zmineskip=0pt plus0pt minus0pt
\mathchardef\mineMM=20000
\newinsert\footins
\def\footnote#1{\let\minesf\empty % parameter #2 (the text) is read later
  \ifhmode\edef\minesf{\spacefactor\the\spacefactor}\/\fi
   #1\minesf\vfootnote{#1}}
\def\vfootnote#1{\insert\footins\bgroup
  \interlinepenalty\interfootnotelinepenalty
  \splittopskip\ht\strutbox % top baseline for broken footnotes
  \splitmaxdepth\dp\strutbox \floatingpenalty\mineMM
  \leftskip\zmineskip \rightskip\zmineskip
\spaceskip\zmineskip \xspaceskip\zmineskip
 \item{#1}\footstrut\futurelet\next\fominet}
\def\fominet{\ifcat\bgroup\noexpand\next \let\next\fmineminet
  \else\let\next\fminet\fi \next}
\def\fmineminet{\bgroup\aftergroup\minefoot\let\next}
\def\fminet#1{#1\minefoot}
\def\minefoot{\strut\egroup}
\def\footstrut{\vbox to\splittopskip{}}
\skip\footins=\bigskipamount % space added when footnote is present
\count\footins=1000 % footnote magnification factor (1 to 1)
\dimen\footins=8in % maximum footnotes per page

%%%%%%%%%%%%%%%%%%%%%%%%%%%%%%%%%%%%%%%%%%%%%%%%%%%%%%%%%%%%%%%%%%%%%%%%%

\overfullrule=0pt
\def \a capo {\hfill \break}

\nopagenumbers
$$\vbox{
\vskip 0.5truecm}$$
\centerline{\bf{Dirac's Observables}}
\centerline{\bf{for}}
\centerline{\bf{Classical Yang-Mills Theory}}
\centerline{\bf {with Fermions}}
\vskip 3truecm
\centerline {\bf Luca Lusanna}
\vskip 1.5truecm
\centerline {Sezione INFN di Firenze}
\centerline {Largo E.Fermi 2 (Arcetri)}
\centerline {50125 Firenze, Italy}
\centerline {e-mail: LUSANNA@FI.INFN.IT}
\vskip 8truecm
\noindent January 1994

hep-th/9402023

\vfill\eject
%%%%%%%%%%%%%%%%%%%%%%%%%%%%%%%%%%%%%%%%%%%%%%%%%%%%%%%%%%%%%%%%%%%%%%%%%%%%%%
%\baselineskip=20pt
\a capo
$$\vbox{
\vskip 1truecm}$$
\centerline
{\bf ABSTRACT}
\bigskip
\noindent

For pure Yang-Mills theory on
Minkowski space-time, formulated in functional spaces where the covariant
divergence is an elliptic operator without zero modes and for a trivial
principal bundle over the fixed-time Euclidean space with a compact,
semisimple,
connected and simply connected structure Lie group, a Green function for the
covariant divergence has been found. It allows to solve the first class
constraints associated with Gauss' laws and to identify a
connection-dependent coordinatization of the trivial principal bundle. In a
neighbourhood of the global identity section, by using canonical coordinates
of first kind on the fibers, one has a symplectic implementation of the
Lie algebra of the small gauge transformations generated by Gauss' laws
and one can make a generalized Hodge decomposition of the gauge potential
one-forms based on the BRST operator.
This decomposition singles out a pure gauge background
connection (the BRST ghost as Maurer-Cartan one-form on the group of gauge
transformations) and a transverse gauge covariant magnetic gauge potential.
After an analogous decomposition of the electric field strength into transverse
and longitudinal part,  Dirac's observables associated with the
transverse electric and magnetic components are identified as their restriction
to the global identity section of the trivial principal bundle. The
longitudinal
part of the electric field can be re-expressed in terms of these electric and
magnetic transverse parts and of the constraints without Gribov ambiguity.
The physical Lagrangian, Hamiltonian, non-Abelian and topological
charges have been obtained
in terms of transverse Dirac's observables, also in presence of fermion fields,
after a symplectic decoupling of the gauge degrees of freedom; one has
an explicit realization of the abstract "Riemannian metric" on the orbit space.
Both the Lagrangian and the Hamiltonian are non-local and
non-polynomial; like in the Coulomb gauge they are not Lorentz invariant, but
the invariance can be enforced on them if one introduces Wigner covariance
of the observables by analyzing the various kinds of
Poincar\'e orbits of the system and by reformulating the theory on suitable
space-like or light-like hypersurfaces, following Dirac. By extending to
classical relativistic field theory the problems associated with the
Lorentz non-covariance of the canonical (presymplectic) center-of-mass for
extended relativistic systems, in the
sector of the field theory with $P^2 > 0$ and $W^2\not= 0$ one identifies a
classical invariant intrinsic unit of length, determined by the Poincar\'e
Casimirs, whose quantum counterpart is the ultraviolet cutoff looked for by
Dirac and Yukawa: it is the Compton wavelength of the field configuration
(in an irreducible Poincar\'e representation) multiplied by the value of its
spin.

\vfill\eject

\def\pagenumbers{\pageno=3\footline={\hss\tenrm\folio\hss}}
\pagenumbers
\noindent
{\bf{1. Introduction}}
\newcount \nfor

\def \form {\global \advance \nfor by 1 \eqno (1.\the\nfor)}
\bigskip

Dirac-Bergmann [1] theory of Hamiltonian constraints and of singular
Lagrangians lies at the basis of most of the special and general
relativistic theories relevant to physics [2]; also Newtonian mechanics
and Newtonian gravities can be reformulated in this framework [3]. The basic
structure behind Dirac-Bergmann approach is presymplectic geometry [4]. In
a series of papers [5] I have revisited Dirac-Bergmann theory, starting, at
the Lagrangian level, with an extended version of the second Noether
theorem allowing an equal treatment of first and second class constraints:
the projection to phase space of the resulting Noether identities
reproduces the Dirac-Bergmann algorithm. A first outcome of this
analysis was the realization that, when $1^{st}$-class constraints are present,
both the Euler-Lagrange and Hamilton equations do not determine the "gauge"
part of the respective trajectories. One has to introduce as many parameters
as $1^{st}$-class constraints to describe the "gauge directions" in the
chosen formalism (either the second order Lagrangian one or the first order
phase space one) and to consider the $1^{st}$-class constraints as
Hamiltonians for the evolution in these parameters. This multiparametric
(also called "multitemporal", because in the case of both nonrelativistic and
relativistic particle mechanics the parameters are affine time coordinates
for the particle world-lines) description produces a local parametrization of
the original variables in terms of the gauge ones by solving the equations
defining the transformation properties under infinitesimal gauge
transformations
in both formalisms (see Ref.[6] for a first attempt of this kind). The next
step was to realize the relevance of the Shanmugadhasan canonical
transformations [7,8], based on Lie theory of function groups and providing
a way for finding local Darboux bases for presymplectic manifolds in the
finite-dimensional setting: given a set of $1^{st}$- ,$(\phi_a)$, and
$2^{nd}$-class, $(\chi_{\alpha},{\tilde \chi}_{\alpha})$, constraints, there
always exist local canonical transformations such that locally the
presymplectic
constraint manifold is described by the vanishing of a subset, $(P_a)$,
of the new momenta (local Abelianization of the $1^{st}$-class constraints;
the conjugate variables, $(Q_a)$, are the Abelian gauge variables adapted to
this Abelianization) and of a subset of pairs, $(P_{\alpha}, Q_{\alpha})$,
of the new canonical coordinates (canonical form of the $2^{nd}$-class
constraints adapted to the Abelianization of the $1^{st}$-class ones). The
remaining coordinates, $(Q_A,P_A)$, are Dirac's observables, also commuting
with the gauge variables $(Q_a)$; $(Q_a,Q_A,P_A)$ form a local Darboux basis
for the constraint presymplectic manifold defined by $\phi_a=\chi_{\alpha}=
{\tilde \chi}_{\alpha}=0$. Since, in general, given a singular Lagrangian
its associated Hamiltonian formulation does not admit any "global"
Shanmugadhasan canonical transformation, the problem arises whether the
singular Lagrangians describing physical systems do have at least one such
global canonical transformation. This would allow to make a global separation
between Dirac's observables and the set of $1^{st}$-class constraints with the
associated gauge variables and of $2^{nd}$-class constraints without
introducing gauge-fixings and with no residual gauge freedom, and to
reformulate the dynamics globally only in terms of Dirac's observables
both at the Lagrangian and at the Hamiltonian level (this is a strong
requirement: the physical or reduced phase space must be the cotangent
bundle over a physical or reduced configuration space) [9,10].
When this is not possible the model has to be considered pathological already
at the classical level, unless one finds a physical interpretation for the
gauge variables (or at least, when the reduced phase space is not a cotangent
bundle, one renounces to a Lagrangian description).
Let us remember that there is a strict interconnection
between this approach and the applicability of the BRST method [11].

Three remarks are needed at this point. Firstly, one has to distinguish between
$1^{st}$-class constraints associated with gauge transformations and
$1^{st}$-class constraints associated with diffeomorphisms: only in the first
case it must always be possible to get a description in terms of global
observables; in the second case Dirac's observables have the meaning of
independent Cauchy data after a convention about what are space and/or
time and, like in general relativity, one does not expect that they are
globally defined, except maybe for systems defined in the flat Minkowski
space-time. Secondly, when Grassmann-valued Dirac's observables are
present, only even functions of them can, in some sense, be associated with
physically measurable quantities. Thirdly, the theorems at the basis of the
existence of the Shanmugadhasan canonical transformation have not been extended
to the infinite-dimensional case, especially to classical field theory. One
of the main obstacles is that in field theory some of the constraints are
partial differential equations, which only in suitable function spaces are
of the elliptic type: generically, there is the problem of the possible
existence of zero modes of these elliptic operators and of their physical
interpretation. This may give origin to problems like the Gribov ambiguity
and the associated impossibility of global gauge-fixings, which implies the
non-existence of global Dirac's observables.

Notwithstanding that, it is reasonable to try to see whether either local or
global Shanmugadhasan canonical transformations can also be defined in these
cases, at least for string and gauge theories. If globally defined Dirac's
observables exist in these cases, from the bosonic ones among them should be
possible to extract the classical basis of the "local observables" (here
local means restricted to a compact submanifold of the Minkowski
space-time, so that these observables can be measured with an apparatus
contained inside such submanifold) of the algebraic approach to quantum field
theory [12] in the still open case of gauge theories (long range interactions)
[13].

After the understanding of a model of two spinless relativistic particles with
an action-at-a-distance interaction instantaneous in the
center-of-mass frame and of its quantization [14] and after a preliminary study
of the Nambu string with light-cone coordinates [15], a global Shanmugadhasan
canonical transformation was found for the Nambu string [16], in which both
the Abelianized $1^{st}$-class constraints and the Abelian conjugated gauge
variables were globally defined Lorentz scalar quantities (even if in this case
one has a diffeomorphism group and not a gauge one, the result relies on the
integrability of the system; let us remark that a global canonical basis of
Dirac's observables has still to be extracted from the known redundant
observables due to the still lacking control on the spin dynamics).
The lesson in both cases was that the source of
globality was the global momentum map associated with the conservation of the
Poincar\'e generators: to get the result one has to analyze the content in
Poincar\'e orbits of the model (and the associated stratification of the
constraint manifold) and to use, as an intermediate step, a canonical
transformation to center-of-mass and relative variables with the relative
variables adapted to the type of Poincar\'e orbits under study, so that the
relative variables have the associated Wigner covariance. It turns out that
in special relativistic extended systems Dirac's observables have
Wigner covariance properties, with the exception of the three positions
among the six observables associated with the center-of-mass: the three
canonical positions (the classical basis of Newton-Wigner-like position
operators) are not the Euclidean components of a four-vector.

The next step is to look for similar results for the classical
Yang-Mills (YM) gauge theory. In the Abelian electromagnetic case the problem
had already been solved by Dirac himself [17] (see also Ref.[18] and the
string variant of Ref.[19]) also in presence of fermions. Apart from the
problem
of Lorentz covariance, Dirac found global observables after the solution
of the $1^{st}$-class constraint associated with Gauss' law and the
introduction of a conjugated non-local gauge variable. At the basis of this
result there was a choice of the Green function for the divergence operator
and the Hodge decomposition of one-forms on the Euclidean space, which
could also be interpreted as the solution of the multitemporal equations.

The aim of this paper is to extend these results to the non-Abelian case.

In Section 2 the needed notations and elements of differential geometry
for pure YM theory over Minkowski space-time
in the case of a trivial principal bundle with a compact, semisimple,
connected, simply connected Lie structure group are introduced. After
a choice of spatial boundary conditions for the gauge potentials
and the gauge transformations, the Lagrangian and Hamiltonian formulation
are developed, with special emphasis on the symplectic structure, the
analysis of the $1^{st}$-class constraints and the boundary conditions.

In Section 3 the Hamiltonian group of gauge transformations is analyzed.

In Section 4 the fermions are included in the description.

In Section 5 the Abelian case is revisited and it is shown how to obtain
the decoupling of the physical degrees of freedom in both the Lagrangian and
Hamiltonian approaches.

In Section 6, after a short review of the problems at the basis of the
Gribov ambiguity and on its dependence from the choice of the functional space
of gauge potentials, it is shown that in a suitable function space [20], in
which the covariant divergence is an elliptic operator without zero modes,
one can evaluate explicitly the Green function associated with the covariant
divergence. This allows to solve the $1^{st}$-class constraints connected with
Gauss' laws, to find the Green function for the Faddeev-Popov operator and,
in the case of transverse vector gauge potentials, the one for the square
of the covariant derivative. Therefore in this function space the Gribov
ambiguity is absent.

In Section 7 a connection dependent coordinatization of the trivial principal
bundle, suggested by the previous Green functions, is introduced; it uses
canonical coordinates of first kind on the group manifold of the structure
group. Then a solution of the multitemporal equations for the vector gauge
potential is given: it is decomposed in the sum of a pure gauge part (the
BRST ghost as the Maurer-Cartan one-form on the group of gauge transformations)
and a gauge-covariant part (the source of the magnetic field strength),
whose transversality properties are determined by means of a generalized
Hodge decomposition based on the BRST operator. Then the canonical variables
of the gauge sector of YM theory are determined.

In Section 8 the electric field strengths (the YM momenta) are decomposed in
a transversal gauge-covariant part and in a longitudinal one, which is then
re-expressed in terms of the transverse and gauge variables by means of the
Green function of the Faddeev-Popov operator. Then the multitemporal equations
for the gauge-covariant transverse vector gauge potential and electric
field strength are solved and global transverse Dirac's observables are
determined as their restriction to the global identity cross section of the
trivial principal bundle. The extension to fermions of the Abelian results
is given. This completes the determination of the Shanmugadhasan canonical
transformation for YM theory. It is argued that with nontrivial
principal bundles one cannot define global Dirac's observables in the
Shanmugadhasan sense.

In Section 9 the non-linear and non-local physical Lagrangian and
Hamiltonian of pure YM theory are determined; the Green function of the
square of the covariant derivative is needed for this derivation and
the final result is an explicit realization of the abstract
"Riemannian metric on the space of connections" introduced in Refs. [21].
Also the non-Abelian and topological charges are expressed in terms of Dirac's
observables. One obtains the determination of the classical physical Lagrangian
and Hamiltonian of QCD, after the global decoupling of the gauge degrees of
freedom connected with the small gauge transformations.

In Section 10 a sketch of how to remedy the lack of Lorentz covariance is
given. Moreover an analysis of the implications of the noncovariance of the
canonical center-of-mass positions, which are Dirac's observables and
should exist also for field configurations, is done and it is shown how this
could provide a definition of an intrinsic ultraviolet cutoff. This can be
a first tool for the understanding of how to quantize and regularize nonlocal
field theories.

In the Conclusions the open problems of the relativistic presymplectic approach
to classical string and gauge theories are presented.

\vfill\eject

%\bigskip
\noindent
{\bf{2. Definitions and Basic Properties of YM Theory}}
\newcount \nfor

\def \form {\global \advance \nfor by 1 \eqno (2.\the\nfor)}
\bigskip

Let us consider YM theory [22] in the case of a trivial principal bundle
$P(M^4,G)=M^4\times G$
over the flat Minkowski space-time $M^4$, whose coordinates are
$x^{\mu}$ and whose metric is $\eta_{\mu\nu}=\eta^{\mu\nu}=(+---)$,
and with structure group a compact (so that $\pi_2(G)=0$, where $\pi_k(G)$
denote the homotopy groups of G), simple (semisimple), connected,
simply connected ($\pi_1(G)=0$) Lie group G, with compact real simple Lie
algebra $g$ (in the semisimple case $g$ is the direct sum of simple Lie
algebras, so that G is a direct product of simple Lie groups). Therefore G
can be SU(n), Spin(n) (the universal covering group of SO(n)) for $n\geq 5$,
Sp(n), $G_2$, $F_4$, $E_8$ and the universal covering spaces of $E_6$, $E_7$;
for these non-Abelian G one has $\pi_1(G)=\pi_2(G)=0$, $\pi_3(G)=Z$.
For G connected, simply connected the
center $Z_G=\lbrace a\in G | ab=ba  \, for\, every\, b\in G\rbrace$, a closed
Abelian
Lie subgroup of G, is discrete (for SU(n): $Z_{SU(n)}=Z_n$) and the Lie algebra
$g$ has a vanishing center $Z_g=\lbrace u\in g | \lbrack u,t\rbrack =0\, for\,
every\, t\in g\rbrace$.

See Refs.[23] for reviews of YM theory and Refs.[24] for
the needed background in differential geometry.

We shall choose a 3+1 splitting of $M^4$; the treatment of Lorentz covariance
will be done later on in Section 10.
Since we are not giving boundary conditions at $x^o\rightarrow \pm \infty$,
because we will be working in the fixed-time Hamiltonian formalism, the
bundle P has to be viewed as the union over $x^o=ct$ of fixed-time trivial
principal bundles $P^t(R^3,G)=(R^3\times \lbrace x^o\rbrace )\times G\sim
R^3\times G$ with the same structure group but with base
manifold the Euclidean space $R^3$ associated with the 3+1 splitting;
certain boundary conditions at spatial infinity will be assigned to the
connections and gauge transformations over $P^t$.
Analogously, at the Lagrangian level, the action will be considered in a
fixed time interval $\triangle x^o=x^o_f-x^o_i$. A priori only
$P^{\triangle x^o}=\cup_{x^o\in \triangle x^o}P^t$
is well defined in every 3+1 splitting:
since we are going to assume that the Poincar\'e group is globally
implemented, this implies that in the limit $x^o_f\rightarrow \infty$,
$x^o_i\rightarrow -\infty$, in every 3+1 splitting, the original
trivial principal bundle P is recovered; but we do not make any
assumption on the $x^o$ asymptotic behaviour of the connections and
gauge transformations over P, since this is a dynamical issue. Therefore
rather than P, the actual geometrical structure we are going to use
in what follows is $P^{'}=\cup_{-\infty < x^o < +\infty}P^t$. Let us remark
that in this way we immediately break conformal invariance, but this is
justified by the following reasons: i) the conformal group cannot be
globally implemented on the Minkowski space-time; ii) the Casimirs of
the conformal group do not have a physical interpretation like the
Casimirs of the Poincar\'e group; iii) conformal invariance is in any case
broken by quantization and regularization.

Since our principal bundles are trivial ($\pi : P(M^4,G)\rightarrow M^4$ is
the canonical projection; the local coordinates of $p\in P$ are $p=(x,a)$,
$x\in M^4$, $a\in G$; the free right action of G on P is $R^P:P\times G
\rightarrow P$, $(p,b)\mapsto R^P(p,b)=p\cdot b=(x,ab)$),
they admit global cross sections
from the base manifold to the bundle manifold and instead of speaking of
connections on either P or $P^t$, we can use global gauge potentials, i.e.
their pullback to the base manifold by means of global cross sections.
While a global cross
section $\sigma_M:M^4\rightarrow P$ associates a gauge 4-potential
${}^{\sigma_M}A_{a\mu}(x)$ on $M^4$ (a is an index in the adjoint
representation of $g$) to each connection on P (i.e. to each connection
1-form $\omega^{\cal A}$ on P;  ${}^{\sigma_M}A_{a\mu}(x){\hat T}^adx^{\mu}=
\sigma^{*}_M\omega^{\cal A}$, where $\sigma^{*}$ is the pull-back of the
differential forms on P to those on $M^4$ and ${\hat T}^a$, a=1,..,dimG, a
basis of generators of $g$ in the adjoint representation of $g$ on $g$),
a connection on $P^t$
gives rise to a vector gauge potential ${}^{\sigma_R}{\vec A}_a(\vec x,x^o)$,
depending on the parameter $x^o$, on $R^3$ by means of a global cross section
$\sigma_R:R^3\rightarrow P^t$: in $P^t(R^3,G)$ the $A^o_a(\vec x,x^o)$'s do
not have an interpretation like gauge potentials, but instead are independent
fields.
The boundary conditions at spatial infinity for the fields and the gauge
transformations, in each 3+1 splitting, will be chosen in such a way to
have finite Poincar\`e generators and covariant non-Abelian charges;
subsequently they will be refined to exclude the Gribov ambiguity.
Higgs fields and monopole solutions, which require special
boundary conditions, are not considered in this paper.

If, in each point p of the trivial principal bundle P, we consider the
tangent space $T_pP$, then its subspace $V_pP\subset T_pP$ of vertical vectors
tangent to the fiber G through p is intrinsically defined; moreover, by using
the right action $R^P$ of G on P, with each element $t\in g$ can be associated
a fundamental vector field $X_t$, which is vertical in each $p\in P$: $X_t{|}_p
\in V_pP$. Instead, there is no intrinsic notion of horizontability. To
introduce such a notion one defines a connection ${\cal A}$ on P (described
by a differential 1-form $\omega^{\cal A}$ on P): it implies the existence,
in a differentiable way and covariantly with respect to the action $R^P$,
of a direct sum decomposition $T_pP=V_pP\oplus H_p^{\cal A}P$, in which
$H_p^{\cal A}P$ is the ${\cal A}$-horizontal subspace; vectors in $H_p
^{\cal A}P$ are said ${\cal A}$-horizontal. Then with each curve x(s),
$s\in [0,1]$, in the base manifold, through each point p=p(0) in the fiber
over x(0), one can associate a unique ${\cal A}$-lift, i.e. a curve p(s)
starting at p(0), projected to x(s) by the canonical projection $\pi : P
\rightarrow M^4$ and with all its tangent vectors ${\cal A}$-horizontal: one
says that p(1) has been obtained by parallel transport of p(0) along p(s)
or, more in general, that the fiber $\pi^{-1}(x(0))$ has been parallel
displaced along the curve x(s) in the base manifold (in this way one can
compare directions in different fibers). Instead of fibers one can parallel
transport other geometrical objects $\phi (x)$ (like matter fields, which are
cross sections of bundles associated with P, whose fibers are representation
spaces of G and with the same base manifold and structure group) defined on
the base manifold along a path $\gamma \subset M^4$: if ${}^{\sigma}A_{\mu}$
is the gauge potential associated with the connection ${\cal A}$ by means of
a global cross section $\sigma :M^4\rightarrow P$ (${}^{\sigma}A_{\mu}dx^{\mu}=
\sigma^{*}\omega^{\cal A}$), one defines the non-integrable
(or path-dependent) Wu-Yang phase [22b] or parallel transporter $V(\gamma ;a)=
Pexp\int_{\gamma}{}^{\sigma}A_{\mu}dx^{\mu}$ (P denotes path ordering); the
gauge transformation properties (2-1), (2-8) (see later on),
of $A_{\mu}$ are equivalent to
the requirement that $\phi (x+dx)$ and $V(x+dx\leftarrow x;A)\phi (x)$ have
the same transformation properties (integral formulation of YM theory); in
this way one also defines the covariant derivative of $\phi (x)$ as the
difference between $\phi (x+dx)$ and $V(x+dx\leftarrow x;A)\phi (x)$, the
parallel transport of $\phi$ from x to x+dx (the covariant derivative (2-5),
see later on, is the ${\cal A}$-lift of the derivative in the base manifold).
By means of the exterior covariant differentiation of the connection 1-form
$\omega^{\cal A}$ one gets the curvature 2-form $\Omega^{\cal A}$ on P, whose
pull-back by means of a cross section gives the field strength 2-form
$F=F_{\mu\nu}dx^{\mu}\wedge dx^{\nu}=2\sigma^{*}\Omega^{\cal A}$ on $M^4$.
When $\Omega^{\cal A}=0$, the connection is flat, i.e. integrable: there
exist global maximal integral manifolds (${\cal A}$-horizontal cross sections),
whose tangent space in each point is the ${\cal A}$-horizontal subspace
$H_p^{\cal A}P$ in that point. When $\Omega^{\cal A}\not= 0$, the connection
is not integrable and such maximal integral submanifolds of P do not exist
(notwithstanding that ${\cal A}$-horizontal cross sections may exist). This is
due to the existence of the holonomy group of the connection: if we consider
the ${\cal A}$-lift of a loop $\alpha$ in $M^4$ (x(s), $s\in [0,1]$, x(0)=
x(1)) through p(0) with $\pi (p(0))=x(0)$, then, in general, one has that
$p(1)=p(0)\cdot a\not= p(0)$, $a\in G$, $\pi (p(1))=x(0)$; the set of all the
group elements a of G obtained in this way by considering all possible
(contractible) loops through x(0) is the (restricted) holonomy group
$\Phi^{\cal A}(x(0))$; when P is connected like in our case, for every $x\in
M^4$ one has $\Phi^{\cal A}(x)\approx \Phi^{\cal A}$, i.e. all the holonomy
groups are isomorphic. When $\Phi^{\cal A}=G$, the connection is said
"fully irreducible": this implies that every two points in P can be joined by
a ${\cal A}$-horizontal curve in P, i.e. the holonomy bundle $P^{\cal A}(p)$
(the set of points of P which can be joined to p by a ${\cal A}$-horizontal
curve) coincide with P for every $p\in P$, $P^{\cal A}(p)=P$.

We shall use the passive interpretation of gauge transformations as
changes of global cross sections at fixed given connection
$\omega^{\cal A}$: if ${\cal G}$
denotes the group of such gauge transformations and $\sigma :M^4\rightarrow
P$ is a global cross section, then $U\in {\cal G}$ is a mapping $U:M^4
\rightarrow G$ such that $\sigma_U(x)=\sigma (x)\cdot U(x)$defines a new global
cross section $\sigma_U$  (the group operation in G generates a pointwise
product for the U's) and for each given connection on P the gauge
potential on $M^4$ transforms as ${}^{\sigma}A(x)\rightarrow {}^{\sigma_U}A(x)=
U^{-1}(x){}^{\sigma}A(x)U(x)+U^{-1}(x)dU(x)$. The Lie algebra $g_{\cal G}$ of
${\cal G}$ is the space of all infinitesimal gauge transformations (see Eqs.
(2-8) later on) parametrized by $g$-valued functions $\xi :M^4\rightarrow g$,
$\xi (x)=\xi_a(x){\hat T}^a$ ; it is a Lie algebra under pointwise Lie bracket
of $g$. For trivial principal bundles
$P(M^4,G)$, ${\cal G}$ is formally isomorphic to the group of vertical
automorphisms $Aut_VP$ of P (which is well defined also for non-trivial
principal bundles, in which only local groups ${\cal G}_x$ can be defined in
neighbourhoods of each point $x\in M^4$, because global cross sections and
gauge potentials do not exist;
$Aut_VP$ is a group under composition and hence a subgroup of the
group of smooth diffeomorphisms of P; if $f\in Aut_VP$, then the automorphism
$f:P\rightarrow P$ is such that its projection on $M^4$ is the identity map
on $M^4$) and to its two isomorphic copies
$GauP=\lbrace \gamma :P\rightarrow G | \gamma (p\cdot a)=a^{-1}
\gamma (p)a\, for\, every\, p\in P,a\in G\rbrace$ of smooth maps from P in G
and $\Gamma (AdP)$ of the global
cross sections of the adjoint group (not principal) bundle, denoted as
$AdP=P{\times}_{Ad}
G$ [an associated bundle of P over $M^4$ with standard fiber the vector space
of
the adjoint representation of G on $g$ and with G
acting on itself with the adjoint (not free) action]. Since in our case $Z_G$
is discrete, the gauge transformations in the center $Z_{\cal G}$ of ${\cal G}$
are constantly equal to the elements of $Z_G$ (so that $Z_{\cal G}\sim Z_G$):
$U\in Z_{\cal G}$, $U:M^4\rightarrow Z_G$, $x\mapsto U(x)=a\in Z_G$ for every
$x\in M^4$. In the case of $P^t$, a priori we have the group of gauge
transformations ${\cal G}_{x^o}$, which contains the fixed-time gauge
transformations of ${}^{\sigma}\vec A(\vec x,x^o)$ [i.e. the mappings $U:R^3
\times \lbrace x^o\rbrace \rightarrow G$, $(\vec x,x^o)\mapsto U(\vec x,x^o)$];
but  then, after the phase space analysis, ${\cal G}_{x^o}$ is enlarged to a
group ${\cal G}^{'}_{x^o}$, which may be larger than ${\cal G}$ (see later on
Eqs.(2-36)). However, since ${\cal G}_{x^o}$ is assumed independent from $x^o$
and since we have ${\cal G}\subset {\cal G}^{'}_{x^o}$ for every $x^o$, we
shall speak of ${\cal G}$ (the true invariance group of the action; ${\cal G}
^{'}_{x^o}$ is the quasi-invariance group) also in the case of $P^t$.

Let us remark that a priori in $Aut_VP$ and in GauP there could be a
subgroup iso-\break
morphic to the structure group G (with our trivial principal
bundles it exists): $G_G=\lbrace f_a\in Aut_VP |
f_a(p)=p\cdot a\, for\, every\, p\in P,\, i.e. \gamma_{f_a}(p)=a\in G\, for\,
\gamma_{f_a}\in GauP\rbrace$, so that the resulting active action of gauge
transformations on P is a right action. The induced active action on gauge
potentials corresponds to vary the connection $\omega^{\cal A}$ taking fixed
the cross section $\sigma$: ${}^{\sigma}A(x)\rightarrow {}^{\sigma}A^f(x)=
U_f^{(\sigma )}(x){}^{\sigma}A(x)U_f^{(\sigma )-1}(x)+U_f^{(\sigma )}(x)d
U_f^{(\sigma )-1}(x)$. From $U^{-1}(x){}^{\sigma}A^f(x)U(x)+U^{-1}(x)dU(x)=
{}^{\sigma_U}A^f(x)=U^{(\sigma_U)}_f(x)[U^{-1}(x){}^{\sigma}A(x)U(x)+U^{-1}(x)
dU(x)]U_f^{(\sigma_U)-1}(x)+U_f^{(\sigma_U)}(x)dU_f^{(\sigma_U)-1}(x)$,
one obtains $U^{(\sigma )}_f(x)U(x)=U(x)$\break
$U_f^{(\sigma_U)}(x)$: in this sense
the active and passive interpretations commute and one can think to the passive
action on P as resulting from a left action $\sigma_U(x)=\sigma (x)U(x)=
(k_U\cdot \sigma )(x)$, $k_U\in G$.
When $G_G$ exists like in our case, then in the group
${\cal G}$ there is a subgroup ${\cal G}_G$ isomorphic
to $G_G\sim G$ (if $U\in {\cal G}_G$, then $U(x)=a\in G$ for every $x\in M^4$)
with a left action on P.

To our understanding there is a gap between the abstract mathematical
description of gauge transformations with $Aut_VP$, GauP and $\Gamma (AdP)$
(to be made rigorous by the use of some Sobolev space, see Section 6) and the
physicist description by means of ${\cal G}$; we shall try to clarify this
point in what follows (see Section 3)
and to show that a more careful choice of the functional
spaces is needed to have the two descriptions implying the wanted physical
results.

In the passive interpretation ${\cal G}$ describes the freedom to parametrize
a given connection on P in terms of the gauge potentials on $M^4$ associated
with all global cross sections $\lbrace \sigma :M^4\rightarrow P\rbrace
\in \Gamma (P)$ ($\Gamma (P)$ is the space of cross sections of P);
in Section 7 we shall define a connection-dependent foliation
of the trivial P, whose leaves are special connection-dependent global cross
sections (a family of special cross sections of this kind is called a "gauge"
in Ref.[23e]): this will reduce the redundancy of the description of P by means
of $\Gamma (P)$.

We shall use the following notations and conventions. The
derivatives with respect to contravariant, $x^{\mu}$, or covariant, $x_{\mu}$,
coordinates of $M^4$ are $\partial_{\mu}={ {\partial}\over {\partial x^{\mu}}
}$
, $\partial ^{\mu}= { {\partial}\over {\partial x_{\mu}} }=(\partial^o;\vec
\partial =\lbrace \partial^i\rbrace =-\vec \nabla )$, where by definition the
3-dimensional gradient is $\vec \nabla =\lbrace \nabla^i\rbrace =\lbrace
\partial_i\rbrace =\lbrace -\partial^i\rbrace =-\vec \partial$. For all the
other Euclidean 3-vectors we have $\vec v=\lbrace v^i\rbrace$ and $v^{\mu}=
(v^o;\vec v)$. The d'Alambertian operator is
$\bar \sqcup =\partial^{\mu}\partial
_{\mu}=\partial_o^2-{\vec \partial}^2=\partial_o^2-{\vec \nabla}^2$. For the
totally antisymmetric Levi-Civita tensor $\epsilon_{\mu\nu\rho\sigma}=-\epsilon
^{\mu\nu\rho\sigma}$, we assume $\epsilon^{0123}=\epsilon^{123}=\epsilon_{123}
=1$, where $\epsilon^{ijk}=\epsilon_{ijk}$ is the 3-dimensional antisymmetric
tensor. The summation convention on repeated indices is assumed.
With regards to units, we shall use natural units
$[c]=[\hbar ]=1$ , so that from $[c]=
[lt^{-1}]$ and $[\hbar ]=[ml^2t^{-1}]$ one gets $[l]=[t]=[x^o]=
[m^{-1}]$; in these units the electric charge $e$ is adimensional since
$e^2=\alpha \hbar c$ ($\alpha$ is the fine structure constant); then the
dimensions of the gauge potential and of the field strength are
$[A]=[m^{1/2}t^{-1/2}]=[l^{-1}]$ and $[F]=[l^{-2}]$ (when one does the
rescaling $A={\sl g}\tilde A$, the coupling constant ${\sl g}$, replacing $e$
in the non-Abelian case, is also adimensional in these units);
the dimensions  of fermion fields is are
$[\psi]=[m^{1/2}l^{-1/2}t^{-1/2}]=[l^{-3/2}]$.
The natural units are used at the classical level for the sake of simplicity;
we shall argue later on that for a YM field configuration belonging to a
time-like
irreducible Poincar\'e representation the quantum unit $\hbar$ could be
replaced
by the classical invariant intrinsic (i.e. field configuration dependent) unit
$\sqrt{-W^2/P^2}=|\, {\hat {\vec S}}\, |$, where $W^2$ and $P^2$ are the
Poincar\'e Casimirs and ${\hat {\vec S}}$ is the rest-frame Thomas spin of the
configuration.

Since the Lie algebra $g$ is (semi)simple and compact, we shall use completely
antisymmetric structure constants $c_{abc}$ (the Jacobi identity is
$c_{abd}c_{dce}+c_{cad}c_{dbe}+c_{bcd}c_{dae}=0$ and the Killing-Cartan metric
is $g_{ab}={({\hat T}^a,{\hat T}^b)}_g=Tr({\hat T}^a{\hat T}^b)=
-{1\over 2}\delta_{ab}$) and the  following notations and normalizations

$$\lbrack t^a,t^b\rbrack =c_{abc}t^c,\quad \lbrack T^a,T^b\rbrack =c_{abc}T^c,
\quad Tr\, T^aT^b=-{1\over 2}\delta_{ab}{ {dim\, \rho}\over {dim\, G}}
{ {C_2(\rho )}\over {C_2(ad)} },\quad T^{a\dagger}=-T^a,$$

$$({\hat T}^c)_{ab}=c_{cab},\quad Tr\, {\hat T}^a {\hat T}^b=-{1\over 2}
\delta^{ab}.$$

\noindent both for the abstract generators $t^a$ and for the generators $T^a$
of any matrix representation $\rho$, with ${\hat T}^a$ denoting the generators
of the adjoint representation $ad$ of $g$ on $g$. Here
$C_2(j)=-\delta_{ab}T^aT^b$ is a normalized form of the second
degree Casimir element $C_2=-g_{ab}t^at^b$ [23i].
Let us remark that often in the physical literature one uses hermitean
generators ${\cal T}^a=iT^a$, so that one has $[{\cal T}^a,{\cal T}^b]=ic_{abc}
{\cal T}^c$.

In $M^4$ the configuration variables are the gauge potentials
$A(x)=A_{a\mu}(x){\hat T}^adx^{\mu}$ (we drop the superscript denoting the
dependence on the cross section),
which are globally defined since $M^4$ is topologically trivial
 and $P(M^4,G)$ is trivial.
The gauge potentials $A_{\mu}(x)$ have the following transformation properties
under  (also called 2nd kind) gauge transformations, i.e. under the smooth
mappings  $U:M^4\rightarrow G$, $x^{\mu}\mapsto U(x)$

$$
\eqalign{
A_{\mu}(x) \mapsto A^U_{\mu}(x)&=U^{-1}(x)A_{\mu}(x)U(x)+U^{-1}(x)
{\partial}_{\mu}U(x)=\cr
                                 &=A_{\mu}(x)+U^{-1}(x)({\partial}_{\mu}
U(x)+\lbrack A_{\mu}(x),U(x)\rbrack )\cr}
\form
$$

\noindent We see that every gauge potential is invariant under the gauge
transformations belonging to the center $Z_{\cal G}$ of ${\cal G}$.
Therefore the group of gauge transformations
does not act freely on the space of connections (or on the associated space of
gauge potentials): $Z_{{\cal G}}\sim Z_G$ is the stability subgroup of a
generic connection (according to the chosen function space there may be special
connections with larger stability subgroups of ${\cal G}$).

The gauge fields or field strengths are

$$
\eqalign{
&F_{\mu\nu}=F_{a\mu\nu}{\hat T}^a={\partial}_{\mu}A_{\nu}-
{\partial}_{\nu}A_{\mu}+\lbrack A_{\mu},A_{\nu}\rbrack \cr
&F_{a\mu\nu}={\partial}_{\mu}A_{a\nu}-{\partial}_{\nu}A_{a\mu}+
c_{abc}A_{b\mu}A_{c\nu}\cr},
\form
$$

\noindent With the 1-form on $M^4$
$A(x)=A_{\mu}(x)dx^{\mu}$, one gets that the field
strength is described by the 2-form $F(x)={1\over 2}F_{\mu\nu}(x)dx^{\mu}
\wedge dx^{\nu}=dA(x)+{1\over 2}\lbrack A(x),A(x)\rbrack$, where $\lbrack
A(x),A(x)\rbrack =\lbrack A_{\mu}(x),A_{\nu}(x)\rbrack dx^{\mu}\wedge
dx^{\nu}$.

The field strengths  have the following transformation properties under gauge
transformations (they too are invariant under $U\in Z_{\cal G}$ and, according
to the chosen functional space, some of them have larger stability subgroups
of ${\cal G}$):

$$
F_{\mu \nu}(x) \mapsto F^U_{\mu \nu}(x)=U^{-1}(x)F_{\mu \nu}(x)U(x)=
F_{\mu\nu}(x)+U^{-1}(x)\lbrack F_{\mu\nu}(x),U(x)\rbrack
\form
$$

\noindent and satisfy the Bianchi identities

$$
{\hat D}^{(A)}_{\alpha}F_{\beta \gamma}+{\hat D}^{(A)}_{\beta}F_{\gamma
\alpha}+
{\hat D}^{(A)}_{\gamma}F_{\alpha \beta}\equiv 0, \qquad or
\qquad {\hat D}^{(A)}_{\mu}{*}F^{\mu \nu}\equiv 0
\form
$$

\noindent which are the integrability conditions for the existence of $A_{\mu}
(x)$. In Eqs.(2-4) ${*}F^{\mu\nu}={1\over 2}\epsilon^{\mu\nu\alpha\beta}
F_{\alpha\beta}$ is the dual field strength (${*}$ is the Hodge star operator
[24l,o] and $F^{\mu\nu}=-{1\over 2}\epsilon
^{\mu\nu\rho\sigma}{*}F_{\rho \sigma}=-{**}F^{\mu\nu}$),  while
$D^{(A)}_{\mu}$ is the gauge covariant derivative,
whose action on a matrix quantity $M=M_aT^a$ in the representation $\rho$
is (in the first and last lines one has
$A^{(\rho )}_{\mu}=A^{(\rho )}_{a\mu}T^a$)

$$
\eqalign{
&D^{(A^{(\rho )})}_{\mu}M={\partial}_{\mu}M+\lbrack A^{(\rho )}_{\mu},M\rbrack
=
(D^{(A^{(\rho )})}_{\mu})_{ab}M_bT^a,\quad
D^{(A^{(\rho )})}=dx^{\mu}D^{(A)}_{\mu}\cr
&({\hat D}^{(A)}_{\mu})_{ac}=\delta_{ac}{\partial}_{\mu}+c_{abc}A_{b\mu}=
 {(\partial_{\mu}-A_{\mu})}_{ac},\quad
\lbrack \, {({\hat {\vec D}}^{(A)})}_{ac}=\delta_{ac}\vec \partial +
c_{abc}{\vec A}_b\, \rbrack\cr
&D^{(A^{(\rho )})}_{\mu}\psi (x)=(\partial_{\mu} +A^{(\rho )}_{\mu}(x))\psi (x)
;\cr}
\form
$$

\noindent the second line gives its form ${\hat D}^{(A)}_{\mu}$
in the adjoint representation and the third line its action on a matter
field $\psi (x)$, whose transformation properties under gauge
transformations are $\psi (x) \mapsto \psi^U(x)=\rho (U^{-1}(x))\psi (x)=
U^{(\rho )-1}(x)\psi (x)$,
if $\rho$ is the representation of G to which the matter field belongs.
The covariant derivative has the property
$\lbrack D^{(A^{(\rho )})}_{\mu},D^{(A^{(\rho )})}_{\nu}\rbrack M=\lbrack
F^{(\rho )}_{\mu\nu},M\rbrack$. Eq.(2-1) may be written in the form
$A^U_{\mu}(x)=A_{\mu}(x)+U^{-1}(x){\hat D}^{(A)}_{\mu}U(x)$.
Instead on matter fields one has $[D^{(A^{(\rho )})}_{\mu},D^{(A^{(\rho )})}
_{\nu}]=F^{(\rho )}_{\mu\nu}$.

When $A_{\mu}(x)=U^{-1}
(x){\partial}_{\mu}U(x)$ (pure gauge), one has $F_{\mu\nu}(x)=0$.
Viceversa, $F_{\mu\nu}=0$ will imply only the pure gauge solution for
$A_{\mu}$, because in this paper
the base manifold M will be connected ($\pi_o(M)=0$) and simply connected
($\pi_1(M)=0$) and moreover topologically trivial ($\pi_k(M)=0$ for every k,
where $\pi_k(M)$ is the k-th homotopy group of M); when $\pi_1(M)\not= 0$ the
pure gauge (vacuum) solution is not unique [25].

Let us remark that often one uses a different notation: i) $\psi \mapsto
\psi^U=\rho (U)\psi$;
ii) $D^{(A^{(\rho )})}_{\mu}=\partial_{\mu}-A^{(\rho )}_{\mu}$ (on matter
fields) and $D^{(A^{(\rho )})}_{\mu}=\partial_{\mu}-
[A^{(\rho )}_{\mu},.]$ on matrices $M=M_a
T^a$ (${({\hat D}^{(A)}_{\mu})}_{ac}={(\partial_{\mu}+A_{\mu})}_{ac}=
\delta_{ac}\partial_{\mu}-c_{abc}A_{b\mu}$);
iii) $A_{\mu}\mapsto A^U_{\mu}=UA_{\mu}U^{-1}+(\partial_{\mu}U)U
^{-1}$; iv) $F_{\mu\nu}=\partial_{\mu}A_{\nu}-\partial_{\nu}A_{\mu}-[A_{\mu},
A_{\nu}]\mapsto F^U_{\mu\nu}=UF_{\mu\nu}U^{-1}$ (essentially one has to do the
substitutions $U\rightarrow U^{-1}$ and $A\rightarrow -A$).
To use hermitean generators ${\cal T}^a=iT^a$ one has to make the substitution
$A_{\mu}=A_{a\mu}T^a=-iA_{a\mu}{\cal T}^a=-iA^{'}_{\mu}$ in both the notations.

The YM action and Lagrangian density are respectively

$$
\eqalign{
&S=\int_1^2 d^4x{\cal L}=\int_{x^o_1}^{x^o_2}dx^oL,\qquad L=\int d^3x{\cal
L}\cr
&{\cal L}={1\over {2{\sl g}^2}}TrF^{\mu\nu}F_{\mu\nu}=-{1\over {4{\sl g}^2}}
F^{\mu\nu}_aF_{a\mu\nu},\cr}
\form
$$

\noindent where ${\sl g}$ is a dimensionless parameter; when one wants to
interpret it as a coupling constant one does the rescaling $A={\sl g}{\tilde
A}$
and one has
${\cal L}={1\over 2}Tr\, {\tilde F}^{\mu\nu}{\tilde F}_{\mu\nu}$ with
${\tilde F}_{\mu\nu}=\partial_{\mu}{\tilde A}_{\nu}-\partial_{\nu}{\tilde A}
_{\mu}+{\sl g}[{\tilde A}_{\mu},{\tilde A}_{\nu}]={\sl g}^{-1}F_{\mu\nu}$ and
either $D^{(A^{(\rho )})}_{\mu}=\partial_{\mu}+{\sl g}{\tilde A}^{(\rho )}
_{\mu}$ or $D^{(A^{(\rho )})}_{\mu}
=\partial_{\mu}+{\sl g}[{\tilde A}^{(\rho )}_{\mu},.]$.
In the Abelian electromagnetic
case one uses $A=e\tilde A$ (${\sl g}=e$) so that ${\cal L}=
-{1\over 4}{\tilde F}^{\mu\nu}{\tilde F}_{\mu\nu}$ and the covariant derivative
acting on matter fields is $D^{(A)}_{\mu}=\partial_{\mu}+e{\tilde A}_{\mu}$.

The associated Euler-Lagrange equations are (${\buildrel \rm
\circ \over =}$ means evaluated on the extremals of the variational principle;
their functional form is modified by a factor $-g^2$ with respect to the
standard expression to conform with the physical literature notation):

$$
\eqalign{
&L^{\nu}_a=g^2({ {\partial {\cal L}}\over {\partial A_{a\nu}} }-
{\partial}_{\mu}{ {\partial {\cal L}}\over {\partial \partial_{\mu}A_{a\nu}}
})=
{\partial}_{\mu}F_a^{\mu\nu}+c_{abc}A_{b\mu}F_c^{\mu\nu}
{\buildrel \rm \circ \over =} 0\cr
&L^{\nu}={\hat D}^{(A)}_{\mu}F^{\mu\nu}=
{\partial}_{\mu}F^{\mu\nu}+\lbrack A_{\mu},F^{\mu\nu}\rbrack
{\buildrel \rm \circ \over =} 0\cr}
\form
$$

\noindent and, given local variations $\delta x^{\mu}=0$, $\delta_oA_{a\mu}(x)=
{\bar A}_{a\mu}(x)-A_{a\mu}(x)$, one gets

$$\delta_o{\cal L}=
{1\over {{\sl g}^2}} [\delta_oA_{a\mu}
L^{\mu}_a-{\partial}_{\mu}(F_a^{\mu\nu}\delta_oA_{a\nu})]$$.

S is invariant under the gauge transformations (2-1), (2-3),
whose infinitesimal form, when
$U(x)=e^{\alpha (x)}$, $\alpha (x)=\alpha _a(x){\hat T}^a=-{\alpha}^{\dag}
(x)$, is

$$
\eqalign{
&\delta A_{\mu}(x)={\hat D}^{(A)}_{\mu}\alpha (x)={\partial}_{\mu}\alpha (x)+
\lbrack A_{\mu}(x),\alpha (x)\rbrack \cr
&\delta A_{a\mu}(x)={\partial}_{\mu}\alpha_a (x)+c_{abc}
A_{b\mu}(x)\alpha_c(x)={\hat D}^{(A)}_{\mu ab}\alpha_b(x)\cr
&\delta F_{\mu\nu}(x)={\hat D}^{(A)}_{\mu}\delta A_{\nu}(x)-
{\hat D}^{(A)}_{\nu}\delta A_{\mu}(x)=
\lbrack F_{\mu\nu}(x),\alpha (x)\rbrack \cr
&\delta F_{a\mu\nu}(x)=c_{abc}F_{b\mu\nu}(x)\alpha_c(x)\cr}
\form
$$

The invariance of S under the gauge transformations generates the following
Noether identity [5c] (usually one uses the notation $G^{\mu}=j^{\mu}$ for the
conserved Noether current and $G^{\mu}_{1a}=j^{\mu}_a$ for the one associated
with rigid global gauge transformations)

\noindent
$$
\eqalign{
\delta {\cal L}=\delta_o {\cal L}&={ {\partial {\cal L}}\over
{\partial A_{a\mu}} }\delta A_{a\mu}
+{ {\partial {\cal L}}\over {\partial \partial_{\nu}A_{a\mu}} }\delta \partial
_{\nu}A_{a\mu}=\cr
&={1 \over {{\sl g}^2}}L^{\mu}_a\delta A_{a\mu}+\partial_{\nu}
(\pi^{\nu\mu}_a\delta A_{a\mu})\equiv 0\cr
{}&\Downarrow \cr
{\partial}_{\mu}G^{\mu}&\equiv -{1\over {{\sl g}^2}}
\delta A_{a\mu}L^{\mu}_a{\buildrel \rm \circ \over =} 0\cr
{}&G^{\mu}(x)=\pi_a^{\mu\nu}\delta A_{a\nu}=\alpha_a(x)G^{\mu}_{1a}(x)+
{\partial}_{\nu}\alpha_a(x)G^{\mu\nu}_{0a}(x)\cr
{}&G^{\mu\nu}_{0a}(x)=\pi^{\mu\nu}_a(x)=-{1\over {{\sl g}^2}}F_a^{\mu\nu}(x)\cr
{}&G^{\mu}_{1a}(x)=c_{abc}\pi^{\mu\nu}_b(x)A_{c\nu}(x)=
-{1\over {{\sl g}^2}}c_{abc}F_b^{\mu\nu}(x)A_{c\nu}(x)\cr}
\form
$$

\noindent where
$\pi^{\mu\nu}_a={ {\partial {\cal L}}\over { \partial {\partial}_{\mu}
A_{a\nu} } }=-F^{\mu\nu}_a/{\sl g}^2$.

The canonical momenta are

$$
\pi^{\mu}_a=\pi^{0\mu}_a={\sl g}^{-2}F^{\mu 0}_a={\sl g}^{-1}{\tilde
F}^{\mu o}_a
\form
$$

Eqs.(2-9) give rise to the Noether identities ($(\mu \nu )$ means
symmetrization)

$$
\eqalign{
&G^{(\mu \nu )}_{0a}\equiv 0\cr
&{\partial}_{\nu}G^{\nu\mu}_{0a}\equiv -G^{\mu}_{1a}-{1\over {{\sl g}^2}}
L^{\mu}_a={1\over {{\sl g}^2}}(c_{abc}F^{\mu\nu}_bA_{c\nu}
-L^{\mu}_a){\buildrel \rm \circ \over =}{1\over {{\sl g}^2}}c_{abc}
F^{\mu\nu}_bA_{c\nu}\cr
&{\partial}_{\mu}G^{\mu}_{1a}\equiv {1\over {{\sl g}^2}}
c_{abc}A_{b\mu}L^{\mu}_c{\buildrel \rm \circ \over =} 0\cr}
\form
$$

The contracted Bianchi identity, correlating the equations of motion, are
obtained from the last of Eqs.(2-11) by using the second one:

$$
{\hat D}^{(A)}_{\mu}L^{\mu}=
({\partial}_{\mu}L^{\mu}_a+c_{abc}A_{b\mu}L^{\mu}_c)
{\hat T}^a\equiv 0
\form
$$

The subset of Eqs.(2-11) relevant for phase space are

$$
\eqalign{
&\pi^0_a=\pi^{00}_a=G^{(00)}_{0a}\equiv 0\cr
0&\equiv {\partial}_0\pi^0_a\equiv {\vec {\partial}}\cdot {\vec \pi}_a-
  {\partial}_{\nu}G^{\nu 0}_{0a}\equiv \vec \partial \cdot {\vec \pi}_a+
  G^o_{1a}+{1\over {{\sl g}^2}}L^o_a=\cr
 &= {\vec {\partial}}\cdot {\vec \pi}_a+
  c_{abc}{\vec A}_b\cdot {\vec \pi}_c
  +{1\over {{\sl g}^2}}L^0_a{\buildrel \rm \circ \over =}
  {{\hat {\vec D}}^{(A)}}_{ab}\cdot {\vec \pi}_b\cr}
\form
$$

One sees the primary, $\pi^0_a\approx 0$, and secondary (Gauss' laws),
$-{{\hat {\vec D}}^{(A)}}_{ab}\cdot {\vec \pi}_b\approx 0$,
Hamiltonian constraints emerge
from the Noether identities.

The strong improper conservation laws (Appendix H of Ref.[5c];
see also Ref.[13], where they are called super conservation
laws) ${\partial}_{\mu}
V^{\mu}_a\equiv 0$, implied by Eqs.(2-11), identify the strong improper
conserved current (strong continuity equation):

$$V^{\mu}_a=-\partial_{\nu}G^{\nu\mu}_{oa}=
{1\over {{\sl g}^2}}
{\partial}_{\nu}F^{\nu\mu}_a={\partial}_{\nu}
U^{\lbrack \mu \nu \rbrack}_a{\buildrel \rm \circ \over =} -
{1\over {{\sl g}^2}} c_{abc}F^{\mu\nu}
_bA_{c\nu}=G^{\mu}_{1a}$$ with the superpotential $$U_a^{\lbrack \mu \nu
\rbrack}=-{1\over {{\sl g}^2}} F^{\mu\nu}_a=\pi^{\mu\nu}_a.$$ Instead the
Trautman strong conservation law is $${\partial}_{\mu}
(G^{\mu}(x)+{1\over {{\sl g}^2}} \alpha_a(x)L^{\mu}_a(x))\equiv 0$$.

When $\alpha =\alpha_a{\hat T}^a=const.$, one speaks of
"rigid" or "global" (or $1^{st}$ kind)
gauge transformations, which are improper Noether symmetry transformations.
In the Abelian case this implies $\delta A_{\mu}(x)=\partial_{\mu}\alpha =0$;
in the non-Abelian case from Eqs.(2-8) one has $\delta A_{a\mu}(x)={\hat D}
^{(A)}_{\mu ab}\alpha_b=c_{abc}A_{b\mu}(x)\alpha_c$ (i.e. under them $A_{a\mu}$
transforms covariantly as a tensor). Therefore the global gauge transformations
$U=e^{\alpha}$ are covariantly constant ($\partial_{\mu}U=0$) in the Abelian
case (with G=U(1) the adjoint action is trivial and $\Gamma (AdP)$ is globally
spanned by covariantly constant cross sections independent from the choice of
the connection), but not, in general, in the non-Abelian case where ${\hat D}
^{(a)}_{\mu}U=[A_{\mu},U]\not= 0$ (the special solutions of this equation for
certain special connections are the "gauge symmetries" of these connections),
except if U belongs to the center $Z_{\cal G}$ of ${\cal G}$
(but this corresponds in our case to $\alpha_a=0$,because with $Z_{\cal G}
\sim Z_G$ discrete one has $Z_{g_{\cal G}}\sim Z_g=0$,
i.e. the center of the Lie
algebra $g_{\cal G}$ of ${\cal G}$ vanishes).
As a consequence in Ref.[26]
it is stressed that an intrinsic definition, like covariant constancy, of the
$1^{st}$ kind gauge transformations
is lacking in the non-Abelian case, so that the Abelian charges of
fermions are on different level with respect to the non-Abelian charges, which
exist also without charged matter. However, if we define the global (rigid,
$1^{st}$ kind) gauge transformations as that subgroup of ${\cal G}$ under which
all the gauge potentials transform gauge covariantly (like in the Abelian
case),
then this is also a definition of $1^{st}$ kind gauge transformations on
fermions: while the Abelian charges are gauge scalars, the non-Abelian ones,
with appropriate boundary conditions on the gauge transformations (see later
on),
turn out to be well defined gauge covariant quantities [27]. But, like
Ref.[26],
also Ref.[27], following the pattern of the definition of energy in general
relativity, uses the concept of covariantly constant gauge transformations,
${\hat D}^{(A)}_{\mu}U=0$, associated with special connections, to try to
define
the non-Abelian charges; as examples, the magnetic charge of the
t'Hooft-Polyakov monopole and the electric and magnetic charges of the
Julia-Zee dyon are evaluated (however Higgs particles and different boundary
conditions are needed in these cases). A full analysis, with fermions
included, of the Abelian and flavour-like charges associated with a connection
with gauge symmetries (i.e. with a stabilizer subgroup ${\cal G}^{\cal A}$ of
gauge transformations satisfying ${\hat D}^{(A)}_{\mu}U=0$) is given
in Ref.[28].

The associated weak improper conservation laws are ${\partial}_{\mu}G^{\mu}
_{1a}{\buildrel \rm \circ \over =} 0$ with $G^{\mu}_{1a}=-
c_{abc}$\break
$F^{\mu\nu}_bA_{c\nu}/{\sl g}^2$, i.e. the last of Eqs.(2-11). The
(adimensional) weak improper conserved charges $Q_a$ are

$$
\eqalign{
Q_a&= \int d^3xG^0_{1a}=-{1\over {{\sl g}^2}}
c_{abc}\int d^3xF^{0k}_bA_{ck}{\buildrel \circ \over =}Q_a^{(V)}=\cr
&=\int d^3xV^0_a={1\over {{\sl g}^2}}\int
d^3x{\partial}_iF^{io}_a=
\int d\Sigma_kU^{\lbrack ok\rbrack}_a={1\over {{\sl g}^2}}
\int d\Sigma_kF^{k0}_a=
-{1\over {{\sl g}^2}}\int d{\vec \Sigma}\cdot {\vec E}_a\cr}
\form
$$

\noindent We see that the Noether charge $Q_a$ (vanishing in the Abelian case
in absence of charged matter) coincides with the strong improper conserved
charges $Q_a^{(V)}$ (for which the strong continuity equation holds
independently from the equations of motion [13]) only by using the equations
of motion: in Ref.[13] it is the second line of Eqs.(2-14) which are called
Gauss' laws.

$Q_a$ is time independent if ${\vec G}_{1a}$ falls off sufficiently rapidly
at large $|\vec x|$. To get $Q_a\not= 0$ in the chosen gauge one needs
${\vec E}_a=O(r^{-2})$ for $r=|{\vec x}|$ going to infinity; if ${\vec E}_a$
fall off more rapidly, the charges vanish.
In Eq.(2-14) one integrates over a region
$\Omega$ of $R^3$, (subsequently enlarged to $R^3$ if the limit exists) and
$d{\vec \Sigma}$  is the integration measure on its boundary
$\partial \Omega$.
In the Abelian electromagnetic case without fermions this charge
vanishes, while in presence of fermions it measures the fermionic
Abelian (for instance electric)
charge. In the non-Abelian case the fields are charged also in absence of
fermions and the $Q_a$'s, if they exist, are the "color" charges, using the
terminology of QCD.

In Eqs.(2-14) we have introduced the following
notation with non-Abelian electric and
magnetic fields ('$\times$' is the cross-product)

$$
\eqalign{
E^k_a&={\sl g}^2\pi^k_a=F^{k0}_a={\partial}^kA^0_a-{\partial}^0A^k_a+
 c_{abc}A^k_bA^0_c=\cr
&=-{\dot A}^k_a+{\hat D}^{(A)k}_{ab}A^o_b={\sl g}{\tilde E}^k_a,
\qquad {*}F^{ij}_a=\epsilon^{ijk}E^k_a\cr
B^k_a&=-{1\over 2}\epsilon^{kij}F^{ij}_a={*}F^{ko}_a
=-{({\vec \partial}\times {\vec A}_a
)}^k-{1\over 2}c_{abc}{({\vec A}_b\times {\vec A}_c)}^k=\cr
&=-{(\lbrack {{\hat {\vec D}}^{(A)}}_{ac}-{1\over 2}c_{abc}{\vec A}_b\rbrack
\times {\vec A}_c)}^k={\sl g}{\tilde B}^k_a,
\quad F^{ij}_a=-\epsilon^{ijk}B^k_a. \cr
&\vec \partial \cdot {\vec B}_a=-{1\over 2}c_{abc}\vec \partial \cdot ({\vec
A}_b\times {\vec A}_c)\cr}
\form
$$

\noindent According to our conventions the Euclidean vectors are always
defined with upper indices: ${\vec B}_a$ is only defined as an Euclidean vector
and all the lower Euclidean indices of Lorentz tensors have to be raised
with the Minkowski metric before interpret them as real Euclidean indices.

With Eqs.(2-15), we can rewrite [21i] $\cal L$ of Eqs.(2-6) and $L^{\mu}_a
{\buildrel \rm \circ \over =} 0$ of Eqs.(2-7) in the following forms

$$
\eqalign{
{\cal L} &= -{1\over {4{\sl g}^2}}F^{\mu\nu}_aF_{a\mu\nu}=
    {1\over {2{\sl g}^2}}\sum_a({\vec E}^2_a-{\vec B}^2_a)=
    {1\over 2}\sum_a({\vec {\tilde E}}^2_a-{\vec {\tilde B}}^2_a)=\cr
       &={1\over {2{\sl g}^2}}\sum_a\lbrace {({\dot {\vec A}}_a-
         {{\hat {\vec D}}^{(A)}}_{ab}A^o_b)}^2-
         {({\vec \partial}\times {\vec A}_a+{1\over 2}c_{abc}
         {\vec A}_b\times {\vec A}_c)}^2\rbrace \cr}
\form
$$

$$
\eqalign{
L^o_a&= -{{\hat {\vec D}}^{(A)}}_{ab}\cdot {\vec E}_b
={{\hat {\vec D}}^{(A)}}_{ab}\cdot {\dot {\vec A}}_b-{{\hat D}^{(A)2}}_{ab}
A^o_b{\buildrel \rm \circ \over =}0\cr
{\vec L}_a&= -{\hat D}^{(A)o}_{ab}{\vec E}_b+{{\hat {\vec D}}^{(A)}}_{ab}\times
{\vec B}_b=\cr
&={\hat D}^{(A)o}_{ab}(-{{\hat {\vec D}}^{(A)}}_{bc}A^o_c+{\dot {\vec A}}_b)-
{{\hat {\vec D}}^{(A)}}_{ab}\times \lbrack ({{\hat {\vec D}}^{(A)}}_{bc}-
{1\over 2}c_{bdc}{\vec A}_d)\times {\vec A}_c\rbrack
{\buildrel \rm \circ \over =}0\cr}
\form
$$

\noindent where ${\hat D}^{(A)2}_{ab}={{\hat {\vec D}}^{(A)}}_{ac}\cdot
{{\hat {\vec D}}^{(A)}}_{cb}$
is the generalized Laplace-Beltrami operator. Let
us remark that, up to surface terms, i.e. with appropriate boundary
conditions, one has $\int d^3x u_a(x){{\hat {\vec D}}^{(A)}}_{ab}v_b(x)=
\int d^3x[-{{\hat {\vec D}}^{(A)}}_{ba}u_a(x)]v_b(x)$, so that ${{\hat {\vec
D}}^{(A){*}}}_{ab}=-{{\hat {\vec D}}^{(A)}}_{ab}$, where ${{\hat {\vec D}}^{(A)
{*}}}_{ab}$ is the "adjoint" of ${{\hat {\vec D}}^{(A)}}_{ab}$.

One has ${1\over 2}\sum_a{\vec B}^2_a={1\over 4}F_{aij}F_a{}^{ij}=
\lbrack {1\over 2}{\partial}^iA^j_a({\partial}^iA^j_a-
{\partial}^jA^i_a)+c_{abc}{\partial}^iA^j_aA^i_bA^j_c+\break
{1\over 4}c_{abc}c_{auv}A^i_bA^j_cA^i_uA^j_v\rbrack$.
Moreover one has ${1\over 4}{*}F^{\mu\nu}_aF_{a\mu\nu}=-{\vec E}_a\cdot {\vec
B}_a=-{\sl g}^2{\vec \pi}_a\cdot {\vec B}_a$ for the gauge invariant density
of topological charge and, from Eqs.(2-4), the Bianchi identities become
${{\hat {\vec D}}^{(A)}}_{ab}\cdot {\vec B}_b\equiv 0$ and ${\hat D}^{(A)o}
_{ab}B^i_b\equiv \epsilon^{ijk}{\hat D}^{(A)j}_{ab}E^k_b$.

The action (2-6) is quasi-invariant under the Poincar\`e transformations
[23a,g]
($A_{\mu}(x)$ is assumed to be a covariant four-vector):

$$
\eqalign{
&\delta x^{\mu}=-a^{\mu}-\omega^{\mu}{}_{\nu}x^{\nu},\qquad \omega_{\mu\nu}=-
\omega_{\nu\mu}=\partial_{\mu}\delta x^{\nu}\qquad
\partial_{\mu}\delta x^{\mu}=0,\cr
&\delta A_{\mu}(x)\equiv
{\bar A}_{\mu}(x+\delta x)-A_{\mu}(x)=\cr
&=\delta_{o}A_{\mu}(x)+\delta x^{\nu}{\partial}_{\nu}A_{\mu}(x)=
-\omega_{\mu}{}^{\nu}A_{\nu}(x)=-(\partial_{\mu}\delta x^{\nu})A_{\nu}(x)\cr
&\delta_{o} A_{\mu}(x)\equiv {\bar A}_{\mu}(x)-A_{\mu}(x)=\cr
&=-\delta x^{\nu}{\partial}_{\nu}A_{\mu}(x)-({\partial}_{\mu}
\delta x^{\nu})A_{\nu}(x)=-\delta x^{\nu}F_{\nu\mu}(x)-
{\hat D}^{(A)}_{\mu}(\delta x^{\nu}A_{\nu}(x))\cr
&\delta_{o}F_{\mu\nu}(x)={\hat D}^{(A)}_{\mu}\delta_{o}A_{\nu}(x)-
{\hat D}^{(A)}_{\nu}\delta_{o}A_{\mu}(x)\cr}
\form
$$

$$
\eqalign{
\delta {\cal L}&=\delta_o{\cal L}-\partial_{\mu}\Omega^{\mu}=
{1\over {{\sl g}^2}}\delta_oA_{a\mu}L^{\mu}_a-
{\partial}_{\mu}({1\over {{\sl g}^2}}F_a^{\mu\nu}\delta
_oA_{a\nu}+\Omega^{\mu})\equiv 0,\cr
&{\Omega}^{\mu}=-\delta x^{\mu}{\cal L}\cr
\delta_o{\cal L}&={1\over {{\sl g}^2}}
\delta_oA_{a\mu}L^{\mu}_a-{\partial}_{\mu}({1\over {{\sl g}^2}}
F^{\mu\nu}_a\delta_oA_{a\nu})\equiv \partial_{\mu}\Omega^{\mu}
\equiv \partial_{\mu}(\Omega^{\mu}+\partial_{\alpha}X^{\mu\alpha})\cr}
\form
$$

\noindent where a term ${\partial}_{\nu}X^{\mu\nu}$,
$X^{\mu\nu}=-X^{\nu\mu}$, ${\partial}_{\mu}{\partial}_{\nu}X^{\mu\nu}\equiv
0$, with the superpotential $X^{\mu\nu}=
{\sl g}^{-2}F_a^{\mu\nu}A_{a\alpha}\delta x^{\alpha}$,
has been added to $\Omega^{\mu}$. See Refs.[29] for the problem that space-time
symmetry Noether transformations are ambiguous for terms corresponding to gauge
transformations. The conserved Noether current is

$$
\eqalign{
0&{\buildrel \circ \over =}-{1\over {{\sl g}^2}}\delta_oA_{a\mu}L^{\mu}_a
\equiv \partial_{\mu}(\Omega^{\mu}+\partial_{\nu}X^{\mu\nu}+
{1\over {{\sl g}^2}}F^{\mu\nu}_a\delta_oA_{a\nu})\equiv
{\partial}_{\mu}(\theta^{\mu\nu}\delta x_{\nu}-{1\over {{\sl g}^2}}
\delta x^{\nu}A_{a\nu}L^{\mu}_a)\equiv \cr
&\equiv \partial_{\mu}{\tilde \theta}^{\mu\nu}{\buildrel \rm \circ \over =}
-a_{\nu}\partial_{\mu}\theta^{\mu\nu}+{1\over 2}\omega_{\nu\alpha}\partial
_{\mu}{\cal M}^{\mu\nu\alpha}=\partial_{\mu}(\theta^{\mu\nu}\delta x_{\nu})
,\cr}
$$

\noindent where $\theta^{\mu\nu}$ is the standard conserved, symmetric,
trace-free, gauge invariant energy-\break
momentum tensor

$$
\eqalign{
\theta^{\mu\nu}&={\sl g}^{-2}(F_a^{\mu\alpha}
F_{a\alpha}{}^{\nu}+{1\over 4}\eta^{\mu\nu}F_a{}^{\alpha\beta}F_{a\alpha\beta}
) \cr
&\theta^{\mu\nu}=\theta^{\nu\mu},\qquad \theta^{\mu}{}_{\mu}
 =0,\qquad {\partial}_{\mu}\theta^{\mu\nu}{\buildrel \rm \circ \over =} 0\cr}
\form
$$

\noindent and where ${\tilde \theta}^{\mu\nu}=\theta^{\mu\nu}-{\sl g}^{-2}
\delta x^{\nu}A_{a\nu}L^{\mu}_a{\buildrel \circ \over =}\theta^{\mu\nu}$
is a form of the energy-momentum tensor
relevant for phase space and ${\cal M}^{\mu\alpha\beta}=x^{\alpha}
\theta^{\mu\beta}-x^{\beta}\theta^{\mu\alpha}$ is the angular momentum
density tensor.

The conserved Poincar\`e generators $P^{\mu}=\int d^3x\theta^{o\mu}$,
$J^{\mu\nu}=\int d^3x{\cal M}^{o\mu\nu}$, have
the expression ($\vec E={\vec E}_a{\hat T}^a$, $\vec B={\vec B}_a{\hat T}^a$;
$H_c$, $H_D$, $\Gamma_a(x)$, ${\bar P}^i$, ${\bar J}^i$ and ${\bar K}^i$ are
phase space quantities to be defined later on)

\noindent
$$
\eqalign{
P^0&=\int d^3x\theta^{00}(\vec x, x^o){\buildrel \circ \over =}\int d^3x
{\tilde \theta}^{oo}(\vec x,x^o)={\bar P}^o=H_c\equiv \cr
&\equiv \int d^3x({\tilde \theta}^{oo}(\vec x,x^o)+\lambda_{ao}(\vec x,x^o)
\pi^o_a(\vec x,x^o))=H_D,\cr
&\theta^{00}={1\over {2{\sl g}^2}}\sum_a({\vec E}^2_a
+{\vec B}^2_a)={1\over 2}\sum_a({\sl g}^2
{\vec \pi}^2_a+{1\over {{\sl g}^2}}{\vec B}^2_a),\cr
&{\tilde \theta}^{oo}=\theta^{oo}-{\sl g}^{-2}A_{ao}L^o_a=\theta^{oo}-A_{ao}
\Gamma_a,\cr
P^i&=\int d^3x \theta^{0i}(\vec x, x^o){\buildrel \circ \over =}
\int d^3x{\tilde \theta}^{oi}(\vec x,x^o)=-\int d^3x{\vec \pi}_a(\vec x,x^o)
\cdot \partial^i{\vec A}_a(\vec x,x^o)={\bar P}^i,\cr
&\theta^{0i}={1\over {{\sl g}^2}}{({\vec E}_a\times {\vec
B}_a)}^i={({\vec \pi}_a\times {\vec B}_a)}^i\cr
&{\tilde \theta}^{oi}=\theta^{oi}-{\sl g}^{-2}A^i_aL^o_a=\theta^{oi}-A^i_a
\Gamma_a,\cr
J^i&={1\over 2}\epsilon^{ijk}J^{jk}
=\int d^3x\epsilon^{ijk}x^j\theta^{0k}(\vec x, x^o)
={1\over {{\sl g}^2}}\int d^3x{\lbrack {\vec x}
\times ({\vec E}_a\times {\vec B}_a)(\vec x,x^o)\rbrack}^i=\cr
&=\int d^3x{\lbrack \vec x\times ({\vec \pi}_a\times {\vec B}_a)(\vec x,x^o)
\rbrack}^i{\buildrel \circ \over =}\int d^3x \epsilon^{ijk}
x^j{\tilde \theta}^{ok}(\vec x,
x^o)=\cr
&=\epsilon^{ijk}\int d^3x[\pi^j_a(\vec x,x^o)A^k_a(\vec x,x^o)+{\vec
\pi}_a(\vec x,x^o)\cdot x^j\partial^k{\vec A}_a(\vec x,x^o)]=
{\bar J}^i,\cr
K^i&=J^{io}=\int d^3x(x^i\theta^{oo}(\vec x, x^o)-x^o\theta^{oi}(\vec x, x^o))
=\int d^3x x^i\theta^{oo}(\vec x, x^o)-x^o P^i{\buildrel \circ \over =}\cr
&{\buildrel \circ \over =}\int d^3x x^i{\cal H}_D(\vec x,x^o)-x^o{\bar P}^i
={\bar K}^i\cr}
\form
$$

\noindent To obtain the final form of $P^i$, $J^i$, Eqs.(2-15), an integration
by parts and $L^o_a={\sl g}^2\Gamma_a{\buildrel \circ \over =}0$ have been
used. Since $\pi^o_a(\vec x,x^o)\equiv 0$ at the Lagrangian level, this term
has been added to ${\tilde \theta}^{oo}$ to reproduce $H_D=\int d^3x{\cal H}
_D(\vec x,x^o)$ of Eqs.(2-27), see later on.

The Poincar\'e algebra satisfied by this final form ${\bar P}^{\mu}$, ${\bar
J}^{\mu\nu}$ of the generators is (for the definition of Poisson brackets see
the subsequent Eqs.(2-29))

$$
\eqalign{
&\lbrace {\bar P}^{\mu},{\bar P}^{\nu}\rbrace =0\quad \lbrace
{\bar J}^{\mu\nu},{\bar P}^{\alpha}\rbrace
=\eta^{\alpha\nu}{\bar P}^{\mu}-\eta^{\alpha\mu}{\bar P}^{\nu}\cr
&\lbrace {\bar J}^{\mu\nu},{\bar J}^{\alpha\beta}\rbrace =
\eta^{\mu\alpha}{\bar J}^{\nu\beta}+\eta^{\nu\beta}{\bar J}^{\mu\alpha}
-\eta^{\mu\beta}{\bar J}^{\nu\alpha}-
\eta^{\nu\alpha}{\bar J}^{\mu\beta}\cr}
\form
$$

\noindent In particular $\lbrace {\bar J}^{io}={\bar K}^i,{\bar P}^o=H_D
\rbrace ={\bar P}^i$ can be reinterpreted as the time constancy of ${\bar
K}^i$,
$\partial {\bar K}^i/\partial x^o +\lbrace {\bar K}^i,H_D\rbrace =0$.

In what follows we shall restrict ourselves to field configurations for which
the 10 Poincar\`e generators are finite when integrated over all $R^3$, so
to have well defined Poincar\`e Casimirs and to be able to study the
Poincar\`e orbits of the field configurations.
If there are no singularities at finite $\vec x=r(sin\theta cos\phi ,sin
\theta sin\phi , cos\theta )$, a set of sufficient boundary conditions to
get finite $P^{\mu}$, $J^{\mu\nu}$ is

$$
\theta ^{oo}(\vec x,x^o),\, \theta^{oi}(\vec x, x^o)\quad
{\buildrel \longrightarrow \over {r\rightarrow \infty}}\quad
{ {T^{oo (oi)}(\theta ,\phi ,x^o)}\over {r^{4+2\epsilon}} }+
O(r^{-s})
\form
$$

\noindent which can be satisfied in at least the two following ways:

$$
\eqalign{
either\, i)\, &F_{a\mu\nu}(\vec x,x^o)\quad
{\buildrel \longrightarrow \over {r\rightarrow \infty}}\quad
{ {f_{a\mu\nu}(\theta ,\phi ,x^o)}\over {r^{2+\epsilon}} }+
O(r^{-3})\cr
or\, ii)\,&E^i_a(\vec x,x^o)=F^{io}_a(\vec x,x^o)\quad
{\buildrel \longrightarrow \over {r\rightarrow \infty}}\quad
{ {e^i_a(\theta ,\phi ,x^0)}\over {r^{2+\epsilon}} }+O(r^{-3})\cr
&B^i_a(\vec x,x^o)=-{1\over 2}\epsilon^{ijk}F^{jk}_a(\vec x,x^o)\quad
{\buildrel \longrightarrow \over {r\rightarrow \infty}}\quad
{ {b^i_a(\theta ,\phi ,x^o)}\over {r^{3+\epsilon}} }+O(r^{-4})\cr}
\form
$$

\noindent so that we can assume

$$
\eqalign{
either\, i)\, &A_{a\mu}(\vec x,x^o)\quad
{\buildrel \longrightarrow \over {r\rightarrow \infty}}\quad
{ {a_{a\mu}(\theta ,\phi ,x^o)}\over {r^{1+\epsilon}} }+O(r^{-2})\cr
or\, ii)\, &A^i_a(\vec x,x^o)\quad
{\buildrel \longrightarrow \over {r\rightarrow \infty}}\quad
{ {a^i_a(\theta ,\phi ,x^o)}\over {r^{2+\epsilon}} }+O(r^{-3})\cr
&A^o_a(\vec x,x^o)\quad
{\buildrel \longrightarrow \over {r\rightarrow \infty}}\quad
{ {a^o_a(\theta ,\phi ,x^o)}\over {r^{1+\epsilon}} }+O(r^{-2})\cr}
\form
$$

The behaviour of Eqs.(2-24) and (2-25)
requires a restriction on the gauge transformations
$U(x)$, to be valid in every gauge: see later on Eqs.(2-40).
The set i) are the "strong boundary conditions" $\lim_{r\rightarrow \infty}rA_i
(x)=0$
of Ref.[30]: for $\epsilon =0$ we get the "weak boundary conditions", allowing
not only finite Poincar\'e generators in general but also the existence of
monopole solutions (see also [23h]) when Higgs fields are present,
which, however, are not considered in this paper.
The set ii) will turn out to be suited for the Hamiltonian formalism.
In both cases i) and ii) there is the possibility of non-vanishing
non-Abelian charges $Q_a$ for $\epsilon \rightarrow 0$.

 In case i), since, on $M^4$,
$F_{\mu\nu}=0$ implies that the gauge potential is pure gauge globally, i.e.
$A_{\mu}=U^{-1}{\partial}_{\mu}U$, Eqs.(2-25) imply
that for $r\rightarrow \infty$  (and also for $|x^o|\rightarrow \infty$ for
instantons) we should have $A_{\mu}\rightarrow
U_{\infty}^{-1}{\partial}_{\mu}U_{\infty}+
b_{\mu}$, with $b_{\mu}$ small for $r\rightarrow \infty$. In the Euclidean
(instanton) case, a  regular $A_{\mu}(x)$ will give finite Euclidean action and
instanton number. Even if there is not a canonical
procedure for this splitting, due to the cancellation
associated with the pure gauge term in the evaluation of $F_{\mu\nu}$, to
get the behaviour of Eqs.(2-24) we can assume that $b_{\mu}$ is vanishing
faster than Eqs.(2-25) (in the instanton case of Ref.[31]
$A_{\mu}=O(r^{-1})$ and $b_{\mu}=O(r^{-3})$). Under a gauge transformation
with suitable boundary behaviour (like in Eqs.(2-40), see later on) one has
$F_{\mu\nu}\rightarrow F^U_{\mu\nu}=U^{-1}F_{\mu\nu}U
\longrightarrow {}_{r\rightarrow \infty}F_{\mu\nu}$, $A_{\mu}=
U^{-1}_{\infty}{\partial}_{\mu}U_{\infty}+b_{\mu}\rightarrow A^U_{\mu}=
U^{-1}{\partial}_{\mu}U+U^{-1}A_{\mu}U={(U_{\infty}U)}^{-1}{\partial}_{\mu}
(U_{\infty}U)+U^{-1}b_{\mu}U\longrightarrow
{}_{r\rightarrow \infty} {\tilde U}_{\infty}^{-1}{\partial}_{\mu}{\tilde U}
_{\infty}+b_{\mu}$, so that the boundary conditions (2-24), (2-25)
are unaffected by gauge transformations.

Since we work at fixed time, or more generally on arbitrary space-like
hypersurfaces, we do not consider the behaviour of $A_{a\mu}$ and $F_{a\mu\nu}$
along null directions (see in the electromagnetic case the radiation part of
the Lienard-Wiechert potentials and field strengths [32]; see Ref.[33] for
the radiation in the non-Abelian case).
Only in the
analysis of the massless sector, $P^2=0$, of YM theory, where one needs the
reformulation with light-cone variables (see Section 10), this behaviour
would become relevant.

Let us define the Hamiltonian formalism. From Eqs.(2-10) and (2-15) we have

$$
\eqalign{
&\pi^0_a(\vec x,x^o)=0\cr
&\pi^k_a(\vec x,x^o)={1\over {{\sl g}^2}}E^k_a(\vec x,x^o)\cr}
\form
$$

\noindent $\pi^0_a(\vec x,x^o)\approx 0$ are the primary constraints.

The canonical and Dirac Hamiltonians are

$$
\eqalign{
H_c&=\int d^3x\lbrack {1\over 2}\sum_a({\sl g}^2{\vec \pi}^2_a(\vec x,x^o)
+{1\over {{\sl g}^2}}{\vec B}^2_a
[{\vec A}_a(\vec x,x^o)])-A_{ao}(\vec x,x^o)\Gamma_a(\vec x,x^o)\rbrack =\cr
&=P^o-\int d^3xA_{ao}(\vec x,x^o)\Gamma_a(\vec x,x^o)\cr
H_D&=H_c+\int d^3x\sum_a \lambda_{ao}(\vec x,x^o)\pi^0_a(\vec x,x^o)\cr}
\form
$$

\noindent where we have discarded the surface term $\int d^3x{\vec
{\partial}}(A_{ao}{\vec \pi}_a)$,consistently with the boundary conditions
of Eqs.(2-24), (2-25).In Eqs.(2-27)
$\Gamma_a(x)$ is the Hamiltonian version of the Euler-Lagrange equation
$L^0_a$, see Eqs.(2-13) and (2-17), and is given by

$$
\Gamma_a(x)={{\hat {\vec D}}^{(A){*}}}_{ab}\cdot {\vec \pi}_b(x)=
-{{\hat {\vec D}}^{(A)}}_{ab}\cdot {\vec \pi}_b(x)={\partial}_i\pi^i_a+
c_{abc}A_{bi}\pi^i_c\equiv {1\over {{\sl g}^2}}L^o_a{\buildrel \circ \over
=}0
\form
$$

The standard Poisson brackets are

$$
\eqalign{
\lbrace F[A,\pi],G[A,\pi]\rbrace &=\sum_c\int d^3z\lbrack
{ {\delta F}\over {\delta A_{c\alpha}(\vec z)} }
{ {\delta G}\over {\delta \pi^{\alpha}_c(\vec z)} }-
{ {\delta F}\over {\delta \pi^{\alpha}_c(\vec z)} }
{ {\delta G}\over {\delta A_{c\alpha}(\vec z)} }\rbrack \cr
&{}\cr
\lbrace A_{a\mu}(\vec x,x^o),
\pi^{\nu}_b(\vec y,x^o)\rbrace &=\delta^{\nu}_{\mu}
\delta_{ab}\delta^3(\vec x-\vec y)\cr
&{}\cr
\lbrace \pi^i_a(\vec x,x^o),B^j_b(\vec y,x^o)\rbrace &=\epsilon^{ijk}{\hat D}
_{ab}^{(A)k}(\vec x,x^o)\delta^3(\vec x-\vec y)\cr
&{}\cr
\lbrace A^i_a(\vec x,x^o),\Gamma_b(\vec y,x^o)\rbrace &=-{\hat D}^{(A)i}_{ab}
(\vec x,x^o)\delta^3(\vec x-\vec y)=\cr
&=-(\delta_{ab}\partial^i_x+c_{acb}A^i_c(\vec x,x^o))\delta^3(\vec x-\vec y)=
\cr
&=-{(\partial^i_x-A^i(\vec x,x^o))}_{ab}\delta^3(\vec x-\vec y)\cr
\lbrace \pi^i_a(\vec x,x^o),\Gamma_b(\vec y,x^o)\rbrace &=-c_{acb}\pi^i_c
(\vec x,x^o)\delta^3(\vec x-\vec y)={(\pi^i(\vec x,x^o))}_{ab}\delta^3(\vec x
-\vec y)\cr
&\lbrace B^i_a(\vec x,x^o),\Gamma_b(\vec y,x^o)\rbrace =-c_{acb}B^i_c(\vec x,
x^o)\delta^3(\vec x-\vec y),\cr}
\form
$$

\noindent where $\vec \pi (x)={\vec \pi}_a(x){\hat T}^a$
(if $A={\sl g}\tilde A$, then $\pi^{\mu}_a={\tilde
F}^{\mu o}_a/{\sl g}={\tilde \pi}^{\mu}_o/{\sl g}$
and $\lbrace A_{a\mu}(\vec
x,x^o),\pi^{\nu}_b(\vec y,x^o)\rbrace =$\break
$=\lbrace {\tilde A}_{a\mu}(\vec x,x^o),
{\tilde \pi}^{\nu}_b(\vec y,x^o)\rbrace$).

Therefore the Hamiltonian formalism requires the definition of the functional
derivatives and this puts restrictions on the functional space of the gauge
potentials ${\vec A}_a(\vec x,x^o)$, of the electric fields ${\vec \pi}_a
(\vec x,x^o)$ and of the fields $A^o_a(\vec x,x^o)$, $\pi^o_a(\vec x,x^o)$.
Moreover one has to define a suitable space of test functions
to treat the distributions needed for the calculations.

The time constancy of the primary constraints

$$
{d\over {dx^o}}\pi^0_a(\vec x,x^o){\buildrel \rm \circ \over =}
\lbrace \pi^0_a(\vec x,x^o),H_D\rbrace =\Gamma_a(\vec x,x^o)\approx 0
\form
$$

\noindent generates Gauss' laws as secondary constraints; their time
constancy

$$
{d\over {dx^o}}\Gamma_a(\vec x,x^o){\buildrel \rm \circ \over =}
\lbrace \Gamma_a(\vec x,x^o),H_D\rbrace =-c_{abc}A_{bo}(\vec x,x^o)
\Gamma_c(\vec x,x^o)\approx 0
\form
$$

\noindent does not imply any other constraint.

All the constraints are first class

$$
\eqalign{
&\lbrace \pi^0_a(\vec x,x^o),\pi^0_b(\vec y,x^o)\rbrace =
\lbrace \pi^0_a(\vec x,x^o),\Gamma_b(\vec y,x^o)\rbrace =0\cr
&\lbrace \Gamma_a(\vec x,x^o),\Gamma_b(\vec y,x^o)\rbrace =
c_{abc}\Gamma_c(\vec x,x^o)\delta^3(\vec x-\vec y)\approx 0\cr}
\form
$$

\noindent and the Dirac multipliers $\lambda_{ao}(\vec x,x^o)$ remain
arbitrary.
Due to the Jacobi identity the vector fields $X_a(\vec x,x^o)=-\lbrace .,
\Gamma_a(\vec x,x^o)\rbrace$ satisfy the algebra $\lbrack X_a(\vec x,x^o),
X_b(\vec y,x^o)\rbrack =\delta^3(\vec x-\vec y)c_{abc}X_c(\vec x,x^o)$.
The Hamilton equations are ("dot" means time derivative):

$$
\eqalign{
{\dot A}_{ao}(\vec x,x^o)&{\buildrel \rm \circ \over =}
\lbrace A_{ao}(\vec x,x^o),H_D\rbrace =\lambda_{ao}(\vec x,x^o)\cr
{\dot A}^k_a(\vec x,x^o)&{\buildrel \rm \circ \over =}
\lbrace A^k_a(\vec x,x^o),H_D\rbrace =
-{\sl g}^2\pi^k_a(\vec x,x^o)+{\partial}^kA^0_a
(\vec x,x^o)+c_{abc}A^k_b(\vec x,x^o)A^0_c(\vec x,x^o)\cr
{\dot \pi}^0_a(\vec x,x^o)&{\buildrel \rm \circ \over =}
\lbrace \pi^0_a(\vec x,x^o),H_D\rbrace =\Gamma_a(\vec x,x^o)\approx 0\cr
{\dot \pi}^k_a(\vec x,x^o)&{\buildrel \rm \circ \over =}
\lbrace \pi^k_a(\vec x,x^o),H_D\rbrace =\cr
&=-{1\over {{\sl g}^2}}{\partial}^iF_a^{ik}(\vec x,x^o)-{1\over {{\sl g}^2}}
c_{abc}A^i_b(\vec x,x^o)F_c^{ik}(\vec x,x^o)-c_{abc}
A^0_b(\vec x,x^o)\pi^k_c(\vec x,x^o)=\cr
&={1\over {{\sl g}^2}}{\partial}_iF_a^{ik}(\vec x,x^o)+c_{abc}
\lbrack {1\over {{\sl g}^2}}A_{bi}(\vec x,x^o)F_c^{ik}
(\vec x,x^o)-A_{b0}(\vec x,x^o)\pi^k_c(\vec x,x^o)\rbrack \cr}
\form
$$

By using Eqs.(2-10) the second and fourth of these equations become the first
of Eqs.(2-15) and $L^k_a{\buildrel \rm \circ \over =} 0$,
see Eqs.(2-17), respectively.

The Dirac Hamiltonian is a special case of the most general generator of
fixed-time infinitesimal canonical gauge transformations, which has the form

$$
\eqalign{
G_{gt}\lbrack \lambda_{ao},\alpha_a\rbrack
&=\int d^3x\lbrace \lambda_{ao}
(\vec x,x^o)\pi^o_a(\vec x,x^o)-\alpha_a(\vec x,x^o)\Gamma_a(\vec x,x^o)
\rbrace =\cr
&=\int d^3x\lbrace \lambda_{ao}(\vec x,x^o)\pi^o_a(\vec x,x^o)-{\vec \pi}_c
(\vec x,x^o)\cdot {\hat {\vec D}}^{(A)}_{ca}\alpha_a(\vec x,x^o)\rbrace +\cr
&+\int d^3x\vec \partial (\alpha_a(\vec x,x^o){\vec \pi}_a(\vec x,x^o))=\cr
&={\hat G}_{gt}[\lambda_{ao},\alpha_a]+\int d^3x\, \vec \partial \cdot
(\alpha_a(\vec x,x^o){\vec \pi}_a(\vec x,x^o))\approx 0\cr}
\form
$$

Sometimes $G_{gt}$ is used as the extended Dirac Hamiltonian $H_E$, but its
use as Hamiltonian is in general not satisfactory; see Ref.[34].
In this Reference there is also the scheme for doing the gauge-fixing of a
chain of $1^{st}$-class
constraints: in the YM case, first one has to add the gauge-fixing constraints
$\phi_a\approx 0$ to Gauss' laws $\Gamma_a\approx 0$; then ${ {d\phi_a}
\over {dx^o} }=\lbrace \phi_a,H_D\rbrace =\psi_a\approx 0$ are the
gauge-fixing constraints for $\pi^o_a\approx 0$, so that generically the
temporal gauge $A^o_a\approx 0$ is not allowed; then ${ {d\psi_a}\over
{dx^o} }=\lbrace \psi_a,H_D\rbrace \approx 0$ are the equations determining
the Dirac multipliers $\lambda_{ao}$'s. See also Ref.[5c,8].

As shown in Ref.[35] (see also Ref.[2c] and Ref.[36], where it is pointed out
how the class of constraints may change going from a 3+1 splitting with
space-like hypersurfaces to one with light-like ones), in the Hamiltonian
formulation of classical field theory one has to make a choice of the
boundary conditions of the canonical variables $A_{ao}(\vec x,x^o)$,
$\pi^o_a(\vec x,x^o)$, $A_{ai}(\vec x,x^o)$, $\pi^i_a(\vec x,x^o)=
E^i_a(\vec x,x^o)/{\sl g}^2$,
and of the functions $\lambda_{ao}(\vec x,x^o)$ (the Dirac
multipliers) and $\alpha_a(\vec x,x^o)$ (generalizing the $A_{ao}(\vec x,
x^o)$'s in $H_D$), which parametrize the infinitesimal gauge transformations,
to give a meaning to the integrations by parts, to the functional derivatives
and therefore to the Poisson brackets, to the proper gauge transformations.
For the canonical variables the boundary conditions are given by Eqs.(2-24),
(2-25) with $\pi^o_a(\vec x,x^o)$ assumed to have the same behaviour of
$A^o_a(\vec x,x^o)$, its conjugate variable, i.e. $\pi^o_a(\vec x,x^o)
{\rightarrow}_{r\rightarrow \infty}p^o_a(\theta ,\phi ,x^o)/r^{1+\epsilon}+
O(r^{-2})$.
Therefore from Eqs.(2-28) one has $\Gamma_a(\vec x,
x^o)\quad {\buildrel \longrightarrow \over {r\rightarrow \infty}}\quad
\gamma_a(\theta ,\phi ,x^o)/r^{3+\epsilon}+O(r^{-4})$ in both cases i) and
ii). The Dirac multipliers $\lambda_{ao}(\vec x,x^o)$, which
parametrize the infinitesimal fixed-time gauge transformations of
$A_{ao}(\vec x,x^o)$ and behave like ${\dot A}_{ao}(\vec x,x^o)$, see Eqs.
(2-33), are assumed to behave like $A_{ao}(\vec x,x^o)$. Since the parameters
$\alpha_a(\vec x,x^o)$ describe the infinitesimal gauge transformations of
${\vec A}_a(\vec x,x^o)$ (see Eqs.(2-8) and (2-36) later on), we could assume
$\alpha_a(\vec x,x^o){\rightarrow}_{r\rightarrow \infty}$\break
${\tilde \alpha}_a
(\theta ,\phi ,x^o)/r^{1+\epsilon}+O(r^{-2})$: then $\delta {\vec A}_a\sim
r^{-2-\epsilon}$ and in case i) it will fall off more rapidly then ${\vec
A}_a$,
while in case ii) like ${\vec A}_a$ ($\delta {\vec \pi}_a\sim r^{-3-\epsilon}$
from Eqs.(2-8)). However, a more reasonable behaviour for the $\alpha_a$'s is
$r^{-3-\epsilon}$ like the $\Gamma_a$'s.

In this way $\lambda_{ao}(\vec x,x^o)$,
$\alpha_a(\vec x,x^o)$ belong to the dual
space of the space of the $1^{st}$-class constraints $\pi^o_a(\vec x,x^o)$,
$\Gamma_a(\vec x,x^o)$ [35,2c,36] and one has: i) these $\lambda_{ao}$'s and
$\alpha_a$'s, called 'proper' test functions, define the most general smearing
of the constraints with test functions so that the knowledge of $G_{gt}[
\lambda_{ao},\alpha_a]$ for all the proper $\lambda_{ao}$, $\alpha_a$ is
equivalent to the knowledge of $\pi^o_a(\vec x,x^o)$, $\Gamma_a(\vec x,x^o)$
for all $\vec x$; ii) with these $\lambda_{ao}$, $\alpha_a$,
$G_{gt}[\lambda_{ao},\alpha_a]={\hat G}_{gt}[\lambda_{ao},\alpha_a]\approx 0$
is the generator of the 'proper' infinitesimal
gauge transformations [35]. However as we shall see further limitations on the
asymptotic behaviour of the proper test functions will come from the
asymptotic behaviour of gauge transformations, see later on Eqs.(2-40).

Since a generic variation of $G_{gt}$ is given by

$$
\eqalign{
&{\bar \delta}G_{gt}[\lambda_{ao},\alpha_a]
=\int d^3x\lbrace \lambda_{ao}(\vec x)\delta \pi^o_a(\vec x)-\delta {\vec
\pi}_c(\vec x)\cdot {\hat {\vec D}}^{(A)}_{ca}\alpha_a(\vec x)+\cr
&+c_{bca}{\vec \pi}_c(\vec x)\alpha_a(\vec x)\cdot \delta {\vec A}_b(\vec x)
+\int d^3x \vec \partial \cdot [\alpha_a(\vec x)\delta {\vec \pi}_a(\vec x)]
\cr}
\form
$$

\noindent we see that for proper gauge transformations the surface term does
not contribute (as all the surface terms encountered till now), so that the
functional derivatives appearing in the Poisson brackets are well defined
and we get (these equations also define the space of allowed proper variations
of the variables)

$$
\eqalign{
&{\bar \delta}A_{ao}(\vec x,x^o)=
\lbrace A_{ao}(\vec x,x^o),G_{gt}\rbrace =
{ {\delta G_{gt}}\over {\delta \pi^o_a(\vec x,x^o)} }=
\lambda_{ao}(\vec x,x^o)\cr
&{\bar \delta}A_{ai}(\vec x,x^o)=\lbrace A_{ai}(\vec x,x^o),G_{gt}\rbrace =
{ {\delta G_{gt}}\over {\delta \pi^i_a(\vec x,x^o)} }=\cr
&=-c_{abc}\alpha_b(\vec x,x^o)A_{ci}(\vec x,x^o)+{\partial}_i\alpha_a
(\vec x,x^o)={\hat D}^{(A)}_{iab}\alpha_b(\vec x)\cr
&{\bar \delta}\pi^o_a(\vec x,x^o)=\lbrace \pi^o_a(\vec x,x^o),G_{gt}\rbrace =
-{ {\delta G_{gt}}\over {\delta A^o_a(\vec x,x^o)} }=0\cr
&{\bar \delta}\pi^i_a(\vec x,x^o)=\lbrace \pi^i_a(\vec x,x^o),G_{gt}\rbrace =
-{ {\delta G_{gt}}\over {\delta A_{ai}(\vec x,x^o)} }=\cr
&=-c_{abc}\alpha_b(\vec x,x^o)\pi^i_c(\vec x,x^o)=c_{abc}\pi^i_b(\vec x,x^o)
\alpha_c(\vec x,x^o)\cr}
\form
$$

\noindent with $\delta A_{ai}={\bar \delta}A_{ai}$, $\delta \pi^i_a={\bar
\delta}\pi^i_a$, in accord with Eqs.(2-8), and with
$\bar \delta A_{ao}=\lambda_{ao}$ as the Hamiltonian fixed-time
counterpart of $\delta A_{ao}=c_{abc}A_{bo}\alpha_c+{\partial}_o
\alpha_a={\hat D}^{(A)}_{oab}\alpha_b$.
Under the fixed-time gauge transformations $\bar \delta A_{ao}$
and $\bar \delta A_{ai}$, the Lagrangian is only weakly quasi-invariant [5b,c],
 i.e.

$$
\eqalign{
\bar \delta {\cal L}&\equiv -{1\over {{\sl g}^2}} \lbrace
{\partial}_{\mu}\lbrack
(\lambda_{ao}-{\partial}_o
\alpha_a+c_{abc}\alpha_bA_{co})F_a^{o\mu}\rbrack +(\lambda_{ao}-
{\partial}_o \alpha_a+c_{abc}\alpha_bA_{co})L^o_a\rbrace =\cr
&=-{1\over {{\sl g}^2}}\lbrace \partial_{\mu}\lbrack (\lambda_{ao}-{\hat D}
^{(A)}_{oab}\alpha_b)F^{o\mu}_a\rbrack +(\lambda_{ao}-{\hat D}^{(A)}_{oab}
\alpha_b)L^o_a\rbrace \cr},
$$

\noindent where we used Eqs.(2-7), (2-12)
and where $L^o_a\equiv \Gamma_a {\buildrel
\rm \circ \over =}0$, Eqs.(2-28), are the Euler-Lagrange equations independent
from the accelerations. We see that by restricting $\lambda_{ao}$, $\lambda
_{ao}=\delta A_{ao}$, one recovers the standard gauge invariance of ${\cal L}$.

Let us now consider finite gauge transformations U. Since $\delta {\vec A}_a=
\bar \delta {\vec A}_a$ under infinitesimal gauge transformations, one has
$\triangle \vec A=\triangle ({\vec A}_a{\hat T}^a)=U^{-1}\vec AU+U^{-1}\vec
\partial U-\vec A=\bar \triangle \vec A$. Instead, since $\bar \delta A_{ao}=
\lambda_{ao}\not= \delta A_{ao}$, let us write $\bar \triangle A_o=\bar
\triangle (A_{ao}{\hat T}^a)=\Lambda_o=\Lambda_{ao}{\hat T}^a=\triangle A_o+
U^{-1}KU$, $K=K_a{\hat T}^a$, with $\triangle A_o=U^{-1}A_oU+U^{-1}\dot U-A_o$;
$\lambda_{ao}{\hat T}^a$ is the infinitesimal form of $\Lambda_o$. Therefore,
we have to check the gauge invariance of ${\cal L}(x)$ written in the form of
Eqs.(2-16) under the finite gauge transformation $\bar \triangle \vec A$ and
$\bar \triangle A_o=\triangle A_o+U^{-1}KU$. One finds the result $\bar
\triangle \vec E=U^{-1}\vec EU-\vec E=\bar \triangle [g^{-1}({\dot {\vec A}}_a-
{\hat {\vec D}}^{(A)}_{ab}A_{bo}){\hat T}^a]=-g^{-1}U^{-1}({\hat {\vec
D}}^{(A)}
K)U$. Then, $\bar \triangle {\cal L}={1\over {2g^2}}\sum_a\, [{({\vec E}_a-
{\hat {\vec D}}^{(A)}_{ab}K_b)}{}^2-{\vec E}^2_a]={1\over {2g^2}}[-2{\vec E}_a
\cdot {\hat {\vec D}}^{(A)}_{ab}K_b+{({\hat {\vec D}}^{(A)}_{ab}K_b)}{}^2]=
{1\over {2g^2}}[2K_a{\hat {\vec D}}^{(A)}_{ab}\cdot {\vec E}_b-K_a{\hat {\vec
D}}^{(A)}_{ab}\cdot {\hat {\vec D}}^{(A)}_{bc}K_c+\vec \partial \cdot (K_a
{\hat {\vec D}}^{(A)}_{ab}K_b-2{\vec E}_aK_a)]{\buildrel \circ \over =}-{1
\over {2g^2}}K_a{\hat {\vec D}}^{(A)}_{ab}\cdot {\hat {\vec D}}^{(A)}_{bc}K_c$
(modulo surface terms and Gauss' equation of motion). Since in the generic case
the covariant derivative has no zero modes (covariantly constant solutions of
${\hat {\vec D}}^{(A)}_{ab}K_b=0$), one must have either $K_a=0$, i.e. $\bar
\triangle A_o=\triangle A_o$, or $K_a{\buildrel \circ \over =}0$, i.e.
$\bar \triangle A_o{\buildrel \circ \over =}\triangle A_o$. Therefore, the
group of gauge transformations of the theory based on $P^t(R^3\times \lbrace
x^o\rbrace ,G)$ may be larger than the group ${\cal G}$ of $P(M^3,G)$, if we
include quasi-invariance, i.e. $\triangle {\cal L}{\buildrel \circ \over =}
0$ and we accept $K_a{\buildrel \circ \over =}0$; at the level of infinitesimal
gauge transformations this implies $\lambda_{ao}(\vec x,x^o){\buildrel \circ
\over =}\, {\hat D}^{(A)}_{oab}\alpha_b(\vec x,x^o)$, i.e. on the
extremals the parameters $\lambda_{ao}(\vec x,x^o)$ are not independent from
the parameters $\alpha_a(\vec x,x^o)$ (this implies that the $\lambda_{ao}$'s
must be functionally dependent on the gauge potentials for consistency) and
then $\bar \delta {\cal L}{\buildrel \circ \over =}0$.

As shown in Ref.[35], when the functions $\lambda_{ao}(\vec x,x^o)$, $\alpha_a
(\vec x,x^o)$ do not satisfy the previous boundary conditions, $G_{gt}
[\lambda_{ao},\alpha_a]$ becomes a generator of "improper" gauge
transformations
(like the global or rigid ones associated with the color charges $Q_a$, Eqs.
(2-14)). However the functions $\lambda_{ao}(\vec x,x^o)$ (and more in
general all the Dirac multipliers associated with the primary $1^{st}$-class
constraints) are not allowed to be improper, because the Dirac Hamiltonian
$H_D$ must be a proper function, generating a mixture of deterministic
evolution and proper gauge transformations.
In the pure YM case the fact that improper $\lambda_{ao}(\vec x,x^o)$'s
are not allowed, is reflected in the absence of improper conservation laws
associated with them, because they are associated with the primary
constraints. The second Noether theorem [5c] shows that if one has a chain
of $1^{st}$-class constraints, only the last constraint of the chain is
involved in the generation of the improper conservation laws: in this case,
as shown by Eqs.(2-14), only improper $\alpha_a(\vec x,x^o)$, associated with
secondary Gauss' law constraints, are relevant. With improper $\alpha_a
(\vec x,x^o)$, the surface term in Eqs.(2-35) can be written as $\delta
{\cal Q}[\alpha_a]$ with ${\cal Q}[\alpha_a]$ given by

$$
{\cal Q}[\alpha_a]=-
\int d^3x\, \vec \partial \cdot \lbrack \alpha_a(\vec x,x^o){\vec \pi}_a
(\vec x,x^o)\rbrack =- \lim_{r\rightarrow \infty}\int_rd{\vec \Sigma}
\alpha_a(\vec x,x^o)\cdot {\vec \pi}_a(\vec x,x^o)
\form
$$

\noindent For improper 'rigid' or 'global' infinitesimal gauge transformations
$\alpha_a(\vec x,x^o)=\alpha_a=const.$ one has ${\cal Q}[\alpha_a]=
\sum_a\alpha_aQ_a$, see Eqs.(2-14).
The generator of the improper infinitesimal gauge transformations, also
taking into account Eqs.(2-34), is

$$
\eqalign{
&G_{gt}[\lambda_{ao}, \alpha_a]={\hat G}_{gt}[\lambda_{ao}, \alpha_a]-
{\cal Q}[\alpha_a]\approx 0\cr
&\Rightarrow \quad {\hat G}_{gt}[\lambda_{ao},\alpha_a]\approx -
\int d^3x\, {\vec \pi}_a(\vec x,x^o)\cdot {\hat {\vec D}}^{(A)}_{ab}
\alpha_b(\vec x,x^o)\approx \cr
&\approx {\cal Q}[\alpha_a]\cr}
\form
$$

Let us remark that the non-Abelian charges
$Q_a$ are not gauge invariant. Eqs.(2-14) should be defined on a ball
${\Omega}_r$ of radius r with boundary $\partial {\Omega}_r$ to give

$$
\eqalign{
&Q_a(r,x^o)={1\over {{\sl g}^2}}c_{abc}\int_{\Omega_r}d^3x{\vec A}_{b}\cdot
{\vec E}_c{\buildrel \rm \circ \over =}-{1\over {{\sl g}^2}}
\int_{{\Omega}_r}d^3x{\vec \partial}\cdot
{\vec E}_a=-{1\over {{\sl g}^2}}
\int_{\partial {\Omega}_r}d{\vec \Sigma}\cdot {\vec E}_a\cr
&{ {dQ_a(r,x^o)}\over {dx^o} }{\buildrel \rm \circ \over =}
{1\over {{\sl g}^2}}c_{abc}
\int_{{\Omega}_r}d^3x{\partial}_i(F^{i\nu}_bA_{c\nu})={1\over {{\sl g}^2}}
c_{abc}\int_{\partial {\Omega}_r}d\Sigma^i(E^i_bA_{co}-\epsilon^{ijk}B^j_b
A^k_c)\cr}
\form
$$

With Eqs.(2-24), (2-25) one gets $Q_a(x^o)=\lim_{r\rightarrow \infty}
Q_a(r,x^o)$ and $dQ_a(x^o)/dx^o=0$. However, as shown in Ref.[37], if
one performs a finite gauge transformation $U(\vec x,x^o)$, in general in
the new gauge the new Noether charges $Q^U_a$ are not conserved. This is due
to the fact that, since from Eqs.(2-3) one has ${\vec E}_a(\vec x,x^o)
\rightarrow {\vec E}^U_a(\vec x,x^o)=
U^{-1}(\vec x,x^o){\vec E}_a(\vec x,x^o)$\break $U(\vec x,x^o)$,
the flux of ${\vec E}^U_a$ may not vanish asymptotically like
the one of ${\vec E}_a$.As shown in Ref.[23a],
the gauge covariance of the charges $Q_a$ depends on the
$r\rightarrow \infty$ limit of $U(\vec x,x^o)$. Only if we assume
$\lim_{r\rightarrow \infty}U(\vec x,x^o)=U_{\infty}=const.$ independently
from the direction, i.e. if the finite gauge transformations are
"asymptotically rigid", one has a well defined covariance for the $g$-valued
non-Abelian charges:

$$Q=Q_a{\hat T}^a\rightarrow Q^U=Q^U_a{\hat T}^a=
U^{-1}_{\infty}QU_{\infty},$$

\noindent so that if Q=0 ($\epsilon \not= 0$ in Eqs.(2-24),
(2-25)) in one gauge, one has Q=0 in every gauge.
The same conclusion is reached in Ref.[33],
where it is shown that the behaviour $U(\vec x,x^o)$\break
${\longrightarrow}_{r\rightarrow \infty}\quad
a(\theta ,\phi ,x^o)\lbrack 1+O(r^{-1})\rbrack$, preserving Eqs.(2-25),
has to be restricted to $a(\theta ,\phi ,x^o)=const.$ from the
requirement of covariant charges (see also Ref.[23b]).
Let us remark that in the previous discussion we could have used
$\alpha_a=\alpha_a(x^o)$ and $U_{\infty}=U_{\infty}(x^o)$, but we shall
exclude these cases, because they break manifest Lorentz covariance in the
sense that they are not preserved by going from a 3+1 splitting to another
one.

Therefore, to satisfy all the previous requirements,
we shall assume the following boundary conditions for $r\rightarrow \infty$
on the finite
gauge transformations, on the field configurations and on the gauge parameters

$$
\eqalign{
either\, i)&{}\cr
&U(\vec x,x^o)\quad {\buildrel \longrightarrow \over {r\rightarrow \infty}}
\quad U_{\infty}+O(r^{-1}),\qquad U_{\infty}=const.\cr
&\partial_{\mu} U(\vec x,x^o)\quad
{\buildrel \longrightarrow \over {r\rightarrow \infty}}\quad
{(\partial_{\mu}U)}_{\infty}={ {u_{\mu}}\over {r^{1+\epsilon}}}+O(r^{-2})\cr
&A_{a\mu}(\vec x,x^o)\quad {\buildrel \longrightarrow \over {r\rightarrow
\infty}}\quad { {a_{a\mu}}\over {r^{1+\epsilon}}}+O(r^{-2})\cr
&A^U_{a\mu}(\vec x,x^o)\quad
{\buildrel \longrightarrow \over {r\rightarrow \infty}}\quad
U^{-1}_{\infty}{(\partial_{\mu}U)}_{\infty}+U^{-1}_{\infty}A^{\infty}_{a\mu}
U_{\infty}={ {a^U_{a\mu}}\over {r^{1+\epsilon}}}+O(r^{-2})\cr
&\pi^o_a(\vec x,x^o)\quad {\buildrel \longrightarrow \over {r\rightarrow \infty
}}\quad { {p^o_a}\over {r^{1+\epsilon}} }\cr
&F_{a\mu\nu}(\vec x,x^o)\quad {\buildrel \longrightarrow \over {r\rightarrow
\infty}}\quad { {f_{a\mu\nu}}\over {r^{2+\epsilon}}}+O(r^{-3})\cr
&\lambda_{ao}(\vec x,x^o)\quad {\buildrel \longrightarrow \over {r\rightarrow
\infty}}\quad { {\lambda^{\infty}_{ao}}\over {r^{1+\epsilon}}}+O(r^{-2})\cr
&\alpha_a(\vec x,x^o)\quad {\buildrel \longrightarrow \over {r\rightarrow
\infty}}\quad { {\alpha^{\infty}_a}\over {r^{3+\epsilon}}}+O(r^{-4})\cr
&\Gamma_a(\vec x,x^o)\quad {\buildrel \longrightarrow \over {r\rightarrow
\infty}}\quad { {\Gamma^{\infty}_a}\over {r^{3+\epsilon}}}+O(r^{-4})\cr
or\, ii)&{}\cr
&U(\vec x,x^o)\quad {\buildrel \longrightarrow \over {r\rightarrow \infty}}
\quad U_{\infty}+O(r^{-1}),\qquad U_{\infty}=const.\cr
&\vec \partial U(\vec x,x^o)\quad
{\buildrel \longrightarrow \over {r\rightarrow \infty}}\quad
{(\vec \partial U)}_{\infty}={ {\vec u}\over {r^{2+\epsilon}}}+O(r^{-3})\cr
&\partial_oU(\vec x,x^o){\buildrel \longrightarrow \over {r\rightarrow \infty}}
{(\partial_oU)}_{\infty}={ {u_o}\over {r^{1+\epsilon}}}+O(r^{-2})\cr
&A_{ao}(\vec x,x^o)\quad {\buildrel \longrightarrow \over {r\rightarrow
\infty}}
\quad { {a_{ao}}\over {r^{1+\epsilon}} }+O(r^{-2})\cr
&\lambda_{ao}(\vec x,x^o)\quad {\buildrel \longrightarrow \over {r\rightarrow
\infty}}\quad { {\lambda^{\infty}_{ao}}\over {r^{1+\epsilon}}}+O(r^{-2})\cr
&\pi^o_a(\vec x,x^o)\quad {\buildrel \longrightarrow \over {r\rightarrow \infty
}}\quad { {p^o_a}\over {r^{1+\epsilon}} }\cr
&A^i_a(\vec x,x^o)\quad {\buildrel \longrightarrow \over {r\rightarrow \infty}}
\quad { {a^i_a}\over {r^{2+\epsilon}} }+O(r^{-3})\cr
&A^{Ui}_a(\vec x,x^o)\quad
{\buildrel \longrightarrow \over {r\rightarrow \infty}}\quad
U^{-1}_{\infty}{(\partial^iU)}_{\infty}+U^{-1}_{\infty}A^{(\infty )i}_a
U_{\infty}={ {a^{Ui}}\over {r^{2+\epsilon}}}+O(r^{-3})\cr
&B^i_a(\vec x,x^o)\quad {\buildrel \longrightarrow \over {r\rightarrow \infty}}
\quad { {b^i_a}\over {r^{3+\epsilon}} }+O(r^{-4})\cr}
\form
$$

$$
\eqalign{
&\pi^i_a(\vec x,x^o)=E^i_a(\vec x,x^o)/{\sl g}^2
\quad {\buildrel \longrightarrow \over {r\rightarrow \infty}}
\quad { {e^i_a}\over {r^{2+\epsilon}} }+O(r^{-3})\cr
&\alpha_a(\vec x,x^o)\quad {\buildrel \longrightarrow \over {r\rightarrow
\infty}}\quad { {\alpha^{\infty}_a}\over {r^{3+\epsilon}}}+O(r^{-4})\cr
&\Gamma_a(\vec x,x^o)\quad {\buildrel \longrightarrow \over {r\rightarrow
\infty}}\quad { {\Gamma^{\infty}_a}\over {r^{3+\epsilon}}}+O(r^{-4})\cr}
$$

While Eqs.(2-40) i) are useful in the manifestly covariant Lagrangian approach,
Eqs.(2-40) ii) are preferred in the Hamiltonian formalism and we shall see
later on which kind of further restrictions are needed by the study of the
group
of gauge transformations and by the requirement to avoid the Gribov ambiguity.

The pure YM non-Abelian charges are now well
defined and gauge covariant. They are given
as surface integrals, and their algebra cannot be evaluated with the Poisson
brackets, but only with the Dirac brackets associated with some
gauge-fixing of the gauge freedom [35] (see later on Section 9). It is
expected that the $Q_a$'s, when they exist, satisfy the same Lie algebra $g$ of
the structure group G, with the opposite structure constants, since they are
associated with the passive interpretation of the gauge transformations and
therefore with a left action; the structure group G, with its right action,
will commute with the ${\cal G}_G\approx G_G\approx G$
group, with its left action, obtained by
exponentiating the $Q_a$'s. The origin of the $Q_a$'s is in the improper
conservation laws associated with the improper global (rigid, $1^{st}$ kind)
gauge transformations. Now, while the proper gauge transformations map one
solution of the Hamilton equations onto a gauge equivalent one, this is no more
true for the improper ones [35], which, in a sense, could
be called "asymptotic dynamical symmetries" : the value of
the charges $Q_a$ is in general changed ($Q^U_a\not= Q_a$),
only the Casimirs of the realization
of the $Q_a$ are left unchanged, like the invariant $-2{(Q,Q)}_g=
\sum_aQ^2_a$. Therefore one should not go to
the quotient with respect to improper rigid gauge transformations; since in
Section 9 we will find Dirac's observables corresponding to the
non-Abelian charges, this would correspond to go to the quotient with
respect to energy in a conservative system.
See in this connection the surface integrals from the improper conservation
laws at the origin of the asymptotic Poincar\'e group in general relativity
with boundary conditions of asymptotic flatness [38,35]. See also Ref.[39]
for a discussion of surface terms in YM-Higgs theory, with particular emphasis
on the monopole and dyons solutions. In electromagnetism surface terms
connected
with solutions of the Maxwell equations with vanishing total four-momentum (the
$P_{\mu}=0$ orbit of the Poincar\'e group, classical basis of the infrared
problems in absence of a mass gap) have been considered in Ref.[40],
to try to avoid the breaking of Lorentz invariance in the charged sectors of
QED.

In Ref.[33] is also considered the limit
of the charges at null infinity, as required by the study of non-Abelian
radiation from spatially bounded sources (a retarded definition of color
charge is used)
, and it is shown that  $\epsilon =0$ in Eqs.(2-25) is required.
It is suggested that the angle dependence of $a(\theta ,\phi )$ may be
eliminated in this case by imposing special boundary conditions at past or
future null infinity on the gauge configurations. In Ref.[41a], Appendix I,
after a study of the boundary conditions needed in general relativity for the
existence of the asymptotic Poincar\'e generators by means of the spi
framework [41b], the YM case in Minkowski space-time is considered.
It is  noted that both the problem of the angle
dependence of $\lim_{r\rightarrow \infty}U(\vec x,x^o)$ and the problem
of the compatibility of charge conservation and Lorentz boosts
can be solved by requiring as boundary condition the vanishing of the
non-Abelian magnetic field strengths on  the hyperboloid of
unit space-like directions in the tangent space at $i^o$ (the point added at
spatial infinity to get a conformal completion of Minkowski space-time).
This condition leaves Eqs.(2-24), with $\epsilon =0$, unchanged on space-like
surfaces.
See Ref.[26] for a discussion of the problem of localization in manifolds with
boundary (i.e. in finite space-like regions)
of the charges $Q_a$ in pure YM theory and Ref.[42] for the
application of this method to a bag model for hadrons.

Let us remark that as a consequence of Eqs.(2-15) one can introduce a magnetic
charge $Q^{(m)}_a=-(1/2{\sl g})\int d^3x\, c_{abc}\vec \partial \cdot ({\vec
A}_b\times {\vec A}_c)=(1/{\sl g})\int d^3x\, \vec \partial \cdot {\vec B}_a=
(1/{\sl g})\int d\vec \Sigma \cdot {\vec B}_a$, which is gauge covariant as a
surface integral due to Eqs.(2-40); but it is not conserved, because there is
no continuity equation deriving from a Noether symmetry of the action.
This is different from the situation of magnetic monopoles and of their
magnetic charges both in the Abelian and non-Abelian cases (see for instance
the review in Ref.[43]).

Till now we have considered field variables $A_{a\mu}(\vec x.x^o)$,
$F_{a\mu\nu}
(\vec x,x^o)$ on $R^3$ with the boundary conditions of Eqs.(2-40); to them we
will add the fermionic field variables $\psi_{a\alpha}(\vec x,x^o)$ and also
bosonic Higgs fields $\phi_A(\vec x,x^o)$ are added in the standard model of
elementary particles (all of them with suitable boundary conditions on $R^3$).
In general, already at the classical level, these variables have to be
considered as distributions and not as functions: usually one considers
tempered distributions, for which the test function space is the Schwartz
space ${\cal S}$ of infinitely often differentiable functions decreasing as
well as their derivatives faster than any power as $\vec x \rightarrow \infty$
(in this framework the Fourier transforms of tempered distributions are again
tempered distributions); in the spirit of the classical Hamiltonian
formalism we shall not make any hypothesis on the time dependence; we disregard
at this stage the fact that in the naive canonical quantization the
renormalized fields do not have well defined canonical commutation relations
(in the interacting case they are more singular than the unrenormalized
fields),
because the approach with classical Dirac's observables (analogously to
the Coulomb gauge in Abelian theories) leads to non-local and non-linear
actions
and to the need to define different regularization and renormalization
procedures. A first insight could come by the discovery of a regularization of
the Coulomb gauge in the Abelian case with its nonlocal quadrilinear fermionic
self-interaction; then one could try to extend it to the non-Abelian case
and to learn how to translate in this language what is known about the
standard renormalizable (but not gauge invariant) Abelian and non-Abelian
theories. Even if many researchers are satisfied with the present status of
the art in connection with applications to particle physics, the known schemes
of regularization are
totally inadequate to general relativity and therefore to any future approach
realizing a consistent fusion of particle physics and general relativity.

On the other hand in local relativistic quantum field theory one emphasizes the
fact that to compare the theory with experiments one utilizes apparatuses,
which are localized on bounded open regions ${\cal O}$ of Minkowski space-time:
this is the origin of the "field algebra" ${\cal F}({\cal O})$ containing a
subalgebra ${\cal A}({\cal O})$, i.e. the "local algebra of observables"
(gauge invariant quantities not containing charged fermionic fields: the
absence of their strict classical background [unobservable fields] forces one
to the pseudoclassical approach [44]
with Grassmann-valued fermionic fields) [12];
a "local" operator is a polynomial in the original field operators with
functions with support in ${\cal O}$ (in each argument) as coefficients, while
an "almost or quasi local" operator has the coefficients in ${\cal S}$ (for
each argument). See Ref.[12b] for the status of this and of related approaches
based on the concept of localized observables, which approximate the
quasi-local
ones due to the Reeh-Schlieder theorem.
For systems with short-range suitable
interactions and with a mass gap, Abelian conserved charges are connected
with the problem of the superselection sectors of the theory. However it is
not yet clear the relation between the global aspects implied by the assumed
implementation of the Poincar\`e group and the local properties of the theory.
For gauge theories with long-range interactions and with no physical mass gap
due to the presence of massless particles, one does not yet have a well defined
and accepted theory along these lines and uses more hypothetical schemes (see
Refs.[13,45] and the review in Ref.[12b]).
All these problems must have a classical basis in the Hamiltonian
formalism with $1^{st}$-class constraints on Minkowski space-time (relativistic
presymplectic approach) and with a pseudoclassical description of the
fermionic degrees of freedom, which is largely unexplored: the search of a
description only in terms of Dirac's observables (generalizing the Coulomb
gauge approach and taking into account the global aspects of the Poincar\'e
group), could serve to clarify some basic issues and to reopen the
research on how to quantize, regularize and renormalize non-local theories;
naturally, for fermionic variables the real observables must be even functions
of fermionic Dirac's observables and, as we shall see in Section 9, this is
implied by the classical analogue of the superselection rules.

In particular, the charges connected with Gauss' laws give rise to
superselection rules at the quantum level in absence of spontaneous symmetry
breaking (when this mechanism is at work with Higgs fields present, one has
superselection rules for the unbroken generators and charge screening for the
broken ones) [13,45], according to which all local
observables commute with the charges; this means that we cannot observe a
coherent superposition of states belonging to different eigenvalues of the
charges; in the non-Abelian case the superselected sectors are labelled by the
Casimirs of G like $\sum_aQ^2_a$, the charges $Q_a$ are not observables,
the physical states are singlets described by mixtures
with respect to the charge quantum numbers inside a given irreducible
representation of G and charged states cannot be obtained
as local excitations of the vacuum. The classical basis
of the superselection rules could be phrased in the following way: even if
one does not go to the quotient with respect to global (rigid, 1st kind)
gauge transformations, one has to consider a family of (pseudo) classical
theories (labelled by  Casimirs like $\sum_aQ^2_a$), described by those
functions of Dirac's observables. which have vanishing Poisson brackets with
Dirac's observables ${\check Q}_a$ corresponding to the non-Abelian
charges $Q_a$ (see Section 9); each (pseudo) classical theory would be a
(pseudo) classical superselection sector.
Further problems, whose classical background has to be investigated in
in terms of Dirac's observables and of the
Poincar\'e orbit $P^{\mu}=0$ for the infrared problems,
are: i) the absence of the cluster property
(hypothesis of quark confinement with a linearly rising quark-antiquark
potential for increasing space-like separations and absence of their
asymptotic states; in turn this implies an infrared mechanism (infrared
slavery)
connected with the indefinite metric, so that the space-time translations are
not unitary on the light-cone; as a result only the superselected sector with
$Q_a=0$ would survive at the level of composite hadrons) and the possible
existence of variables at infinity giving volume effects [13,45].
ii) the absence of mass gap (separating the time-like Poincar\'e orbits
$P^2 > 0$ from the exceptional one $P^{\mu}=0$),
which already in the simpler Abelian case (no
confinement), makes the situation  extremely
complicated : localized charged states do not strictly
correspond to mass eigenvalues (electrons are "infraparticles" due to the
infrared problem of the Coulomb cloud) and the Lorentz group is spontaneously
broken on them; if we accept non-local charged states with $P^2=m^2$, then the
velocity of the electron gives a superselection rule, so that electrons with
different velocities are described not equivalently (the classical basis of it
is the fact that the separation of the radiation and Coulomb parts of the
Lienard-Wiechert field of an electron is not invariant under boosts).

See Ref.[46a] for the implementation of local fields  at the
classical level at fixed time, if our basic variables, instead of being
tempered distributions, are smeared with test functions of compact support in
$R^3$.  With regards to the charges $Q_a$, in Ref.
[46a] it is stressed that the original idea of Yang and Mills was the absence
of an absolute way of comparing the charge spaces of matter fields (associate
bundles of P) in different points of the base manifold (a connection is
needed for the comparison); if the global ($1^{st}$ kind) gauge
transformations,
associated with the definitions of the $Q_a$'s, exist, they should allow the
definition of privileged global cross sections both for the principal bundle
of pure YM theory and for the associated bundles of matter fields; this would
imply an absolute comparison of the charge spaces of matter fields on different
points of the base manifold, because it would assure the existence of
complete (either local or non-local) gauge-fixings for the action of ${\cal
G}$:
this is, in general, not possible for the Gribov ambiguity, for the
presence of the center $Z_G$ of G or for problems with the boundary conditions
in gauges like the axial one.
With our trivial principal bundles with
topologically trivial base manifold and simply connected fiber G, we will show
that in suitable function spaces one can obtain an almost complete non-local
gauge-fixing (more exactly a global decoupling of the gauge degrees of
freedom from Dirac's observables), which does not fix the action of $Z_G$
(see Ref.[46b] for $SU(3)/Z_3$); this was already emphasized after Eqs.(2-1).
In ref.[46a] there is also a discussion of local gauges for YM theory, in which
one poses a condition on the field strengths $F_{a\mu\nu}$ (like in
Refs.[47], where the role of the Cartan subalgebra of the semisimple Lie
algebra
$g$ is emphasized): it is
shown that they cannot be complete and that there are local gauge singularities
connected with some stability group leaving invariant the condition on the
$F_{a
\mu\nu}$'s.

\vfill\eject

\bigskip\noindent
{\bf{3. The Hamiltonian Group of Gauge Transformations.}}
\newcount \nfor

\def \form {\global \advance \nfor by 1 \eqno (3.\the\nfor)}
\bigskip

Instead of the group ${\cal G}$ of gauge transformations,
in the Hamiltonian formalism one has
to do with fixed-time gauge transformations, whose infinitesimal form is given
by Eqs.(2-36): under them the Lagrangian only is weakly quasi-invariant.
These Hamiltonian gauge transformations form a group ${\bar {\cal G}}$
(a priori ${\bar {\cal G}}_{x^o}$, but we shall assume that all these groups
are isomorphic), which is more general of
${\cal G}$ restricted at $x^o=const.$ at the level of infinitesimal
transformations at least outside the extremals of the action as we have seen
in Section 2: the difference between ${\bar \delta}A_{ao}$
and $\delta A_{ao}$ must be vanishing by using the equations of motion.
With the chosen boundary conditions for every given $x^o$ a
fixed-time Hamiltonian gauge transformation can be considered as a mapping
$\bar U:R^3\times \lbrace x^o\rbrace \rightarrow G$, $(\vec x,x^o)\mapsto
\bar U(\vec x,x^o)$, which must satisfy the same boundary conditions (2-40)
as $U(x)$ for the consistency of the whole scheme, since one wishes to have
independence from the 3+1 splitting.

With the boundary conditions of Eqs.(2-40)ii), the group ${\bar {\cal G}}$ of
all the proper and improper fixed-time gauge transformations coincides with
the group ${\cal G}^c$ of Ref.[48a], which is a subgroup of the more general
group ${\bar {\cal G}}^{'}$ of all the maps $R^3\rightarrow G$, which, however,
does not seem to be physically relevant for what has been said about having
covariant non-Abelian charges. When it exists, like in our case,
the rigid group ${\cal G}_G\sim G_G\sim G$ is a subgroup of ${\bar {\cal G}}$.
In this Section we shall treat the group ${\bar {\cal G}}$ in a naive way,
deferring to the Section 6 some informations about its definition as an
infinite dimensional Hilbert-Lie group.

${\bar {\cal G}}$ has the normal subgroup ${\bar {\cal G}}_{\infty}$,
defined by

$$
\bar U(\vec x,x^o)\quad {\buildrel \longrightarrow \over {r\rightarrow
\infty}}\quad I+O(r^{-1})
\form
$$

\noindent which contains both proper and improper gauge transformations; in the
case of $R^3$ with the boundary conditions (2-40)ii) it defines the group of
gauge transformations associated with the subgroup of $Aut_VP^t$,
which preserve the fibers over the points of the 2-sphere at spatial infinity.

Then there is the normal subgroup ${\bar {\cal G}}^o_{\infty}$ of
${\bar {\cal G}}_{\infty}$, of all the proper and improper gauge
transformations, which are continuously connected with the identity. The
generators of the Lie algebra $g_{{\bar {\cal G}}^o_{\infty} }$  of the proper
infinitesimal gauge transformations in
${\bar {\cal G}}^o_{\infty}$ are the $1^{st}$-class constraints
$\pi^o_a$, $\Gamma_a$ and Eqs.(2-34) give the generic generator ${\hat G}_{gt}$
of the proper infinitesimal gauge transformations if one uses proper
$\lambda^0_a$, $\alpha_a$: the proper gauge transformations contained
in ${\bar {\cal G}}^o_{\infty}$ form a subgroup (or better a $\infty$-
dimensional generalization of a local Lie group,
i.e. an analytic subgroup [49a];
it depends on the choice of the function spaces,
of the topology and of the definition of the infinite dimensional Lie group,
 see Section 6) of the canonical
transformations. Instead Eqs.(2-38) give the generator $G_{gt}$
of the improper infinitesimal gauge transformations: the term ${\hat G}_{gt}
[\lambda_{ao},\alpha_a]\approx {\cal Q}[\alpha_a]$,
with improper $\alpha_a$, of $G_{gt}$ again
generates a canonical transformation; the term ${\cal Q}[\alpha_a]$,
a surface term at spatial infinity, is not a generator
of canonical transformations with the Poisson brackets: only with Dirac
brackets [35] associated with a gauge-fixing it becomes a generator of
canonical transformations on  (non-local) Dirac's observables (see
Section 9). For constant $\alpha_a$'s, i.e. for global or rigid gauge
transformations in ${\cal G}_G\sim G_G\sim G$, one has ${\hat G}_{gt}[0,
\alpha_a]=-c_{acb}\alpha_b\int d^3x\, {\vec \pi}_a(\vec x,x^o)\cdot
{\vec A}_a(\vec x,x^o)\approx {\cal Q}[\alpha_a]=\sum_a\alpha_aQ_a$,
where, from Eq.(2-14), $Q_a$ are the weak improper conserved non-Abelian
charges; this is the classical basis of the group of unitary
transformations $exp(iq{\hat Q})$ used in quantum field theory: in a gauge
theory it is well defined only on the observables and it is the quantization
of the rigid group ${\cal G}_G\sim G_G\sim G$, or better of $G/Z_G$,where
$Z_G$ is the discrete center of G.

In Ref.[50] the condition to have $G_G\sim G$ or $K\subset G\sim G_G$ (with K
a compact, connected Lie subgroup of G)
implemented as "left rigid internal actions" on P are
given (this is the exact definition of $G_G$):
essentially this requires that P must be "reducible" to a principal
$Z_G$-bundle [or $Z_G(K)$-bundle, where $Z_G(K)=\lbrace a\in G\, |\, ak=ka\,
for\, all\, k\in K\rbrace$ is the centralizer of K in G]
and that the rigid internal action is a cross section
of a suitable associated bundle. Since this implies that the transition
functions of P (in a coordinate principal bundle representation) all belong
to $Z_G$, this requirement is satisfied by our trivial principal G-bundle
$P^t$, which can be described with only one coordinate chart, so that
the transition functions are trivial, $\psi_{\alpha\beta}=I$. When this does
not happen,
one has only infinitesimal improper global (rigid) gauge transformations
which do not exponentiate to global finite ones due to some obstruction; in the
Noether theorem one gets invariance only at the algebra level and not at the
group level, so that a global momentum map cannot be defined: this implies that
all or some non-Abelian charges $Q_a$ are locally, and not globally, defined.
With $P^t$ the $Q_a$'s are globally defined, as already said in the discussion
about Ref.[26]; the rigid group ${\cal G}_G\sim G_G\sim G$ has a left rigid
internal
action. Given a connection on $P^t$, it admits G or $K\subset G$ as an internal
rigid symmetry group if the connection is reducible to a connection on the
principal $Z_G$- or $Z_G(K)$-bundle (i.e. the holonomy group $\Phi
^{\cal A}$ of the connection is contained in $Z_G$ or $Z_G(K)$); these
connections are special cases of the reducible connections, which have gauge
symmetries (i.e. stability subgroups ${\cal G}^{\cal A}\supset Z_G$; see
Section 6).

Let us call ${\bar U}^{\lambda }(\vec x,x^o)$ the gauge transformations
generated by $\pi^o_a$ and ${\hat {\bar U}}(\vec x,x^o)$ those generated by
$\Gamma_a$.
Let us remark that usually one ignores the generators $\pi^0_a$ of the Abelian
subalgebra of the Lie algebra of ${\bar {\cal G}}^o_{\infty}$, and only
considers the subgroups ${\hat {\cal G}}$, ${\hat {\cal G}}
_{\infty}$, ${\hat {\cal G}}_{\infty}^o$ associated with Gauss'
laws, because they give rise to a Lie algebra isomorphic to $g$, the Lie
algebra of the structure group G (not to be confused with the algebra of the
non-Abelian charges).
{}From Eqs.(2-40)ii) we deduce the following asymptotic behaviour for
the infinitesimal form of these gauge transformations, when they are proper
gauge transformations:

$$
\eqalign{
&{\bar U}^{\lambda}(\vec x,x^o) \sim I+t^a\lambda_{ao}(\vec x,x^o)\quad
{\buildrel \longrightarrow \over {r\rightarrow \infty}}\quad I+
{ {t^a\lambda^{\infty}_{ao}}\over {r^{1+\epsilon}} }+O(r^{-2})\cr
&{\hat {\bar U}}(\vec x,x^o)\sim I+t^a\alpha_a(\vec x,x^o)\quad
{\buildrel \longrightarrow \over {r\rightarrow \infty}}\quad I+
{ {t^a\alpha_a^{\infty}}\over {r^{3+\epsilon}} }+O(r^{-4})\cr}
\form
$$

\noindent We shall also assume this asymptotic behaviour for the proper
"finite" gauge transformations in ${\bar {\cal G}}_{\infty}$ (and also
in ${\bar {\cal G}}$ when $\bar U\rightarrow const.$): finite transformations
${\hat {\bar U}}$ approaching I in a slower way are "improper" but not rigid.
In this way the new subgroup ${\bar {\cal G}}_{\infty}^{o(P)}$ of
${\bar {\cal G}}_{\infty}^o$ of the "proper" canonical
gauge transformations connected with the identity is defined
(and the previous equations define the behaviour of the Lie algebra
$g_{ {\bar {\cal G}}_{\infty}^{o(P)}}$ of
${\bar {\cal G}}_{\infty}^{o(P)}$; with
improper $\alpha_a$'s one gets the Lie algebra $g_{{\bar {\cal G}}_{\infty}^o}$
of ${\bar {\cal G}}_{\infty}^o$). In this paper
we shall study Dirac's observables with respect to this subgroup
(or better $\infty$-dimensional analogue of a
local Lie group) ${\bar {\cal G}}^{o(P)}_{\infty}$.

Let us now review the two main types of improper gauge transformations,
besides the global ones, connected with the non-Abelian charges in $G/Z_G$,
and the non-global ones in ${\bar {\cal G}}^o_{\infty}$ (which will be
discussed
at the end of this Section), i.e.
i) the gauge transformations with winding number; ii) those belonging to the
center $Z_G$.

i) In electrodynamics G=U(1) and ${\hat {\cal G}}_{\infty}^o={\hat {\cal G}}
_{\infty}$. Instead when G is simple [48], one has

$$
{\hat {\cal G}}_{\infty}/{\hat {\cal G}}_{\infty}^o=Z
\form
$$

\noindent with Z the group of integers under addition. Therefore
$\pi_o({\hat {\cal G}}_{\infty})=Z$, i.e. ${\hat {\cal G}}
_{\infty}$ has as many disconnected components ${\hat {\cal G}}_
{\infty,n}$ as the integers $n\in Z$. This means that if $\bar U\in
{\hat {\cal G}}_{\infty,n}$, then $\bar U$ should be written as
${\bar U}_n(\vec x,x^o)$. ${\bar U}_o$ are the gauge transformations
connected with the identity, also called "small" gauge transformations
( they contain the proper ones previously defined);
instead the ${\bar U}_n$'s with $n\not= 0$ are called the "large" gauge
transformations. For each $n\not= 0$, among the ${\bar U}_n$'s there are the
improper ones ${\bar U}^{(IM)}_n$ not connected with the identity. A
generic ${\bar U}_n$ can be written as ${{\bar U}_1}^n{\bar U}_o$, where
${\bar U}_1$ is a gauge transformation with n=1. With each connected component
${\hat {\cal G}}_{\infty ,n}$ can be associated a pure gauge (flat) gauge
potential (a classical vacuum): if ${\vec A}_o=U^{-1}_o\vec \partial U_o$,
then ${\vec A}_o^{U_n}={(U_oU_n)}^{-1}\vec \partial (U_oU_n)=U^{'-1}_n\vec
\partial U^{'}_n={\vec A}_n$. Therefore the number n introduces a splitting of
the space of gauge potentials on $R^3\times G$ into an infinite countable
number of subsets each one with a flat gauge potential or classical vacuum
as a consequence of the non connectedness of the group of gauge transformations
(for $M\times G$ with $\pi_1(M)\not= 0$ and with Eq.(3-3) valid, one has more
than one flat gauge potential in each subset).

The integer n is called the "winding number" of the gauge transformation.
For G=SU(2) [51] and with $R^3$ compactified to $S^3$,
since the group manifold of SU(2) is also homeomorphic to
$S^3$, $\bar U$ may be considered as a mapping $\bar U:S^3\rightarrow S^3$
and such mappings falls into homotopy classes labelled by an integer
winding number n. An example of gauge transformation with n=1 is
${\bar U}_1(\vec x,x^o)=exp\lbrace f(r)\tau_ix^i/r\rbrace =cos\, f(r)+i\tau_i
x^i/r\, sin\, f(r)$, where $r=|\vec x|$,
$\tau_i$ are the Pauli matrices and $f(0)=0$, $f(\infty )=2\pi$ [51,48],
so that for $r\rightarrow \infty$ one has ${\bar U}_1(\vec x,x^o)\rightarrow
1$, i.e. ${\bar U}_1\in {\hat {\cal G}}_{\infty}$.
Z is the group generated by the coset ${\bar U}_1{\hat {\cal G}}^o
_{\infty}$.In the case of $R^3$ with the boundary conditions (2-40)
we still have the winding number, because
among the mappings $\bar U:R^3\rightarrow S^3$ there is a subset which may be
identified with  those of the type $\bar U:S^3\rightarrow S^3$ (which are
differentiable also in the added point to infinity, but this is not
relevant for homotopy, where only continuity is considered). Other examples
of gauge transformations in ${\hat {\cal G}}_{\infty ,1}$ are: i) ${\bar U}_1
=-exp{ {i\pi \vec x\cdot \vec \tau}\over {\sqrt{{\vec x}^2+\lambda^2}} }
{\rightarrow}_{r\rightarrow \infty}I$; ii) ${\bar U}_1(\vec x,x^o)=-{
{{(\lambda
\vec \tau +i\vec x)}^2}\over {{\vec x}^2+\lambda^2}
}{\rightarrow}_{r\rightarrow
\infty}I$. The generalization of these ${\bar U}_1(\vec x,x^o)$ to simple Lie
groups as G=SU(3) can be done by using the SU(2) subgroups of G.

Since any connected Lie group is the quotient of the direct product of simple
and Abelian Lie groups by discrete Abelian groups (which can be trivial)[49],
Eq.(3-3) can be generalized to arbitrary Lie groups. In Ref.[25a,b,52] it
is shown that  for G non Abelian and simple, Eq.(3-3) and the winding number
derive from $\pi_1(\lbrack S^3\rightarrow G\rbrack )=\pi_3(G)=Z$, where
$\lbrack S^3\rightarrow G\rbrack$ is the set of the base point preserving
maps from $S^3$ onto G. Then the result is extended to the case in which
$R^3$ or $S^3$ are replaced by an arbitrary compact, connected, orientable
3-manifold $\Sigma$: if the principal G-bundle is trivial (this implies the
triviality of the cohomology group $H^2(\Sigma ,\pi_1(G)$)
and $\pi_1(\Sigma )=0$, the group Z is
replaced by the group $Hom(\lbrack \Sigma \rightarrow G\rbrack ,U(1))$ of the
homomorphisms of $\lbrack \Sigma \rightarrow G\rbrack$ onto U(1); for the
case $\pi_1(\Sigma )\not= 0$ see Refs.[25a,b].

To find the winding number of a gauge transformation $\bar U$ in the case of
G=SU(2) and $R^3$ compactified to $S^3$ with radius $R_{S^3}$,
one has to evaluate the volume integral (see for instance [43,23b,53]):

$$\eqalign{
n[\bar U]&={1\over {24\pi^2}}\int_{S^3,R_{S^3}\rightarrow \infty}
d^3x \epsilon^{ijk}Tr\lbrack ({\bar U}^{-1}(\vec x,x^o)
{\partial}^i{\bar U}(\vec x,x^o))\cr
&({\bar U}^{-1}(\vec x,x^o){\partial}^j{\bar U}
(\vec x,x^o))({\bar U}^{-1}(\vec x,x^o)
{\partial}^k{\bar U}(\vec x,x^o))\rbrack \cr}
\form
$$

\noindent For G=SU(2) the integrand of Eq.(3-4) is the invariant Haar
measure on SU(2) [31]; the integrand of $n[\bar U]$ is the Jacobian of the
mapping $\bar U:S^3\rightarrow SU(2)\sim S^3$, a quantity which relates the
surface elements on the two spheres $S^3$ [31,54] and each winding of a $S^3$
onto the other contributes a unit to $n[\bar U]$. The result that $n[\bar U]$
is an integer remains valid for every compact simple Lie group [SO(n), $n\geq
5$; SU(n), $n\geq 2$; Sp(n), $n\geq 1$; $E_6,E_7,E_8,F_4,G_8$] by evaluating
the $n[\bar U]$ of the mapping $\bar U:S^3\rightarrow a\, SU(2)\, subgroup\,
of\, G$ [see appendix of [54]].
More in general the winding number is connected with the degree of
proper surjective maps $U:M\rightarrow G$, when M and G are compact, oriented
manifolds of the same dimension [55].

All the subsets of gauge potentials labelled by the winding number can be
defined on the same trivial principal G-bundle $R^3\times G$ or $S^3\times G$,
since all the principal G-bundles over $R^3$ and $S^3$ are trivial (in fact the
second Chern characteristic class for G=SU(n) or the second Pontryagin class
for G=O(n), which classifies the not equivalent principal G-bundles on a given
base manifold [24b], vanishes in these cases). The same is true for $R^4\times
G$ (Euclidean YM theory), but not for principal G-bundles over $S^4$ like
$S^4\times G$ (compactified Euclidean YM theory): all the gauge transformations
on $S^4\times G$ have zero winding number; in the case of $S^4$ and with,
for instance, G=SU(n), the second Chern characteristic class $\gamma_2$
(a gauge invariant closed 4-form on $S^4$, belonging to the cohomology group
$H^4(S^4,R)$ and function of the field strength; $\gamma_2=0$ on $R^3$ and
$S^3$ because there are no 4-forms on them) classifies all the not equivalent
principal G-bundles $P_n(S^4,G)$
over $S^4$ [24b] and, since the integral of $\gamma_2$ over
$S^4$ is an integer coinciding with the winding number of the gauge potential
used to evaluate $\gamma_2$ [31,54], this implies that all the gauge
transformations on the principal G-bundle $P_n(S^4,G)$
over $S^4$ with $\int_{S^4}\gamma_2
d^4x=n$ have winding number n (and also a nontrivial topology due to the
Gribov ambiguity, even for n=0), so that every connection on such a bundle
has all associated gauge potentials in only one set. For $S^4$ the
"instanton number" (Chern or Pontryagin number) is defined by

$$\eqalign{
n_I&=\int_{S^4}\gamma_2={1\over {16\pi^2}}\int d^4x\, Tr(F^E_{\mu\nu}(x){*}
F^{E\mu\nu}(x))=-{1\over {32\pi^2}}\int d^4x
F^E_{a\mu\nu}(x){*}F^{E\mu\nu}_a(x)
\cr
&={1\over {8\pi^2}}\int d^4x\, {\vec E}^E_a(x)\cdot {\vec B}^E_a(x)
=\int d^4x\, \nu [x;A]\cr}
\form
$$

\noindent where the index E denotes Euclidean. In the case of $R^4$ the same
formula holds; for both $S^4$ and $R^4$ the Euclidean behaviour $F^E_{\mu\nu}
(x){\rightarrow}_{R\rightarrow \infty}1/R^2$ ($R^2=x_E^2+{\vec x}^2$, $x_E=i
x^o$) assures a finite Euclidean action and $\int_{S^4}\gamma_2$ finite, but
it is not known what other supplementary hypotheses are needed to get an
integer n (i.e. for $S^4$ to get a well defined principal G-bundle over $S^4$)
[24b,54].

In the compactified Euclidean $S^4$ theory the identification of $n_I$ with the
winding number n [31,54] is based on the observation that the quantity $\nu
[x;A^E]$ in Eq.(3-5) is a topological density insensitive to local deformations
of $A^E_{\mu}$ due to the Bianchi identities (2-4): $\delta \nu [x;A^E]/\delta
A^E_{a\mu}(y)=(1/8\pi^2){({\hat D}^{(A)}_{\nu}{*}F^{E\mu\nu})}_a\delta^4(x-y)
\equiv 0$. Thus  $\nu [x;A^E]$ depends only on the large scale properties of
$A^E_{\mu}$, as it is evident from its being a divergence of a topological
current $K^{\mu}[x;A^E]$ [54], and gauge potentials can be grouped in classes
labelled by the value of the topological charge.

$$
\eqalign{
&\nu [x;A^E]={1\over {8\pi^2}}{\vec E}^E_a(x)\cdot {\vec B}^E_a(x)=
{\partial}_{\mu}K^{\mu}[x;A^E]\cr
&K^{\mu}[x;A^E]=-{ 1\over {16\pi^2} }\epsilon^{\mu\alpha\beta\gamma}
\sum_aA^E_{a\alpha}(x)({\partial}_{\beta}A^E_{a\gamma}(x)+{1\over 3}
c_{abc}A^E_{b\beta}(x)A^E_{c\gamma}(x))=\cr}
\form
$$

\vfill\eject

$$
=-{ 1\over {32\pi^2} }\epsilon^{\mu\alpha\beta\gamma}\sum_aA^E_{a\alpha}(x)
(F^E_{a\beta\gamma}(x)-{1\over 3}c_{abc}A^E_{b\beta}(x)A^E_{c\gamma}(x))
$$

\noindent Therefore one has ($S^3$ is the sphere at infinity; the limit exists
if $\int_{S^4}d^4x\, \nu [x;A^E]$ converges):

$$n_I=\int_{S^4}d^4x\, \nu [x;A^E]=\int_{S^4}d^4x\, \partial_{\mu}K^{\mu}[x;
A^E]=lim_{R_{S^3}\rightarrow \infty}\int_{S^3}d^3\sigma_{\mu}K^{\mu}[x;A^E]
\form
$$

Since $F^E_{\mu\nu}(x){\rightarrow}_{R\rightarrow \infty}0$
at least as $1/R^2$ to have the Euclidean action
finite, this implies that $A^E_{\mu}(x){\rightarrow}_{R\rightarrow \infty}
U^{E-1}_{\infty}(x)\partial_{\mu}U^E_{\infty}(x)\sim O(1/R)$ and one gets
($\vec y$ is a coordinate system for $S^3$) by using the antisymmetric
properties of the indices

$$\eqalign{
n_I&=-{1\over {24\pi^2}}\int_{S^3,R_{S^3}\rightarrow \infty}
d^3y \epsilon^{ijk}Tr\lbrack (U^{E-1}_{\infty}(\vec y)
{\partial}^iU^E_{\infty}(\vec y))\cr
&(U^{E-1}_{\infty}(\vec y){\partial}^jU^E_{
\infty}(\vec y))(U^{E-1}_{\infty}(\vec y)
{\partial}^kU^E_{\infty}(\vec y))\rbrack \cr}
\form
$$

\noindent By comparison with Eq.(3-4), one finds $n_I=n[U^E_{\infty}]$, and
one has $n[U^{E-1}_{\infty}]=-n[U^E_{\infty}]$. In
the Euclidean $S^4$ YM theory, one has the positivity condition $Tr\int d^4x\,
{(F^E\pm {*}F^E)}^2=Tr\int d^4x\, 2(F^{E2}\pm F^E{*}F^E)\geq 0$ and one gets
$Tr\int d^4x\, F^{E2}\geq | Tr\int d^4x\, F^E{*}F^E | =16\pi^2n$, so that for
the Euclidean action one has $S^E=(1/2g^2)Tr\int d^4x\, F^{E2}\geq
8\pi^2n/g^2$;
the Euclidean action is minimized when $F^E_{\mu\nu}=\pm {*}F^E_{\mu\nu}$, i.e.
the self-dual or antiself-dual field strengths are the finite-action solutions
to the classical Euclidean YM theory and for them $\int_{S^4}d^4x\, \nu
[x;A^E]$
is absolutely convergent, so that $n_I$ is well defined. Let us remark that
$S^E[A^E]$ is not a topological invariant, since $\delta S^E[A^E]/\delta A^E
_{a\mu}(x)=g^{-2}L^{\mu}_a(x)=g^{-2}{\hat D}^{(A^E)}_{\nu}F^{E\nu\mu}_a(x)
{\buildrel \circ \over =}0$, which only vanishes on the extremals (but $F^E=
\pm {*}F^E$ are extremals due to the Bianchi identity). The fact that $n_I$ is
an integer is a consequence of the Atiyah-Singer index theorem for compact
manifolds; on them it relates the number of zero eigenvalues of the covariant
derivative $D^{(A)}_{\mu}$ to the topological charge of the gauge potential
A contained in it [54].
Therefore in the $S^4$ theory one has a vacuum gauge potential (and also a flat
connection) $A^{E(n)}_{\mu}=U^{E-1}_n\partial_{\mu}U^E_n$ and a different
principal G-bundle over $S^4$ for each value of the instanton=winding number;
instanton solutions, like the previous one, are interpreted at the Euclidean
quantum level as corresponding to tunnelling between the different vacuum
states
corresponding to the $A^{E(n)}_{\mu}$ [51]. This is usually done in the path
integral formalism in the Coulomb gauge $A^E_o(x)=0$ (which still leaves the
gauge freedom of U's such that $\partial_oU=0$) by choosing ${\vec A}^E=0$ for
$x_E\rightarrow -\infty$ and ${\vec A}^E=U^{E-1}_n\vec \partial U^E_n$ for
$x_E\rightarrow \infty$ [for instance with $U^E_1(x)=exp\lbrace i\pi \vec \tau
\cdot \vec x/{(R^2+\lambda^2)}^{1/2}\rbrace$ for n=1], i.e. $F^E_{\mu\nu}=0$
for $x_E\rightarrow \pm \infty$; these two asymptotic vacua are interpolated
by an instanton solution with instanton number n, rewritten in the Coulomb
gauge [strictly speaking the two vacua are flat connections in two different
not equivalent principal G-bundles over $S^4$ and the instanton
solution connects
these two asymptotic bundles]; since the instanton solution is a minimum of the
Euclidean action (i.e. it is the classical extremal path for immaginary time
$x^o=ix_E$), one has the interpretation of the instanton solution as a
tunnelling through the potential barrier between different vacuum states $| n
>$ (see Refs.[53,43] for a detailed discussion). Let us remark that in Ref.
[56] for G=SU(2) and for gauge potentials with an O(3) gauge symmetry, the fact
that the Coulomb gauge has the Gribov ambiguity (see later on in Section 6 its
connection with the gauge symmetries of certain connections) is interpreted as
implying a discontinuity in the time evolution of the gauge potential.
As vacuum states $| n >$ corresponding to different topological winding
(instanton) numbers are separated by finite-energy barriers and there are
tunnellings between these states, one expects the true vacuum state ($\theta$
vacuum) to be a suitable superposition of these $| n >$ states. Since there are
quantum gauge transformations ${\hat U}_1$, leaving the quantum Hamiltonian
invariant, such that ${\hat U}_1| n > =| n+1 >$, the $\theta$-vacuum can be
defined as an eigenstate of ${\hat U}_1$: ${\hat U}_1| \theta >=e^{i\theta}
| \theta >$, $| \theta > = \sum_ne^{-in\theta}| n >$ (${\hat U}_1$
can be interpreted as the
translation operator, $| \theta >$ as the Bloch wave, $\theta$ as the conserved
pseudomomentum of a quantum-mechanical problem of periodic potential). Each
value of $\theta$ defines a physically not equivalent sector and each sector
does not communicate with the others, so that there is no a priori method to
determine $\theta$. From the study of the vacuum-to-vacuum ($| \theta >
\rightarrow | \theta^{'} >$) transition amplitude, it turns out that it is
the sum over the winding number n of terms, each one defining the path
integral over the connections with winding number n of an effective action
of the type [43,23]:

$$\eqalign{
S^E_{eff}&=\int d^4x\, {\cal L}_{\theta}(x)=S^E_{YM}[A_{(n)}]+\theta n=\int
d^4x[{\cal L}(x)+\theta \nu [x;A^E]=\cr
&=\int d^4x\lbrace -{1\over {4g^2}}F^{E\mu\nu}_{a(n)}(x)F^E_{a(n)\mu\nu}(x)
-{{\theta} \over {32\pi^2}}F^{E\mu\nu}_{a(n)}(x){*}F^E_{a(n)\mu\nu}(x)\rbrace
\cr}
\form
$$

\noindent with $\theta$ an arbitrary parameter. ${\cal L}_{\theta}$ is
gauge equivalent to ${\cal L}$ in the sense that the new term is a total
divergence : the Euler-Lagrange equations do not change and the new canonical
momenta ${\pi}_{\theta}{}_a^i=E^i_a-(1/8\pi^2)\theta B^i_a$,
$\pi_{\theta}{}^o_a=0$, generate the same Gauss' laws, because
from the Bianchi identities one has ${{\hat {\vec D}}^{(A)}}_{ab}\cdot
{\vec B}_b\equiv 0$.

Let us now consider $\int d^4x\, \nu [x;A]=\int d^4x\, \partial_{\mu}K^{\mu}
[x;A]$ on Minkowski space-time. Let us evaluate it in a volume ${\cal V}$
whose boundary is $\partial {\cal V}=\Sigma_1-\Sigma_2+\Xi$, where $\Sigma_1$
and $\Sigma_2$ are two space-like surfaces and $\Xi$ is a boundary at spatial
infinity (for instance the portion of a space-like hyperboloid at spatial
infinity between $\Sigma_1$ and $\Sigma_2$). Since $K^{\mu}[x;A]{\rightarrow}
_{r\rightarrow \infty}$\break $O(1/r^3)$, then
$\int_{\Xi}d^3\sigma_{\mu}K^{\mu}
[x;A]=0$. Therefore, when $\Sigma_1$ and $\Sigma_2$ are space-like hyperplanes,
one gets:

$$\int_{\cal V}d^4x\, \nu [x;A]={1\over {8\pi^2}}\int_{\cal V}d^4x\, {\vec E}_a
(x)\cdot {\vec B}_a(x)=(\int_{\Sigma_1}-\int_{\Sigma_2})d^3x\, K^o[x;A]=
W[\vec A{|}_{\Sigma_1}]-W[\vec A{|}_{\Sigma_2}]
\form
$$

\noindent where $W[\vec A]$ is the (non conserved) "charge" associated with
the topological current \break $K^{\mu}[x;A]$

$$
\eqalign{
W[\vec A]&=\int d^3x\, K^o[x;\vec A]=\cr
&={ 1\over {8\pi^2}}\epsilon^{ijk}\int d^3x\, Tr(A^i(\vec x,x^o)
{\partial}^jA^k(\vec x,x^o)+{2\over 3}A^i(\vec x,x^o)A^j(\vec x,x^o)
A^k(\vec x,x^o))\cr}
\form
$$

$W[\vec A]$ is called, by analogy with the Euclidean case, the "winding
number" of the gauge potential [23b] [when space-time is 3-dimensional, $d^3x
\rightarrow dx^od^2x$, it is called the Chern-Simon action (see for instance
Ref.[48b])]. If we evaluate $W[\vec A]$ for a pure gauge (flat) gauge potential
$\vec A(x)={\bar U}^{-1}(x)\vec \partial \bar U(x)$, like for Eqs.(3-8) we
get $W[{\bar U}^{-1}\vec \partial \bar U]=n[\bar U]$ with $n[\bar U]$ of Eq.
(3-4); that is for G=SU(2) we get the integer giving the winding number
of the gauge transformation $\bar U$ and, moreover, under a gauge
transformation
${\bar U}_n(\vec x,x^o)$ with winding number n we get

$$
W[{\bar U}^{-1}_n{\vec A}{\bar U}_n-{\bar U}^{-1}_n{\vec
\partial}{\bar U}_n]=W[\vec A]+n[{\bar U}_n]=W[\vec A]+n
\form
$$

\noindent Therefore $W[\vec A]$ is a quantity invariant under ${\bar {\cal G}}
_{\infty}^o$, but not under ${\bar {\cal G}}_{\infty}$.

If $F_{a\mu\nu}(x)=0$ on $\Sigma_1$ and $\Sigma_2$, one has $A_{\mu}(x){|}
_{\Sigma_1}={\bar U}^{-1}_{\Sigma_1}(x)\partial_{\mu}{\bar U}_{\Sigma_1}(x)$
and $A_{\mu}(x){|}_{\Sigma_2}={\bar U}^{-1}_{\Sigma_2}(x)\partial_{\mu}
{\bar U}_{\Sigma_2}(x)$ and from Eq.(3-11) one obtains

$$\triangle n=
n[{\bar U}_{\Sigma_1}]-n[{\bar U}_{\Sigma_2}]={1\over {8\pi^2}}\int_{\cal V}
d^4x\, {\vec E}_a(x)\cdot {\vec B}_a(x)
\form
$$

In Refs.[57] it is investigated the problem of which field configurations could
provide the vacuum tunnelling
in Minkowski space-time, of which kind of potential
barrier exists in this space-time and to what extent vacuum tunnelling can be
understood as a solution to the Minkowski YM equations. In Ref.[57b] for
G=SU(2)
, it is noted that if Eqs.(3-13) give $\triangle n\not= 0$, then there must be
some region between $\Sigma_1$ and $\Sigma_2$ where $F_{\mu\nu}\not= 0$; but
since the energy density is positive definite and vanishes only if $F_{\mu\nu}
=0$, the existence of such a region with $F_{\mu\nu}\not= 0$ can only occur
during a tunnelling process. While it is impossible to change the winding
number of a classical vacuum ($F_{\mu\nu}=0$ on $\Sigma_1$ and $\Sigma_2$
implies $F_{\mu\nu}=0$ in ${\cal V}$), the winding number of a vacuum quantum
state may be changed at the quantum level due to quantum fluctuations: this
can be understood as a tunnelling in some collective mode of the gauge
potential configuration (see for instance [57a] for the collective coordinate
method). In [57b] the rules for obtaining the maximal tunnelling amplitude
for a single collective mode are given in the Coulomb gauge and it is shown
that the vacuum tunnelling is consistent with finding a first quantized
solution
to the field equations only for this mode.
In particular, the maximal tunnelling configuration with
$\vec E\parallel \vec B$ can be obtained from an Euclidean self-dual instanton
solution by a suitable parametrization of the variable $x_E$ of the Euclidean
theory in terms of $x^o$, which is different from the definition of the
Euclidean time $x_E=ix^o$, being connected with the definition of the previous
collective mode.

As a consequence of this discussion, Eq.(3-9) is also used in Minkowski
space-time to take into account the effects of the winding number as a
topological contribution to the determination of the $\theta$-vacuum;

$$
S_{eff}=S_{YM}+\theta \, \int d^4x\, \nu [x;A]=S_{YM}+{{\theta}\over {8\pi^2}}
\int d^4x\, {\vec E}_a(x)\cdot {\vec B}_a(x)
\form
$$

See also Ref.[58], whose results generalize the previous ones on the winding
number in Minkowski space-time to every
compact semisimple G: when $A_{\mu}$ is assumed to approach $U^{-1}{\partial}
_{\mu}U$, pure gauge, at spatial infinity (a sphere $S^2$),
one finds that the gauge potentials
fall into homotopy classes corresponding to the relative homotopy classes
of pure gauges, i.e. of the gauge transformations U.

In conclusion, in the case of $P^t$, the gauge potentials fall into classes
$\lbrace {\vec A}_n(\vec x,x^o)\rbrace$, labelled by the winding number
$n\in Z$, with a gauge orbit of flat gauge potentials in each class [see also
Section 7, Eqs.(7-22)].

ii) Let us now consider the global (rigid) gauge transformations belonging to
the center $Z_G$ of G.
Since our structure group G has a discrete center $Z_G$, one can name
"centrality" the transformation law under $Z_G$ of gauge potentials and
matter fields, in analogy to the "color triality" in QCD with G=SU(3) and $Z_G=
Z_3$; in this case the elements of $Z_3$ are $I_m=e^{2\pi im/3}I=exp[2\pi im
\lambda_8/\sqrt{3}]$, m=0,1,2, where $\lambda_8={\hat H}_8$ lies in a Cartan
subalgebra $\hat h$ of $g=su(3)$ whose unit lattice is $\lbrace 0\rbrace$ and
for which the subset of $\hat h$ identifying $Z_3$
is $k(su(3))=\lbrace 0, {\hat H}_8, 2{\hat H}_8\rbrace$.
The gauge potentials $A_{\mu}=A_{a\mu}{\hat T}^a$ are in the
adjoint representation ad of $g$
(${(ad_{A_{a\mu}t^a})}_{bc}={(A_{a\mu}[t^a,.])}
_{bc}=A_{a\mu}{({\hat T}^a)}_{bc}=A_{a\mu}c_{abc}$); now the adjoint
representation Ad of G on $g$ is not faithful: it is a faithful representation
only of $G/Z_G$. But, since $Z_G$ is a normal subgroup of G and $G\sim \cup
_{\hat a\in G/Z_G}\lbrace \hat a\cdot Z_G\rbrace=\cup_{\hat a\in G/Z_G}\lbrace
\hat a\cdot I,\hat a\cdot I_1,..,\hat a\cdot I_d\rbrace$, $d=dimZ_G-1$, this
means that $A_{\mu}$ ($\vec A$) is equivalent to the gauge potential of a
connection on the not simply connected trivial principal $(G/Z_G)$-bundle
$M^4\times (G/Z_G)$ ($R^3\times (G/Z_G)$). Let us consider a fiber G of
$M^4\times G$ ($R^3\times G$); since the group manifold of G is a compact
connected simply connected manifold, it is also a globally symmetric
Riemannian space [49b], whose geodesic exponential map Exp coincides with the
exponential map exp associated with the one-parameter subgroups of G, i.e.
geodesics through $I\in G$ are one-parameter subgroups. As a compact
semisimple Lie group, G has a surjective exponential map $exp:g\rightarrow G$
and, as a Riemannian manifold with a bi-invariant Riemannian metric induced by
the Killing-Cartan form $-{(.,.)}_g$, G is geodesically complete;
but the points in
$Z_G\subset G$, except the identity, are singular points of G and the
geodesics emanating from I have $I_m\in Z_G$, $m=1,..,d$, as points of
focusing (i.e. these points are conjugated to I in the sense of the geodesic
deviation equation, which is the Jacobi equation for Riemannian manifolds).
Let us recall [49] that for a compact connected semisimple classical
Lie group G, an analytic Cartan subgroup $H_G$ of G, whose Lie algebra is
a Cartan subalgebra $\hat h\subset g^C$ of $g^C$ (the complexification of $g$),
is also a maximal torus
(maximal Abelian subgroup of G); the "diagram" of G is the set $D(G)=\lbrace
h\in \hat h\, |\, \alpha (h)\in 2\pi iZ\, for\, some\, \alpha \in \triangle
\rbrace \subset \hat h\subset g^C$ ($\triangle$ is the set of roots) of
elements of $g$ which defines the "singular set" of G, $S=\psi [(G/H_G)\times
D(G)]$ where $\psi$ is the surjective map $\psi :(G/H_G)\times \hat h
\rightarrow G$, $(aH_G,h)\mapsto exp(Ad_ah)$, $a\in G$, $h\in \hat h$ ($Ad_a:
t\mapsto ata^{-1}$ is the adjoint representation of  G on $g$); in the points
$a\in S$ the differential $d\cdot exp$ of the exponential mapping $g\rightarrow
G$ ($d\cdot exp :g\rightarrow g$  after the identification $g\sim Tg$) is
singular, $det(d\cdot exp)=0$, i.e. each singular point of G belongs to more
than one maximal torus of G, while a regular point of G belongs only to one
maximal torus; the unit lattice ${\hat h}_I=\lbrace h\in \hat h\, |\, exp\, h=I
\rbrace$ is a subset of D(G), which reduces to ${\hat h}_I=\lbrace 0\rbrace$
if $\pi_1(G)=0$ like in our case; $k(g)=\lbrace h\in \hat h\, |\, exp\, h\in
Z_G\rbrace =\lbrace h\in \hat h\, |\, \alpha (h)\in 2\pi iZ\, for\, each\,
\alpha \in \triangle \rbrace \supset {\hat h}_I$ is the subset of D(G)
identifying $Z_G$; it is $k(g)=\cup_{m=0}^{d-1}k_m(g)$, $d=dimZ_G-1$, with
$k_m(g)=\lbrace h\in \hat h\, |\, exp\, h=I_m\in Z_G\rbrace$ and $k_o(g)=
{\hat h}_I$; if $\tilde G$ is the universal covering group of G ($\pi_1
(\tilde G)=0$), which has the same Lie algebra as G, and ${\hat {\tilde h}}_I$
its unit lattice, then one has $\pi_1(G)\sim {\hat h}_I/{\hat {\tilde h}}_I$
and
$Z_{\tilde G}=k(g)/{\hat {\tilde h}}_I$: for G simply connected, $\pi_1(G)=0$,
one has $\tilde G=G$ and $k(g)=D(G)$. Since
$d\, exp$ is singular at $I_m\in Z_G$, $m\not= 0$, $exp:g\rightarrow G$ cannot
be a diffeomorphism in a neighbourhood of every $I_m\in Z_G$; on the other
hand,
since exp is surjective, one has $I_m=exp\, h$, $h\in k(g)=D(G)=\lbrace h\in
g\,
|\, exp\, h\in Z_G\rbrace$; if $\phi^{(m)}_t(s)$ is an analytic
one-parameter subgroup of G (with the form $exp(s\tau_at^a)$ in a normal
neighbourhood of $0\in g$ with $\tau_a$ the canonical coordinates of first
kind of the point $exp(\tau_at^a)\in G$), which satisfies $lim_{s\rightarrow
\infty}\phi^{(m)}_t(s)=exp\, h=I_m\in Z_G$, every its parametrization near
$I_m$ of the form $exp\, f(s)h$ cannot be analytic (usually the exponential
representations of the elements of compact semisimple Lie groups have the
parameters only continuous or differentiable, never analytic).

Let  us consider
a connection 1-form $\omega^{\cal A}$ on $M^4\times G$ ($R^3\times G$); let
$\sigma_{(m)}:M^4\rightarrow I_m\in Z_G$ ($R^3\rightarrow I_m$), $m=1,..,d=
dimZ_G-1$, ($\sigma_{(o)}=\sigma_I$, the identity cross section) be the d
global cross sections passing through the same element $I_m\in Z_G$ in each
fiber; let $A^{(m)}_{\mu}=\sigma^{*}_{(m)}\omega^{\cal A}$ (${\vec A}^{(m)}=
\sigma^{*}_{(m)}\omega^{\cal A}$), m=0,1,..,d, be the d+1=$dimZ_G$ associated
gauge potentials with $A^{(o)}=A$ and let us look for a gauge transformation
U such that $A^{(m)}=U^{-1}A^{(o)}U+U^{-1}dU$ ($A^{(m)}=A^{(m)}_{\mu}dx^{\mu}$
or $={\vec A}^{(m)}\cdot d\vec x$); as this is equivalent to $\sigma_{(m)}(x)=
\lbrace x,I_m\rbrace =\sigma_I(x)\cdot U(x)=\lbrace x,U(x)\rbrace$, one gets
$U(x)=I_m$ and $A^{(m)}=A^{(o)}$. This implies that each connection gives rise
to $dimZ_G$ identical gauge potentials $A^{(m)}$ which cannot be connected by
proper gauge transformations, because $U=I_m\in {\bar {\cal G}}$ is neither in
${\bar {\cal G}}_{\infty}$ nor in ${\bar {\cal G}}_{\infty}^o$; this is a
consequence of the fact that the adjoint representation of G is a faithful
representation of the not simply connected, connected, compact, semisimple
Lie group $G/Z_G$ and not of its simply connected universal covering group G;
in other words any connection on $M^4\times G$ ($R^3\times G$) in the adjoint
representation can be reduced to a connection on $M^4\times (G/Z_G)$
($R^3\times (G/Z_G)$), the one with m=0.
Instead of the YM action $S_{YM}[A]=S_{YM}[A^{(o)}]=S_{YM}[A^{(m)}]$, m=0,1,..,
d, due to gauge invariance, one could use the more symmetric form emphasizing
the role of the center $Z_G$,

$$S_{YM}={1\over d}\sum_{m=0}^{d-1}S_{YM}[A^{(m)}]={1\over d}\sum_{m=0}^{d-1}
S^{(m)}_{YM}.$$

The gauge potentials $A^{(m)}$ were introduced also in the approach of
Ref.[59],
in which confinement (in the sense of charge screening) is connected with
destructive interference in the path integral in the temporal gauge over the
multiply connected group $SU(n)/Z_n$ (taking into account the Haar measure
for global effects from the group manifold) among homotopically not equivalent
trajectories corresponding to the various $A^{(m)}$ [$\pi_1(SU(n)/Z_n)=Z_n$];
this would imply a centrality also for gluons in the theory with the covering
group SU(n), which, like quarks,
would have a charge screening. However, as it
appears from the previous construction, there is no centrality connecting the
$A^{(m)}$ in the gauge group ${\bar {\cal G}}$ in contrast to the fermion case
(see Section 4); to introduce it one should enlarge ${\bar {\cal G}}$.

Also in the approach of Ref.[60] the center $Z_G$ plays a crucial role in
studying the duality of electric and magnetic fluxes and the confinement,
Higgs and Coulomb phases. However in these papers one considers Euclidean tori,
$M=T^3\, or\, T^4$, arising from periodic boundary conditions in a box, so that
$\pi_1(M)\not= 0$; therefore there are extra topologically nontrivial gauge
transformations (the twisted ones) and not
equivalent pure gauges (vacua) for each
value of the winding number, so that a more complicated analysis is needed at
the classical level.

As a consequence of this discussion, if the group of proper gauge
transformations ${\hat {\cal
G}}^{o(P)}_{\infty}$ can, in some sense, be defined as the analogous in the
infinite dimensional case of a local Lie group (or analytic Lie subgroup) of
${\hat {\cal G}}$, its gauge transformations cannot connect in an analytic way
$\sigma_{(o)}=\sigma_I$ to the other $\sigma_{(m)}$, m=1,..,d. If in some
topological sense ${\hat {\cal G}}^{o(P)}_{\infty}$ is closed and normal in
${\hat {\cal G}}$, then ${\hat {\cal G}}/{\hat {\cal G}}^{o(P)}_{\infty}\sim
G\times Z$ (modulo the nonrigid improper gauge transformations, see
later on) are the only
gauge transformations which can connect $\sigma_I$ to one $\sigma_{(m)}$ and
transform ${}^{\sigma_I}A$ into ${}^{\sigma_{(m)}}A={}^{\sigma_I}A$.
Let us now choose a
gauge transformation ${\bar U}_1$ with winding number one (there is no
canonical
choice); then with the d+1 privileged global cross sections
$\sigma_{(m)}=\sigma
_{o(m)}$, m=0,1,..,d, (and to the gauge potentials $A^{(m)}_o={}^{\sigma
_{o(m)}}A$) one can associate d+1 privileged global cross sections
$\sigma_{1(m)}(\vec x,x^o)={\bar U}_1(\vec x,x^o)\sigma_{o(m)}(\vec x,
x^o)$ (and gauge potentials $A^{(m)}_1={}^{\sigma_{1(m)}}A$) which can
be said to carry winding number one; the same can be done for each value of the
winding number $n\in Z$. This will imply that the method of construction of
Dirac's observables, which privileges ${\hat {\cal G}}^{o(P)}_{\infty}$
and $\sigma
_I=\sigma_{o(I)}$, can be repeated around each $\sigma_{n(m)}$, m=0,1,..
,d, $n\in Z$, by considering the effect of the gauge transformations generated
by Gauss' laws on each gauge potential $A^{(m)}_n$. Therefore final
Dirac's observables will have two indices, one for the winding number and the
other labelling the elements of $Z_G$. As we have already said, it is not
physically reasonable to go to the quotient with respect to the residual gauge
transformations $G\times Z\sim {\hat {\cal G}}/{\hat {\cal
G}}^{o(P)}_{\infty}$;
only the gauge transformations associated with Gauss' laws are associated
with unphysical degrees of freedom. Since the original action is invariant
(or quasi-invariant) under all the gauge transformations, the final action
$S_D$ for Dirac's observables will have the form

$$
S_D=\sum_n\sum_m[S_{D(n)}^{(m)}+\theta Q^{(m)}_{T(n)}],
\form
$$

\noindent where $Q_{T(n)}^{(m)}$ is the topological contribution to the
$\theta$-vacuum from gauge potentials with winding number n (see Eq.(3-14))
and obtained from the connection on $P^t$ by means of the global cross
section constantly equal to the element $I_m\in Z_G$, namely $\sigma_{n(m)}$,
to maintain the invariance under all the gauge transformations. See also
Section 9.

$${}$$

Let us come back to ${\hat {\cal G}}$. The group ${\cal G}_G={\hat
{\cal G}}/{\hat {\cal G}}_{\infty}$ describes the "improper global (rigid,
1st kind)
gauge transformations" or "asymptotic dynamical symmetries": in Ref.[48] it
is called the "internal symmetry group". In electrodynamics on $R^3$,
where G=U(1), it is shown in Ref.[48a] that ${\hat {\cal G}}/{\hat {\cal G}}
_{\infty}={\hat {\cal G}}/{\hat {\cal G}}^o_{\infty}=U(1)$ and that the
associated
charge Q is the electric charge, which vanishes in absence of charged fermions.
In the non Abelian case with $R^3$ as the spatial slice (or with a
3-dimensional
$\Sigma$ with $\pi_1(\Sigma )=0$) one can show that

$${\hat {\cal G}}/{\hat {\cal G}}_{\infty}={\cal G}_G\sim G
\form
$$

\noindent with the rigid group ${\cal G}_G$
isomorphic to $G_G\sim G$ (the charges $Q_a$ of Eqs.(2-14)), but now Eqs.(3-3)
holds. It can be shown [48a] that $\tilde G= {\hat
{\cal G}}/{\hat {\cal G}}_{\infty}=({\hat {\cal G}}/
{\hat {\cal G}}_{\infty}^o)/({\hat {\cal G}}_{\infty}/
{\hat {\cal G}}_{\infty}^o)=({\hat {\cal G}}/
{\hat {\cal G}}_{\infty}^o)/Z$. Therefore, since the elements of Z
commute with ${\hat {\cal G}}/{\hat {\cal G}}_{\infty}^o$,
one gets that the "asymptotic dynamical symmetry group"
${\hat {\cal G}}/{\hat {\cal G}}_{\infty}^o$
is a trivial central extension of G by Z:

$${\hat {\cal G}}/{\hat {\cal G}}_{\infty}^o={\cal G}_G\times Z
\sim G\times Z.
\form
$$

\noindent In
QCD G=SU(3): at the quantum level the irreducible unitary representations of
SU(3) with generators $Q_a$ account for colour ; instead the unitary
irreducible
representations of Z, which are in one-to-one correspondence with the points
of $S^1$, associate to $n\in Z$ $e^{in\theta}$, where $\theta$ is the QCD
$\theta$-parameter labelling not equivalent vacua $|\, \theta \, >$.
In $G\times Z$ are contained the gauge transformations valued in $Z_G$, the
center of G; when $\pi_1(G)=0$, as in our case,
($Z_G$ discrete and $Z_{\cal G}\sim Z_G$) they are global (rigid) improper ones
; with $\pi_1(G)\not= 0$ the gauge transformations in $Z_{\cal G}$ can be
dependent on the  base point and $Z_{\cal G}\supset Z_G$.

Let us remark that in more complicated situations, like with 't'Hooft-
Polyakov monopoles [61], one has to restrict the boundary conditions (2-40)
to leave invariant the $r\rightarrow \infty$ limit of the solution for the
Higgs field. This changes the group ${\hat {\cal G}}$ and, when
G=SO(3) or SU(2) (spontaneously broken to $G^{'}$=U(1)), the asymptotic
dynamical symmetry group ${\hat {\cal G}}/{\hat {\cal G}}^o
_{\infty}$ is the additive group $R^1$ of real numbers (non trivial
central extension of U(1) by Z)[62], and not $U(1)\times Z$. Since $R^1$
does not contain U(1) as a subgroup, we see that
${\hat {\cal G}}/{\hat {\cal G}}^o_{\infty}$ may not contain
the original G [48a].

Let us now consider  the Lie algebra $g_{\bar {\cal G}}$ of the group of gauge
transformations ${\bar {\cal G}}$. One of the ambiguous points in all its
formulations is which kind of "improper non rigid" gauge transformations are
allowed, that is which choice of the functional space of the "parameters"
$\xi_A(\vec x,x^o)=\lbrace \lambda_{ao}(\vec x,x^o), \alpha_a(\vec x,x^o)
\rbrace$
of the infinitesimal gauge transformations (i.e. which
choice of $g_{\bar {\cal G}}$) is physically reasonable among all the
possibilities left open by the abstract definitions of the original gauge
transformations on the principal bundle ($Aut_VP$, GauP, $\Gamma (AdP)$)
with the generic associated Sobolev spaces as we shall see in Section 6. It
seems that the only physically relevant concepts are: i) "proper gauge
transformations" with the phase space infinitesimal generators given by the
first class constraints; ii) "global or rigid improper gauge transformations"
associated with the non-Abelian charges via the first Noether theorem
aspects contained in the second Noether theorem; iii) global or rigid improper
gauge transformations in $Z_G$; iv) gauge transformations with nonvanishing
winding number $n\in Z$. Let $\xi_A^{(PR)}(\vec x,x^o)$,
satisfying Eqs.(2-40)ii), and $\xi_A^{(R)}$ be the parameters associated with
proper and rigid improper gauge transformations respectively. We shall assume
that the functional space $W_g$ needed to define $g_{\bar {\cal G}}$ is some
refinement of ordinary Sobolev spaces which contains only "parameters" $\xi_A
(\vec x,x^o)$ of the form

$$\eqalign{
\lambda_{ao}(\vec x,x^o)&=\lambda_{ao}^{(PR)}(\vec x,x^o){\rightarrow}_{r
\rightarrow \infty}0,\cr
\alpha_a(\vec x,x^o)&=\alpha_a^{(R)}+\alpha_a^{(PR)}(\vec x,x^o){\rightarrow}
_{r\rightarrow \infty}\,\, \alpha_a^{(R)}\cr
or\, &=\alpha_a^{(PR)}(\vec x,x^o){\rightarrow}
_{r\rightarrow \infty}\,\, 0.\cr}
\form
$$

\noindent so that there are no non-rigid improper gauge transformations between
the rigid ones $\alpha^{(R)}_a(\vec x,x^o)$ and the proper ones $\alpha^{(PR)}
_a(\vec x,x^o)$ [or at least they tend to zero for $r\rightarrow \infty$ faster
than $\alpha^{(PR)}_a(\vec x,x^o)$].

\noindent According to this ansatz these $\xi_A(\vec x,x^o)$ are the most
general allowed non rigid improper gauge transformations; for $r\rightarrow
\infty$ they tend to the rigid limit with a rate determined by Eqs.(2-40)ii).
At the level of the groups ${\hat {\cal G}}$, ${\hat {\cal G}}_{\infty}$ and
${\hat {\cal G}}^o_{\infty}\supset {\hat {\cal G}}^{o(P)}_{\infty}$ with
${\cal G}_G={\hat {\cal G}}/{\hat {\cal G}}_{\infty}\sim G$, this implies that

$$
\eqalign{
{\hat {\cal G}}&\equiv G\times {\hat {\cal G}}_{\infty},\quad \quad
\quad {\hat {\cal G}}^o_{\infty}={\hat {\cal G}}^{o(P)}_{\infty}\subset
{\hat {\cal G}}_{\infty}\cr
{\bar {\cal G}}&\equiv G\times {\bar {\cal G}}_{\infty},\quad \quad \quad {\bar
{\cal G}}^o_{\infty}={\bar {\cal G}}^{o(P)}_{\infty}\subset {\bar {\cal G}}
_{\infty},\cr}
\form
$$

\noindent due to Eqs.(3-18) ($\lambda_{ao}=\lambda^{(PR)}_{ao}$).

All the improper gauge transformations are in

$$
{\bar {\cal G}}/{\bar {\cal G}}^{o(P)}_{\infty}\sim G\times Z,
\form
$$

\noindent As we shall see in Section 6, only ${\bar {\cal G}}
^{o(P)}_{\infty}$ can be given the structure of Hilbert-Lie groups with our
boundary conditions (2-40)ii) and (3-18); if $R^3$ is compactified to $S^3$,
${\bar {\cal G}}^{o(P)}_{\infty}={\bar {\cal G}}_{*}$, where ${\bar {\cal G}}
_{*}\subseteq {\cal G}_{*}$ is induced by the group of pointed gauge
transformations ${Aut_VP^t}_{*}$ which leaves fixed
the fiber over a point (here the compactification point at infinity to be
added to $R^3$ to get $S^3$): in general ${\cal G}_{*}$ is a closed subgroup
of the Hilbert-Lie group ${\cal G}^{(HL)}$.
Finally one denotes ${\tilde {\cal G}}$ the quotient of ${\cal G}$ with respect
to its center $Z_{\cal G}\sim Z_G$ (in our case), ${\tilde {\cal G}}={\cal G}/
Z_{\cal G}$, so that one has

$$
{\tilde {\bar {\cal G}}}={\bar {\cal G}}/Z_G,\quad \quad {\tilde {\bar {\cal
G}}}/{\bar {\cal G}}^{o(P)}_{\infty}\sim (G/Z_G)\times Z.
\form
$$

With regard to the gauge transformations carrying winding number (${\hat {\cal
G}}_{\infty}/{\hat {\cal G}}^o_{\infty}=$\break
$={\hat {\cal G}}_{\infty}/{\hat {\cal
G}}^{o(P)}_{\infty}=Z=\pi_o({\hat {\cal G}})$), one has to face an
infinite dimensional version of the problem of non connected Lie groups G,
whose component $G_o$ connected with the identity is a normal subgroup of G and
for which $G/G_o=H$ with H a finite group (in our case H=Z, a discrete not
finite but denumerable group). For finite dimensional Lie groups G one says
that G is an extension of $G_o$ by H; the possible types of extensions are
classified in Ref.[63] in the
case that $G_o$ is a semisimple Lie group; there, it is also given a method
for finding all extensions in the class denoted natural extensions if $G_o$
is a simple Lie group.
One should need an infinite dimensional generalization of the concept of
character $\chi$ (a representation ${\cal I}$ of G into the group of
automorphisms of $G_o$).
Since a gauge transformation of winding number n, $U_n\in
{\hat {\cal G}}_{\infty}$, can be represented as $U_n={(U_1)}^nU_o$, $U_o\in
{\hat {\cal G}}^o_{\infty}={\hat {\cal G}}^{o(P)}_{\infty}$, for some $U_1\in
{\hat {\cal G}}_{\infty}$ with winding number one, one could think to the
map $U_1\mapsto {\cal I}_{U_1}\in Aut\, {\hat {\cal G}}^{o(P)}_{\infty}$
(${\cal I}_{U_1}U_o=U_1U_oU_1^{-1}$ is an automorphism of ${\hat {\cal G}}^{o
(P)}_{\infty}$)as defining a homomorphism ${\hat {\cal
G}}_{\infty}\rightarrow Aut\, {\hat {\cal G}}^{o(P)}_{\infty}$; since ${\hat
{\cal G}}_{\infty}/{\hat {\cal G}}^{o(P)}_{\infty}=Z$, this homomorphism
induces
a character $\chi$, i.e. a homomorphism $\chi : Z\rightarrow Aut\, {\hat {\cal
G}}^{o(P)}_{\infty}/Int\, {\hat {\cal G}}^{o(P)}_{\infty}$. Here $Int\, {\hat
{\cal G}}^{o(P)}_{\infty}$ is the group of inner automorphisms of ${\hat {\cal
G}}^{o(P)}_{\infty}$, for instance $U^{'}_o\mapsto U_oU_o^{'}U_o^{-1}$ with
$U_o, U_o^{'}\in {\hat {\cal G}}^{o(P)}_{\infty}$; for Lie groups G, $Aut\,
G_o/Int\, G_o$ is isomorphic to a finite subgroup $V(G_o)$ of $Aut\, G_o$ such
that $V(G_o)\cap Int\, G_o=I_{Aut\, G_o}$; in the infinite dimensional case
of a Hilbert-Lie group
the nature of $Aut\, {\hat {\cal G}}^{o(P)}_{\infty}/Int\, {\hat {\cal G}}
^{o(P)}_{\infty}$ is unknown. If one could give a sounded basis to a similar
generalization, one could try to identify which kind of character $\chi$ is
connected with the winding number and may be able to define the
extension ${\hat {\cal G}}_{\infty}$ of ${\hat {\cal G}}^o_{\infty}={\hat
{\cal G}}^{o(P)}_{\infty}$ of character $\chi$ as a semidirect product
(in analogy to the case of Lie groups [63])

$${\hat {\cal G}}_{\infty}={\hat {\cal G}}^{o(P)}_{\infty}{\times}_{\chi}\,\,
Z,
\form
$$

\noindent so that

$${\hat {\cal G}}=G\times ({\hat {\cal G}}^{o(P)}_{\infty} {\times}_{\chi}\,\,
Z),\quad \quad {\hat {\cal G}}^o_{\infty}={\hat {\cal G}}^{o(P)}_{\infty}.
\form
$$

This construction resembles the approach of Ref.[52]; since our group ${\bar
{\cal G}}_{\infty}$ is the phase space analogue of the group
${\cal G}_{*}$ of pointed gauge transformations
and our ${\bar {\cal G}}^{o(P)}_{\infty}$ corresponds to ${\cal G}
_{{*}o}$, the group of pointed gauge transformations homotopic to the identity,
the reduction of the
space of connections ${\cal C}$ with respect to ${\cal G}_{{*}o}$ will
create a simply connected orbit space, which is the universal covering space
of the orbit space ${\cal C}/{\cal G}_{*}$, which has $\pi_1({\cal C}/
{\cal G}_{*})=\pi_o({\cal G}_{*})=Z$ (in general, as we shall see in Section 6,
the orbit space ${\cal C}/{\cal G}_{*}$ is a stratified manifold with a
complicated topology and the higher $\pi_n({\cal C}/{\cal G}_{*})$ are
connected with the Gribov ambiguity [nontriviality of the principal ${\cal G}
_{*}$-bundle ${\cal C}$] [21a].

\vfill\eject

\bigskip\noindent
{\bf{4. YM Theory with Fermions.}}
\newcount \nfor

\def \form {\global \advance \nfor by 1 \eqno (4.\the\nfor)}
\bigskip

Let us now introduce the "pseudo-classical" fermionic Dirac field
[2c] $\psi (x)=\lbrace \psi_{a\alpha}
(x)\rbrace$: it is a complex 4-component spinor ($\alpha=1,\dots ,4$) for each
value of $a$; moreover it carries a representation $\rho$ of the structure
group G of P ($T^a$ are the matrices of the generators of $g$ in this
representation); finally the fields $\psi (x)$ together with their complex
conjugated $\bar \psi (x)=\psi^{\dagger}(x)\gamma_o$ form a
$(8\cdot dim\, g)$-dimensional Grassmann algebra
$\psi_{a\alpha}(x)\psi_{b\beta}
(y)+\psi_{b\beta}(y)\psi_{a\alpha}(x)=\psi_{a\alpha}(x){\bar \psi}_{b\beta}(y)+
{\bar \psi}_{b\beta}(y)\psi_{a\alpha}(x)={\bar \psi}_{a\alpha}(x){\bar \psi}
_{b\beta}(y)+{\bar \psi}_{b\beta}(y){\bar \psi}_{a\alpha}(x)=0$,
so that ${[\bar \psi (x)\psi (x)]}^2=\sum_{a\not= b,\alpha \not= \beta}
{\bar \psi}_{a\alpha}(x)\psi_{a\alpha}(x){\bar \psi}_{b\beta}(x)\psi_{b\beta}
(x)$.
The 4-dimensional Dirac matrices ${\gamma}_{\mu}$ are defined by
${\gamma}_{\mu}
{\gamma}_{\nu}+\gamma_{\nu}\gamma_{\mu}=2\eta_{\mu\nu}$,
$\gamma^{\dagger}_{\mu}=
\gamma_o\gamma_{\mu}\gamma_o$, $\gamma_5=\gamma^5=-i\gamma_o\gamma_1\gamma_2
\gamma_3$, $\gamma_5\gamma_{\mu}+\gamma_{\mu}\gamma_5=0$; if we introduce
$\beta =\gamma_o$ and $\vec \alpha =\gamma_o\vec \gamma$ we have $\alpha_i
\alpha_j+\alpha_j\alpha_i=\alpha_i\beta +\beta \alpha_i=0$ ($i\not= j$),
$\alpha^2=\beta^2=1$.

In the fiber bundle language the "pseudo-classical" Dirac fields are cross
sections of an associated bundle $E(M^4,G,F^{\rho},P)$ to
$P(M^4,G)$,with standard fiber the previous Grassmann algebra $F^{\rho}$.

The free Lagrangian density, in which $\psi (x)$ and $\bar \psi (x)$ have to
be considered as independent variables, is

$$
\eqalign{
{\cal L}_F(x)&={i\over 2}\lbrack \bar \psi (x)\gamma^{\mu}{\partial}_{\mu}
\psi (x)-({\partial}_{\mu}\bar \psi (x))\gamma^{\mu}\psi (x)\rbrack -m\bar
\psi (x)\psi (x)=\cr
&=i\bar \psi (x)\gamma^{\mu}{\partial}_{\mu}\psi (x)-m\bar \psi (x)\psi (x)
-{\partial}_{\mu}\lbrack {i\over 2}\bar \psi (x)\gamma^{\mu}\psi (x)\rbrack
\cr}
\form
$$

For other choices of the Lagrangian density see Ref.[2c]. The interaction
Lagrangian density with a YM gauge potential is ($A^{(\rho)}_{\mu}(x)=
A_{a\mu}(x)T^a$; $A_{\mu}^{(\rho )}(x)dx^{\mu}$ is the pullback of the
connection 1-form in the $\rho$-representation)

\noindent
$$
{\cal L}_I(x)=iTr\lbrack \bar \psi (x)\gamma^{\mu}A^{(\rho)}_{\mu}(x)\psi
(x)\rbrack =i{\bar \psi}_a(x)\gamma^{\mu}A_{c\mu}(x){(T^c)}_{ab}\psi_b(x)
{\buildrel \rm {def} \over =}i\bar \psi (x)\gamma^{\mu}A^{(\rho)}_{\mu}(x)
\psi (x)
\form
$$

The total Lagrangian density can be written in the form with the minimal
coupling

\noindent
$$
\eqalign{
{\cal L}_T(x)&={\cal L}_{YM}(x)+{\cal L}_F(x)+{\cal L}_I(x)=\cr
          &-{1\over {4{\sl g}^2}}F^{\mu\nu}_a(x)F_{a\mu\nu}(x)+i\bar \psi
(x)\gamma^{\mu}(\partial_{\mu}+A^{(\rho)}_{\mu}(x))\psi (x)-m\bar \psi (x)
\psi (x)-\cr
          &-\partial_{\mu}[{i\over 2}\bar \psi (x)\gamma^{\mu}\psi (x)]=\cr}
\form
$$

$$
\eqalign{
          &=-{1\over {4{\sl g}^2}}F^{\mu\nu}_a(x)F_{a\mu\nu}(x)+i\bar \psi
(x)\gamma^{\mu}D_{\mu}^{(A^{(\rho)})}\psi (x)-m\bar \psi (x)
\psi (x)-\cr
          &-\partial_{\mu}[{i\over 2}\bar \psi (x)\gamma^{\mu}\psi (x)]=\cr
          &=-{1\over {4{\sl g}^2}}F^{\mu\nu}_a(x)F_{a\mu\nu}(x)+{i\over 2}
\bar \psi (x)\gamma^{\mu}(\partial_{\mu}+A^{(\rho)}_{\mu}(x))\psi (x)-\cr
          &-{i\over 2}\bar \psi (x)(\overleftarrow {
          \partial_{\mu}-A^{(\rho)}_{\mu}(x)})
          \gamma^{\mu}\psi (x)-m\bar \psi (x)\psi (x)=\cr
          &=-{1\over {4{\sl g}^2}}F^{\mu\nu}_a(x)F_{a\mu\nu}(x)+{i\over 2}
\bar \psi (x)\gamma^{\mu}D_{\mu}^{(A^{(\rho)})}\psi (x)+{i\over 2}(D_{\mu}
^{(A^{(\rho)}){*}}\bar \psi (x))\gamma^{\mu}\psi (x)\cr
          &-m\bar \psi (x)\psi (x)\cr}
$$

\noindent where $D_{\mu}^{(A^{(\rho )})*}=-(\partial_{\mu}-A^{(\rho )}_{\mu}
(x))$.

It is gauge invariant under the gauge transformations (2-1), if the Dirac
fields have the following transformation properties ($\rho$ is a unitary
representation of G; the group of gauge transformations ${\cal G}$
now becomes ${\cal G}^{(\rho )}$, to denote the change of representation)

$$
\eqalign{
&\psi (x)\mapsto \psi^U(x)=\rho (U^{-1}(x))\psi (x)=U^{(\rho)-1}(x)\psi (x)
\simeq \psi (x)-\alpha^{(\rho )} (x)\psi (x)\cr
&\bar \psi (x)\mapsto {\bar \psi}^U(x)=\bar \psi (x)U^{(\rho)-1\dagger}(x)=
\bar \psi (x)U^{(\rho)}(x)\simeq \bar \psi (x)+\bar \psi (x)\alpha^{(\rho )}
(x)\cr}
\form
$$

\noindent where we have also given the form of the infinitesimal gauge
transformations ($U^{(\rho )}(x)\simeq I+\alpha^{(\rho )} (x)$, $\alpha^{(
\rho )}(x)=\alpha_a(x)T^a=-\alpha^{(\rho )\dagger}(x)$).

They imply

\noindent
$$
\eqalign{
&\bar \psi (x)\psi (x)\mapsto \bar \psi (x)\psi (x),\qquad
\bar \psi (x) \gamma^{\mu}\psi (x) \mapsto \bar \psi (x)\gamma^{\mu}\psi (x)\cr
&D_{\mu}^{(A^{(\rho)})}\psi^U(x)=\lbrack \partial_{\mu}+U^{(\rho)-1}(x)
A^{(\rho)}_{\mu}(x)U^{(\rho)}(x)+U^{(\rho)-1}(x)\partial_{\mu}U^{(\rho)}(x)
\rbrack U^{(\rho)-1}(x)\psi (x)=\cr
&=U^{(\rho)-1}(x)D_{\mu}^{(A^{(\rho)})}\psi (x)\cr}
\form
$$

\noindent showing that $D_{\mu}^{(A^{(\rho)})}$ is the covariant derivative on
the associated bundle E.

Now the improper gauge transformations in the center $Z_{{\cal G}^{(\rho )}}
\sim Z_G$ of ${\cal G}^{(\rho )}$  act non-trivially on
$\psi (x)$: if $U^{(\rho )}\in Z_{{\cal G}^{(\rho )}}$, then $U^{(\rho )}
(x)=U^{(\rho )}=I_m^{(\rho )}=\lambda_mI^{(
\rho )}=exp(f_mH^{(\rho )})$, m=0,1,...,$d=dimZ_G-1$, with $I^{(\rho )}_m\in
Z_G$ and with $H^{(\rho )}$ in a Cartan subalgebra of g. For G=SU(3), $Z_G=Z_3$
and one speaks of triality: $U^{(\rho )(m)}=I^{(\rho )}_m
=exp(-2\pi im/3)I^{(\rho )}=exp(-2\pi im\lambda_8/
\sqrt{3})$, m=0,1,2; $\lambda_8$ is a Gell-Mann matrix. In general for the
action of the centrality on the fermionic fields one has
($\psi (x)=\psi^{(o)}(x)$):

$$\psi^{U^{(\rho )(m)}}(x)=U^{(\rho )(m)-1}\psi (x)=e^{{2\pi im}\over d}\psi
(x)=e^{{2\pi imQ_z}\over d}\psi (x)=\psi^{(m)}(x)
\form
$$

\noindent where $Q_z$ ($Q_z\psi (x)=\psi (x)$) is the center charge.
In QCD let $\rho$ be the triplet fundamental representation
of quarks and ${\rho}^{*}$ that of antiquarks; in a tensor product of N $\rho$
and $N^{*}$ $\rho^{*}$ (corresponding to N quarks and $N^{*}$ antiquarks)
the centrality would be $Q_z=(N-N^{*})(mod\, dimZ_G)$, .
Since $A^{(\rho )(m)}_{\mu}(x)=U^{(\rho )(m)-1}A^{(\rho )(o)}_{\mu}(x)U^{(\rho
)(m)}+U^{(\rho )(m)-1}\partial_{\mu}U^{(\rho )(m)}=A^{(\rho )(o)}_{\mu}(x)=
A^{(\rho )}_{\mu}(x)$, with $U^{(\rho )(m)}=I_m\in Z_G$, due to gauge
invariance
one could write for the total action ($d=dimZ_G-1$)

$$S_T[A,\psi ]=S_T[A^{(o)},\psi^{(o)}]=S_T[A^{(m)},\psi^{(m)}]={1\over d}
\sum_{m=0}^{d-1}S_T[A^{(m)},\psi^{(m)}].
\form
$$

In the Abelian case we shall use as antihermitean generator for U(1), $T^o=-i$,
so that we obtain the Lagrangian density ($A_{\mu}=e{\tilde A}_{\mu}$)

$$
\eqalign{
{\cal L}_T(x)&=-{1\over 4}{\tilde F}^{\mu\nu}(x){\tilde F}_{\mu\nu}(x)+\bar
\psi
(x)\gamma^{\mu}(i\partial_{\mu}+e{\tilde A}_{\mu}(x))\psi (x)-m\bar \psi (x)
\psi (x)-\cr
&-\partial_{\mu}({i\over 2}\bar \psi (x)\gamma^{\mu}\psi (x))=\cr
&=-{1\over {4e^2}}F^{\mu\nu}F_{\mu\nu}+\bar \psi
(x)\gamma^{\mu}(i\partial_{\mu}
+A_{\mu}(x))\psi (x)-m\bar \psi (x)\psi (x)-\cr
&-\partial_{\mu}({i\over 2}\bar \psi (x)\gamma^{\mu}\psi (x))\cr}
\form
$$

The Euler-Lagrange equations for the gauge potentials and the Dirac fields are

$$
\eqalign{
L^{T\nu}_a(x)&={\sl g}^2 ({ {\partial {\cal L}_T(x)}\over {\partial
A_{a\nu}(x)} }-\partial_{\mu}{ {\partial {\cal L}_T(x)}\over
{\partial \partial_{\mu}A_{a\nu}(x)} })=\cr
&=\partial_{\mu}F^{\mu\nu}_a(x)+c_{abc}A_{b\mu}(x)F^{\mu\nu}_c(x)+i{\sl g}^2
\bar \psi (x)\gamma^{\nu}T^a\psi (x)=\cr
&={\hat D}^{(A)}_{\mu}F^{\mu\nu}_a(x)+{\sl g}^2J^{\nu}_a(x)
{\buildrel \rm \circ \over =}0\cr
L_{\psi}(x)&={ {\partial {\cal L}_T(x)}\over {\partial \psi (x)} }-\partial
_{\mu}{ {\partial {\cal L}_T(x)}\over {\partial \partial_{\mu}\psi (x)} }=\cr
&=\bar \psi (x)[i(\overleftarrow
{\partial_{\mu}-A^{(\rho)}_{\mu}(x)})\gamma^{\mu}
+m]=\bar \psi (x)[-i\overleftarrow {D_{\mu}^{(A^{(\rho )})*}}\gamma^{\mu}+m]
{\buildrel \rm \circ \over =}0\cr
L_{\bar \psi}(x)&={ {\partial {\cal L}_T(x)}\over {\partial \bar \psi (x)} }-
\partial_{\mu}{ {\partial {\cal L}_T(x)}\over
{\partial \partial_{\mu}\bar \psi (x)} }=\cr
&=[i\gamma^{\mu}(\partial_{\mu}+A^{(\rho)}_{\mu}(x))-m]\psi (x)=
[i\gamma^{\mu}D_{\mu}^{(A^{(\rho )})}-m]\psi (x)
{\buildrel \rm \circ \over =}0\cr}
\form
$$

\noindent $J_a^{\mu}$ is the matter current and we have

$$
\eqalign{
J_a^{\mu}(x)&=i\bar \psi (x)\gamma^{\mu}T^a\psi (x),\quad {\hat J}^{\mu}(x)=
J_a^{\mu}(x){\hat T}^a\cr
&{\sl g}^2{\hat D}^{(A)}_{\mu}{\hat J}^{\mu}(x){\buildrel \rm \circ \over =}
-{\hat D}^{(A)}_{\mu}{\hat D}^{(A)}_{\nu}F^{\nu\mu}(x)=\cr
&=-{1\over 2}[{\hat D}^{(A)}_{\mu},{\hat D}^{(A)}_{\nu}]F^{\nu\mu}(x)=
-{1\over 2}[F_{\mu\nu}(x),F^{\nu\mu}(x)]=0\cr}
\form
$$

The total action is quasi-invariant under the Poincar\'e transformations
(2-18) and $\delta_o\psi (x)=-\delta x^{\nu}\partial_{\nu}\psi (x)+{i\over 4}
\partial_{\alpha}\delta x_{\beta}\sigma^{\alpha\beta}\psi (x)=a^{\nu}\partial
_{\nu}\psi (x)+{1\over 2}\omega_{\alpha\beta}(x^{\alpha}\partial^{\beta}-
x^{\beta}\partial^{\alpha}+{i\over 2}\sigma^{\alpha\beta})\psi (x)$,
$\delta_o\bar \psi (x)=-\delta x^{\nu}\partial_{\nu}\bar \psi (x)-{i\over 4}
\partial_{\alpha}\delta x_{\beta}\bar \psi (x)\sigma^{\alpha\beta}=
a^{\nu}\partial_{\nu}\bar \psi (x)+{1\over 2}\omega_{\alpha\beta}\bar \psi
(x)(x^{\alpha}\overleftarrow {\partial^{\beta}}-x^{\beta}\overleftarrow
{\partial^{\alpha}}-{i\over 2}\sigma^{\alpha\beta})$, with
$\sigma^{\alpha\beta}={i\over 2}\lbrack \gamma^{\alpha},\gamma^{\beta}\rbrack$,
$\lbrack \sigma^{\alpha\beta},\gamma^{\mu}\rbrack =2i(\gamma^{\alpha}\eta
^{\mu\beta}-\gamma^{\beta}\eta^{\mu\alpha})$:

$$
\eqalign{
\delta {\cal L}_T&=\delta_o{\cal L}_T-\partial_{\mu}\Omega_T^{\mu}
\equiv 0,\qquad \Omega_T^{\mu}=-\delta x^{\mu}{\cal L}_T\cr
\delta_o{\cal L}_T(x)&={1\over {{\sl g}^2}}\delta_oA_{a\mu}L^{T\mu}_a+
\delta_o\bar \psi L_{\bar \psi}-L_{\psi}\delta_o\psi +\partial_{\mu}\lbrace
-{1\over {{\sl g}^2}}F_a^{\mu\nu}\delta_oA_{a\nu}+{i\over 2}\lbrack \bar \psi
\gamma^{\mu}\delta_o\psi -\cr
&-\delta_o\bar \psi \gamma^{\mu}\psi \rbrack \rbrace
\equiv \partial_{\mu}\Omega^{\mu}\equiv \partial_{\mu}(\Omega_T^{\mu}+\partial
_{\nu}X^{\mu\nu})\cr}
\form
$$

\noindent with $X^{\mu\nu}={\sl g}^{-2}F^{\mu\nu}_aA_{a\alpha}\delta
x^{\alpha}$
as without fermions.

The conserved Noether current is

$$
\eqalign{
0&{\buildrel \rm \circ \over =}-{1\over {{\sl g}^2}}\delta_oA_{a\mu}L^{T\mu}_a-
\delta_o\bar \psi L_{\bar \psi}+L_{\psi}\delta_o\psi\equiv \cr
&\equiv \partial_{\mu}(\Omega_T^{\mu}+\partial_{\nu}X^{\mu\nu}+
{1\over {{\sl g}^2}}F^{\mu\nu}_a\delta_oA_{a\nu}-{i\over 2}[\bar \psi \gamma
^{\mu}\delta_o\psi -\delta_o\bar \psi \gamma^{\mu}\psi ])\equiv \cr
&\equiv \partial_{\mu}(\theta_T^{\mu\nu}\delta x_{\nu}+{1\over 8}\partial
_{\alpha}\delta x_{\beta}\bar \psi (\gamma^{\mu}\sigma^{\alpha\beta}+
\sigma^{\alpha\beta}\gamma^{\mu})\psi-{1\over {{\sl g}^2}}\delta x^{\nu}
A_{a\nu}L^{T\mu}_a-{1\over 2}\delta x^{\mu}(\bar \psi L_{\bar \psi }-
L_{\psi}\psi ))\equiv \cr
&\equiv \partial_{\mu}({\tilde \theta}_T^{\mu\nu}\delta
x_{\nu}+{1\over 8}\partial
_{\alpha}\delta x_{\beta}\bar \psi (\gamma^{\mu}\sigma^{\alpha\beta}+
\sigma^{\alpha\beta}\gamma^{\mu})\psi ){\buildrel \rm \circ \over =}\cr
&{\buildrel \rm \circ \over =}\partial_{\mu}({\tilde \theta}_T^{\mu\nu}\delta
x_{\nu}+{1\over 8}\partial
_{\alpha}\delta x_{\beta}\bar \psi (\gamma^{\mu}\sigma^{\alpha\beta}+
\sigma^{\alpha\beta}\gamma^{\mu})\psi ){\buildrel \circ \over =}\cr
&{\buildrel \circ \over =}-a_{\nu}\partial_{\mu}\theta_T^{\mu\nu}+{1\over 2}
\omega_{\alpha\beta}\partial_{\mu}{\cal M}_T^{\mu\alpha\beta}\cr}
\form
$$

\noindent with the energy-momentum tensors ($\theta^{\mu\nu}$ is given
in Eqs.(2-20)) $\theta_T^{\mu\nu}$, ${\tilde \theta}_T^{\mu\nu}$ and the
angular momentum density tensor given by:

$$
\eqalign{
\theta_T^{\mu\nu}&=\theta^{\mu\nu}+{i\over 2}\lbrack \bar \psi
\gamma^{\mu}D^{(A^{(\rho )})\nu}\psi +\bar \psi \overleftarrow
{D^{(A^{(\rho )})* \nu}}\gamma^{\mu}\psi \rbrack \cr
{\tilde \theta}_T^{\mu\nu}&=\theta_T^{\mu\nu}-{1\over {{\sl g}^2}}
A^{\nu}_aL^{T\mu}_a-{1\over 2}\eta^{\mu\nu}(\bar \psi L_{\bar \psi}-
L_{\psi}\psi ){\buildrel \circ \over =}\theta_T^{\mu\nu}\cr
{\cal M}_T^{\mu\alpha\beta}&=x^{\alpha}\theta_T^{\mu\beta}-x^{\beta}
\theta_T^{\mu\alpha}+{1\over 4}\bar \psi (\gamma^{\mu}\sigma^{\alpha\beta}+
\sigma^{\alpha\beta}\gamma^{\mu})\psi \cr}
\form
$$

\noindent The symmetric energy-momentum tensor is ${1\over 2}(\theta_T^{\mu
\nu}+\theta_T^{\nu\mu})$ [23g].

The conserved Poincar\'e generators are ($\sigma^i={1\over 2}\epsilon^{ijk}
\sigma^{jk}$; $\sigma^{io}\gamma^o+\gamma^o\sigma^{io}=0$; $H_{DT}$ is given
by Eq.(4-20) below):

$$
\eqalign{
P^0&=\int d^3x\theta_T^{00}(\vec x, x^o){\buildrel \circ \over =}\int d^3x
{\tilde \theta}_T^{oo}(\vec x,x^o)={\bar P}^o\equiv H_{DT},\cr
&\theta_T^{00}={1\over {2{\sl g}^2}}\sum_a({\vec E}^2_a
+{\vec B}^2_a)+{i\over 2}(\bar \psi \gamma^oD^{(A^{(\rho )})o}\psi +\bar \psi
\overleftarrow {D^{(A^{(\rho )})* o}}\gamma^o\psi )\cr
&{\tilde \theta}_T^{oo}=\theta_T^{oo}-A_{ao}\Gamma^T_a-[{i\over 2}(\bar \psi
\gamma^{\alpha}D_{\alpha}^{(A^{(\rho )})}\psi +\bar \psi \gamma^{\alpha}
\overleftarrow { D_{\alpha}^{(A^{(\rho )})*} }\psi )-m\bar \psi \psi ]=\cr
&={1\over {2{\sl g}^2}}\sum_a({\vec E}^2_a+{\vec B}^2_a)+{i\over 2}(\bar \psi
\vec \gamma \cdot {\vec D}^{(A^{\rho )})}\psi +\bar \psi \vec \gamma \cdot
\overleftarrow { {\vec D}^{(A^{\rho )})*} }\psi )+m\bar \psi \psi -
A_{ao}\Gamma^T_a\cr
P^i&=\int d^3x \theta_T^{0i}(\vec x, x^o){\buildrel \circ \over =}\int d^3x
{\tilde \theta}_T^{oi}(\vec x,x^o)=\cr
&=-\int d^3x ({\vec \pi}_a(\vec x,x^o)\cdot \partial^i{\vec A}_a(\vec x,x^o)-
{i\over 2}\psi^{\dagger}(\vec x,x^o)(\partial^i-\overleftarrow {\partial^i})
\psi (\vec x,x^o)={\bar P}^i,\cr
&\theta_T^{0i}={1\over {{\sl g}^2}}{({\vec E}_a\times {\vec
B}_a)}^i+{i\over 2}(\bar \psi \gamma^oD^{(A^{(\rho )})i}\psi +\bar \psi
\overleftarrow {D^{(A^{(\rho )})* i}}\gamma^o\psi )\cr
&{\tilde \theta}_T^{oi}=\theta_T^{oi}-A^i_a\Gamma^T_a=\vec \partial \cdot
(A^i_a{\vec \pi}_a)-{\vec \pi}_a\cdot \partial^i{\vec A}_a+{i\over 2}\psi
^{\dagger}(\partial^i-\overleftarrow {\partial^i})\psi \cr
J^i&={1\over 2}\epsilon^{ijk}J^{jk}
=\int d^3x\epsilon^{ijk}(x^j\theta_T^{ok}(\vec x, x^o)+{1\over 2}\psi^{\dagger}
(\vec x,x^o)\sigma^{jk}\psi (\vec x,x^o)) =\cr
&=\int d^3x({1\over {{\sl g}^2}}{\lbrack {\vec x}
\times ({\vec E}_a\times {\vec B}_a)(\vec x,x^o)
\rbrack}^i+\psi^{\dagger}(\vec x,x^o)
\sigma^i\psi (\vec x,x^o) ){\buildrel \circ \over =}\cr
&{\buildrel \circ \over =}\int d^3x (\epsilon^{ijk}x^j{\tilde \theta}_T^{ok}
(\vec x,x^o)+\psi^{\dagger}(\vec x,x^o)\sigma^i\psi (\vec x,x^o))=\cr
&=\int d^3x\lbrace -\epsilon^{ijk}\lbrack \pi^j_a(\vec x,x^o)A^k_a(\vec x,x^o)
+{\vec \pi}_a(\vec x,x^o)\cdot x^j\partial^k{\vec A}_a(\vec x,x^o)-\cr
&-{i\over 2}
\psi^{\dagger}(\vec x,x^o)(x^j\partial^k-\overleftarrow {\partial^k}x^j)\psi
(\vec x,x^o)\rbrack+\psi^{\dagger}(\vec x,x^o)\sigma^i\psi (\vec x,x^o)
\rbrace ={\bar J}^i\cr
K^i&=J^{io}=\int d^3x(x^i\theta_T^{oo}(\vec x, x^o)-x^o\theta_T^{oi}(\vec x,
x^o)\cr
&=\int d^3x x^i\theta_T^{oo}(\vec x, x^o)-x^o P^i{\buildrel \circ \over =}
\int d^3x x^i{\tilde \theta}_T^{oo}(\vec x,x^o)-x^o{\bar P}^i\cr}
\form
$$

For their convergence
one has to impose the following boundary conditions (also ensuring
the possibility to discard the surface terms in the integration by parts)
besides Eqs.(2-40):

$$
\psi (\vec x,x^o)\rightarrow_{r\rightarrow \infty} { {\chi}\over {r^{3/2+
\epsilon}} }+O(r^{_2})
\form
$$

\noindent with $\chi$ independent from the direction and the time as in
Eqs.(2-40).

The canonical momenta for the Dirac fields are

$$
\eqalign{
&\pi_{a\alpha}(x)={ {\partial {\cal L}_T}\over {\partial ({\partial}_o\psi
_{a\alpha}(x))} }=-{i\over 2}{(\bar \psi (x)\gamma_o)}_{a\alpha}\cr
&{\bar \pi}_{a\alpha}(x)={ {\partial {\cal L}_T}\over {\partial ({\partial}_o
{\bar \psi}_{a\alpha}(x))} }=-{i\over 2}{(\gamma_o\psi (x))}_{a\alpha}\cr}
\form
$$

\noindent and the standard Poisson brackets for the odd variables $\psi
_{a\alpha}(x)$, $\pi_{a\alpha}(x)$, ${\bar \psi}_{a\alpha}(x)$, ${\bar \pi}
_{a\alpha}(x)$ are

$$
\eqalign{
&\lbrace \psi_{a\alpha}(\vec x,x^o),\pi_{b\beta}(\vec y,x^o)\rbrace =-\delta
_{ab}\delta_{\alpha\beta}\delta^3(\vec x-\vec y)\cr
&\lbrace {\bar \psi}_{a\alpha}(\vec x,x^o),{\bar \pi}_{b\beta}(\vec y,x^o)
\rbrace =-\delta_{ab}\delta_{\alpha\beta}\delta^3(\vec x-\vec y)\cr}
\form
$$

We have now the primary constraints

$$
\eqalign{
&\pi^o_a(x)\approx 0\cr
&\chi_{a\alpha}(x)=\pi_{a\alpha}(x)+{i\over 2}{(\bar \psi (x)\gamma_o)}
_{a\alpha}\approx 0\cr
&{\bar \chi}_{a\alpha}(x)={\bar \pi}_{a\alpha}(x)+{i\over 2}{(\gamma_o
\psi (x))}_{a\alpha}\approx 0\cr}
\form
$$

\noindent with the only non-vanishing Poisson brackets:

$$
\lbrace \chi_{a\alpha}(\vec x,x^o),{\bar \chi}_{b\beta}(\vec y,x^o)\rbrace
=-i\delta_{ab}{(\gamma^o)}_{\beta\alpha}\delta^3(\vec x-\vec y)
\form
$$

The total Dirac Hamiltonian is

$$
\eqalign{
H_{DT}&=\int d^3x\lbrace {1\over 2}\sum_a({\sl g}^2{\vec \pi}^2_a(\vec x,x^o)+
{1 \over {{\sl g}^2}}{\vec B}^2_a(\vec x,x^o))+\cr
   &+{i\over 2}[\bar \psi (\vec x,
x^o)\vec \gamma \cdot (\vec \partial +{\vec A}^{(\rho)}(\vec x,x^o))\psi (\vec
x
,x^o)-\bar \psi (\vec x,x^o)(\overleftarrow {\vec \partial -
{\vec A}^{(\rho)}(\vec x,x^o)})\cdot \vec \gamma \psi (\vec x,x^o)]+\cr
   &+m\bar \psi (\vec x,x^o)\psi (\vec x,x^o)-
A_{ao}(\vec x,x^o)\Gamma^T_a(\vec x,x^o)+\lambda_{ao}(\vec x,x^o)\pi^o_a(\vec
x,
x^o)+\cr
   &+\chi_{a\alpha}(\vec x,x^o)\mu_{a\alpha}(\vec x,x^o)+{\bar \mu}_{a\alpha}
(\vec x,x^o){\bar \chi}_{a\alpha}(\vec x,x^o)\rbrace .\cr}
\form
$$

\noindent In Eqs.(4-20) $\Gamma^T_a$ is given by

$$
\Gamma^T_a(x)=\Gamma_a(x)+i\bar \psi (x)\gamma^oT^a\psi (x)=
{1\over {{\sl g}^2}}L^o_a{\buildrel \circ \over =}0
\form
$$

\noindent with

$$
\eqalign{
\lbrace \Gamma^T_a(\vec x,x^o),\chi_{b\alpha}(\vec y,x^o)\rbrace &=-i
{\bar \psi}_{b\beta}(\vec x,x^o){(\gamma^o)}_{\beta\alpha}T^a\delta^3(\vec x
-\vec y)\cr
\lbrace \Gamma^T_a(\vec x,x^o),{\bar \chi}_{b\alpha}(\vec y,x^o)\rbrace &=i
T^a{(\gamma^o)}_{\alpha\beta}\psi_{b\beta}(\vec x,x^o)\delta^3(\vec x-
\vec y)\cr}
\form
$$

The time constancy of the primary constraints implies:

$$
\eqalign{
{\dot \pi}^o_a(\vec x,x^o)&{\buildrel \circ \over =}\lbrace \pi^o_a(\vec x,
x^o),H_{DT}\rbrace =\Gamma^T_a(\vec x,x^o)\approx 0\cr
{\dot \chi}_{a\alpha}(\vec x,x^o)&{\buildrel \circ \over =}\lbrace \chi
_{a\alpha}(\vec x,x^o),H_{DT}\rbrace =i{\bar \mu}_{a\beta}(\vec x,x^o)
{(\gamma^o)}_{\beta\alpha}-\cr
&-{(\bar \psi(\vec x,x^o)[i(\overleftarrow {\vec \partial
-{\vec A}^{(\rho)}(\vec x,x^o)})\cdot \vec \gamma -m])}_{a\alpha}\approx 0\cr
{\dot {\bar \chi}}_{a\alpha}(\vec x,x^o)&{\buildrel \circ \over =}\lbrace
{\bar \chi}_{a\alpha}(\vec x,x^o),H_{DT}\rbrace =-i{(\gamma^o)}_{\alpha\beta}
\mu_{a\beta}(\vec x,x^o)-\cr
&-{([i\vec \gamma \cdot (\vec \partial +{\vec A}^{(\rho)}(\vec x,x^o))+m])}_{a
\alpha}\approx 0\cr}
\form
$$

One obtains Gauss' laws  $\Gamma^T_a(x)\approx 0$ as secondary constraints
and the determination of the Dirac multipliers $\mu_{a\alpha}(x)$, ${\bar \mu}
_{a\alpha}(x)$: therefore $\chi_{a\alpha}(x)\approx 0$, ${\bar \chi}_{a\alpha}
(x)\approx 0$ are pairs of $2^{nd}$-class constraints. One has:

$$
\eqalign{
\mu_{a\alpha}(x)&\approx i{(\gamma^o[i\vec \gamma \cdot (\vec \partial +
{\vec A}^{(\rho)}(x))+m]\psi (x))}_{a\alpha}\cr
{\bar \mu}_{a\alpha}(x)&\approx
-i {(\bar \psi (x)[i(\overleftarrow {\vec \partial -
{\vec A}^{(\rho)}(x)})\cdot \vec \gamma -m]\gamma^o)}_{a\alpha}\cr}
\form
$$

The time constancy of Gauss' laws implies

$$
\eqalign{
{\dot \Gamma}^T_a(\vec x,x^o)&{\buildrel \circ \over =}\lbrace \Gamma^T_a
(\vec x,x^o),H_{DT}\rbrace =-c_{abc}A_{bo}(\vec x,x^o)\Gamma^T_c(\vec
x,x^o)-\cr
&-i{(\bar \psi (\vec x,x^o)\gamma^o)}_{b\alpha}T^a\mu_{b\alpha}(\vec x,x^o)
+i{\bar \mu}_{b\alpha}(\vec x,x^o)T^a{(\gamma^o\psi (\vec x,x^o))}_{b\alpha}
\approx 0\cr}
\form
$$

\noindent by using $\Gamma^T_a(x)\approx 0$ and Eqs.(4-24); therefore, no
more constraints are implied.

The non-vanishing Poisson brackets of the algebra of the constraints $\pi^o_a
(x)\approx 0$, $\Gamma^T_a(x)\approx 0$, $\chi_{a\alpha}(x)\approx 0$,
${\bar \chi}_{a\alpha}(x)\approx 0$ are given by Eqs.(4-19), (4-22) and by

$$
\lbrace \Gamma^T_a(\vec x,x^o),\Gamma^T_b(\vec x,x^o)\rbrace =c_{abc}\Gamma
_c(\vec x,x^o)\delta^3(\vec x-\vec y)\approx -ic_{abc}\bar \psi (x)\gamma^o
T^a\psi (x)\not= 0
\form
$$

\noindent
$\chi_{a\alpha}(x)\approx 0$ and ${\bar \chi}_{a\alpha}(x)\approx 0$ are
$2^{nd}$-class constraints, while the $1^{st}$-class ones are $\pi^o_a(x)
\approx 0$ and

$$
{\tilde \Gamma}^T_a(x)=\Gamma^T_a(x)-(\chi_{b\gamma}(x)T^a\psi_{b\gamma}(x)+
{\bar \psi}_{b\gamma}(x)T^a{\bar \chi}_{b\gamma}(x))\approx 0
\form
$$

\noindent because

$$
\eqalign{
\lbrace {\tilde \Gamma}^T_a(\vec x,x^o),{\tilde \Gamma}^T_b(\vec y,x^o)
\rbrace &=c_{abc}{\tilde \Gamma}^T_c(\vec x,x^o)\delta^3(\vec x-\vec y)
\approx 0\cr
\lbrace {\tilde \Gamma}^T_a(\vec x,x^o),\chi_{b\alpha}(\vec y,x^o)\rbrace
&=\chi_{b\alpha}(\vec x,x^o)T^a\delta^3(\vec x-\vec y)\approx 0\cr
\lbrace {\tilde \Gamma}^T_a(\vec x,x^o),{\bar \chi}_{b\alpha}(\vec y,x^o)
\rbrace&=-T^a{\bar \chi}_{b\alpha}(\vec x,x^o)\delta^3(\vec x-\vec y)\approx
o\cr}
\form
$$

The second class constraints are eliminated by introducing the Dirac
brackets, so that $\chi_{a\alpha}(x)\equiv {\bar \chi}_{a\alpha}(x)\equiv
0$ hold in Dirac strong sense. One eliminates the conjugated variables ${\bar
\psi}_{a\alpha}(x)$ and ${\bar \pi}_{a\alpha}(x)$; however, instead of working
with $\psi_{a\alpha}(x)$ and $\pi_{a\alpha}(x)$, it is simpler [2c] to use
$\psi_{a\alpha}(x)$ and ${\bar \psi}_{a\alpha}(x)\equiv 2i{(\pi (x)\gamma^o)}
_{a\alpha}$. The Dirac brackets of $\psi (x)$ and $\bar \psi (x)$ are

$$
{\lbrace \psi_{a\alpha}(\vec x,x^o),{\bar \psi}_{b\beta}(\vec
y,x^o)\rbrace}^{*}
=-i\delta_{ab}{(\gamma^o)}_{\alpha\beta}\delta^3(\vec x-\vec y)
\form
$$

\noindent with all the others vanishing. Now ${\tilde \Gamma}_a^T(x)\equiv
\Gamma^T_a(x)\approx 0$ and

$$
{\lbrace \Gamma^T_a(\vec x,x^o),\Gamma^T_b(\vec y,x^o)\rbrace}^{*}=c_{abc}
\Gamma^T_c(\vec x,x^o)\delta^3(\vec x-\vec y)
\form
$$

Instead of ${\lbrace .,.\rbrace}^{*}$ from now on
we shall use the notation $\lbrace .,.\rbrace$ also for the Dirac brackets.

The Dirac Hamiltonian strongly becomes

$$
\eqalign{
H_{DT}&=\int d^3x\lbrace {1\over 2}\sum_a({\sl g}^2{\vec \pi}^2_a(\vec x,x^o)+
{1 \over {{\sl g}^2}}{\vec B}^2_a(\vec x,x^o))+\cr
   &+{i\over 2}[\bar \psi (\vec x,
x^o)\vec \gamma \cdot (\vec \partial +{\vec A}^{(\rho)}(\vec x,x^o))\psi (\vec
x
,x^o)-\bar \psi (\vec x,x^o)(\overleftarrow {\vec \partial -
{\vec A}^{(\rho)}(\vec x,x^o)})\cdot \vec \gamma \psi (\vec x,x^o)]+\cr
   &+m\bar \psi (\vec x,x^o)\psi (\vec x,x^o)-
A_{ao}(\vec x,x^o)\Gamma^T_a(\vec x,x^o)+\lambda_{ao}(\vec x,x^o)\pi^o_a(\vec
x,
x^o)\rbrace \cr
&=\int d^3x\lbrace {1\over 2}\sum_a({\sl g}^2{\vec \pi}^2_a(\vec x,x^o)+
{1 \over {{\sl g}^2}}{\vec B}^2_a(\vec x,x^o))+\cr
&+i\psi^{\dagger}(\vec x,x^o)\vec \alpha \cdot (\vec \partial +{\vec A}_a(\vec
x,x^o)T^a)\psi (\vec x,x^o)+m\psi^{\dagger}(\vec x,x^o)\psi (\vec x,x^o)-\cr
&-A_{ao}(\vec x,x^o)\Gamma^T_a(\vec x,x^o)+\lambda_{ao}(\vec x,x^o)\pi^o_a
(\vec x,x^o)-{i\over 2}\vec \partial \cdot [\psi^{\dagger}(\vec x,x^o)\vec
\alpha \psi (\vec x,x^o)]\rbrace \cr}
\form
$$

Let us remark that from the primary constraints $\chi_{a\alpha}(x)\approx 0$,
${\bar \chi}_{a\alpha}(x)\approx 0$ one can build the generator
$G=\int d^3x[\chi_{a\alpha}(\vec x,x^o)\epsilon (\vec x,x^o)+{\bar \epsilon}
_{a\alpha}(\vec x,x^o){\bar \chi}_{a\alpha}(\vec x,x^o)]$ of the
generalized gauge transformations $\delta \psi_{a\alpha}(\vec x,x^o)=
\lbrace \psi_{a\alpha}(\vec x,x^o),G\rbrace =-\epsilon_{a\alpha}(\vec x,x^o)$,
$\delta {\bar \psi}_{a\alpha}(\vec x,x^o)=\lbrace {\bar \psi}_{a\alpha}(\vec x,
x^o),G\rbrace ={\bar \epsilon}_{a\alpha}(\vec x,x^o)$ ($\epsilon_{a\alpha}(x)$
and ${\bar \epsilon}_{a\alpha}(x)$ are Grassmann valued arbitrary spinors).
Under them we get

$$
\eqalign{
\delta {\cal L}_T(x)&=\delta \psi (x){ {\partial {\cal L}_T}\over {\partial
\psi (x)} }+\delta \partial_{\mu}\psi (x){ {\partial {\cal L}_T(x)}\over
{\partial \partial_{\mu}\psi (x)} }+\cr
&\delta \bar \psi (x){ {\partial {\cal L}_T(x)}\over{\partial \bar \psi (x)} }
+\delta \partial_{\mu}\bar \psi (x){ {\partial {\cal L}_T(x)}\over
{\partial \partial_{\mu} \bar \psi (x)} }=\cr
&=\delta \bar \psi (x)L_{\bar \psi}(x)-L_{\psi}(x)\delta \psi (x)+
\partial_{\mu}(\delta \bar \psi (x) { {\partial {\cal L}_T(x)}\over
{\partial \partial_{\mu}\bar \psi (x)} }-{ {\partial {\cal L}_T(x)}\over
{\partial \partial_{\mu}\psi (x)} }\delta \psi (x))\equiv \cr
&\equiv \bar \epsilon (x)L_{\bar \psi}(x)+L_{\psi}(x)\epsilon (x)+
\partial_{\mu}[-{i\over 2}(\bar \epsilon (x)\gamma^{\mu}\psi (x)+
\bar \psi (x)\gamma^{\mu}\epsilon (x))]{\buildrel \circ \over =}\cr
&{\buildrel \circ \over =}\partial_{\mu}[-{i\over 2}(\bar \epsilon (x)
\gamma^{\mu}\psi (x)+\bar \psi (x)\gamma^{\mu}\epsilon (x))]\cr}
\form
$$

This is the "weak quasi-invariance" (quasi-invariance modulo $1^{st}$-order
Euler-Lagrange equations), which characterize singular Lagrangians generating,
at the Hamiltonian level, primary $2^{nd}$-class constraints [5c]. The
Noether identities in this case are trivial

$$
G^{\mu}(x)=\bar \epsilon (x)({ {\partial {\cal L}_T(x)}\over
{\partial \partial_{\mu} \bar \psi (x)} }+{i\over 2}\gamma^{\mu}\psi (x))+
({ {\partial {\cal L}_T(x)}\over {\partial \partial_{\mu}\psi (x)} }+
{i\over 2}\bar \psi (x)\gamma^{\mu})\epsilon (x)\equiv 0
\form
$$

\noindent but as expected [5c] for $\mu =0$ they reproduce the primary
$2^{nd}$-
class constraints

$$
G^o(x)=\bar \epsilon (x)\bar \chi (x)+\chi (x) \epsilon (x)\equiv 0
\form
$$

Following Ref.[5c] the Noether identities associated with the $1^{st}$-class
constraints $\pi^o_a(x)\approx 0$, ${\tilde \Gamma}^T_a(x)\approx 0$ (replacing
Eqs.(2-9) in presence of fermions), generating the standard infinitesimal
gauge transformations $\delta A_{\mu}(x)={\hat D}^{(A)}_{\mu}\hat \alpha (x)$,
$\delta \psi (x)=-\alpha^{(\rho )} (x)\psi (x)$, $\delta \bar \psi (x)=\bar
\psi (x)\alpha^{(\rho )} (x)$ ($\hat \alpha (x)=\alpha_a(x){\hat T}^a$,
$\alpha^{(\rho )} (x)=\alpha_a(x)
T^a$; for $K=A_{a\mu},\psi,\bar \psi$ one has $\delta K(\vec x,x^o)=\lbrace
K(\vec x,x^o),G\rbrace$ with $G=\int d^3x[{({\hat D}^{(A)o}\hat \alpha (\vec x,
x^o))}_a\pi^o_a(\vec x,x^o)-\alpha_a(\vec x,x^o){\tilde \Gamma}^T_a(\vec x,
x^o)]$ ) under which the action $S_T=\int d^3x {\cal L}_T(x)$ is invariant, are

$$
\eqalign{
\delta {\cal L}_T(x)&={1\over {{\sl g}^2}}L^{\mu}_a(x)\delta A_{a\mu}(x)+
\delta \bar \psi (x)L_{\bar \psi}(x)-L_{\psi}(x)\delta \psi (x)+\cr
&+\partial_{\nu}[\pi^{\nu\mu}_a(x)\delta A_{a\mu}(x)+\delta \bar \psi (x)
{ {\partial {\cal L}_T(x)}\over {\partial \partial_{\nu}\bar \psi (x)} }
-{ {\partial {\cal L}_T(x)}\over {\partial \partial_{\nu}\psi (x)} }\delta
\psi (x)]\equiv 0\cr
&\Downarrow \cr
\partial_{\mu}G^{\mu}(x)&\equiv -{1\over {{\sl g}^2}}\delta A_{a\mu}(x)L^{\mu}
_a(x)+L_{\psi}(x)\delta \psi (x)-\delta \bar \psi (x)L_{\bar \psi}(x)
{\buildrel \circ \over =}0\cr
G^{\mu}(x)&=\alpha_a(x)G^{\mu}_{1a}(x)+\partial_{\nu}\alpha_a(x)G^{\mu\nu}
_{0a}(x)=\cr
&\pi_a^{\mu\nu}(x)\delta A_{a\nu}(x)-{i\over 2}(\delta \bar \psi
(x)\gamma^{\mu}
\psi (x)-\bar \psi (x)\gamma^{\mu}\delta \psi (x))\cr
&G^{\mu\nu}_{0a}(x)=-{1\over {{\sl g}^2}}F^{\mu\nu}_a(x)\cr
&G^{\mu}_{1a}(x)=-{1\over {{\sl g}^2}}c_{abc}F^{\mu\nu}_b(x)A_{c\nu}(x)-
i\bar \psi (x)\gamma^{\mu}T^a\psi (x)\cr}
\form
$$

\noindent so that

$$
\eqalign{
&G^{(\mu\nu )}_{0a}\equiv 0\cr
&\partial_{\nu}G^{\nu\mu}_{0a}\equiv -G^{\mu}_{1a}-{1\over {{\sl g}^2}}L^{\mu}
_a={1\over {{\sl g}^2}}(c_{abc}F_b^{\mu\nu}A_{c\nu}-L^{\mu}_a)+i\bar \psi
\gamma^{\mu}T^a\psi\cr
&\partial_{\mu}G^{\mu}_{1a}\equiv {1\over {{\sl g}^2}}c_{abc}A_{b\mu}L^{\mu}_c
+L_{\psi}T^a\psi +\bar \psi T^aL_{\bar \psi}{\buildrel \circ \over =}0\cr}
\form
$$

\noindent and the contracted Bianchi identities become

$$
{\hat D}^{(A)}_{\mu}L^{\mu}-{\sl g}^2{\hat T}^a(L_{\psi}T^a\psi +\bar \psi
T^aL_{\bar \psi})\equiv 0
\form
$$

Eqs.(2-13) are replaced by

$$
\eqalign{
\pi^o_a&=G^{(oo)}_{0a}\equiv 0\cr
0&\equiv \partial_o\pi^o_a\equiv \vec \partial \cdot \vec \pi_a-\partial_{\nu}
G^{\nu o}_{0a}\equiv \vec \partial \cdot \vec \pi_a+G^o_{1a}+{1\over {{\sl g}^2
}}L^o_a=\cr
&=\vec \partial \cdot \vec \pi_a+c_{abc}{\vec A}_b\cdot {\vec \pi}_c+i\bar \psi
\gamma^oT^a\psi +{1\over {{\sl g}^2}}L^o_a=-\Gamma^T_a+{1\over {{\sl g}^2}}
L^o_a{\buildrel \rm \circ \over =}-\Gamma^T_a\cr}
\form
$$

\noindent consistently with Eqs.(4-21) (see also Eqs.(2-28)).

Then as before we have

$$
V^{\mu}_a=-\partial_{\nu}G^{\nu\mu}_{0a}={1\over {{\sl g}^2}}\partial_{\nu}
F^{\nu\mu}_a=\partial_{\nu}U_a^{[\mu\nu ]}{\buildrel \circ \over =}-
{1\over {{\sl g}^2}}c_{abc}F_b^{\mu\nu}A_{c\nu}+i\bar \psi \gamma^{\mu}T^a
\psi
\form
$$

\noindent with $U_a^{[\mu\nu ]}=-F_a^{\mu\nu}/g^2$.

{}From $\partial_{\mu}G^{\mu}_{1a}{\buildrel \circ \over =}0$
we get the improper conserved charge

$$
\eqalign{
Q_a^T&=
\int d^3x G^o_{1a}=-{1\over {{\sl g}^2}}c_{abc}\int d^3x F^{ok}_bA_{ck}
+i\int d^3x \bar \psi \gamma^oT^a\psi {\buildrel \circ \over =}\cr
&{\buildrel \circ \over =}Q_a^{(V)T}=\int d^3x V^o_a=-
{1\over {{\sl g}^2}}\int d^3x \vec \partial \cdot {\vec E}_a=
-{1\over {{\sl g}^2}}\int d\vec \Sigma \cdot {\vec E}_a \cr}
\form
$$

In the Abelian electromagnetic case $\partial_{\mu}G^{\mu}_{1a}{\buildrel
\circ \over =}0$ is the conservation of the electric current and $eQ=e\int
d^3x \psi^{\dagger} \psi$. Let us remark that, in this pseudoclassical
background of QED, Q is an even quantity
bilinear in Grassmann variables: this shows the intrinsically quantum
character of the charge, like in the pseudoclassical description of
particles with either spin or inner degrees of freedom (charges) [44].
In the one-particle quantum Dirac equation, with the normalization
$\int d^3x{\hat \psi}^{\dagger}{\hat \psi}=1$ for the wave function
$\hat \psi$, one has eQ=e.
In the non-Abelian case one has
${\sl g}Q^T_a={\sl g}\int d^3x \psi^{\dagger}(i
T^a)\psi -{\sl g}{({\hat T}^a)}_{bc}\int d^3x {\tilde F}_b^{ok}{\tilde A}
_{ck}$; this again shows the quantum character of the charges: the
fermionic part is a bilinear in Grassmann variables, while the bosonic
part is a number, which, however, depends on the matrix elements of the
generators ${\hat T}^a$ in the adjoint representation (see also the approach
of Ref.[64]).

Let us remark that gauge fields were introduced in physics starting from
global symmetries associated with a Lie group G
of fermion fields (Eqs.(4-5) with U and $\alpha$
independent from x) and looking for a theory with the global symmetry
extended to a local one to realize the idea that the charges can be defined
independently in each space-time point (gauge principle). This led to the
introduction of the gauge fields via the covariant derivative (the minimal
coupling) and to the identification of the gauge fields with connections
on a principal G-bundle so to introduce a concept of parallel transport to
compare charges in one point with charges in another one. Thus the fermion
part $Q^F_a$ of $Q_a$ must be well defined under global (first kind) gauge
transformations, since this is the starting point of the construction; the
delicate point is a sound definition of $Q^{({\cal A})}_a=Q_a-Q^F_a$. Instead
in the mathematical description with fiber bundles, gauge fields as connections
on a principal G-bundle are the starting building block of the theory and
matter fields are a subsequent ingredient of it connected with suitable
associated bundles.

\vfill\eject

\bigskip\noindent
{\bf{5. The Abelian Case}}
\newcount \nfor

\def \form {\global \advance \nfor by 1 \eqno (5.\the\nfor)}
\bigskip

Let us review the theory of electromagnetism, whose description with
a trivial principal G-bundle involves the structure group $G=U(1)$. The gauge
transformations associated with Gauss' law
have the form $U(x)=e^{i\alpha (x)}$, with $\alpha (\vec x,x^o)
\quad {\longrightarrow}_{r\rightarrow \infty} \quad const.$
for improper ones and $\alpha (\vec x,x^o)\quad
{\longrightarrow}_{r\rightarrow \infty} \quad { {a}\over
{r^{3+\epsilon}} }+O(r^{-4})$ for proper ones.

The gauge potentials ${\tilde A}_{\mu}(x)=A_{\mu}(x)/e$
associated with a connection $\omega^{\cal A}$ on the trivial
principal bundle $P(M^4,G)$ satisfy the boundary conditions
${\tilde A}_{\mu}(\vec x,x^o)\quad {\longrightarrow}
_{r\rightarrow \infty}  \quad { {a_{\mu}}\over
{r^{1+\epsilon}} }+O(r^{-2})$, while for the associated field strengths
${\tilde F}_{\mu\nu}(x)={\partial}_{\mu}{\tilde A}_{\nu}(x)-{\partial}_{\nu}
{\tilde A}_{\mu}(x)$ we have ${\tilde F}_{\mu\nu}(\vec x,x^o)\quad
{\longrightarrow}_{r\rightarrow
\infty} \quad { {f_{\mu\nu}}\over {r^{2+\epsilon}} }+O(r^{-3})$. The Bianchi
identities are ${\partial}_{\mu}{*}{\tilde F}^{\mu\nu}(x)\equiv 0$.

The electric and magnetic fields are ${\tilde E}^i(x)={\tilde F}^{io}(x)$ and
${\tilde B}^i(x)=-{1\over 2}\epsilon^{ijk}{\tilde F}^{jk}(x)=-\epsilon^{ijk}
{\partial}^jA^k(x)$. The Lagrangian density is ${\cal L}(x)=-{1\over 4}
{\tilde F}_{\mu\nu}(x){\tilde F}^{\mu\nu}(x)={1\over 2}({\vec {\tilde E}}^2(x)-
{\vec {\tilde B}}^2(x))={1\over 2}\lbrace {\lbrack {\dot {\vec {\tilde A}}}(x)-
{\vec \partial}{\tilde A}_o(x)\rbrack }{}^2-{\lbrack {\vec \partial}\times
{\vec {\tilde A}}(x)\rbrack}^2\rbrace$ and the canonical momenta are
$\pi^o(x)=e\pi^o(x)=0$,
${\tilde \pi}^i(x)=e\pi^i(x)={\tilde E}^i(x)$; the Poisson brackets are
\break $\lbrace A_{\mu}(\vec x,x^o),
\pi^{\nu}(\vec y,x^o)\rbrace =\lbrace {\tilde A}_{\mu}(\vec
x,x^o),{\tilde \pi}^{\nu}(\vec y,x^o)\rbrace =\eta^{\nu}_{\mu}\delta^3(\vec x-
\vec y)$ so that $\lbrace {\tilde A}^i(\vec x,x^o),{\tilde \pi}^j(\vec y,x^o)
\rbrace \break
=-\delta^{ij}\delta^3(\vec x-\vec y)$).  The Euler-Lagrange equations
and the contracted Bianchi identities are ($\triangle =-{\vec \partial}^2 $):

$$
\eqalign{
{\tilde L}^{\mu}(x)&=e^{-1}L^{\mu}(x)={\partial}_{\nu}{\tilde F}^{\nu\mu}(x)=
{\buildrel {-}\over \sqcup}{\tilde A}^{\mu}(x)
-{\partial}^{\mu}{\partial}_{\nu}{\tilde A}^{\nu}(x){\buildrel \rm \circ \over
=}0\cr
&{\tilde L}^o(x)=e^{-1}L^o(x)=-{\vec \partial}\cdot {\vec {\tilde E}}(x)=
{\vec \partial}\cdot {\dot {\vec {\tilde A}}}(x)+\triangle A^o(x){\buildrel
\rm \circ \over =}0\cr
&{\vec {\tilde L}}(x)=e^{-1}\vec L(x)=-{\dot {\vec {\tilde E}}}(x)+{\vec
\partial}\times {\vec {\tilde B}}(x)={\partial}_o(-{\vec \partial}{\tilde A}^o
(x)+{\dot {\vec {\tilde A}}}(x))-\cr
&-{\vec \partial}\times
\lbrack {\vec \partial}\times {\vec {\tilde A}}(x)\rbrack
{\buildrel \circ \over =}0\cr
{\partial}_{\mu}{\tilde L}^{\mu}&(x)\equiv 0\cr}
\form
$$

The $1^{st}$-class constraints are

$$
\eqalign{
{\tilde \pi}^o(x)&\approx 0\cr
\tilde \Gamma (x)&=e\Gamma (x)=-\vec \partial \cdot {\vec {\tilde \pi}}(x)=-
\vec \partial \cdot {\vec {\tilde E}}(x)={\tilde L}^o(x)\approx 0\cr}
\form
$$

\noindent and the Dirac Hamiltonian is

$$
H_D=\int d^3x\lbrace {1\over 2}\lbrack {\vec {\tilde \pi}}^2(x)+{\vec {\tilde
B}}^2(x)\rbrack -{\tilde A}_o(x)\tilde \Gamma (x)+{\tilde \lambda}_o(x)
{\tilde \pi}^o(x)\rbrace
\form
$$

We see immediately that ${\tilde A}_o(x)$, ${\tilde \pi}^o(x)$ are a pair of
conjugated gauge canonical variables. A second such pair is formed by
${\tilde \eta} (x)$, $\tilde \Gamma (x)$,where $\tilde \eta (x)$ describes the
longitudinal degrees of freedom of the gauge potential. To find $\tilde \eta
(x)$, let us first find the solutions of Gauss'
law. With our boundary conditions, the Beltrami-Laplace operator $\triangle =
-{\vec \partial}^2$ has no zero modes and there are no harmonic forms in the
Hodge decomposition of 1-forms  [24l] on $R^3$, so that its inverse
${\triangle}
^{-1}$ is well defined; one has

$$
{1\over \triangle }\delta^3(\vec x-\vec y)=c(\vec x-\vec y)={ {-1}\over
{4\pi |\vec x-\vec y|} },\quad \quad \triangle c(\vec x-\vec y)=\delta^3(\vec
x-\vec y)
\form
$$

\noindent This implies that the following distribution [17] is well defined

$$
\eqalign{
&c^i(\vec x-\vec y)={\partial}_x^ic(\vec x-\vec y)={ {{\partial}_x^i}\over
\triangle }\delta^3(\vec x-\vec y)= { {{(x-y)}^i}\over {4\pi {|\, \vec x-
\vec y\, |}^3} }\cr
&{\vec \partial}_x\cdot {\vec c}(\vec x-\vec y)=-{\vec \partial}_y\cdot
{\vec c}(\vec x-\vec y)=-\delta^3(\vec x-\vec y)\cr}
\form
$$

\noindent and it serves to solve the equation ${\vec \partial}\cdot \vec f
(\vec x)=-\rho (\vec x)$:

$$
f^i(\vec x)=\int d^3yc^i(\vec x-\vec y)\rho (\vec y)+(\delta^{ij}+{ {{\partial}
^i{\partial}^j}\over \triangle })g^j(\vec x)
\form
$$

\noindent with $\vec g(\vec x)$ an arbitrary function; one has $(\delta^{ij}+
\partial^i\partial^j/\triangle )f^j(\vec x)=(\delta^{ij}+\partial^i\partial^j
/\triangle )g^j(\vec x)$.

Then the solution of Gauss' law is

$$
\eqalign{
{\tilde \pi}^i_{\perp}(\vec x,x^o)&={\tilde E}^i_{\perp}(\vec x,x^o)=
(\delta^{ij}+{ {{\partial}^i
{\partial}^j}\over {\triangle} }){\tilde \pi}^j(\vec x,x^o)=\cr
&=\int d^3y\lbrack \delta^{ij}\delta^3(\vec x-\vec y)+c^i(\vec x-\vec y)
{\partial}^j_y\rbrack {\tilde \pi}^j(\vec y,x^o)\cr}
\form
$$

\noindent Here ${\vec {\tilde E}}_{\perp}(\vec x,x^o)$ is the transverse
electric field and we have

$$
{\vec {\tilde \pi}}(x)={\vec {\tilde E}}(x)={\vec {\tilde \pi}}_{\perp}(x)+
{ {\vec \partial}\over {\triangle}}\tilde \Gamma (x)
\form
$$

Let us recall (see for instance [24l,o]) the Hodge decomposition of forms
on a differentiable manifold. Let (M,g) be an oriented, pseudo-Riemannian
manifold with non-degenerate (0,2) tensor field g and with a
metric volume form $\mu$; if $p\in M$, g(p) defines an inner product on
$T_pM$ (the tangent space at p) of signature (r,s), r+s=m=dimM, and index
$i_g=s$: for s=0 one has a Riemannian manifold and for s=1 a Lorentz one;
g induces an inner product on all tensor spaces and therefore also on $\Lambda
^k(M)$ (k=0,1,..,m), the space of all differential k-forms on M: if $\alpha ,
\beta \in \Lambda^k(M)$, then $(\alpha ,\beta )\mapsto g(\alpha ,\beta )\in R$.
The "Hodge star operator" on (M,g), ${*} : \Lambda^k(M)\rightarrow
\Lambda^{m-k}
(M)$, sends $\beta \in \Lambda^k(M)$ ($0\leq k\leq m$) into ${*}\beta \in
\Lambda^{m-k}(M)$ such that $\alpha \wedge {*}\beta=g(\alpha ,\beta )\mu$ for
every $\alpha \in \Lambda^k(M)$; one has ${**}\beta ={(-)}^{i_g+k(m-k)}\beta$,
${*}\mu={(-)}^{i_g}$, ${*}1=\mu$. For 2-forms on Minkowski space-time $M^4$ one
has
${**}=-{(-)}^{k(4-k)}$; on $R^3$ and $S^3$ one has ${**}={(-)}^{k(3-k)}$.
The "codifferential" $\delta$ on (M,g) is a linear map $\delta :\Lambda^k(M)
\rightarrow \Lambda^{k-1}(M)$ defined as

$\delta ={(-)}^{i_g+mk+m+1}{*}d{*}$,

\noindent where d is the exterior differential on M; one has $\delta f=0$ on
functions $f\in \Lambda^o(M)$. Like the property $d^2=0$ allows to define
closed ($d\alpha =0$) and exact ($\alpha =d\beta$) forms, the resulting
property $\delta^2=0$ of the codifferential leads to "coclosed" ($\delta
\alpha =0$) and "coexact" ($\alpha =\delta \beta$) forms. Given a vector field
$X^i\partial_i$ on M, one defines a 1-form $X_idx^i$ by means of $X_i=g_{ij}
X^j$ ($g_{ij}=g(\partial_i,\partial_j)$): then $\delta (X_idx^i)=0$ implies
$div(X^i\partial_i)=0$, i.e. $\partial_iX^i(x)=0$. When (M,g) is an oriented,
closed (i.e. compact and without boundary), Riemannian manifold ($i_g=s=0$),
the "Laplace-Hodge-deRahm operator" $\triangle :\Lambda^k(M)\rightarrow
\Lambda^k(M)$ is defined by $\triangle = d\, \delta +\delta \, d$ (on functions
one has $-\triangle ={\nabla}^2=div\circ grad$, the classical Laplace-Beltrami
operator). An $L^2$-inner product on $\Lambda^k(M)$ is defined by the map
$< \, ,\, > =\Lambda^k(M)\times \Lambda^k(M)\rightarrow R$, $< \, \alpha ,
\beta >=\int_M \alpha \wedge {*}\beta =\int_M g(\alpha ,\beta )\mu$,
$\alpha ,\beta \in \Lambda^k(M)$. For $\alpha ,\beta \in \Lambda^k(M)$,
$\gamma \in \Lambda^{k+1}(M)$, one has : i) $< d\alpha ,\gamma > = < \alpha ,
\delta \beta >$ ($\delta$ is the adjoint of d); ii) $< \triangle \alpha ,
\beta > = < \alpha ,\triangle \beta >$ ($\triangle$ is self-adjoint); iii)
$\triangle \alpha =0$ if and only if $d\alpha =\delta \alpha =0$: $\alpha$
is said to be a "harmonic k-form" (it is both closed and coclosed). The
set ${\cal H}^k=\lbrace \alpha \in \Lambda^k(M) |\triangle \alpha =0\rbrace$
is a subspace of $\Lambda^k(M)$. The "Hodge decomposition theorem" asserts
that ${\cal H}^k$ is finite-dimensional (it is isomorphic to the k-th
de Rahm cohomology space $H^k(M,R)$) and that there is the following direct
sum decomposition into orthogonal (with respect to $< \, ,\, >$) subspaces

$\Lambda^k(M)=d\Lambda^{k-1}(M)\oplus \delta \Lambda^{k+1}(M)\oplus {\cal
H}^k$

\noindent so that for any k-form $\alpha$ one has $\alpha =d\beta +\delta
\gamma +\theta$. The proof uses the theory of elliptic operators and this
requires the compactness of M. In our case, $R^3$, the chosen boundary
conditions allow to avoid the compactification to $S^3$ and
to apply the theorem: one has $\Lambda^1(R^3)=d\Lambda^o
(R^3)\oplus \delta \Lambda^2(R^3)$, since ${\cal H}^1$ is void.
When $\pi_1(M)\not= 0$, like in the ideal Bohm-Aharonov setting, ${\cal H}^1$
contains extra physical degrees of freedom of the gauge potential (whose
conjugated variables are the harmonic forms in the Hodge decomposition of
${\vec {\tilde \pi}}(x)\cdot d\vec x$); this seems to imply a classical basis
for the description of the observable quantum Bohm-Aharonov phase, which will
be explored elsewhere.
Therefore the Hodge decomposition of the 1-form
$\vec A(\vec x,x^o)\cdot d\vec x$ on
$R^3$ with our boundary conditions is given by

$$
{\vec {\tilde A}}(x)\cdot d\vec x=d\tilde \eta (x)+{\vec {\tilde D}}(x)\cdot
d\vec x
\form
$$

\noindent where ${\vec {\tilde D}}(x)\cdot d\vec x$ is a coexact 1-form:
$\delta ({\vec {\tilde D}}(x)\cdot d\vec x)=0$; this implies ${\vec \partial}
\cdot {\vec {\tilde D}}(x)=0$, i.e. that ${\vec {\tilde D}}(\vec x,x^o)$ is
divergence free. Therefore ${\tilde D}^i(x)={\tilde A}^i_{\perp}
(x)=(\delta^{ij}+{ {{\partial}^i{\partial}^j}\over {\triangle} })A^j(x)$
is the transverse part of the gauge potential. From the Hodge decomposition
we get

$$
\eqalign{
{\vec {\tilde A}}(\vec x,x^o)&={\vec \partial}\tilde \eta (\vec x,x^o)+
{\vec {\tilde A}}_{\perp}(\vec x,x^o)\cr
\tilde \eta (\vec x,x^o)&=-{1\over {\triangle} }{\vec \partial}\cdot
{\vec {\tilde A}}(\vec x,x^o)=
-\int d^3y\vec c(\vec x-\vec y)\cdot {\vec {\tilde A}}(\vec y,x^o)=\cr
&=\int d^3y{\vec \partial}_yc(\vec x-\vec y)\cdot {\vec {\tilde A}}(\vec
y,x^o)=-\int d^3yc(\vec x-\vec y){\vec \partial}_y\cdot {\vec {\tilde A}}
(\vec y,x^o)\cr}
\form
$$

\noindent where an integration by parts (allowed by our boundary conditions)
has
been done to get the last line.

Since we have

$$
\lbrace \tilde \eta (\vec x,x^o),\tilde \Gamma (\vec y,x^o)\rbrace =
-\delta^3(\vec x-\vec y),
\form
$$

\noindent
we see that $\tilde \eta (x)$ is the looked for longitudinal gauge variable.
We also have

$$
\lbrace .,\tilde \Gamma (\vec y,x^o)\rbrace =-{ {\delta}\over {\delta \tilde
\eta (\vec y,x^o)} }
\form
$$

\noindent and

$$
{ {\delta \tilde \eta (\vec x,x^o)}\over {\delta {\tilde A}^i(\vec y,x^o)} }=
-c^i(\vec x-\vec y)
\form
$$

Therefore the two canonically conjugated pairs of physical variables, spanning
the reduced phase space, are contained in Dirac's observables ${\vec
{\tilde A}}_{\perp}(x)$, ${\vec {\tilde \pi}}_{\perp}(x)={\vec {\tilde E}}
_{\perp}(x)$, which satisfy

$$
\eqalign{
&\lbrace {\vec {\tilde A}}_{\perp}(\vec x,x^o),\tilde K(\vec y,x^o)\rbrace =0
\quad \quad \tilde K(x)={\tilde A}_o(x),{\tilde \pi}^o(x),\tilde \eta (x),
\tilde \Gamma (x)\cr
&\lbrace {\vec {\tilde \pi}}_{\perp}(\vec x,x^o),\tilde K(\vec y,x^o)\rbrace
=0\cr}
\form
$$

$$
\lbrace {\tilde A}^i_{\perp}(\vec x,x^o),{\tilde \pi}^j_{\perp}(\vec y,x^o)
\rbrace=-(\delta^{ij}+{ {{\partial}^i{\partial}^j}\over {\triangle} })
\delta^3(\vec x-\vec y)
\form
$$

Since from Eqs.(5-8) and from ${\tilde B}^i(x)=-\epsilon^{ijk}{\partial}^j
{\tilde A}^k(x)=-\epsilon^{ijk}{\partial}^j{\tilde A}_{\perp}^{k}(x)$ one has

$$
{\vec {\tilde \pi}}^2(x)={\vec {\tilde \pi}}^2_{\perp}(x)+2{\vec {\tilde \pi}}
_{\perp}(x)\cdot { {\vec \partial} \over {\triangle} }\tilde \Gamma (x)+
{\lbrack { {\vec \partial}\over {\triangle} }\tilde \Gamma (x)\rbrack }^2
\form
$$

\noindent the Dirac Hamiltonian (5-3) becomes

$$
\eqalign{
H_D&=H_{\perp}+\int d^3x\lbrace {1\over 2}{\lbrack { {\vec \partial}\over
{\triangle} }\tilde \Gamma (x)\rbrack}^2+{\vec {\tilde \pi}}_{\perp}(x)\cdot
{ {\vec \partial}\over {\triangle} }\tilde \Gamma (x)-{\tilde A}_o(x)\tilde
\Gamma (x)+{\tilde \lambda}_o(x){\tilde \pi}^o(x)\rbrace =\cr
&\equiv H_{\perp}+\int d^3x\lbrack -{\tilde A}_o(x)\tilde \Gamma (x)+{\tilde
\lambda}_o(x){\tilde \pi}^o(x)\rbrack \approx H_{\perp}\cr}
\form
$$

$$
H_{\perp}=\int d^3x{1\over 2}\lbrack {\vec {\tilde \pi}}^2_{\perp}(x)+
{\vec {\tilde B}}^2(x)\rbrack
\form
$$

To get the second line of Eqs.(5-17) we have done an allowed integration by
parts and we have discarded a term quadratic in the constraints: $"\equiv "$
means here "strong equality" in the sense of Dirac.

We see that $H_{\perp}$ is the physical (i.e. Dirac's observable) Hamiltonian
on the reduced phase space. There the Hamilton equations are

$$
\eqalign{
&{\dot {\tilde A}}^i_{\perp}(\vec x,x^o){\buildrel \rm \circ \over =}\lbrace
{\tilde A}^i_{\perp}(\vec x,x^o),H_{\perp}\rbrace =-{\tilde \pi}^i_{\perp}
(\vec x,x^o)=-{\tilde E}^i_{\perp}(\vec x,x^o)\cr
&{\dot {\tilde \pi}}^i_{\perp}(\vec x,x^o)={\dot {\tilde E}}^i_{\perp}(\vec x,
x^o){\buildrel \rm  \circ \over =}\lbrace {\tilde \pi}^i_{\perp}(\vec x,x^o),
H_{\perp}\rbrace =\triangle {\tilde A}^i_{\perp}(\vec x,x^o)\cr}
\form
$$

Therefore the effective equations of motion are

$$
\bar
\sqcup {\tilde A}^i_{\perp}(x)=({\partial}_o^2+\triangle ){\tilde A}^i_{\perp}
(x){\buildrel \rm \circ \over =}0
\form
$$

\noindent which are the Euler-Lagrange equations deriving from the non Lorentz
covariant Lagrangian density (see Section 10 for the problem of manifest
Lorentz covariance replaced by manifest Wigner covariance):

$$
{\cal L}_{\perp}(x)={1\over 2}({\dot {\vec {\tilde A}}}^2_{\perp}(x)-{\vec
{\tilde B}}^2[{\vec {\tilde A}}_{\perp}(x)])=
{1\over 2}\lbrack {\dot {\vec {\tilde A}}}^2_{\perp}(x)-{({\vec \partial}
\times {\vec {\tilde A}}_{\perp}(x))}^2\rbrack
\form
$$

Eqs.(5-21) can be obtained from ${\cal L}(x)={1\over 2}\lbrace {\lbrack
{\dot {\vec {\tilde A}}}(x)-\vec \partial {\tilde A}^o(x)\rbrack }^2-
{\lbrack \vec \partial \times {\vec {\tilde A}}_{\perp}(x)\rbrack }^2\rbrace =
{1\over 2}\lbrace {\lbrack \vec \partial (\dot \eta (x)-{\tilde A}^o(x))+
{\dot {\vec {\tilde A}}}^2_{\perp}(x)\rbrack }^2-{\lbrack \vec \partial \times
{\vec {\tilde A}}_{\perp}(x)\rbrack }^2\rbrace {\buildrel \circ \over =}
{\cal L}_{\perp}(x)$ by using the first of Eqs.(5-22) below.

That ${\tilde A}_o(x)$ and $\tilde \eta (x)$ are the natural gauge variables
can also be seen by recovering Eqs.(5-20) from Eqs.(5-1) without introducing
gauge-fixings. Actually by using Eqs.(5-10) we can rewrite ${\tilde L}^o(x)
{\buildrel \rm \circ \over =}0$ and ${\vec {\tilde L}}(x){\buildrel \rm \circ
\over =}0$ as

$$
\eqalign{
&{\tilde L}^o(x)=\tilde \Gamma (x)=
\triangle \lbrack {\tilde A}_o(x)-{\dot {\tilde \eta}} (x)
\rbrack {\buildrel \rm \circ \over =}0\cr
&{\vec {\tilde L}}(x)=
\bar \sqcup {\vec {\tilde A}}_{\perp}(x)-{\partial}^o{\vec
\partial}\lbrack {\tilde A}_o(x)-{\dot {\tilde \eta}}(x)\rbrack
{\buildrel \rm \circ \over =}0\cr}
\form
$$

\noindent which imply

$$
\bar \sqcup
{\vec {\tilde A}}_{\perp}(x)={\vec {\tilde L}}(x)+{\partial}^o{ {\vec
\partial}\over {\triangle} }{\tilde L}^o(x){\buildrel \rm \circ \over =}0
\form
$$

$$
{\tilde A}_o(x){\buildrel \rm \circ \over =}{\dot {\tilde \eta}}(x)=-
{1\over {\triangle} }{\vec \partial}\cdot {\dot {\vec {\tilde A}}}(x).
\form
$$

In absence of charged fermions, the natural gauge-fixing to Gauss' law,
following the scheme of Ref.[34], is $\tilde \eta (x)\approx 0$, i.e. the
Coulomb gauge. Then its time constancy

$$
{\dot {\tilde \eta}}(\vec x,x^o){\buildrel \rm \circ \over =}\lbrace \tilde
\eta (\vec x,x^o),H_D\rbrace ={\tilde A}_o(\vec x,x^o)-{1\over {\triangle} }
\tilde \Gamma (\vec x,x^o)\approx {\tilde A}_o(\vec x,x^o)\approx 0
\form
$$

\noindent implies the temporal or Weyl gauge ${\tilde A}_o(x)\approx 0$, and
its time constancy

$$
{\dot {\tilde A}}_o(\vec x,x^o){\buildrel \rm \circ \over =}\lbrace {\tilde A}
_o(\vec x,x^o),H_D\rbrace ={\tilde \lambda}_{o}(\vec x,x^o)\approx 0
\form
$$

\noindent determines the Dirac multiplier. Therefore in absence of charged
fermions the Coulomb and the temporal gauges are compatible and we get a
subcase of the Lorentz gauge ${\partial}^{\mu}{\tilde A}_{\mu}(x)\approx 0$.

{}From Eqs.(5-17) we see that the Hamiltonian for the decoupled gauge variables
${\tilde A}_o(x)$, ${\tilde \pi}^o(x)$, $\tilde \eta (x)$, $\tilde \Gamma (x)$
is $H_G=\int d^3x\lbrack -{\tilde A}_o(x)\tilde \Gamma (x)+{\tilde \lambda}_{o}
(x){\tilde \pi}^o(x)\rbrack$, whose Hamilton equations are
${\dot {\tilde A}}_o(x){\buildrel \rm \circ \over =}{\tilde \lambda}_o(x)$,
${\dot {\tilde \pi}}^o(x){\buildrel \rm \circ \over =}\tilde \Gamma (x)\approx
0$, ${\dot {\tilde \eta}}(x){\buildrel \rm \circ \over =}{\tilde A}_o(x)$,
${\dot {\tilde \Gamma}}(x){\buildrel \rm \circ \over =}0$,
consistently with Eqs.(5-25), (5-26).

To clarify the comparison with the Lagrangian approach using covariant gauges,
let us consider how  the Lorentz gauge $\partial^{\mu}A_{\mu}(x)=0$ can be
implemented at the Hamiltonian level. Since ${\dot {\tilde A}}_o(x){\buildrel
\circ \over =}{\tilde \lambda}_o(x)$, one sees that the Lorentz gauge-fixing
constraint is

$$
0\approx {\tilde \lambda}_o+\triangle \tilde \eta {\buildrel \circ \over =}
\partial^{\mu}{\tilde A}_{\mu}
$$

\noindent that is the gauge-fixing constraint depends on the Dirac multiplier
(see Ref.[5c] for this kind of constraints).
Its time constancy, i.e. the gauge-fixing constraint for ${\tilde \pi}^o(x)
\approx 0$ , is

$$
{d\over {dx^o}}({\tilde \lambda}_o+\triangle \tilde \eta ){\buildrel \circ
\over =}{ {\partial {\tilde \lambda}_o}\over {\partial x^o} }+\lbrace
\triangle \tilde \eta ,H_D\rbrace ={\dot {\tilde \lambda}}_o+\triangle
{\tilde A}_o{\buildrel \circ \over =}\bar
\sqcup {\tilde A}_o\approx 0
$$

Therefore ${\tilde A}_o$ is not fixed. Also ${\tilde \lambda}_o$ is not fixed,
because

$$
{d\over {dx^o}}({\dot {\tilde \lambda}}_o+\triangle {\tilde A}_o){\buildrel
\circ \over =}{\ddot {\tilde \lambda}}_o+\triangle {\tilde \lambda}_o=\bar
\sqcup {\tilde \lambda}_o\approx 0
$$

We find the residual gauge freedom of the Lorentz gauge $A_{\mu}\rightarrow
A_{\mu}+\partial_{\mu}\Lambda$ with $\bar \sqcup \Lambda =0$.

In Ref.[21i], following Refs.[21a-h], it is shown that the orbit space, i.e.
the reduced configuration space in which the reduced Lagrangian density (5-21)
is defined (an orbit is the set of gauge equivalent gauge potentials described
by ${\vec {\tilde A}}_{\perp}(x)$), is endowed with a natural Riemannian
metric, whose Levi-Civita connection and curvature tensor can be evaluated.
Actually in analogy to a non-relativistic particle of mass one moving in a
curved space  described by the Lagrangian $L={1\over 2}g_{ij}(\vec x){\dot x}^i
{\dot x}^j-V(\vec x)$, where the infinitesimal squared distance in that space
is given by $ds^2=g_{ij}(\vec x)dx^idx^j$, with an allowed integration by parts
we can rewrite Eqs.(5-21) as

$$
{\cal L}_{\perp}(x)={1\over 2}\lbrack {\dot {\tilde A}}^i_{\perp}(x)
{\dot {\tilde A}}^i_{\perp}(x)-{(\vec \partial \times {\vec {\tilde A}}_{\perp}
(x))}^2\rbrack
={1\over 2}\lbrack {\dot {\tilde A}}^i(x){\hat P}^{ij}_{\perp}(\vec x)
{\dot {\tilde A}}^j(x)
-{({\vec \partial}\times {\vec {\tilde A}}_{\perp}(x))}^2\rbrack
\form
$$

\noindent where ${\hat P}^{ij}_{\perp}(\vec x)$ is the projection operator

$$
{\hat P}^{ij}_{\perp}(\vec x)=
\delta^{ij}+{ {{\partial}^i{\partial}^j}\over {\triangle} },
\quad {\hat P}^{ij}_{\perp}{\hat P}^{jk}_{\perp}={\hat P}^{ik}_{\perp}
\form
$$

{}From Eqs.(5-10), we see that the longitudinal modes ${\vec \partial}\tilde
\eta (x)$ contained in ${\vec {\tilde A}}(x)$ are eigenvectors of
${\hat P}^{ij}_{\perp}(\vec x)$ with eigenvalue
zero. This suggests a natural way to define an infinitesimal squared distance
in the functional space $\lbrace {\vec {\tilde A}}(x)\rbrace$ of the vector
gauge potentials

$$
\delta s^2 =\int d^3x \delta A^i(\vec x){\hat P}^{ij}_{\perp}(\vec x)
\delta A^j(\vec x)=
\int d^3x \delta {\tilde A}^i_{\perp}(\vec x){\hat P}^{ij}_{\perp}(\vec x)
\delta {\tilde A}^j_{\perp}(\vec x)
\form
$$

\noindent where $\delta A(\vec x)=A_1(\vec x)-A_2(\vec x)$.
This metric depends on the orbit ${\vec {\tilde A}}_{\perp}(\vec x)$, because
it does not depend on the longitudinal part ${\vec \partial}\tilde \eta (\vec
x)$, i.e. it is gauge invariant: the distance between two gauge equivalent
gauge potentials, i.e. having the same transverse part, is zero. Therefore we
have:
i) in the functional space $\lbrace {\vec {\tilde A}}(x)=\vec \partial \tilde
\eta (x)+{\vec {\tilde A}}_{\perp}(x)\rbrace$ it is possible to define slices
$\lbrace {\vec {\tilde A}}_{\perp}(x)\rbrace$ (isomorphic to the orbit space)
orthogonal to the gauge orbit $\lbrace \vec \partial \tilde \eta (x)\rbrace$;
ii) Eq.(5-29) defines a metric on the orbit space.
Since the metric (5-29) does not depend on the vector gauge potential, it is
constant, so that the functional space of gauge potentials in the Abelian
theory is flat.

Let us now consider the coupling of the electromagnetic potential to Dirac
fermions, assumed to behave as $\psi (\vec x,x^o)\rightarrow \chi r^{-3/2+
\epsilon}+O(r^{-2})$ for $r\rightarrow \infty$ like in Eqs.(4-15).
{}From Section 4, Eqs.(4-8), we have the Lagrangian density

$$
\eqalign{
{\cal L}(x)&=\bar \psi (x)\gamma^{\mu}(i{\partial}_{\mu}+e{\tilde A}_{\mu}(x))
\psi (x)-m\bar \psi (x)\psi (x)-{1\over 4}{\tilde F}_{\mu\nu}(x){\tilde F}
^{\mu\nu}(x)-\cr
&-\partial_{\mu}[{i\over 2}\bar \psi (x)\gamma^{\mu}\psi (x)]\cr}
\form
$$

\noindent implying the Euler-Lagrange equations

$$
\eqalign{
&{\partial}_{\mu}{\tilde F}^{\mu\nu}(x)=
\bar \sqcup {\tilde A}^{\nu}(x)-{\partial}
^{\nu}({\partial}_{\mu}{\tilde A}^{\mu}(x))=-e\bar \psi (x)\gamma^{\nu}\psi
(x)+
{\tilde L}^{T\nu}{\buildrel \rm \circ \over =}
-e\bar \psi (x)\gamma^{\nu}\psi (x)\cr
&L_{\psi}(x)=\bar \psi (x)[\overleftarrow {(i\partial_{\mu}-e{\tilde A}_{\mu}
(x))}\gamma^{\mu}+m]{\buildrel \circ \over =}0\cr
&L_{\bar \psi}(x)=[\gamma^{\mu}(i\partial_{\mu}+e{\tilde A}_{\mu}(x))-m]
\psi (x){\buildrel \circ \over =}0\cr}
\form
$$

The canonical momenta of the various fields and the primary constraints are
unchanged (see Section 4), and the secondary constraint given by Gauss'
law becomes

$$
{\tilde \Gamma}^T(x)={\tilde L}^{To}(x)=
-{\vec \partial}{\vec {\tilde \pi}}(x)+e\psi^{\dagger}(x)\psi (x)
\approx 0
\form
$$

\noindent with ${\tilde \Gamma}^T(\vec x,x^o){\rightarrow}_{r\rightarrow
\infty}\, const. \, r^{-(3+\epsilon )}$ like in Eqs.(2-40) due to Eq.(4-15).

By eliminating the fermionic momenta and by using the Dirac brackets (4-29)
(or $\lbrace \psi_{\alpha}(\vec x,x^o),\psi^{\dagger}_{\beta}(\vec y,x^o)
\rbrace =-i\delta_{\alpha\beta}\delta^3(\vec x-\vec y)$),
we are led to the following Dirac Hamiltonian in terms of $\beta =\gamma^o$,
$\vec \alpha =\gamma^o\vec \gamma$ (after an integration by parts; see Eqs.
(4-20))

$$
\eqalign{
H_D&=\int d^3x\lbrace \psi^{\dagger}(x)\vec \alpha \cdot (i{\vec \partial}
+e{\vec {\tilde A}}(x))\psi (x)+m\psi^{\dagger}(x)\beta \psi (x)+{1\over 2}
\lbrack {\vec {\tilde \pi}}^2(x)+{\vec {\tilde B}}^2(x)\rbrack -\cr
&-{\tilde A}_o(x){\tilde \Gamma}^T(x)+{\tilde \lambda}_o(x){\tilde \pi}^o(x)
\rbrace{|}_{x=(\vec x,x^o)}\cr}
\form
$$

Since $\lbrace \psi_{\alpha}(\vec x,x^o),{\tilde \Gamma}^T (\vec y,x^o)\rbrace
=-ie\psi_{\alpha}(\vec x,x^o)\delta^3(\vec x-\vec y)$, by using Eqs.(5-11), one
obtains [17] the following Dirac's observables for the fermion

$$
\eqalign{
{\tilde \psi}_{\alpha}(x)&=e^{-ie\tilde \eta (x)}\psi_{\alpha}(x)\cr
{\tilde \psi}^{\dagger}_{\alpha}(x)&=\psi^{\dagger}_{\alpha}(x)e^{ie\tilde
\eta (x)}\cr}
\form
$$

$$
\lbrace {\tilde \psi}_{\alpha}(\vec x,x^o),{\tilde \Gamma}^T(\vec y,x^o)\rbrace
=\lbrace {\tilde \psi}^{\dagger}_{\alpha}(\vec x,x^o),{\tilde \Gamma}^T(\vec y,
x^o)\rbrace =0
\form
$$

\noindent Their only non vanishing Dirac brackets are

$$
\lbrace {\tilde \psi}_{\alpha}(\vec x,x^o),{\tilde \psi}^{\dagger}_{\beta}
(\vec y,x^o)\rbrace =-i\delta_{\alpha\beta}\delta^3(\vec x-\vec y)
\form
$$

These Dirac's observables are not measurable quantities, because they are
Grassmann odd: only Grassmann even combinations of them, like the bilinears,
are measurable in the sense of Ref.[44].

Following the same procedure used to get Eq.(5-18), the final physical
Hamiltonian depending on Dirac's observables ${\vec {\tilde A}}_{\perp}(x)$,
${\vec {\tilde \pi}}_{\perp}(x)={\vec {\tilde E}}_{\perp}(x)$, $\tilde \psi (x)
$, ${\tilde \psi}
^{\dagger}(x)$, whose brackets are given by Eqs.(5-15) and (5-36), is

$$
\eqalign{
H_{\perp}&=\int d^3x\lbrace i{\tilde \psi}^{\dagger}(\vec x,x^o)\vec
\alpha \cdot {\vec \partial}\tilde \psi (\vec x,x^o)+m{\tilde \psi}^{\dagger}
(\vec x,x^o)\beta \tilde \psi (\vec x,x^o)+\cr
&+e{\tilde \psi}^{\dagger}(\vec x,x^o)\vec \alpha \cdot {\vec {\tilde A}}
_{\perp}(\vec x,x^o)\tilde \psi (\vec x,x^o)+{1\over 2}\lbrack {\vec {\tilde
\pi}}^2_{\perp}(\vec x,x^o)+{\vec {\tilde B}}^2[{\vec {\tilde A}}_{\perp}
(\vec x,x^o)]\rbrack +\cr
&+{{e^2}\over 2}
{\lbrack { {\vec \partial}\over {\triangle} }({\tilde \psi}^{\dagger}(\vec
x,x^o)\tilde \psi (\vec x,x^o))\rbrack}^2 \rbrace \cr}
\form
$$

\noindent where $c(\vec x-\vec y)$ is given by Eq.(5-4).
${\vec {\tilde A}}_{\perp}(x)$, ${\vec {\tilde E}}(x)$ describe the radiation
field and there is a minimal coupling of the Dirac's observable fermion field
to  ${\vec {\tilde A}}_{\perp}(x)$.
Instead the last term describes the non-local Coulomb self-interaction of
Dirac's observable fermion field.

In the derivation of Eq.(5-37) one has used Eq.(5-16) and

$$
\eqalign{
{\lbrack { {\vec \partial}\over {\triangle} }\tilde \Gamma (x)\rbrack}^2&=
{\lbrack { {\vec \partial}\over {\triangle} }({\tilde \Gamma}^T(x)-e\psi
^{\dagger}(x)\psi (x))\rbrack}^2={\lbrack { {\vec \partial}\over {\triangle}}
{\tilde \Gamma}^T(x)\rbrack}^2-\cr
&-2e\lbrack { {\vec \partial}\over {\triangle}}
{\tilde \Gamma}^T(x)\rbrack \cdot { {\vec
\partial}\over {\triangle} }({\tilde \psi}^{\dagger}(x)\tilde \psi (x))+
e^2{\lbrack { {\vec \partial}\over {\triangle} }({\tilde \psi}^{\dagger}(x)
\tilde \psi (x))\rbrack}^2\cr}
\form
$$

\noindent so that, after an allowed integration by part, the Dirac Hamiltonian
is

$$
\eqalign{
H_T&=\int d^3x\lbrace i{\tilde \psi}^{\dagger}(\vec x,x^o)\vec
\alpha \cdot {\vec \partial}\tilde \psi (\vec x,x^o)+m{\tilde \psi}^{\dagger}
(\vec x,x^o)\beta \tilde \psi (\vec x,x^o)+\cr
&+e{\tilde \psi}^{\dagger}(\vec x,x^o)\vec \alpha \cdot {\vec {\tilde A}}
_{\perp}(\vec x,x^o)\tilde \psi (\vec x,x^o)+{1\over 2}\lbrack {\vec {\tilde
\pi}}^2_{\perp}(\vec x,x^o)+{\vec {\tilde B}}^2[{\vec {\tilde A}}_{\perp}
(\vec x,x^o)]\rbrack \rbrace +\cr
&+\int d^3x\lbrace {1\over 2}{\lbrack { {\vec \partial}\over {\triangle}}
{\tilde \Gamma}^T(\vec x,x^o)\rbrack}^2-
e\lbrack { {\vec \partial}\over {\triangle}}
{\tilde \Gamma}^T(\vec x,x^o)\rbrack \cdot
{ {\vec \partial}\over {\triangle} }({\tilde \psi}^{\dagger}(\vec x,x^o)
\tilde \psi (\vec x,x^o))+\cr
&+{{e^2}\over 2}{\lbrack { {\vec \partial}\over {\triangle} }
({\tilde \psi}^{\dagger}(\vec x,x^o)\tilde \psi (\vec x,x^o))\rbrack}^2-
{\tilde A}_o(\vec x,x^o){\tilde \Gamma}^T(\vec x,x^o)+{\tilde \lambda}_o
(\vec x,x^o){\tilde \pi}^o(\vec x,x^o)\rbrace = \cr
&=H_{\perp}
+\int d^3x\lbrace -\lbrack {\tilde A}_o(\vec x,x^o)+e{1\over {\triangle}}
({\tilde \psi}^{\dagger}(\vec x,x^o)\tilde \psi (\vec x,x^o))\rbrack {\tilde
\Gamma}^T(\vec x,x^o)+\cr
&+{\tilde \lambda}_o(\vec x,x^o){\tilde \pi}^o(\vec x,x^o)+{1\over 2}
{\lbrack { {\vec \partial}\over {\triangle} }{\tilde \Gamma}^T(\vec x,x^o)
\rbrack}^2\rbrace \cr}
\form
$$

\noindent with $H_{\perp}$ of Eq.(5-37); in $H_{\perp}$ the self-energy term
can also be written as

$$
\eqalign{
&e^2\int d^3x {\lbrack { {\vec \partial}\over {\triangle} }({\tilde \psi}
^{\dagger}(\vec x,x^o)\tilde \psi (\vec x,x^o))\rbrack}^2=\cr
&=e^2\int d^3xd^3y
\lbrack {\tilde \psi}^{\dagger}(\vec x,x^o)\tilde \psi (\vec x,x^o)\rbrack
c(\vec x-\vec y)\lbrack {\tilde \psi}^{\dagger}(\vec y,x^o)
\tilde \psi (\vec y,x^o)\rbrack =\cr
&=-{{e^2}\over {4\pi}}\int d^3xd^3y{ {[{\tilde \psi}^{\dagger}(\vec
x,x^o)\tilde
\psi (\vec x,x^o)][{\tilde \psi}^{\dagger}(\vec y,x^o)\tilde \psi (\vec
y,x^o)]}
\over {| \vec x-\vec y |} }\cr}
\form
$$

Note that with the boundary conditions $f(\vec x,x^o)\rightarrow const. r^{-
(3+\epsilon )}$ ($f={\tilde \Gamma}_a^T, {\tilde \psi}^{\dagger}\tilde \psi$),
these functions are square integrable and therefore tempered distributions,
so that their Fourier transforms are well defined;
for two of such functions f,g the following integration by parts
is well defined
$\int d^3xf(\vec x,x^o){1\over {\triangle}}g(\vec x,x^o)=\int { {d^3p}\over
{(2\pi )^3} }{ {f(\vec p,x^o)g(\vec p,x^o)}\over {{\vec p}^2} }=
\int d^3xg(\vec x,x^o){1\over {\triangle}}f(\vec x,x^o)$.

Let us remark that now the Coulomb gauge $\tilde \eta (\vec x,x^o)\approx 0$ is
not compatible with the temporal gauge  ${\tilde A}_o(\vec x,x^o)\approx 0$,
because

$$
\eqalign{
&{\dot {\tilde \eta}}(\vec x,x^o){\buildrel \rm \circ \over =}\lbrace
\tilde \eta (\vec x,x^o),H_D\rbrace \approx {\tilde A}_o(\vec x,x^o) +
{ {e}\over {\triangle} }{\tilde \psi}^{\dagger}(\vec x,x^o)
\tilde \psi (\vec x,x^o)\approx 0\cr
&\partial^o\lbrack {\tilde A}_o(\vec x,x^o)+{e\over {\triangle}}({\tilde
\psi}^{\dagger}(\vec x,x^o)\tilde \psi (\vec x,x^o))\rbrack \approx \cr
&\approx \lambda_o
(\vec x,x^o)+{e\over {\triangle}}\lbrack {\tilde \psi}^{\dagger}(\vec x,x^o)
(i\overleftarrow  {\partial}+i\overrightarrow
{\partial})\cdot \vec \alpha \tilde \psi (\vec x,x^o)\rbrack \approx 0\cr}
\form
$$

To decouple completely the physical degrees of freedom from the gauge ones,
one has to do a final canonical transformation to a new set (what is lacking
is the extraction of a covariant canonical basis from ${\vec {\tilde
A}}_{\perp}
(x)$, ${\vec {\tilde \pi}}_{\perp}(x)$):

$$
\eqalign{
&{\tilde V}_o(x)=\triangle {\tilde A}_o(x)+e{\tilde \psi}
^{\dagger}(x)\tilde \psi (x)\cr
&{\tilde \Pi}^o(x)={1\over {\triangle}}{\tilde \pi}^o(x)\approx 0\cr
&{}\cr
&\tilde \eta (x)\cr
&{\tilde \Gamma}^T(x)\approx 0\cr}
\form
$$

$$
\eqalign{
&{\vec {\tilde A}}_{\perp}(x)\cr
&{\vec {\tilde \pi}}_{\perp}(x)\cr
&{}\cr
&{\hat {\tilde \psi}}(x)=e^{ -{{ie}\over {\triangle}}{\tilde \pi}^o(x)}
\tilde \psi (x)\approx \tilde \psi (x)\cr
&{\hat {\tilde \psi}}^{\dagger}(x)=e^{ {{ie}\over {\triangle}}{\tilde \pi}^o
(x)}{\tilde \psi}^{\dagger}(x)\approx {\tilde \psi}^{\dagger}(x)\cr}
\form
$$

\noindent because, since $\lbrace {\tilde \psi}^{\dagger}(\vec x,x^o)\tilde
\psi (\vec x,x^o),{\tilde \psi}^{\dagger}(\vec y,x^o)
\tilde \psi (\vec y,x^o)\rbrace =0$, we have

$$
\eqalign{
&\lbrace {\hat {\tilde \psi}}(\vec x,x^o),{\hat {\tilde \psi}}^{\dagger}(\vec
y,x^o)\rbrace -i\delta_{\alpha\beta}\delta^3(\vec x-\vec y)\cr
&\lbrace {\tilde V}_o(\vec x,x^o),{\tilde \Pi}_o(\vec y,x^o)\rbrace =
\delta^3(\vec x-\vec y)\cr
&\lbrace {\hat {\tilde \psi}}_{\alpha}(\vec x,x^o),{\tilde V}_o(\vec y,x^o)
\rbrace =0\cr
&\lbrace {\hat {\tilde \psi}}^{\dagger}_{\alpha}(\vec x,x^o),{\tilde V}_o
(\vec y,x^o)\rbrace =0\cr}
\form
$$

Now we have

$$
H_D={\hat H}_{\perp}+H_G
\form
$$

\noindent with ${\hat H}_{\perp}$ the same as $H_{\perp}$ but as a function of
${\hat {\tilde \psi}}$ and with the gauge Hamiltonian $H_G$ given by
(the fermionic term in the second line comes from re-expressing $H_{\perp}$ in
terms of ${\hat {\tilde \psi}}$ to get ${\hat H}_{\perp}$)

$$
\eqalign{
H_G&=\int d^3x\lbrace -({1\over {\triangle}}{\tilde V}_o(\vec x,x^o)){\tilde
\Gamma}^T(\vec x,x^o)+{1\over 2}
{\lbrack {{\vec \partial}\over {\triangle}}{\tilde \Gamma}^T
(\vec x,x^o)\rbrack}^2+\cr
&+\lbrack {\tilde \lambda}_o(\vec x,x^o)\triangle -e{\hat {\tilde \psi}}
^{\dagger}(\vec x,x^o)\vec \alpha {\hat {\tilde \psi}}(\vec x,x^o)\cdot {\vec
\partial}\rbrack {\tilde \Pi}^o(\vec x,x^o)\rbrace =\cr
&=\int d^3x\lbrace -({1\over {\triangle}}{\tilde V}_o(\vec x,x^o)){\tilde
\Gamma}^T(\vec x,x^o)+{1\over 2}
{\lbrack {{\vec \partial}\over {\triangle}}{\tilde
\Gamma}^T(\vec x,x^o)\rbrack}^2+\cr
&\lbrack \triangle {\tilde \lambda}_o(\vec x,x^o)+e{\vec \partial}
\cdot ({\hat {\tilde \psi}}^{\dagger}(\vec x,x^o)
\vec \alpha {\hat {\tilde \psi}}
(\vec x,x^o))\rbrack {\tilde \Pi}^o(\vec x,x^o)\rbrace =\cr
&=\int d^3x\lbrace -({1\over {\triangle}}{\tilde V}_o(\vec x,x^o)){\tilde
\Gamma}^T(\vec x,x^o)+{1\over 2}
{\lbrack {{\vec \partial}\over {\triangle}}{\tilde \Gamma}^T
(\vec x,x^o)\rbrack}^2+\cr
&+{\hat \lambda}_o(\vec x,x^o){\tilde \Pi}^o(\vec x,x^o)\rbrace \cr}
\form
$$

\noindent after an allowed integration by parts and a redefinition of the
arbitrary Dirac multiplier. The analogue of the temporal gauge is now
${\tilde V}_o(x)\approx 0$.

{}From Eqs.(5-31) instead of Eqs.(5-22) one gets

$$
\eqalign{
{\tilde L}^{To}(x)&=\triangle \lbrack {\tilde A}_o(x)-{\dot {\tilde \eta}}(x)
\rbrack +e\psi^{\dagger}(x)\psi (x)=\cr
&=\triangle \lbrack {\tilde A}_o(x)-{\dot {\tilde \eta}}(x)\rbrack
+e{\tilde \psi}^{\dagger}(x)\tilde \psi (x){\buildrel \rm \circ \over =}0\cr
{\vec {\tilde L}}^T(x)&=
\bar \sqcup {\vec {\tilde A}}_{\perp}(x)-{\partial}^o{\vec
\partial}\lbrack {\tilde A}_o(x)-{\dot {\tilde \eta}}(x)\rbrack
+e\psi^{\dagger}(x)\gamma^o\vec \gamma \psi (x)=\cr
&=\bar \sqcup {\vec {\tilde A}}_{\perp}(x)-{\partial}^o{\vec
\partial}\lbrack {\tilde A}_o(x)-{\dot {\tilde \eta}}(x)\rbrack
+e{\tilde \psi}^{\dagger}(x)\gamma^o\vec \gamma {\tilde \psi}(x)
{\buildrel \rm \circ \over =}0\cr
L_{\bar \psi}(x)&=e^{ie\tilde \eta (x)}\lbrace \gamma^o\lbrack i\partial_o-e
(-{\tilde A}_o(x)+{\dot {\tilde \eta}}(x))\rbrack +\gamma^k(i\partial_k+e
{\tilde A}_{\perp k}(x))-m\rbrace \tilde \psi (x){\buildrel \circ \over =}0\cr
L_{\psi}(x)&={\tilde \psi}^{\dagger}(x)\gamma^o\lbrace \lbrack
i\overleftarrow {\partial_o}+
e\overleftarrow {(-{\tilde A}_o(x)+{\dot {\tilde \eta}}(x))}\rbrack \gamma^o+
\overleftarrow {(i\partial_k-e{\tilde A}_{\perp k}(x))}\gamma^k+m\rbrace
e^{-ie\tilde \eta (x)}{\buildrel \circ \over =}0\cr}
\form
$$

\noindent which imply

$$
\eqalign{
&{\tilde A}_o(x){\buildrel \circ \over =}{\dot {\tilde \eta}}(x)-{e\over
{\triangle}}({\tilde \psi}^{\dagger}(x)\tilde \psi (x))=
-{1\over {\triangle}}\vec \partial \cdot {\dot {\vec {\tilde A}}}(x)-{e\over
{\triangle}}({\tilde \psi}^{\dagger}(x)\tilde \psi (x))\cr
&\Rightarrow { 1\over {\triangle} }{\tilde V}_o(x){\buildrel \circ \over =}
{\dot {\tilde \eta}}(x)\cr}
\form
$$

\noindent so that the physical equations of motion, also deducible from the
Hamilton equations associated with Eqs.(5-37), are (the difference between
$\tilde \psi$ and ${\hat {\tilde \psi}}$ is here irrelevant)

$$
\eqalign{
\bar \sqcup {\tilde A}^i_{\perp}(x)&{\buildrel \rm \circ \over =}-e{\tilde
\psi}
^{\dagger}(x)
\gamma^o\gamma^i\tilde \psi (x)+e{\partial}^o{ {{\partial}^i}\over
{\triangle} }\lbrack {\tilde \psi}^{\dagger}(x)\tilde \psi (x)\rbrack =\cr
&=-e\int d^3y\lbrace \delta^3(\vec x-\vec y){\tilde \psi}^{\dagger}(\vec y,x^o)
\alpha^i\tilde \psi (\vec y,x^o)-\cr
&-c^i(\vec x-\vec y){\tilde \psi}^{\dagger}
(\vec y,x^o)(\vec \partial + \overleftarrow {\vec \partial})\cdot \vec \alpha
\tilde \psi (\vec y,x^o)\rbrace \cr
i\partial_o\tilde \psi (x)&{\buildrel \circ \over =}\lbrack \vec \alpha \cdot
(i\vec \partial +e{\vec {\tilde A}}_{\perp}(x))+\beta m+{{e^2}\over
{\triangle}}
({\tilde \psi}^{\dagger}(x)\tilde \psi (x))\rbrack \tilde \psi (x)\cr
i\partial_o{\tilde \psi}^{\dagger}(x)&{\buildrel \circ \over =}{\tilde \psi}
^{\dagger}(x)\lbrack \overleftarrow {i\vec \partial -e{\vec {\tilde A}}_{\perp}
(x))}\cdot \vec \alpha -\beta m-{{e^2}\over {\triangle}}({\tilde
\psi}^{\dagger}
(x)\tilde \psi (x))\rbrack \cr
&\Rightarrow \, i\partial_o\lbrack {\tilde \psi}^{\dagger}(x)\tilde \psi (x)
\rbrack {\buildrel \circ \over =}{\tilde \psi}^{\dagger}(x)(i\overleftarrow
{\partial}+i\overrightarrow {\partial})\cdot \vec \alpha \tilde \psi (x)\cr}
\form
$$

To find the associated Lagrangian by means of the inverse Legendre
transformation we need the first half of the Hamilton equations for $H_{\perp}$
of Eqs.(5-37):

$$
\eqalign{
{\dot {\tilde A}}^i_{\perp}(\vec x,x^o)&{\buildrel \circ \over =}-(\delta^{ij}+
{ {\partial^i\partial^j}\over {\triangle} }){\tilde \pi}^j_{\perp}(\vec
x,x^o)=-
{\tilde \pi}^i_{\perp}(\vec x,x^o)\cr
i{\dot {\tilde \psi}}(\vec x,x^o)&=\lbrack \vec \alpha \cdot (i\vec \partial +e
{\vec {\tilde A}}_{\perp}(\vec x,x^o))+\beta m+{{e^2}\over {4\pi} }\int d^3y
{ {{\tilde \psi}^{\dagger}(\vec y,x^o)\tilde \psi (\vec y,x^o)} \over {|\vec x
-\vec y|} }\rbrack \tilde \psi (\vec x,x^o)\cr}
\form
$$

Having ${\vec {\tilde \pi}}_{\perp}(x)=-{\dot {\vec {\tilde A}}}_{\perp}(x)$
and taking into account second class constraints for the Dirac fields, see
Eqs.(4-18), one gets

$$
\eqalign{
{\cal L}^T_{\perp}(x)&={1\over 2}\lbrack {\dot {\tilde A}}^i_{\perp}(x)
{\dot {\tilde A}}^i_{\perp}(x)-{(\vec \partial \times {\vec {\tilde A}}_{\perp}
(x))}^2\rbrack +\cr
&+{\tilde \psi}^{\dagger}(x)\lbrace i\partial_o-{1\over 2}
{{e^2}\over {4\pi} }\int d^3y
{ {{\tilde \psi}^{\dagger}(\vec y,x^o)\tilde \psi (\vec y,x^o)}\over {| \vec x
-\vec y |} }-\vec \alpha \cdot (i\vec \partial +e{\vec {\tilde A}}_{\perp}
(x))-\cr
&-\beta m\rbrace \tilde \psi (x)
-\partial_{\mu}\lbrack {i\over 2}{\bar {\tilde \psi}}(x)\gamma^{\mu}\tilde
\psi (x)\rbrack \cr}
\form
$$

Indeed, from Eqs.(5-30) and (5-48) one has

$$
\eqalign{
{\cal L}^T(x)&={1\over 2}\lbrace {\lbrack \vec \partial (\dot \eta (x)-{\tilde
A}^o(x))+{\dot {\vec {\tilde A}}}_{\perp}(x)\rbrack }^2-{\lbrack \vec \partial
\times {\vec {\tilde A}}_{\perp}(x)\rbrack }^2\rbrace +\cr
&+{\tilde \psi}^{\dagger}(x)\lbrack i\partial^o-e(\dot \eta (x)-{\tilde A}^o
(x))\rbrack \tilde \psi (x)-{\tilde \psi}^{\dagger}(x)\vec \alpha \cdot
\lbrack i\vec \partial +e{\vec {\tilde A}}_{\perp}(x)\rbrack \tilde \psi
(x)-\cr
&-m{\tilde \psi}^{\dagger}(x)\beta \tilde \psi (x)-\partial_{\mu}\lbrack
{i\over 2}{\tilde \psi}(x)\gamma^o\gamma^{\mu}\tilde \psi (x)\rbrack
{\buildrel \circ \over =}\cr
&{\buildrel \circ \over =}{1\over 2}\lbrace {\lbrack {\dot {\vec {\tilde A}}}
_{\perp}(x)-e{ {\vec \partial}\over {\triangle} }{\tilde \psi}^{\dagger}(x)
\tilde \psi (x)\rbrack }^2-{\lbrack \vec \partial \times {\vec {\tilde A}}
_{\perp}(x)\rbrack }^2\rbrace +\cr
&+{\tilde \psi}^{\dagger}(x)\lbrack i\partial^o-{{e^2}\over {\triangle}}
({\tilde \psi}^{\dagger}(x)\tilde \psi (x))-\vec \alpha \cdot (i\vec \partial
+e{\vec {\tilde A}}_{\perp}(x))-m\beta \rbrack \tilde \psi (x)-\cr
&-\partial_{\mu}\lbrack {i\over 2}{\tilde \psi}^{\dagger}(x)\gamma^o
\gamma^{\mu}\tilde \psi (x)\rbrack ={\cal L}^T_{\perp}(x)\cr}
\form
$$

Let us consider the Abelian version of the energy-momentum tensor and of the
\break Poincar\`e generators of Eqs.(4-14) and let us look at their
decomposition into Dirac's observables and gauge part. By remembering that
${\vec {\tilde \pi}}={\vec {\tilde \pi}}_{\perp}-e{ {\vec \partial}\over
{\triangle} }({\hat {\tilde \psi}}^{\dagger}{\hat {\tilde \psi}})+
{ {\vec \partial}\over {\triangle} }{\tilde \Gamma}^T$, by using Eqs.(5-8),
(5-10), (5-34), (5-42), (5-43), (5-45), and by making integrations by parts,
one obtains:

$$
\eqalign{
{\bar P}^o&\equiv H_T={\hat H}_{\perp}+H_G\cr
{\bar P}^i&={\bar P}^i_{\perp}+{\bar P}^i_G+{\bar P}^i_{G\perp}\approx
{\bar P}^i_{\perp}\cr
{\bar J}^i&={\bar J}^i_{\perp}+{\bar J}^i_G+{\bar J}^i_{G\perp}\approx
{\bar J}^i_{\perp}\cr
{\bar K}^i&={\bar K}^i_{\perp}+{\bar K}^i_G+{\bar K}^i_{G\perp}\approx
{\bar K}^i_{\perp}\cr}
\form
$$

\noindent where

$$
\eqalign{
{\hat H}_{\perp}&=\int d^3x\lbrace i{\hat {\tilde \psi}}^{\dagger}(\vec x,x^o)
\vec \alpha \cdot {\vec \partial}{\hat {\tilde \psi}}(\vec x,x^o)+
m{\hat {\tilde \psi}}^{\dagger}(\vec x,x^o)\beta {\hat {\tilde \psi}}
(\vec x,x^o)+\cr
&+e{\hat {\tilde \psi}}^{\dagger}(\vec x,x^o)\vec \alpha \cdot {\vec {\tilde
A}}
_{\perp}(\vec x,x^o){\hat {\tilde \psi}}(\vec x,x^o)+{1\over 2}\lbrack
{\vec {\tilde \pi}}^2_{\perp}(\vec x,x^o)+{\vec {\tilde B}}^2[{\vec {\tilde A}}
_{\perp}(\vec x,x^o)]\rbrack +\cr
&+{{e^2}\over 2}{\lbrack { {\vec \partial}\over {\triangle} }
({\hat {\tilde \psi}}^{\dagger}(\vec x,x^o)
{\hat {\tilde \psi}}(\vec x,x^o))\rbrack}^2 \rbrace \cr
{\bar P}^i_{\perp}&=\int d^3x \lbrack -{\vec {\tilde \pi}}_{\perp}(\vec x,x^o)
\cdot \partial^i{\vec {\tilde A}}_{\perp}(\vec x,x^o)+{i\over 2}{\hat {\tilde
\psi}}^{\dagger}(\vec x,x^o)(\partial^i-\overleftarrow {\partial^i}){\hat
{\tilde \psi}}(\vec x,x^o)\rbrack \cr
{\bar J}^i_{\perp}&=\int d^3x\lbrace -\epsilon^{ijk}\lbrack {\tilde \pi}^j
_{\perp}(\vec x,x^o){\tilde A}^k_{\perp}(\vec x,x^o)+{\vec {\tilde
\pi}}_{\perp}
(\vec x,x^o)\cdot x^j\partial^k{\vec {\tilde A}}_{\perp}(\vec x,x^o)-\cr
&-{i\over
2}{\hat {\tilde \psi}}^{\dagger}(\vec x,x^o)(x^j\partial^k-\overleftarrow
{\partial^k}x^j){\hat {\tilde \psi}}(\vec x,x^o)\rbrack +
{\hat {\tilde \psi}}^{\dagger}(\vec x,x^o)\sigma^i{\hat {\tilde \psi}}(\vec
x,x^o)\rbrace \cr
{\bar K}^i_{\perp}&=\int d^3x x^i\lbrace {1\over 2}\lbrack {\vec {\tilde \pi}}
^2_{\perp}(\vec x,x^o)+{\vec {\tilde B}}^2(\vec x,x^o)\rbrack +{1\over 2}
\lbrack {\hat {\tilde \psi}}^{\dagger}(\vec x,x^o)\vec \alpha \cdot (i\vec
\partial +e{\vec {\tilde A}}_{\perp}(\vec x,x^o)){\hat {\tilde \psi}}(\vec x,
x^o)-\cr
&-{\hat {\tilde \psi}}^{\dagger}(\vec x,x^o)\vec \alpha \cdot \overleftarrow
{(i\vec \partial -e{\vec {\tilde A}}_{\perp}(\vec x,x^o))}{\hat {\tilde \psi}}
(\vec x,x^o)\rbrack
+m{\hat {\tilde \psi}}^{\dagger}(\vec x,x^o){\hat {\tilde \psi}}(\vec
x,x^o)+\cr
&+{ {e^2}\over 2}{\lbrack { {\vec \partial}\over {\triangle} }({\hat {\tilde
\psi}}(\vec x,x^o){\hat {\tilde \psi}}(\vec x,x^o)\rbrack}^2-e{\vec {\tilde
\pi}}_{\perp}(\vec x,x^o)\cdot { {\vec \partial}\over {\triangle} }({\hat
{\tilde \psi}}^{\dagger}(\vec x,x^o){\hat {\tilde \psi}}(\vec x,x^o))
\rbrace -x^o{\bar P}^i_{\perp}\cr}
\form
$$

$$
\eqalign{
H_G&=\int d^3x\lbrace -({1\over {\triangle}}{\tilde V}_o(\vec x,x^o)){\tilde
\Gamma}^T(\vec x,x^o)+{1\over 2}
{\lbrack {{\vec \partial}\over {\triangle}}{\tilde \Gamma}^T
(\vec x,x^o)\rbrack}^2+
{\hat \lambda}_o(\vec x,x^o){\tilde \Pi}^o(\vec x,x^o)\rbrace \cr
{\bar P}^i_G&=-\int d^3x {\tilde \Gamma}^T(\vec x,x^o)\partial^i\tilde
\eta (\vec x,x^o)\cr
{\bar J}^i_G&=-\epsilon^{ijk}\int d^3x {\tilde \Gamma}^T(\vec x,x^o)x^j
\partial^k\tilde \eta (\vec x,x^o)\cr
{\bar K}^i_G&=\int d^3x x^i\lbrack {1\over 2}{\lbrack { {\vec \partial}
\over {\triangle} }{\tilde \Gamma}^T(\vec x,x^o)\rbrack}^2-{1\over {\triangle}}
{\tilde V}^o(\vec x,x^o){\tilde \Gamma}^T(\vec x,x^o)+{\hat {\tilde \lambda}}
(\vec x,x^o){\tilde \Pi}^o(\vec x,x^o)\rbrack -x^0{\bar P}^i_G\cr}
\form
$$

$$
\eqalign{
{\bar P}^i_{G\perp}&=-e\int d^3x {\hat {\tilde \psi}}^{\dagger}(\vec x,x^o)
{\hat {\tilde \psi}}(\vec x,x^o)\partial^i{\tilde \Pi}(\vec x,x^o)\cr
{\bar J}^i_{G\perp}&=-e\int d^3x {\hat {\tilde \psi}}^{\dagger}(\vec x,x^o)
{\hat {\tilde \psi}}(\vec x,x^o)\epsilon^{ijk}x^j\partial^k{\tilde \Pi}^o
(\vec x,x^o)\cr
{\bar K}^i_{G\perp}&=\int \lbrace e{\hat {\tilde \psi}}^{\dagger}(\vec x,x^o)
\alpha^i{\hat {\tilde \psi}}(\vec x,x^o){\tilde \Pi}^o(\vec x,x^o)-\cr
&-\lbrack
{1\over {\triangle}}({\tilde \pi}_{\perp}(\vec x,x^o)+e{ {\partial^i}\over
{\triangle}}({\hat {\tilde \psi}}^{\dagger}(\vec x,x^o)
{\hat {\tilde \psi}}(\vec x,x^o)))\rbrack {\tilde \Gamma}^T(\vec x,x^o)-
x^o{\bar P}^i_{G\perp}\cr}
\form
$$

We shall study elsewhere the implications of the decomposition (5-54), the
terms (5-56) mixing constraints and physical degrees of freedom, and the
problem of the decoupling of the gauge degrees of freedom in the
Lienard-Wiechert potential [32].

\vfill\eject

\bigskip\noindent
{\bf{6. Gribov Ambiguity and the Solution of Gauss' Laws.}}
\newcount \nfor

\def \form {\global \advance \nfor by 1 \eqno (6.\the\nfor)}
\bigskip

Till now we have treated connections on $P^3=P^t=(R^3\times \lbrace x^o\rbrace
)\times G\sim R^3\times G$ or on $P^4=P=M^4\times G$ and the associated
gauge potentials on $R^3$ or $R^4$ (or gauge potentials and electric fields in
the phase space approach) and the group of gauge transformations $Aut_VP\sim
GauP\sim \Gamma (AdP)$ [$P=P^3$ or $P^4$] or ${\cal G}$ (or ${\bar {\cal G}}$
in phase space) disregarding the fact that we have to do with infinite
dimensional manifolds.

Let us define some notations: if $P^3=R^3\times G$ and $P^4=M^4\times G$ are
the
trivial principal G-bundles, let us call ${\hat {\cal C}}^3$ or ${\hat {\cal
C}}
^4$ the respective functional spaces of connections (of connections 1-forms
$\omega^{\cal A}$; see Appendix B) on $P^3$ or $P^4$ respectively, while let
${\cal C}^3$ and ${\cal C}^4$ denote the corresponding functional spaces of
$g$-valued (in the adjoint representation)
gauge potentials $A_{a\mu}(x)dx^{\mu}{\hat T}^a$ or ${\vec A}_a(\vec x,x^o)
\cdot d\vec x{\hat T}^a$; ${\bar {\cal C}}$ will denote the functional space
whose elements are the pairs $\lbrace {\vec A}_a(\vec x,x^o)\cdot d\vec x{\hat
T}^a,{\vec \pi}_a(\vec x,x^o)\cdot d\vec x{\hat T}^a=g^{-2}{\vec E}_a(\vec
x,x^o
)\cdot d\vec x{\hat T}^a\rbrace$ used in the phase space approach (it is the
cotangent bundle of ${\cal C}^3$); $\lbrace A^o_a(\vec x,x^o){\hat T}^a,\pi^o_a
(\vec x,x^o){\hat T}^a\rbrace$ form the remaining variables to obtain the full
phase space ${\bar {\cal C}}_T$ of YM theory once ${\bar {\cal C}}$ is given
(${\bar {\cal C}}_T$ is the cotangent bundle of ${\cal C}^4$);
while $Aut_VP\sim GauP\sim \Gamma (AdP)$  is the group
of gauge transformations on the principal G-bundle $P=P^3$ or $P^4$ acting on
${\hat {\cal C}}^3$ and ${\hat {\cal C}}^4$ respectively, ${\cal G}
^3$ and ${\cal G}^4$ are the groups of gauge transformations acting on the
functional spaces ${\hat {\cal C}}^3$ and ${\hat {\cal C}}^4$ respectively;
finally ${\hat {\cal G}}$ and ${\bar {\cal G}}$ will denote the phase space
groups of gauge transformations acting on ${\bar {\cal C}}$ and ${\bar {\cal
C}}_T$ respectively (see Section 2 and 3).
The associated Lie algebras will be denoted
$aut_VP\sim gauP\sim \Gamma (adP)$ ($P=P^3$ or $P^4$), $g_{\cal G}^{3\, or\,
4}$, $g_{\hat {\cal G}}$, $g_{\bar {\cal G}}$. In the case of phase space one
has ${\bar {\cal G}}^o_{\infty}\subset {\bar {\cal G}}_{\infty}\subset {\bar
{\cal G}}$. In the case of $R^3$, $R^4$ ($M^4$ after the rotation to
immaginary time) compactified to
$S^3$, $S^4$ one defines ${\cal G}^{3\, or\, 4}_{*}$
as the subgroup of gauge transformations in ${\cal G}^{3\, or\, 4}$ which
leaves the fiber G over a given point of $P^{3\, or\, 4}$ fixed [induced by the
group ${(Aut_VP)}_{*}\sim \Gamma (AdP)_{*}\sim GauP_{*}$] and usually this
point is the point at infinity which compactifies $R^{3\, or\, 4}$ to $S^{3\,
or\, 4}$; instead the group of gauge transformations module its center
$Z_{\cal G}\sim Z_G$ is denoted ${\tilde {\cal G}}^{3\, or\, 4}={\cal G}
^{3\, or\, 4}/Z_G$ [on ${\hat {\cal C}}^{3\, or\, 4}$ one has $\tilde \Gamma
(AdP)=\Gamma (AdP)/Z_{\Gamma (AdP)}$ and $\tilde {Gau}P=GauP/Z_G$] and
analogously in phase space one defines ${\tilde {\bar {\cal G}}}$.

${\hat {\cal C}}^4$ and ${\hat {\cal C}}^3$ are affine spaces [21b,c];
fixed an origin
${\cal A}_o\in {\hat {\cal C}}^{4\, or\, 3}$, i.e. a "background connection",
one has ${\hat {\cal C}}^{4\, or\, 3}=\lbrace {\cal A}_o+\hat a\, |\, \hat
a\in C^{\infty}(\Lambda^1(P^{4\, or\, 3})\otimes
adP)\rbrace$, i.e. they become vector spaces isomorphic
to $C^{\infty}(\Lambda^1(P^{4\, or\, 3})\otimes adP)$,
the smooth equivariant $g$-valued 1-forms on $P^{4\, or\, 3}$
($\Lambda^1(P)$ denotes the Grassmann algebra of 1-forms on P).
${\hat {\cal C}}^{4\, or\, 3}$ is an infinite dimensional Riemannian manifold
with the tangent space at the background connection $T_{{\cal A}_o}{\hat {\cal
C}}^{4\, or\, 3}$ identified with $C^{\infty}(\Lambda^1(P^{4\, or\, 3})\otimes
g)$. In ${\cal C}^{4\, or\, 3}$ every connection in ${\hat {\cal C}}^{4\, or\,
3}$ is described by the system of gauge potentials (the gauge orbit of the
given
connection) associated with the smooth (global in our case) cross sections from
$M^4$ ($R^3$) to $P^4$ ($P^3$); the $g$-valued gauge potentials are elements of
$C^{\infty}(\Lambda^1(M)\otimes g)$, $M=M^4$ or $R^3$, once a background  gauge
potential has been fixed. The (weak) Riemannian structure on ${\hat {\cal C}}
^{4\, or\, 3}$ is induced from the one in ${\cal C}^{4\, or\, 3}$; if $A=\sigma
^{*}\omega^{{\cal A}_o+\tilde a}=A_o+\tilde A$, $\tilde A\in C^{\infty}(\Lambda
^1(M)\oplus g)$, and $A^{'}=\sigma^{*}\omega
^{{\cal A}_o+{\tilde a}^{'}}=A_o+{\tilde A}^{'}$, the inner product inducing
a (weak) Riemannian metric is $< \tilde A,{\tilde A}^{'} > \, =\, \int_M
{(\tilde A\, \cdot \, ,\, \wedge \, {*}{\tilde A}^{'})}_g$. Here ${*}$ is the
Hodge star operation needed to construct the volume form on $M=M^4$ or $R^3$;
${(.,.)}_g$ is the Killing-Cartan form on $g$; $\cdot \,$ denotes the
Riemannian
metric on TM ($\eta_{\mu\nu}$ for $M^4$; $\delta^{ij}$ for $R^3$); the
integration over M requires suitable boundary conditions [they usually imply
the
possibility to compactify either $R^4$ ($M^4$ after the rotation to immaginary
time) to $S^4$ as in the Euclidean theory or of $R^3$ to $S^3$]. Let us remark
that compactness of M, besides its relevance for instantons when $M=S^4$, is a
form of "volume cutoff".

With regard to the gauge transformations on the principal G-bundles, it is
convenient to use the formulation with $\Gamma (AdP)$, $P=P^4$ or $P^3$, i.e.
the space of $C^{\infty}$ cross sections of the associated bundle
$AdP=P{\times}
_G\, G$, which acts on ${\hat {\cal C}}=
{\hat {\cal C}}^{4\, or\, 3}$, ${\hat {\cal C}}
\times \, \Gamma (AdP)\rightarrow {\hat {\cal C}}$. To introduce suitable
function space topologies one completes the space of smooth cross sections
to a Hilbert space $\Gamma^{2,s+1}(AdP)$ in some Sobolev norm ${\| .\|}_{2,s}$
with $s\, >\, {1\over 2}dimM$ [21b,c] ($s\, >\, 2$ in $M^4$, $s\, >\, 3/2$ in
$R^3$) acting on a Sobolev space ${\hat {\cal C}}^{2,s}$ of connections;
this condition is also necessary to have the possibility to
define gauge-fixings selecting only one point from each gauge orbit in ${\cal
C}^{4\, or\, 3}$ [65a]: without it any functional hypersurface in ${\cal
C}^{4\, or\, 3}$ would be intersected by any gauge orbit in a dense set of
points, while with $s\, >\, {1\over 2}dimM$ one at most has the well known
Gribov ambiguity, which, however, may be absent in suitable functional spaces
as we shall see. In this way one can show [21b,c] that the group of gauge
transformations ${\cal G}$ becomes an $\infty$-dimensional Hilbert-Lie group
${\cal G}^{2,s+1}$ (${\bar {\cal G}}^{2,s+1}$ in phase space) [66] with a well
defined action on the space ${\hat {\cal C}}^{2,s}$ of gauge potentials
(Eqs.(2-1) show that gauge transformations need one extra order of
differentiation that the gauge potentials).

Let us now give a classification of the types of connections ([21a]; see
[67] for more details; it does not seem that there is a final
universally accepted classification in the literature); we shall use a rough
classification by putting all connections in three classes. However, before
that, let us define two important stability subgroups of gauge transformations
connected with the realization with gauge potentials $A\in {\cal C}$ of
a given connection ${\cal A}\in {\hat {\cal C}}$:

\noindent 1) "Stability subgroup or gauge symmetries of a connection" [68b,
67]. It is the subgroup ${\cal G}^{\cal A}$ of ${\cal G}^{2,s}$ which
leaves fixed a given gauge potential associated with the given connection
${\cal A}$:

$$A^U=A+U^{-1}D^{(A)}U=A.
\form
$$

\noindent Therefore a gauge transformation U belongs to ${\cal G}^{\cal A}$ if
and only if it is covariantly constant with respect to ${\cal A}$, i.e.
$D^{(A)}U=0$; for a generic connection with our kind of structure group G
($\pi_1(G)=0$ so that $Z_G$ is discrete and $Z_g=0$), the only solutions of
these equations are constant gauge transformations belonging to the center of
the gauge group (isomorphic to the discrete $Z_G$ in our case), which, for
such connections, is also isomorphic to $Z_{\Phi^{\cal A}}$ (the center of the
holonomy group $\Phi^{\cal A}$ of the connection, which is isomorphic
to a subgroup of $Z_G$): $U\in Z_{\cal G}\sim Z_G
\sim Z_{\Phi^{\cal A}}$; in these cases one speaks of trivial stability
subgroup
${\cal G}^{\cal A}=Z_{\Phi^{\cal A}}=Z_G$. However, there can exist special
connections (it depends on the chosen functional space; with the Sobolev
space ${\cal C}^{2,s}=W^{2,s}(M,R)$
they can exist) for which ${\cal G}^{\cal A}\supset Z_G=Z_{\Phi^{\cal
A}}$ strictly and it can be shown [67] that in this case ${\cal G}^{\cal
A}$ is equal to the centralizer in G of the holonomy group, ${\cal G}^{\cal A}=
Z_G(\Phi^{\cal A})=\lbrace a\in G\, |\, ab=ba\,\, for\, each\, b\in
\Phi^{\cal A}\rbrace$, i.e.

$${\cal G}^{\cal A}=\lbrace U\in {\cal G}^{2,s}\, |\, D^{(A)}U=0\rbrace =
Z_G(\Phi^{\cal A})\supset Z_G= Z_{\Phi^{\cal A}}.
\form
$$

\noindent The Lie algebra of ${\cal G}^{\cal A}$ will be denoted $g^{\cal A}$
and one has ${\cal G}^{\cal A}\sim G$ if and only if $\Phi^{\cal A}\subseteq
Z_G$ (i.e. the connection is invariant under the rigid internal action of G,
as said in Section 3).
When a connection admits gauge symmetries, it has been shown [68]
that the space of solutions of YM equations is not a
manifold, but there are conical singularities ("cone over cone" structure)
in correspondence of solutions associated with these special connections
(these singularities appear as spurious solutions of the Jacobi equations
associated with YM equations); these conical singularities are therefore also
present in the YM constraint hypersurface in phase space, which contains
the solutions to YM equations. Since ${\cal G}^{\cal A}=Z_G(\Phi^{\cal A})$,
it follows that the corresponding stability subgroup $\Gamma^{\cal A}(AdP)$
is defined by the condition that $\tilde \gamma \in \Gamma^{\cal A}(AdP)$
if and only if $[\tilde \gamma ,b]=0$ for all $b\in \Phi^{\cal A}$. To get the
result that only conical singularities are present it is needed the theory of
the momentum map [69] extended to the second Noether theorem [68a,b$\gamma$].

\noindent 2) "Stability subgroup of the curvature of a connection" [70 and
references therein].
It is the subgroup ${\cal G}^{\Omega}$ of ${\cal G}^{2,s}$ which leaves fixed
the field strength derived from a gauge potential associated with a given
connection ${\cal A}$ with curvature $\Omega$, i.e. such that

$$F^U=F+U^{-1}[F,U]=F;
\form
$$

\noindent clearly one has ${\cal G}^{\Omega}\supseteq {\cal G}^{\cal A}$. For
a generic connection one has ${\cal G}^{\Omega}={\cal G}^{\cal A}=Z_{\Phi^{\cal
A}}=Z_G$. But there are special connections (without or with gauge symmetries)
for which ${\cal G}^{\Omega}\supset {\cal G}^{\cal A}\supseteq Z_G=
Z_{\Phi^{\cal A}}$; in these cases one has the problem of "gauge copies"
[70a]: there exist different gauge potentials $A_{(i)}$ giving rise to the
same field strength F. Only when the structure group G is simply connected
($\pi_1(G)=0$, so that $Z_G$ is discrete), like in our case, all these gauge
potentials lie on the same gauge orbit, i.e. $A_{(i)}=A_{(j)}+U^{-1}_{ij}
D^{(A_{(j)})}U_{ij}$; when $\pi_1(G)\not= 0$, there exist gauge not equivalent
families of $A_{(i)}$'s in connection with the homotopically not equivalent
families of paths in the group manifold of G (the gauge orbit of a given
connection has disjoined components related to this fact). For these
connections one has

$$
{\cal G}^{\Omega}=\lbrace U\in {\cal G}^{2,s}\, |\, [F,U]=0\rbrace \supseteq
{\cal G}^{\cal A}=Z_G(\Phi^{\cal A})\supseteq Z_G\supseteq Z_{\Phi^{\cal A}}.
\form
$$

\noindent Its Lie algebra will be denoted $g^{\Omega}$.

We can now divide the connections on a connected, simply connected principal
G-bundle P (like in our case with $\pi_1(G)=\pi_1(M)=0$) in the following
three classes ($\Phi^{\cal A}$ is the holonomy group of the connection, whose
Lie algebra is denoted $g_{\Phi^{\cal A}}$):

\noindent i) "Fully irreducible connections, ${\cal A}\in {\hat {\cal C}}_{fir}
$, $A\in {\cal C}_{fir}$". In this case one has

$$\eqalign{
\Phi^{\cal A}&=G,\quad \quad g_{\Phi^{\cal A}}=g,\quad \quad P^{\cal A}=P,\cr
{\cal G}^{\Omega}&={\cal G}^{\cal A}=Z_G(\Phi^{\cal A})=Z_G=Z_{\Phi^{\cal A}}
\cr}
\form
$$

\noindent and there are neither gauge copies nor gauge symmetries; here
$P^{\cal
A}$ is the holonomy bundle, i.e. the set of points of P connected by parallel
transport [24a].

\noindent ii) "Irreducible connections, ${\cal A}\in {\hat {\cal C}}_{ir}$,
$A\in {\cal C}_{ir}$". In this case the holonomy group is an irreducible
matrix subgroup [21a] of the structure group G (more in general [67]
$\Phi^{\cal A}$ is a "not closed" subgroup of G, so that the holonomy bundle
is immersed, but not embedded in P; we include in this class also the "weakly
irreducible connections" in ${\cal C}_{wir}$ for which one has $\Phi^{\cal A}
\subset G$ but ${\bar {\Phi^{\cal A}}}=G$, where the bar means set closure: in
general, one has ${\cal C}_{fir}\subseteq {\cal C}_{ir}\subseteq {\cal C}_{wir}
\subseteq {\cal C}_{gen}\subseteq {\cal C}$, where ${\cal C}_{gen}$ means only
the condition $Z_G(\Phi^{\cal A})=Z_G$), $\Phi^{\cal A}\subset G$. In this case
there are gauge copies (but not gauge symmetries), and one can show [70a,71]
that there exists a basis of the Lie algebra $g$ of G such that one has
the vector space sum

$$g=g_{\Phi^{\cal A}}+g^{\Omega}
\form
$$

\noindent and that one has

$${\cal G}^{\Omega}\supset {\cal G}^{\cal A}=Z_G(\Phi^{\cal A})=Z_{\Phi^{\cal
A}}=Z_G.
\form
$$

Let us remark that if P is a nontrivial SU(2) bundle over M and if for the
integer-valued second cohomology group one has
$H^2(M,Z)=0$, then ${\cal C}_{ir}={\cal C}$.

\noindent iii) "Reducible connections, ${\cal A}\in {\hat {\cal C}}_{red}$,
$A\in {\cal C}_{red}$". In this case the connection ${\cal A}$ on P is
reducible to a connection on a subbundle Q of P with structure group
$Z_G(\Phi^{\cal A})$ and one has ($\oplus$ means direct sum of Lie algebras):

$$\eqalign{
\Phi^{\cal A}&\subset G,\quad \quad g=g_{\Phi^{\cal A}}\oplus Z_g(g_{\Phi^{\cal
A}}),\cr
{\cal G}^{\Omega}&\supset {\cal G}^{\cal A}=Z_G(\Phi^{\cal A})\supset Z_G=
Z_{\Phi^{\cal A}},\cr}
\form
$$

\noindent where $Z_g(g_{\Phi^{\cal A}})=\lbrace t\in g\, |\, ad_tv=[t,v]=0\,\,
for\, each\, v\in g_{\Phi^{\cal A}}\rbrace$ is the centralizer in $g$ of
$g_{{\Phi}^{\cal A}}$. In this case one has both gauge copies and gauge
symmetries.
A "reducible" connection ${\cal A}$ would split the associated vector bundle
E to P with standard fiber a representation space of G into a direct sum of two
line bundles trivial over M.

In Refs.[21l,a,b,c] it has been shown that there are
three well defined $C^{\infty}$
actions, two of which are free actions, of the group of gauge transformations
or of some of its subgroups (these results are valid for G=SU(n) and
$M=S^{3\, or\, 4}$; instead of stating the results for ${\hat {\cal C}}$ and
$Aut_VP\sim \Gamma (AdP)\sim GauP$ [valid even when P is non-trivial], we
shall use ${\cal C}$ and ${\cal G}$ with both only locally defined when P is
non-trivial):\hfill\break

\noindent
i) ${\cal C}^{2,s}\times \, {\cal G}^{2,s+1}\rightarrow {\cal C}^{2,s}$;\hfill
\break
ii) ${\cal C}^{2,s}\times {\cal G}_{*}^{2,s+1}\rightarrow {\cal C}^{2,s}\,\,$
(free action);\hfill\break
iii) ${\cal C}^{2,s}_{ir}\times \, {\tilde {\cal G}}^{2,s+1}\rightarrow
{\cal C}_{ir}^{2,s}\,\, $ (free action);

\noindent where ${\cal C}^{2,s}_{ir}$ is the space of gauge potentials
associated with irreducible connections. One has

$${\cal G}_{*}^{2,s+1}
\subset {\tilde {\cal G}}^{2,s+1}\subset {\cal G}^{2,s+1}
\form
$$

One can define three different orbit spaces:

\noindent i) ${\cal N}^{2,s}={\cal C}^{2,s}/{\cal G}^{2,s+1}$\hfill\break

\noindent with ${\cal G}^{2,s+1}$ not acting freely on ${\cal C}^{2,s}$ due to
the nontriviality of the stability group ${\cal G}^{\cal A}=Z_G(\Phi^{\cal A})$
for reducible connections. The $C^{\infty}$ topology of the orbit space
${\cal N}^{2,s}$ is the topology discussed in Refs.[72a,b]
in connection with the group of diffeomorphisms Diff(M). In short, the
topology of ${\cal N}$ is determined by the local "slices" ${\cal S}_{\cal A}
=\lbrace A+\tilde A\, |\,  A\in {\cal C}^{2,s},\, \tilde A\in C^{\infty}
(\Lambda^1(M)\otimes g),\, D^{(A){*}} \tilde A=-D^{(A)}\tilde A=0
\rbrace$, which are "generalized local
Coulomb gauges" (for $A=0$ one gets the standard Coulomb gauge $\partial
\cdot \tilde A=0$); since ${\cal S}_{\cal A}$ is orthogonal (with respect to
the
Riemannian metric) to the gauge orbit ${\cal O}_{\cal A}$ through $A$,
${\cal S}_{\cal A}$ does not intersect ${\cal O}_{\cal A}$ near $A$
except at $A$; ${\cal N}^{2,s}$ is a metrizable Hausdorff space.
${\cal C}^{2,s}({\cal N}^{2,s},{\cal G}^{2,s+1})$ is not
a principal bundle. If a continuous gauge $s:{\cal N}^{2,s}\rightarrow
{\cal C}^{2,s}$ were to exist, then $s{|}_{{\cal N}^{2,s}_{ir}}:
{\cal N}^{2,s}_{ir}\rightarrow {\cal C}^{2,s}_{ir}$ would give a cross section
of the principal bundle ${\cal C}^{2,s}_{ir}({\cal N}^{2,s}_{ir},{\tilde {\cal
G}}^{2,s+1})$ [see iii) later on].
In Refs.[21a,l] it has been shown that no
continuous gauge (i.e. global cross section)
$s:{\cal N}^{2,s}\rightarrow {\cal C}^{2,s}$ exists for $M=S^4\, or\, S^3$ ;
therefore there is the Gribov ambiguity [73,30].

\noindent ii) ${\cal N}_{*}^{2,s}={\cal C}^{2,s}/{\cal G}_{*}^{2,s+1}$,\hfill
\break

\noindent where ${\cal G}_{*}^{2,s+1}\subset {\cal G}^{2,s+1}$
is the subgroup of gauge transformations
which leaves fixed the fiber over a point $x_o\in M$ (usually $\infty$ in
$R^3$ or $R^4$; this is the case for ${\bar {\cal G}}^{2,s+1}_{\infty}$
with the boundary conditions of Eqs.(2-40) in
the Hamiltonian formalism), i.e. $U(x_o)=I$; ${\cal G}_{*}^{2,s+1}$ acts freely
and properly on ${\cal C}^{2,s}$ (i.e. without fixed points),
which turns out to be a
principal fiber bundle ${\cal C}^{2,s}({\cal N}_{*}^{2,s},{\cal
G}_{*}^{2,s+1})$
over ${\cal N}_{*}^{2,s}$ with structure group ${\cal G}_{*}^{2,s+1}$,
which differs from ${\cal G}^{2,s+1}$ for the action of the compact group G:
${\cal G}^{2,s+1}/{\cal G}_{*}^{2,s+1}=G$.
The Lie algebra of ${\cal G}_{*}^{2,s+1}$ consists of elements of
$g^{2,s+1}_{\cal G}$ which vanish at $x_o$. In Refs.[21a,l] it is shown that
the principal bundle ${\cal C}^{2,s}({\cal N}^{2,s}_{*},{\cal G}
_{*}^{2,s+1})$ is not trivial, so that it has no global cross sections;
again we get the Gribov ambiguity.

\noindent iii) ${\cal N}_{ir}^{2,s}={\cal C}_{ir}^{2,s}/{\tilde {\cal G}}
^{2,s+1}\subset {\cal N}^{2,s}$,\hfill\break

\noindent
where ${\tilde {\cal G}}^{2,s+1}={\cal G}^{2,s+1}/Z_G$ acts freely
and properly on ${\cal C}_{ir}^{2,s}$,
which therefore is a principal bundle over ${\cal N}^{2,s}_{ir}$ with structure
group ${\tilde {\cal G}}^{2,s+1}$. ${\cal G}_{*}^{2,s+1}$ is a normal
subgroup of ${\tilde {\cal G}}^{2,s+1}$ ,
but the principal bundle ${\cal C}_{ir}^{2,s}({\cal N}^{2,s}_{ir},{\tilde
{\cal G}}^{2,s+1})$ "cannot be reduced" to ${\cal C}^{2,s}
({\cal N}_{*}^{2,s},{\cal G}_{*}^{2,s+1})$.
The $C^{\infty}$ topology of the orbit space ${\cal N}^{2,s}_{ir}$ is the
topology discussed in Ref.[71a,b; 21a,b,c].
${\cal N}^{2,s}_{ir}$ is an infinite
dimensional $C^{\infty}$ Riemannian manifold,
whose tangent space at ${\cal O}_{A_{ir}}$ is isomorphic to the slice
${\cal S}_{{\cal A}_{ir}}=\lbrace A_{ir}+\tilde A | A_{ir}\in
{\cal C}^{2,s}_{ir},\, \tilde A\in C^{\infty}(\Lambda^1(M)\otimes g),\,
D^{(A_{ir}){*}}\tilde A=0\rbrace$; the inner product is $\int_M{({\tilde A}_1
\wedge {*}{\tilde A}_2)}_g$.
The tangent space at the gauge orbit ${\cal O}_{A_{ir}}$
at $A_{ir}$ is $\lbrace D^{(A_{ir})}{\tilde A}^{'} \, |\, {\tilde A}^{'} \in
C^{\infty}(\Lambda^o(M)\otimes g)\rbrace$ and it is orthogonal to the
slice ${\cal S}_{{\cal A}_{ir}}$: $\, \int_M(\tilde A\wedge$\break $ {*}D^{(A
_{ir})}{\tilde A}^{'} ){}_g=\int_M\, {(D^{(A_{ir}){*}}\tilde A
\wedge {*}{\tilde A}
^{'} )}_g=0$ by definition of ${\cal S}_{{\cal A}_{ir}}$.
In Ref.[21c] it is shown that ${\cal N}^{2,s}_{ir}$ is a Hilbert manifold [i.e.
it is a manifold whose structure is given by a system of neighbourhoods
covering ${\cal N}^{2,s}_{ir}$, homeomorphic to open sets in a Hilbert space
(model space); the Hilbert space is ${\cal H}^{2,s}(\Lambda^1(M)\otimes g)$,
the kernel of $D^{(A){*}}$ in $\Gamma^{2,s}(\Lambda^1(M)\otimes g)$; a slice
${\cal S}_{A_{ir}}$ is one of its open sets].
One has that ${\cal C}^{2,s}_{ir}$ is an "open dense" set with trivial homotopy
groups [$\pi_j({\cal C}^{2,s}_{ir})=0$]
in ${\cal C}^{2,s}$
and ${\cal N}^{2,s}_{ir}$ is an "open dense" set in ${\cal N}^{2,s}$.
The complement of ${\cal N}^{2,s}_{ir}$ in ${\cal N}^{2,s}$, the "gauge orbits
of reducible potentials", is a "stratified closed, nowhere dense variety";
therefore, since the action of ${\cal G}^{2,s+1}$ on ${\cal C}^{2,s}$ is not
free, ${\cal N}^{2,s}$ is not a
manifold. Standard methods in elliptic analysis allow us to conclude that there
are local cross sections, the slices ${\cal S}_{{\cal A}_{ir}}$, whose $C^
{\infty}$ topology determines the topology of ${\cal N}^{2,s}_{ir}$.
The closed set $C={\cal C}^{2,s}-{\cal C}^{2,s}_{ir}$ is a "stratified set",
i.e. the union of disjoint submanifolds of ${\cal C}^{2,s}$. If $A\in C$, one
can find an  infinite dimensional subspace of the tangent space to $A$ in
${\cal C}^{2,s}$ "orthogonal to C" at $A$. As a result there is a neighbourhood
V of $A$ in ${\cal C}^{2,s}$ such that $\pi_j({\cal C}^{2,s}-{\cal C}^{2,s}\cap
 C)=0$. The set C is a "bifurcation set" [see Refs.[72] for the study of the
stratification and of its connection with the existence of the various types
of stability subgroups ${\cal G}^{\cal A}$ associated with the gauge
symmetries of certain connections and Refs.[74,28] for the G-space
reinterpretation of the stratification].
Since one can show [21a] that $\pi_j({\tilde {\cal G}}^{2,s+1})\not= 0$ for
some j for $M=S^{4\, or\, 3}$, one has ${\hat {\cal C}}_{ir}^{2,s}\not=
{\cal N}^{2,s}_{ir}\times {\tilde {\cal G}}^{2,s+1}$, so that
the principal bundle ${\cal C}^{2,s}_{ir}({\cal N}
^{2,s}_{ir},{\tilde {\cal G}}^{2,s+1})$ has no flat connection, i.e. it is
nontrivial and has no global cross section and we get the Gribov ambiguity.
See Refs.[21a-h] for a discussion of the Riemannian geometry of ${\cal N}^{2,s}
_{ir}$ in analogy to Eqs(5-29) of the Abelian case.

Therefore all three Sobolev spaces ${\cal C}^{2,s}({\cal N}^{2,s},{\cal
G}^{2,s+1})$, ${\cal C}^{2,s}({\cal N}_{*}^{2,s},{\cal G}_{*}^{2,s+1})$
and \break
${\cal C}_{ir}^{2,s}({\cal N}_{ir}^{2,s},{\tilde {\cal G}}^{2,s+1})$
for G=SU(n) and $M=S^{4\, or\, 3}$, are not trivial principal bundles (due to
the non-vanishing higher homotopy groups of ${\cal G}^{2,s+1}$, ${\cal G}_{*}
^{2,s+1}$ and ${\tilde {\cal G}}^{2,s+1}$ respectively), do not admit global
cross sections [namely global slices or horizontal subspaces (with respect to a
connection on the various bundles ${\cal C}$; this is not true in the
Abelian case as shown by the Coulomb gauge) through a given background
gauge potential $A_o$, which intersect the gauge orbit through $A_o$
(orthogonally with respect to the weak Riemannian structure on the ${\cal
C}$'s) and all other gauge orbits only once] and the orbit spaces have
nontrivial topology. The study of the nontrivial topology of the orbit spaces
is
at the basis of the topological theory of the anomalies, in which one takes
into account the global properties of the group ${\cal G}$ of gauge
transformations; instead the algebraic (or BRST) theory takes into account the
local properties of the algebra $g_{\cal G}$ of ${\cal G}$ (with the residual
gauge freedom of infinitesimal gauge transformations after a BRST gauge-fixing
replaced with the degrees of freedom of the BRST ghosts [5e]), avoiding the
not under control problem to rebuild the infinite-dimensional Hilbert-Lie
group ${\cal G}$ from its Lie algebra $g_{\cal G}$ (see for instance
Refs.[75]).

For $M=S^3$ the same result extends to the three associated phase spaces (see
for instance Ref.[76]): the three corresponding constraint submanifolds are
not trivial principal bundles over the associated reduced phase spaces (phase
space orbit spaces of Dirac's observables), namely there are no global Dirac's
gauge-fixings.

Moreover, in contrast to the Polyakov string [65b],
it is also impossible to make an
incomplete gauge fixing leaving a compact group of residual gauge
transformations due to the fact that the homotopy type of the group of gauge
transformations is different from that of finite dimensional Lie groups [65a].

This phenomenon is due to the fact that in ordinary Sobolev spaces $W^{2,s}(M,
R)$, $s > {1\over 2}dim\, M$, $M=S^3$ or $M=R^3$ with the boundary conditions
(2-40), there can exist connections with non-trivial stability subgroups
${\cal G}^{\cal A}$ of the gauge potentials (gauge symmetries) and/or
${\cal G}^{\Omega}$ of the field strength (gauge copies), creating the
stratifications of the spaces ${\cal C}^{2,s}$ (and of their orbit
spaces) and $T^{*}{\cal C}^{2,s}$. As shown in Refs.[73,30], the gauge
orbits in these functional spaces have a nontrivial behaviour (complicated by
the winding number) and intersect many times every hyperplane in ${\cal
C}^{2,s}$ (or in $T^{*}{\cal C}^{2,s}$) supposed to correspond to a
gauge-fixing, like $\vec \partial \cdot {\vec A}_a(x)=0$ in the Coulomb
gauge, so that one has Gribov copies. For instance, in the Coulomb gauge
hyperplane there is a region $\Gamma$ around the pure gauge background
connection where the determinant of the Faddeev-Popov operator (connected with
the Poisson brackets of the first-class constraints and of the gauge-fixings)
is positive. On the boundary $\Gamma^{*}$ (the first Gribov horizon) of this
so called first Gribov region $\Gamma$,
the Faddeev-Popov operator develops zero
modes; outside $\Gamma^{*}$ there are other Gribov regions with Gribov copies.
As a consequence one cannot invert relevant operators like the Faddeev-Popov
one and find the associated Green functions [77]. By using more sophisticated
functional spaces for ${\cal C}$ in the Euclidean setting (Euclidean
torus) and by studying the stationarity of a functional F (the Hilbert norm of
the gauge transformed $A^U$ as a function of U, if the background connection is
a pure gauge described by the zero gauge potential) over a gauge orbit in
${\cal C}$ (in this way one takes into account only the stability
subgroups ${\cal G}^{\cal A}$, but not ${\cal G}^{\Omega}$), whose local
minima are connected with the Coulomb gauge, it is shown in Refs.[78] that
$\Gamma$ is a bounded convex region, that every gauge orbit intersects
$\Gamma$,
that the gauge orbits of gauge potentials with gauge symmetries (nontrivial
${\cal G}^{\cal A}$) intersect the Gribov horizon $\Gamma^{*}$, but moreover
that there can be Gribov copies also inside $\Gamma$. It is then shown that the
absolute minima of the functional F determine a fundamental modular region
$\Lambda$ without Gribov copies, which however can be present on the boundary
$\Lambda^{*}$ of $\Lambda$ (no way is known to find $\Lambda^{*}$). $\Lambda
^{*}$ lies inside $\Gamma^{*}$, but may have points of contact with
$\Gamma^{*}$
in correspondence of gauge potentials with gauge symmetries (nontrivial
${\cal G}^{\cal A}$). In Ref.[79] it is argued that Gribov copies are present
in
the part of $\Lambda^{*}$ not in contact with $\Gamma^{*}$ and that the
nontrivial homotopy properties of the gauge group and of the orbit spaces [21a]
can be rediscovered by identifying all the sets of points on $\Lambda^{*}$
which are Gribov copies of each other. It is plausible, even if not proved,
that
the gauge potentials ${\vec A}_a$ inside each set of Gribov copies on
$\Lambda^{*}$ would correspond to gauge copies on the same gauge orbit ($\pi_1
(G)=0$) giving the same field strength ${\vec B}_a$ (nontrivial ${\cal G}
^{\Omega}$) and also the same electric  fields (canonical momenta); probably
one should reformulate the approach of Ref.[78] for the gauge orbits in
phase space so to take into account ${\cal G}^{\cal A}$ and ${\cal G}^{\Omega}$
simultaneously (see later on); however the situation could be more complicated
and the connection, if any, between part of the Gribov copies and the gauge
copies has still to be clarified (see for instance the results in Ref.[80]).
Also the general relevance of the Gribov ambiguity for physics, for instance
for the problem of confinement, is far from clear (see for instance Ref.[81]).
In any case it is clear that this phenomenon depends on the choice of the
functional space ${\cal C}$ of gauge potentials and that, if we could
have only fully irreducible connections, ${\cal C}_{fir}$, all these
problems would disappear.

Let us analyze in more detail
the Coulomb gauge-fixing in YM theory on $P^{t}=(R^3\times
\lbrace x^o\rbrace )\times G$ with the boundary conditions of Eqs.(2-40)ii)
by choosing the flat background connection described by the zero gauge
potential; the configuration space of gauge potentials ${\vec A}_a$ is
${\cal C}^3$, while the phase space is $T^{*}{\cal C}={\cal C}^3\times {\cal
E}^3$ ($A^o_a$ and $\pi^o_a$ can be ignored in this discussion), where ${\cal
E}^3$ is the space of electric fields ${\vec \pi}_a=g^{-2}{\vec E}_a$ (they
are momenta and not field strengths on $P^{t}$; instead the field strengths
are the magnetic fields ${\vec B}_a$); ${\cal G}^3$ is the group of fixed
time gauge transformations $U(\vec x,x^o)$ (see the discussion in Sections 2
and 3);
${\cal G}^{\cal A}$ is the stability subgroup of ${\vec A}_a(\vec x,x^o)\in
{\cal C}^3$ (gauge symmetries of ${\vec A}_a$); the stability group ${\cal G}
^{\Omega}$ of the YM theory on $P^4$ defined by $[U,F]=0$, i.e. $[U,\vec E]=
[U,\vec B]=0$, gives rise to a simultaneous stability subgroup ${\cal G}
^{\Omega}\subset {\cal G}^3$ of the momenta ${\vec \pi}_a$ and of the field
strengths ${\vec B}_a$ of the YM theory on $P^{t}$. On $P^{t}$ a priori one
could define two independent stability subgroups: i) ${\cal G}^{\Omega E}
\supseteq {\cal G}^{\Omega}$, ${\cal G}^{\Omega E}\subset {\cal G}^3$, of the
momenta ${\vec \pi}_a=g^{-2}{\vec E}_a$ defined by $[U,{\vec E}_a]=0$; ii)
${\cal G}^{\Omega B}\supseteq {\cal G}^{\Omega}$, ${\cal G}^{\Omega B}\subset
{\cal G}^3$, of the $P^{t}$ field strengths ${\vec B}_a$ (only it would be
responsible of the gauge copies in the $P^{t}$ YM theory); and one has
${\cal G}^{\Omega E}\cap {\cal G}^{\Omega B}={\cal G}^{\Omega}$. However, due
to
the implementation of the Poincar\'e group, as already said, we look at the
$P^{t}$ YM theory as a time slice of the $P^4=M^4\times G$ YM theory, in which
, however, no hypothesis is done on the time asymptotism of both connections
and
gauge transformations. Therefore, we shall consider only ${\cal G}^{\Omega}$.
Now, while ${\cal G}^{\cal A}$ is the stability subgroup of a point ${\vec A}_a
\in {\cal C}^3$, $({\cal G}^{\cal A},{\cal G}^{\Omega})$ is the stability
subgroup of a point $({\vec A}_a,{\vec \pi}_a)\in T^{*}{\cal C}^3={\cal C}^3
\times {\cal E}^3$. Therefore like the various possible types of subgroups
${\cal G}^{\cal A}$ generate the stratification of ${\cal C}^3$ (and of its
reduced orbit space) [72,74,28],
so the various possible types of subgroups ${\cal G}
^{\Omega}$ will generate a stratification of ${\cal E}^3$, so that $T^{*}{\cal
C}^3={\cal C}^3\times {\cal E}^3$ will have a double stratification structure
a priori, of which there is no trace in the literature; therefore, a priori
one expects a doubling of the Gribov ambiguity in phase space in the sense of
Ref.[21a] ( the constraint set, supposed to be a submanifold of $T^{*}{\cal
C}^3$, will be a nontrivial bundle on the reduced phase space of Dirac's
observables); however, since ${\cal G}^{\Omega}$ is also the stability subgroup
of the field strength ${\vec B}_a$ and since, as we shall see, it too is
responsible of the Gribov ambiguity, it is better to speak only of a
generalized
Gribov ambiguity (it would be needed a re-examination of the problem,
in particular of the stratification aspects both from the differential
geometry and functional analysis points of view). When the chosen functional
space is such that ${\cal G}^{\cal A}=Z_G$ one has ${\cal C}^3={\cal
C}^3_{ir}$,
but if ${\cal G}^{\Omega}\supset Z_G$ one has ${\cal E}^3\not= {\cal E}^3_{ir}$
(${\cal E}^3_{ir}$ denotes the space of electric fields with trivial ${\cal G}
^{\Omega}$),
so that $T^{*}{\cal C}^3_{ir}={\cal C}^3\times {\cal E}^3\not= {(T^{*}{\cal C}
^3)}_{ir}={\cal C}^3_{ir}\times {\cal E}_{ir}$; only when fully irreducible
connections alone , for which ${\cal G}^{\Omega}={\cal G}^{\cal A}=Z_G$, are
allowed, one has ${\cal E}^3={\cal E}^3_{ir}$ and then ${\cal C}^3={\cal C}^3
_{fir}$ and $T^{*}{\cal C}^3_{fir}={\cal C}^3_{fir}\times {\cal E}^3_{ir}$.

For a given ${\vec A}(\vec x,x^o)={\vec A}_a(\vec x,x^o){\hat T}^a$ one has the
Coulomb gauge  condition

$${\vec \partial}\cdot {\vec A}_a(\vec x,x^o)=0,\quad \quad \Rightarrow
{\vec A}_a(\vec x,x^o)={\vec A}_{a\perp}(\vec x,x^o).
\form
$$

\noindent Gribov copies will exist if there exist gauge transformations $U
(\vec x,x^o)$ such that the transformed gauge potential ${\vec A}^U_{a\perp}$
is still transverse

$$\eqalign{
{\vec A}^U_{\perp}(\vec x,x^o)&={\vec A}_{\perp}(\vec x,x^o)+U^{-1}(\vec x,
x^o){\hat {\vec D}}^{({\vec A}_{\perp})}U(\vec x,x^o)\cr
\vec \partial \cdot {\vec A}^U_{\perp}(\vec x,x^o)&=\vec \partial \cdot [U
^{-1}(\vec x,x^o){\hat {\vec D}}^{({\vec A}_{\perp})}U(\vec x,x^o)]=0;\cr}
\form
$$

\noindent if $U=e^{\alpha}$, this last equation becomes

$$
\vec \partial \cdot {\vec A}^U_{\perp}(\vec x,x^o)=U^{-1}(\vec x,x^o)[K({\vec
A}_{\perp})\alpha (\vec x,x^o)]U(\vec x,x^o)=0
\form
$$

\noindent where K(.) denotes the Faddeev-Popov operator

$$\eqalign{
K(\vec A)&=-\vec \partial \cdot {\hat {\vec D}}^{({\vec A})}=\triangle \, 1+
[\lbrace {\vec A}_a\cdot \vec \partial +\vec \partial \cdot {\vec A}_a\rbrace
{\hat T}^a,.]\cr
K({\vec A}_{\perp})&=-\vec \partial \cdot {\hat {\vec D}}^{({\vec A}_{\perp})}
=\triangle \, 1+[{\vec A}_{a\perp}\cdot \vec \partial {\hat T}^a,.]\cr}
\form
$$

\noindent When the elliptic operator K has zero modes
(their existence depends on the chosen functional space), in particular when
$K({\vec A}_{\perp})\alpha =0$, then one has Gribov copies.

If we go back to the global form (6-11) of this equation (whose solutions
do not imply a priori the vanishing of the Faddeev-Popov determinant and,
therefore, can give origin to Gribov copies not on the Gribov horizons),
we immediately see:

\noindent i) every connection on $R^3\times G$ with a nontrivial stability
subgroup ${\cal G}^{\cal A}\supset Z_G$, has Gribov copies associated with the
gauge symmetries U solutions of ${\hat {\vec D}}^{({\vec A}_{\perp})}U(\vec x,
x^o)=0$ since for these U's one has $\vec \partial \cdot {\vec A}^U_{\perp}
=0$;\hfill\break
ii) for every connection on $R^3\times G$ with a nontrivial stability subgroup
${\cal G}^{\Omega}\supset Z_G$ for its field strength ${\vec B}_a$, the gauge
transformations $U\in {\cal G}^{\Omega}$ are solutions of the equation $[U,
\vec B]=0$; by using the Bianchi identity ${\hat {\vec D}}^{({\vec A})}
\cdot \vec B\equiv 0$, the defining equation $[U,\vec B]=0$ can be rewritten
as $[U^{-1}{\hat {\vec D}}^{({\vec A})}U,\cdot \, \vec B]=0$ where $U^{-1}
{\hat {\vec D}}^{({\vec A})}U$ is an element ${\vec t}^{\Omega}(\vec x,x^o;
\vec A)$ of the Lie algebra $g^{\Omega}$ of ${\cal G}^{\Omega}$ valued in the
space of functions $\vec g(\vec x,x^o)$ on $R^3$; therefore any such element
of $g^{\Omega}$ which, for ${\vec A}_a={\vec A}_{a\perp}$, is a transverse
function on $R^3$, i.e. $\vec \partial \cdot {\vec t}^{\Omega}(\vec x,x^o;
{\vec A}_{\perp})=0$, gives a Gribov copy being a solution of Eq.(6-11).

Therefore on $P^t=R^3\times G$ with the boundary conditions (2-40)ii) we
expect to have:

\noindent i) Let ${\cal A}\in {\hat {\cal C}}^3_{fir}$, namely ${\vec A}_a\in
{\cal C}^3_{fir}$ is the gauge potential of a fully irreducible connection so
that its stability subgroup ${\cal G}^{\cal A}$ and the stability subgroup
${\cal G}^{\Omega}$ of its field strength $\vec B$ are trivial: ${\cal G}
^{\Omega}={\cal G}^{\cal A}=Z_G$ (neither gauge copies nor gauge symmetries).
Then its gauge orbit ${\cal O}_A$ intersects the transverse functional
hyperplane $\vec \partial \cdot {\vec A}_a=0$ only once so that on the gauge
orbit there is only one transverse gauge potential ${\vec A}_{a\perp}$ for the
given connection and there are no Gribov copies.\hfill\break
ii) Let ${\cal A}\in {\hat {\cal C}}^3_{ir}$, namely
${\vec A}_a\in {\cal C}^3_{ir}$
is the gauge potential of an irreducible connection with gauge copies but no
nontrivial gauge symmetry: ${\cal G}^{\Omega}\supset {\cal G}^{\cal A}=Z_G$.
In this case the gauge orbit has as many intersections with the transverse
hyperplane as gauge copies of the given gauge potential and some of these
Gribov copies can be inside the Gribov horizon $\Gamma^{*}$; inside it,
without points of contact, there is the fundamental modular region $\Lambda$
and on its boundary $\Lambda
^{*}$ there are sets of points which are gauge copies connected by gauge
transformations induced by those $U\in {\cal G}^{\Omega}\supset Z_G$
which are not solutions of Eq.(6-11); these U's should also be responsible
for the Gribov copies which lie in between the boundary $\Lambda^{*}$
of the fundamental region and the Gribov horizon $\Gamma^{*}$,
even if we do not have an explicit demonstration, and of the Gribov copies
on $\Lambda^{*}$ (see however Ref.[80]).\hfill\break
iii) Let ${\cal A}\in {\hat {\cal C}}^3_{red}$, namely ${\vec A}_a\in
{\hat {\cal C}}
^3_{red}$ is the gauge potential of a reducible connection with both gauge
copies and gauge symmetries: ${\cal G}^{\Omega}\supset {\cal G}^{\cal A}
\supset Z_G$. In this case we have all the possible Gribov copies, the
Gribov horizon $\Gamma^{*}$
has points connected with the gauge symmetries and points
connected with the gauge copies for those $U\in {\cal G}^{\Omega}$ giving
solutions of Eq.(6-11), the fundamental modular region boundary $\Lambda^{*}$
has points of tangency with the Gribov horizon $\Omega^{*}$
in connection with the gauge symmetries $U\in {\cal G}^{\cal A}$ and on the
remaining part of the boundary $\Lambda^{*}$ there are Gribov copies
in the sense of gauge copies induced by those $U\in {\cal G}^{\Omega}$
not giving solutions of Eq.(6-11) as in ii).

${}$

The conclusion is that one has to look for more sophisticated functional
spaces than the Sobolev spaces, to see whether Gribov copies and Gribov
horizon can be avoided, so that the whole space is the fundamental
modular region.
In Ref.[20] it is assumed that the gauge potentials and the
electric field strengths belong to the following weighted Sobolev spaces
(see Ref.[82])

$$\eqalign{
{\vec A}_a&\in W^{p,s-1,\delta +1}(R^3,R),\quad \quad i.e.\,\,\,
{\cal C}^3={\cal C}^{3;p,s-1,\delta +1}\, \Rightarrow {\vec B}_a\in W^{p,
s-2,\delta +2},\cr
{\vec \pi}_a&\in W^{p,s-1,\delta +1},\quad \quad i.e.\,\,\,
{\cal E}^3={\cal E}^{3;p,s-1,\delta +1}\cr
&p > 3,\quad \quad s\geq 3,\quad \quad 0\leq \delta \leq 1-{3\over p}\cr}
\form
$$

\noindent In this space one has the asymptotic behaviours

$${\vec A}_a(\vec x,x^o), {\vec \pi}_a(\vec x,x^o)\,
{\rightarrow}_{r\rightarrow
\infty}\, O(r^{-3/2+\epsilon})
\form
$$

\noindent so that both of them are square integrable. In this way one has that
all the Poincar\'e generators (2-21) are convergent well defined real valued
functions on $T^{*}{\cal C}^{3;p,s-1,\delta +1}$ and that the non-Abelian
charges are convergent on the constraint subset of $T^{*}{\cal C}^{3;p,s-1,
\delta +1}$ defined by $\Gamma_a(\vec x,x^o)
\approx 0$. Let us remark that in the YM theory on $P^4=M^4\times G$, one
should
take also $\vec \pi =g^{-2}\vec E$ belonging to the same space as $\vec B$ (for
instance this is done in the first half of Ref.[20] when discussing Gribov
ambiguity); here, due to the absence of requirements on time asymptotism, there
is a difference between $P^4$ and $\cup_{x^o}(R^3\times \lbrace x^o\rbrace )
\times G$.

Eqs.(6-14) are a weakening of Eqs.(2-40)ii), which, however, have to be
mantained with $\epsilon =0$ if one wishes to have neither exploding nor
vanishing non-Abelian charges. Consistency
requires a group ${\cal G}^{3;p,s,\delta}$ of gauge transformations acting on
this space of gauge potentials, but to get Eqs.(3-18) and (3-19) and in
particular ${\cal G}^{3;o}_{\infty}={\cal G}^{3;o(P)}_{\infty}$
(absence of non-rigid improper gauge transformations) one needs to restrict
strongly ${\cal G}^{3;p,s,\delta}$ (whose strongest allowed behaviour is
$r^{-(1
+\epsilon )}$) to a subgroup ${\check {\cal G}}$ whose
Hilbert-Lie algebra $g_{\check {\cal G}}$ contains only
functions such that $\alpha_a(\vec x,x^o)\sim r^{-(5/2+\epsilon )}$ or
$r^{-(3+\epsilon )}$. If this restriction can be defined in a consistent way,
then the conclusions of Section 3 can be made rigorous and the constraint
(presymplectic) submanifold $\bar \gamma$ in phase space would become $\bar
\gamma ={\bar \gamma}_R\times {\bar {\cal G}}^{3;o(P)}_{\infty}$, where
${\bar \gamma}_R$ is the reduced phase space (Hamiltonian orbit space):
$\bar \gamma$ would be a trivial principal bundle and ${\bar \gamma}_R$
would still contain the structures connected with the winding number $n\in Z$,
with the non-Abelian charges and with the center $Z_G$.

For every ${\vec A}_a\in {\cal C}^{3;p,s-1,\delta +1}$, the covariant
derivative define the map ${\hat
{\vec D}}^{(\vec A)}:W^{p,s,\delta }$\break $(R^3,R)
\rightarrow W^{p,s-1,\delta +1}(R^3,R)$, whose formal $L^2$ adjoint is
${\hat {\vec D}}^{(\vec A){*}}=-{\hat {\vec D}}^{(\vec A)}$, so that one has
$\Gamma_a=-{\hat {\vec D}}^{(\vec A)}_{ab}\cdot {\vec \pi}_b=
{\hat {\vec D}}^{(\vec A){*}}_{ab}\cdot {\vec \pi}_b\approx 0$.
Then following the method used in general relativity [83], one can think
to a decomposition in a covariantly transverse part plus the covariant
derivative of something

$$\eqalign{
{\vec \pi}_a(\vec x,x^o)&={\vec \pi}_{a,D\perp}(\vec x,x^o)+{\hat {\vec D}}
^{(\vec A)}_{ab} \phi_b(\vec x,x^o)\cr
&{}\cr
&{\hat {\vec D}}^{(\vec A)}_{ab}\cdot {\vec \pi}_{b,D\perp}(\vec x,x^o)=-{\hat
{\vec D}}^{(\vec A){*}}_{ab}\cdot {\vec \pi}_{b,D\perp}(\vec x,x^o)\equiv 0.\cr
&{}\cr
&{\hat {\vec D}}^{(\vec A){*}}_{ab}\cdot {\vec \pi}_b(\vec x,x^o)=\Gamma_a(\vec
x,x^o)={\hat {\vec D}}^{(\vec A){*}}_{ab}\cdot {\hat {\vec D}}^{(\vec A)}_{bc}
\phi_c(\vec x,x^o)=-{\triangle^{(\vec A)}}_{ab}\phi_b(\vec x,x^o).\cr}
\form
$$

In Ref.[20] it is shown that

$$\triangle^{(\vec A)}=-{\hat {\vec D}}^{(\vec A){*}}\cdot {\hat {\vec D}}
^{(\vec A)}={\hat {\vec D}}^{(\vec A)}\cdot {\hat {\vec D}}^{(\vec A)}:W^{p,s,
\delta}(R^3,R)\rightarrow W^{p,s-2,\delta +2}(R^3,R)
\form
$$

\noindent is an isomorphism; namely it is a second order elliptic operator,
which is a continuous map with closed range and a priori a finite dimensional
kernel; than it is shown that $ker \triangle^{(\vec A)}=0$ by showing that
the equation ${\hat {\vec D}}^{(\vec A)}_{ab}\phi_b(\vec x,x^o)=0$ has only the
solution $\phi_a=0$ so that $\triangle^{(\vec A)}$ is an isomorphism; this also
shows that in these functional spaces there are no gauge symmetries for the
gauge potentials in ${\cal C}^{3;p,s-1,\delta +1}$ [the generators U
of the gauge symmetries cannot vanish asymptotically and thus cannot belong
to $W^{p,s,\delta}(R^3,R)$]: i.e. ${\cal C}_{red}^{3;p,s-1;\delta +1}=0$,
${\cal C}^{3;p,s-1,\delta +1}={\cal C}_{ir}^{3;p,s-1,\delta +1}$ and there is
no stratification due to the ${\cal G}^{\cal A}$'s.

Therefore one obtains

$$\eqalign{
\phi_a(\vec x,x^o)&=-{1\over {\triangle^{(\vec A)}} }(\vec x,x^o)\,
\Gamma_b(\vec x,x^o)\cr
&{}\cr
\pi^i_{a,D\perp}(\vec x,x^o)&={(1-{\hat {\vec D}}^{(\vec A)}
{1\over {\triangle^{(\vec A)}} }{\hat {\vec D}}^{(\vec A)})}^{ij}_{ab}(\vec x
,x^o)\, \pi^j_b(\vec x,x^o)={\cal P}^{(\vec A)ij}_{ab}(\vec x,x^o)
\pi^j_b(\vec x,x^o)\cr}
\form
$$

\noindent where

$${\cal P}^{(\vec A)ij}_{ab}(\vec x,x^o)=\delta^{ij}\delta_{ab}-[{\hat D}^{(
\vec A)i}_{ac}{1\over {\triangle^{(\vec A)}} }_{cd}{\hat D}^{(\vec A)j}_{db}
](\vec x,x^o)
\form
$$

\noindent is a globally defined projection operator onto the covariantly
constant transverse space for each ${\vec A}_a\in {\cal C}^{3;p,s-1,\delta
+1}$.

In Ref.[20] it is also shown that the constraint subset of $T^{*}{\cal C}
^{3;p,s-1,\delta +1}\sim {\cal C}^{3;p,s-1,\delta +1}\times {\cal E}^{3;p,s-1,
\delta +1}$ determined by the first class constraints $\Gamma_a\approx 0$ is
a submanifold of \break
$T^{*}{\cal C}^{3;p,s-1,\delta +1}$. If one could
show that ${\cal E}^{3;p,s-1,\delta +1}={\cal E}_{ir}^{3;p,s-1,\delta +1}$,
i.e. ${\cal G}^{\Omega}={\cal G}^{\cal A}=Z_G$ for all connections, then one
would be sure that ${\cal C}^{3;p,s-1,\delta +1}={\cal C}_{fir}^{3;p,s-1,
\delta +1}$: no Gribov ambiguity would be present, ${\cal C}_{fir}^{3;p,s-1,
\delta +1}$ would be a trivial principal bundle over the reduced orbit
space (with structure group ${\cal G}_{\infty}^{3;o(P)}\subset {\cal
G}_{\infty}
^{3;o;p,s,\delta}$) and the constraint submanifold $\Gamma_a\approx 0$ of
$T^{*}{\cal C}_{fir}^{3;p,s-1,\delta +1}={\cal C}_{fir}^{3;p,s-1,\delta +1}
\times {\cal E}_{ir}^{3;p,s-1,\delta +1}$ a trivial principal bundle over the
reduced phase space of Dirac's observables (with structure group
${\bar {\cal G}}_{\infty}^{3;o(P)}\subset {\bar {\cal G}}_{\infty}
^{3;o;p,s,\delta}$).

For YM theory in Minkowski space-time the existence and uniqueness theorems
were obtained in Refs.[84a,b,c]: for Cauchy data $(\vec A(\vec x,0),\vec
\pi (\vec x,o))$ in the Sobolev space $T^{*}{\cal C}={\cal C}\times {\cal E}=
W^{2,s}(R^3,R)\times W^{2,s-1}(R^3,R)$, $s\geq 2$, which satisfy the constraint
equations $\Gamma_a(\vec x,0)=0$, there exists a global solution $(\vec A(\vec
x,x^o),\vec \pi (\vec x,x^o))$ of YM equations. On the other hand the
regularity
of the constraint set (i.e. its being a submanifold of the phase space and the
absence of connections with gauge symmetries), were just shown to hold in
the previous weighted Sobolev spaces [20]. In Ref.[84d] the regularity of the
constraint set is proved in $W^{2,2}(R^3,R)\times W^{2,1}(R^3,R)$, for which
one already has the existence and uniqueness theorem; again connections with
gauge symmetries are excluded (an U which is a gauge symmetry cannot approach
the boundary limits so quickly; since the number of independent gauge
symmetries is equal to the number of zero eigenvalues of the elliptic
operator ${\hat {\vec D}}^{(\vec A)}$ [${\hat {\vec D}}^{(\vec A)}\phi =0$],
on the compactification $S^3$ of $R^3$ it would also be the topological
charge (due to the index theorem), which vanishes since all principal bundles
over $S^3$ are trivial; this result extends to $R^3$ with suitable boundary
conditions like Eqs.(2-40));
it is however noted that the assumption of
square integrability is violated in the presence of symmetry breaking and
there,
maybe, the connections with gauge symmetries could play some role. Let us
remark
that the boundary conditions for these existence and uniqueness theorems
exclude configurations containing Coulomb charges ($\sim 1/r^2$)  and also
dipole type waves; see Ref.[84e] for more recent developments.

After the previous results let us try to solve Gauss' law 1st class
constraints \break $\Gamma_a(\vec x,x^o)\approx 0$ of Eqs.(2-28); to this end
we need a Green function ${\vec \zeta}^{(\vec A)}_{ab}(\vec x,\vec y;x^o)$ for
the covariant divergence:

$$\eqalign{
{\hat {\vec D}}^{(\vec A)}_{ab}(\vec x,x^o)\cdot {\vec \zeta}^{(\vec A)}_{bc}
(\vec x,\vec y;x^o)&=\vec \partial\cdot {\vec \zeta}^{(\vec A)}_{ac}(\vec x,
\vec y,x^o)+c_{adb}{\vec A}_d(\vec x,x^o)\cdot {\vec \zeta}^{(\vec A)}_{bc}
(\vec x,\vec y;x^o)\cr
&=-\delta_{ac}\delta^3(\vec x-\vec y)\cr}
\form
$$

If we consider the path-dependent or Wu-Yang nonintegrable phase [22,23f]
joining the point $\vec y\in R^3$ to the point $\vec x\in R^3$
along the segment of straight-line (the geodesic joining $\vec x$ to $\vec y$),
we have the following solution of the previous equation:

$$\eqalign{
{\vec \zeta}^{(\vec A)}_{ab}(\vec x,\vec y;x^o)&=\vec c(\vec x-\vec y)\zeta
^{(\vec A)}_{ab}(\vec x,\vec y;x^o)\cr
\zeta^{(\vec A)}_{ab}(\vec x,\vec y;x^o)&={[\zeta^{(\vec A)}(\vec x,\vec y;x^o)
]}_{ab}
={(P\, e^{\int_{\vec y}^{\vec x}d\vec z\cdot {\vec A}_u(\vec z,x^o)
{\hat T}^u})}_{ab}\cr}
\form
$$

\noindent where $\vec c(\vec x-\vec y)$ is the Green function of the
ordinary divergence in the Abelian case, given by Eqs.(5-5) and satisfying
${\vec \partial}_x\cdot \vec c(\vec x-\vec y)=-\delta^3(\vec x-\vec y)$; in
the Abelian limit, $c_{abc}=0$, one has ${\vec \zeta}^{(\vec A)}
_{ab}(\vec x,\vec y;
x^o)=\delta_{ab}\delta^3(\vec x-\vec y)$. The key point is that the operator
$\vec c(\vec x-\vec y)\cdot {\vec \partial}_x$ is the directional derivative
along the straight-line joining $\vec y$ and $\vec x$. Therefore the solution
of the equations $\Gamma_a(\vec x,x^o)=-{\hat {\vec D}}^{(\vec A)}_{ab}(\vec x,
x^o)\cdot {\vec \pi}_b(\vec x,x^o)$ is

$$\eqalign{
\pi^i_a(\vec x,x^o)&=\pi^i_{a,D\perp}(\vec x,x^o)+\int d^3y\, \zeta^{(\vec A)i}
_{ab}(\vec x,\vec y;x^o)\Gamma_b(\vec y,x^o)\cr
&{}\cr
&{\hat {\vec D}}^{(\vec A)}_{ab}(\vec x,x^o)\cdot {\vec \pi}_{b,D\perp}(\vec x,
x^o)\equiv 0.\cr}
\form
$$

The definition of the path-dependent phase  along a path
$\gamma$ in the case $M=R^3$, ($\gamma :[0,1]\rightarrow R^3$, $s\mapsto
\gamma (s)$) is

$$\eqalign{
&V(\gamma ;\vec A)=P\, e^{\int_{\gamma}\, \vec A(\vec x,x^o)\cdot d\vec x}=P\,
e^{\int_0^1ds\, { {d\vec x(s)}\over {ds}}\cdot \vec A(\vec x(s),x^o)}=\cr
&{}\cr
&=\sum_{n=0}^{\infty}\int_0^1ds_n\, \int_0^{s_n}\, \cdots \int_0^{s_2}ds_1\cr
&{{dx^{i_n}(s_n)}\over {ds_n}}A^{i_n}_{a_n}(\vec x(s_n),x^o)\cdots {{dx^{i_1}(
s_1)}\over {ds_1}}A^{i_1}_{a_1}(\vec x(s_1),x^o){\hat T}^{a_n}\cdots {\hat
T}^{a_1},\cr}
\form
$$

\noindent where P denotes path ordering. The transformation properties of
$V(\gamma ,\vec A)$ under a gauge transformation $U(\vec x,x^o)$ is

$$V(\gamma ;\vec A)\rightarrow V(\gamma ,{\vec A}^U)=U(\gamma (1),x^o)
V(\gamma ,\vec A)U^{-1}(\gamma (0),x^o).
\form
$$

\noindent This implies that, under infinitesimal gauge transformations $U=1+
\alpha_a{\hat T}^a$, $\zeta^{(\vec A)}(\vec x,\vec y;x^o)$ transforms as

$$\eqalign{
\zeta^{(\vec A)}_{ab}(\vec x,\vec y;x^o)&\rightarrow {[(1+\alpha_u(\vec x,x^o)
{\hat T}^u)\zeta^{(\vec A)}(\vec x,\vec y;x^o)(1-\alpha_v(\vec y,x^o){\hat
T}^v)
]}_{ab}\approx \cr
&\approx \zeta^{(\vec A)}_{ab}(\vec x,\vec y;x^o)+\cr
&+{[\alpha_u(\vec x,x^o){\hat T}^u\zeta^{(\vec A)}(\vec x,\vec y;x^o)]}_{ab}-
{[\alpha_u(\vec y,x^o)\zeta^{(\vec A)}(\vec x,
\vec y;x^o){\hat T}^u]}_{ab}=\cr
&=\zeta^{(\vec A)}_{ab}(\vec x,\vec y;x^o)+\cr
&+\int d^3z\, \alpha_u(\vec z,x^o)\lbrack
\delta^3(\vec x-\vec z)c_{uac}\zeta^{(\vec A)}_{cb}(\vec z,\vec y;x^o)-\zeta
^{(\vec A)}_{ac}(\vec x,\vec z;x^o)c_{ucb}\delta^3(\vec z-\vec y)\rbrack =\cr
&=\zeta^{(\vec A)}_{ab}(\vec x,\vec y;x^o)+\int d^3z\, \alpha_u(\vec z,x^o)
\lbrace \zeta^{(\vec A)}_{ab}(\vec x,\vec y;x^o),
\Gamma_u(\vec z,x^o)\rbrace ,\cr}
\form
$$

\noindent where in the last line we have introduced the phase space generator
of the gauge transformation; by comparison we get the following action of
Gauss' law constraints on the bilocal Green function

$$\eqalign{
\lbrace &\zeta^{(\vec A)}_{ab}(\vec x,\vec y;x^o),\Gamma_u(\vec z,x^o)\rbrace
=\cr
&=\delta^3(\vec x-\vec z)c_{uac}\zeta^{(\vec A)}_{cb}(\vec z,\vec y;x^o)-\zeta
^{(\vec A)}_{ac}(\vec x,\vec z;x^o)c_{ucb}\delta^3(\vec z-\vec y).\cr}
\form
$$

Since Eq.(6-16) implies

$$\eqalign{
\pi^i_{a,D\perp}(\vec x,x^o)&=\int d^3y\, [\delta^{ij}\delta_{ab}\delta^3(\vec
x-\vec y)+\zeta^{(\vec A)i}_{ac}(\vec x,\vec y;x^o){\hat D}^{(\vec A)j}_{cb}
(\vec y,x^o)]\pi^j_b(\vec y,x^o)=\cr
&=\int d^3y\, {\cal P}^{(\vec A)ij}_{ab}(\vec x,\vec y;x^o)\pi^j_b(\vec y,x^o)
,\cr}
\form
$$

\noindent
then by using Eqs.(2-29) one gets that ${\vec \pi}_{a,D\perp}(\vec x,x^o)$
behaves as a tensor like ${\vec \pi}_a(\vec x,x^o)$:

$$\lbrace \pi^i_{a,D\perp}(\vec x,x^o),\Gamma_b(\vec y,x^o)\rbrace =-c_{acb}
\pi^i_{c,D\perp}(\vec x,x^o)\delta^3(\vec x-\vec y).
\form
$$

Since the path-dependent phase performs the parallel transport of geometrical
objects belonging to representations of G, along a curve $\gamma$ in $M=R^3$,
associated with the gauge potential $A\in {\cal C}^{3;p,s-1 ,\delta +1}$, its
${\cal A}$-horizontal lift to $P^t=R^3\times G$ is the parallel
transporter on P along an ${\cal A}$-horizontal lift $\gamma^{\cal A}_{p_o}
(s)$ [starting at any point $p_o\in \pi^{-1}(\gamma (0))\subset P^t$ of the
fiber over $\gamma (0)$] of $\gamma (s)$ associated with the connection 1-form
$\omega^{\cal A}$ describing the given connection ${\cal A}\in {\hat {\cal
C}}^{3;p,s-1,\delta +1}$; if the parallel transporter on M is globally defined
for any pair of points in M, then the parallel transporter in $P^t$ will be
globally defined on $P^t$, and this can happen only if all the points of
$P^t$ can be connected by parallel transport; but this implies that the
holonomy bundle of the connection coincides with $P^t$, $P^{\cal A}=P^t$, and
this may happen only
is the connection is completely irreducible, ${\cal A}\in {\hat {\cal C}}^{3;
p,s-1,\delta +1}_{fir}$, i.e. $\Phi^{\cal A}=G$. Therefore the global
existence of the Green function ${\vec \zeta}^{(\vec
A)}_{ab}(\vec x,\vec y;x^o)$ of
the covariant divergence, which is assured from being the covariant divergence
an elliptic operator without zero modes in the weighted Sobolev space
$W^{p,s-1,\delta +1}(R^3,R)$ of Ref.[20], is synonimus of the fact that in such
spaces only fully irreducible connections are present, just those for which the
Gribov ambiguity is absent so that ${\cal C}^{3;p,s-1,\delta +1}={\cal C}_{fir}
^{3;p,s-1,\delta +1}$ is a trivial principal bundle over the orbit space
${\cal N}_{fir}^{3;p,s-1,\delta +1}$ with fiber the group of gauge
transformations ${\cal G}^{p,s,\delta}$. But then, for what has been said
previously, also ${\cal E}^{3;p,s-1,\delta +1}={\cal E}_{ir}^{3;p,s-1,\delta
+1}
$ since ${\cal G}^{\Omega}={\cal G}^{\cal A}=Z_G$ and there is no Gribov
ambiguity.

We shall now evaluate other Green functions, which will be useful later on.

First of all let us consider the following operator which coincides with
the Faddeev-Popov operator on transverse gauge potentials

$$\eqalign{
{\tilde K}_{ab}(\vec A)(\vec x,x^o)&=-{\hat {\vec D}}^{(\vec A)}(\vec x,x^o)
\cdot {\vec \partial}_x=-{\vec \partial}_x\cdot {\hat {\vec D}}^{(\vec A)}
(\vec x,x^o)-c_{abc}({\vec \partial}_x\cdot {\vec A}_c(\vec x,x^o))=\cr
&=K(\vec A)-c_{abc}{\vec \partial}_x\cdot {\vec A}_c(\vec x,x^o)\cr
{\tilde K}_{ab}({\vec A}_{\perp})(\vec x,x^o)&=K({\vec A}_{\perp})(\vec x,x^o)
\cr
&{}\cr
{\tilde K}_{ab}(\vec A)(\vec x,x^o)\delta^3(\vec x-\vec y)&={\tilde K}_{ab}
(\vec A)(\vec x,\vec y;x^o)=-{\hat {\vec D}}^{(\vec A)}(\vec x,x^o)\triangle_x
\cdot { {{\vec \partial}_x}\over {\triangle_x} }\delta^3(\vec x-\vec y)=\cr
&=-{\hat {\vec D}}^{(\vec A)}(\vec x,x^o)\triangle_x \cdot \vec c(\vec x-\vec
y)
;\cr}
\form
$$

\noindent its Green function satisfies

$$\eqalign{
&G^{(\vec A)}_{\tilde K,ab}(\vec x,\vec y;x^o)={({\tilde K}^{-1})}_{ab}(\vec A)
(\vec x,x^o)\delta^3(\vec x-\vec y)\cr
&{}\cr
&{\tilde K}_{ab}(\vec A)(\vec x,x^o)G^{(\vec A)}_{\tilde K,bc}(\vec x,\vec y;
x^o)=\delta_{ac}\delta^3(\vec x-\vec y).\cr}
\form
$$

The second operator is

$$\eqalign{
Z^{(\vec A)}(\vec x,x^o)&=I+\vec A(\vec x,x^o)\cdot { {{\vec \partial}_x}\over
{\triangle_x} }=I+{\hat T}^c{\vec A}_c(\vec x,x^o)\cdot { {{\vec \partial}_x}
\over {\triangle_x} }=\tilde K (\vec A)(\vec x,x^o){1\over {\triangle_x}}\cr
&{}\cr
Z^{(\vec A)}_{ab}(\vec x,x^o)&=(\delta_{ab}\triangle_x+c_{abc}{\vec A}_c(\vec x
,x^o))\cdot {\vec \partial}_x){1\over {\triangle_x}}=-(\delta_{ab}{\vec
\partial}_x+c_{acb}{\vec A}_c(\vec x,x^o))\cdot {1\over {\triangle_x}}\cr}
\form
$$

\noindent and for its Green function one has

$$\eqalign{
&Z^{(\vec A)}_{ab}(\vec x,x^o)\delta^3(\vec x-\vec y)=Z^{(\vec A)}_{ab}(\vec x,
\vec y;x^o)=-{\hat {\vec D}}^{(\vec A)}_{ab}(\vec x,x^o)\cdot \vec c(\vec x-
\vec y)\cr
&{}\cr
&G^{(\vec A)}_{Z,ab}(\vec x,\vec y;x^o)={(Z^{-1})}^{(\vec A)}_{ab}(\vec x,x^o)
\delta^3(\vec x-\vec y)=\triangle_x\, G^{(\vec A)}_{\tilde K,ab}(\vec x,\vec y;
x^o)\cr
&{}\cr
&Z^{(\vec A)}_{ab}(\vec x,x^o)G^{(\vec A)}_{Z,bc}(\vec x,\vec y;x^o)=-{\hat
{\vec D}}^{(\vec A)}_{ab}(\vec x,x^o)\cdot \int d^3z\, \vec c(\vec x-\vec z)
G^{(\vec A)}_{Z,bc}
(\vec z,\vec y;x^o)=\delta_{ac}\delta^3(\vec x-\vec y)\cr}
\form
$$

{}From Eq.(6-20) we obtain

$${ {{\vec \partial}_x}\over {\triangle_x}}\, G^{(\vec A)}_{Z,ab}(\vec x,\vec
y;
x^o)=\int d^3z\, \vec c(\vec x-\vec z)G^{(\vec A)}_{Z,ab}(\vec z,\vec y;x^o)=
{\vec \zeta}^{(\vec A)}_{ab}(\vec x,\vec y;x^o).
\form
$$

Therefore we get the following solutions for the two Green functions

$$\eqalign{
G^{(\vec A)}_{Z,ab}(\vec x,\vec y;x^o)&={(Z^{-1})}_{ab}^{(\vec A)}(\vec x,x^o)
\delta^3(\vec x-\vec y)=[\triangle_x{({\tilde K}^{-1})}_{ab}(\vec A)](\vec x,
x^o)\delta^3(\vec x-\vec y)=\cr
&=-{\vec \partial}_x\cdot {\vec \zeta}^{(\vec A)}_{ab}(\vec x,\vec y;x^o)\cr
&{}\cr
G^{(\vec A)}_{\tilde K,ab}(\vec x,\vec y;x^o)&={({\tilde K}^{-1})}_{ab}(\vec A)
(\vec x,x^o)\delta^3(\vec x-\vec y)=\cr
&=-{ {{\vec \partial}_x}\over {\triangle_x} }\cdot {\vec \zeta}^{(\vec A)}_{ab}
(\vec x,\vec y;x^o)=-\int d^3z\, \vec c(\vec x-\vec z)\cdot {\vec \zeta}^{(\vec
A)}_{ab}(\vec z,\vec y;x^o)\cr}
\form
$$

Let us remark that from the last equation we obtain the result that the
inverse of the Faddeev-Popov $K({\vec A}_{\perp})(\vec x,x^o)$ exists in
these weighted Sobolev space (no zero modes) for transverse gauge potentials.

The next operator is the one of Eq.(6-17)

$$\eqalign{
\triangle^{(\vec A)}_{ab}(\vec x,x^o)&={\hat {\vec D}}^{(\vec A)}_{ac}(\vec x,
x^o)\cdot {\hat {\vec D}}^{(\vec A)}_{cb}(\vec x,x^o),\quad \quad \triangle
^{(\vec A)}=\triangle^{(\vec A){*}}\cr
&{}\cr
\triangle^{(\vec A)}_{ab}(\vec x,\vec y;x^o)&=\triangle^{(\vec A)}_{ab}(\vec x,
x^o)\delta^3(\vec x-\vec y)\cr
&{}\cr
G^{(\vec A)}_{\triangle ,ab}(\vec x,\vec y;x^o)&={(\triangle^{-1})}^{(\vec A)}
_{ab}(\vec x,x^o)\delta^3(\vec x-\vec y)\cr
&{}\cr
\triangle^{(\vec A)}_{ab}(\vec  x,x^o)\, G^{(\vec A)}_{\triangle ,bc}(\vec x,
\vec y;x^o)&=\int d^3z\, \triangle^{(\vec A)}_{ab}(\vec x,\vec z;x^o)\, G^{(
\vec A)}_{\triangle ,bc}(\vec z,\vec y;x^o)=\cr
&={\hat {\vec D}}^{(\vec A)}_{au}(\vec x,x^o)\cdot \lbrack {\hat {\vec D}}^{(
\vec A)}_{ub}(\vec x,x^o)\, G^{(\vec A)}_{\triangle ,bc}(\vec x,\vec y;x^o)
\rbrack =\delta_{ac}\delta^3(\vec x-\vec y)\cr
&{}\cr
G^{(\vec A)}_{\triangle ,uv}(\vec y,\vec z;x^o)&=\int d^3x\, G^{(\vec A)}
_{\triangle ,ua}(\vec y,\vec x;x^o)\triangle^{(\vec A)}_{ab}(\vec
x,x^o)G^{(\vec
A)}_{\triangle ,bv}(\vec x,\vec z;x^o)=\cr
&=\int d^3x\, [\triangle^{(\vec A)}_{ba}(\vec x,x^o)G^{(\vec A)}_{\triangle ,
ua}(\vec y,\vec x;x^o)]G^{(\vec A)}_{\triangle ,bv}(\vec x,\vec z;x^o),\cr
&{}\cr
\triangle^{(\vec A)}_{ac}(\vec x,x^o)G^{(\vec A)}_{\triangle ,bc}(\vec y,\vec
x;x^o)&={\hat {\vec D}}^{(\vec A)}_{au}(\vec x,x^o)\cdot [{\hat {\vec
D}}^{(\vec
A)}_{uc}(\vec x,x^o)G^{(\vec A)}_{\triangle ,bc}(\vec y,\vec x;x^o)]=\cr
&=\delta_{ab}\delta^3(\vec x-\vec y),\cr}
\form
$$

\noindent which for $c_{abc}=0$ becomes $\triangle^{(\vec A)}_{ab}(\vec x^o)
\rightarrow -\triangle_x$. By using Eqs.(6-33), (6-34) one obtains

$$\eqalign{
&{\hat {\vec D}}^{(\vec A)}_{ab}(\vec x,x^o)\, G^{(\vec A)}_{\triangle ,bc}
(\vec x,\vec y;x^o)=-{\vec \zeta}^{(\vec A)}_{ac}(\vec x,\vec y;x^o)=\cr
&=-{ {{\vec \partial}_x}\over {\triangle_x} }[{(Z^{-1})}^{(\vec A)}_{ac}(\vec
x,
x^o)\delta^3(\vec x-\vec y)]=-{\vec \partial}_x{({\tilde K}^{-1})}{(\vec A)}
_{ac}(\vec x,x^o)\delta^3(\vec x-\vec y)\cr
&{}\cr
&{\hat {\vec D}}^{(\vec A)}_{ac}(\vec x,x^o)G^{(\vec A)}_{\triangle ,bc}(\vec
y,
\vec x;x^o)=-{\vec \zeta}^{(\vec A)}_{ab}(\vec x,\vec y;x^o)\cr
&{}\cr
&G^{(\vec A)}_{\tilde K,ac}(\vec x,\vec y;x^o)={({\tilde K}^{-1})}_{ac}(\vec
A)(\vec x,x^o)\delta^3(\vec x-\vec y)=-{ {{\vec \partial}_x}\over {\triangle_x
} }\cdot {\vec \zeta}^{(\vec A)}_{ac}(\vec x,\vec y;x^o)\cr
&{}\cr
&G^{(\vec A)}_{Z,ac}(\vec x,\vec y;x^o)
={(Z^{-1})}^{(\vec A)}_{ac}(\vec x,x^o)\delta^3(\vec x-\vec y)=[\triangle_x\,
{({\tilde K}^{-1})}_{ac}(\vec A)](\vec x,x^o)\delta^3(\vec x-\vec y)=\cr
&=-[{\tilde K}_{ab}(\vec A)(\vec x,x^o)+c_{abv}({\vec \partial}_x\cdot {\vec
A}_v(\vec x,x^o))]\, G^{(\vec A)}_{\triangle ,bc}(\vec x,\vec y;x^o)=\cr
&-K_{ab}(\vec A)(\vec x,x^o)\, G^{(\vec A)}_{\triangle ,bc}(\vec x,
\vec y;x^o)={\vec \partial}_x\cdot {\hat {\vec D}}^{(\vec A)}_{ab}(\vec x,x^o)
\, G^{(\vec A)}_{\triangle ,bc}(\vec x,\vec y;x^o)=\cr
&=-{\vec \partial}_x\cdot {\vec \zeta}^{(\vec A)}_{ac}(\vec x,\vec y;x^o)\cr}
\form
$$

\noindent and

$$\eqalign{
[\delta_{ab}+&{({\tilde K}^{-1})}_{au}(\vec A)(\vec x,x^o)c_{ubv}
({\vec \partial}_x\cdot {\vec A}_v(\vec x,x^o))]\, G^{(\vec A)}_{\triangle ,bc}
(\vec x,\vec y;x^o)=\cr
&=-[{({\tilde K}^{-1})}_{au}(\vec A)\, \triangle_x\, {({\tilde K}
^{-1})}_{uc}(\vec A)](\vec x,x^o)\delta^3(\vec x-\vec y)\cr}
\form
$$

For transverse potentials $\vec A={\vec A}_{\perp}$ ($\vec \partial \cdot
{\vec A}_{\perp}\equiv 0$), one gets

$$-\triangle^{({\vec A}_{\perp})}_{ab}(\vec x,x^o)={[K({\vec A}_{\perp})
\, {1\over {\triangle_x} }\, K({\vec A}_{\perp})](\vec x,x^o)}_{ab}
\form
$$

\noindent so that we get the following solution for the Green function of Eqs.
(6-35) in the case of transverse potentials

$$\eqalign{
G^{({\vec A}_{\perp})}_{\triangle ,ab}(\vec x,\vec y;x^o)&={(\triangle^{-1})}
^{({\vec A}_{\perp})}_{ab}(\vec x,x^o)\delta^3(\vec x-\vec y)=\cr
&=-[{(K^{-1})}_{au}({\vec A}_{\perp})\, \triangle_x{(K^{-1})}_{ub}({\vec A}
_{\perp})](\vec x,x^o)\delta^3(\vec x-\vec y)=\cr
&=-\int d^3z\, G^{({\vec A}_{\perp})}_{K,au}(\vec x,\vec z;x^o)\, G^{({\vec A}
_{\perp})}_{Z,ub}(\vec z,\vec y;x^o)=\cr
&=-\int d^3z\, {1\over {\triangle_x}}{\vec \partial}_x\cdot {\vec
\zeta}^{({\vec
A}_{\perp})}_{au}(\vec x,\vec z;x^o){\vec \partial}_z\cdot {\vec \zeta}^{({\vec
A}_{\perp})}_{ub}(\vec z,\vec y;x^o)=\cr
&=-\int d^3z_1d^3z_2\, \vec c(\vec x-{\vec z}_1)\cdot {\vec \zeta}^{({\vec A}
_{\perp})}_{au}({\vec z}_1,{\vec z}_2;x^o)\, {\vec \partial}_{z_2}\cdot {\vec
\zeta}_{ub}^{({\vec A}_{\perp})}({\vec z}_2,\vec y;x^o)\cr}
\form
$$

Since for transverse potentials $A_{\perp}$ one has $\tilde K({\vec A}_{\perp})
=K({\vec A}_{\perp})$ and $K({\vec A}_{\perp})=K^{*}({\vec A}_{\perp})$, one
also gets

$$\eqalign{
G^{({\vec A}_{\perp})}_{\triangle ,ab}(\vec x,\vec y;x^o)&=
-\int d^3z\, G^{({\vec A}_{\perp})}_{K,au}(\vec x,\vec z;x^o)\, G^{({\vec A}
_{\perp})}_{Z,ub}(\vec z,\vec y;x^o)=\cr
&=-\int d^3z\, G^{({\vec A}_{\perp})}_{K,ua}(\vec z,\vec x;x^o)\, G^{({\vec A}
_{\perp})}_{Z,ub}(\vec z,\vec y;x^o)=\cr
&=-\int d^3z {1\over {\triangle_z}}{\vec \partial}_z\cdot {\vec \zeta}^{({\vec
A}_{\perp})}_{ua}(\vec z,\vec x;x^o){\vec \partial}_z\cdot {\vec \zeta}
^{({\vec A}_{\perp})}_{ub}(\vec z,\vec y;x^o)=\cr}
\form
$$

Since from Eq.(6-20) we have

$$\eqalign{
{\vec \partial}_x\cdot {\vec \zeta}^{(\vec A)}_{ab}(\vec x,\vec y;x^o)&=-\delta
_{ab}\delta^3(\vec x-\vec y)-c_{auc}{\vec A}_u(\vec x,x^o)\cdot {\vec \zeta}
_{cb}^{(\vec A)}(\vec x,\vec y;x^o)=\cr
&=-\delta_{ab}\delta^3(\vec x-\vec y)+{({\hat T}^u)}_{ac}{\vec A}_u(\vec x,x^o)
\cdot {\vec \zeta}^{(\vec A)}_{cb}(\vec x,\vec y;x^o)\cr}
\form
$$

\noindent we obtain the following forms of the three Green functions by using
Eqs.(6-21) and (6-35)

$$\eqalign{
G^{(\vec A)}_{\tilde K,ab}(\vec x,\vec y;x^o)&={({\tilde K}^{-1})}_{ab}(\vec A)
(\vec x,x^o)\delta^3(\vec x-\vec y)=\cr
&=-\int d^3z\, [{\vec \partial}_xc(\vec x-\vec z)]\cdot [{\vec \partial}_zc
(\vec z-\vec y)]\zeta^{(\vec A)}_{ab}(\vec z,\vec y;x^o)\cr}
\form
$$

$$\eqalign{
G^{(\vec A)}_{Z,ab}(\vec x,\vec y;x^o)&={(Z^{-1})}_{ab}^{(\vec A)}(\vec x,x^o)
\delta^3(\vec x-\vec y)=\cr
&=\delta_{ab}\delta^3(\vec x-\vec y)-{(\vec A(\vec x,x^o))}_{ac}\cdot [{\vec
\partial}_xc(\vec x-\vec y)]\zeta^{(\vec A)}_{cb}(\vec x,\vec y;x^o)\cr}
\form
$$

\noindent and, after an integration by parts, we have

$$\eqalign{
&G^{({\vec A}_{\perp})}_{\triangle ,ab}({\vec y}_1,{\vec y}_2;x^o)={(\triangle
^{-1})}^{({\vec A}_{\perp})}_{ab}({\vec y}_1,x^o)\delta^3({\vec y}_1-{\vec y}
_2)=\cr
&{}\cr
&=\delta_{ab}\, c({\vec y}_1-{\vec y}_2)-2\int d^3z\, c({\vec y}_1-\vec z)
[{\vec \partial}_zc(\vec z-{\vec y}_2)]\cdot {({\vec A}_{\perp}(\vec z,x^o))}
_{au}\, \zeta^{({\vec A}_{\perp})}_{ub}(\vec z,{\vec y}_2;x^o)+\cr
&+\int d^3z_1d^3z_2\, c({\vec y}_1-{\vec z}_1)[\partial^h_{z_1}c({\vec z}_1-
{\vec z}_2)]\, [\partial^k_{z_2}c({\vec z}_2-{\vec y}_2)]\cr
&{(A^h_{\perp}({\vec z}_1,x^o))}_{au}\zeta_{uv}^{({\vec A}_{\perp})}({\vec z}_1
,{\vec z}_2;x^o){(A^k_{\perp}({\vec z}_2,x^o))}_{vr}\zeta_{rb}^{({\vec A}
_{\perp})}({\vec z}_2,{\vec y}_2;x^o)\cr}
\form
$$

\noindent where Eq.(6-20) and the following notations are used

$$c(\vec x)={1\over {\triangle} }\delta^3(\vec x)={ {-1}\over {4\pi | \vec x|}}
,\quad \quad \vec c(\vec x)={\vec \partial}c(\vec x)
\form
$$

$${(\vec A(\vec x,x^o))}_{ab}={\vec A}_c(\vec x,x^o){({\hat T}^c)}_{ab}=
c_{abc}{\vec A}_c(\vec x,x^o).
\form
$$

Eq.(6-18) now becomes

$$\phi_a(\vec x,x^o)=-\int d^3z\, G^{(\vec A)}_{\triangle ,ab}(\vec x,\vec z;
x^o)\Gamma_b(\vec z,x^o)
\form
$$

\noindent and the first line of Eqs.(6-16) and Eq.(6-22) agree due to Eqs.(6-
36), i.e.

$$\eqalign{
{\vec \pi}_a(\vec x,x^o)&={\vec \pi}_{a,D\perp}(\vec x,x^o)+\int d^3y\, {\vec
\zeta}^{(\vec A)}_{ac}(\vec x,\vec y;x^o)\Gamma_c(\vec y;x^o)=\cr
&={\vec \pi}_{a,D\perp}(\vec x,x^o)-{\hat {\vec D}}^{(\vec A)}_{ab}(\vec x,x^o)
\int d^3y\, G^{(\vec A)}_{\triangle ,bc}(\vec x,\vec y;x^o)\Gamma_c(\vec y,x^o)
\cr}
\form
$$

\noindent Then Eqs.(6-18) agree with Eq.(6-27), so that the projector (6-19)
can
now be written as

$$\eqalign{
{\cal P}^{(\vec A)\, ij}_{ab}(\vec x,\vec
y;x^o)&=\delta^{ij}\delta_{ab}\delta^3
(\vec x-\vec y)-{\hat D}^{(\vec A)i}_{ad}(\vec x,x^o)G^{(\vec A)}_{\triangle ,
dc}(\vec x,\vec y;x^o){\hat D}^{(\vec A)j}_{cb}(\vec y,x^o)=\cr
&=\delta^{ij}\delta_{ab}\delta^3(\vec x-\vec y)+\zeta^{(\vec A)i}_{ac}(\vec x,
\vec y;x^o){\hat D}^{(\vec A)j}_{cb}(\vec y,x^o)\cr
&{}\cr
\int d^3y\, {\cal P}^{(\vec A)\, ij}_{ab}(\vec x,\vec y;x^o)&{\hat D}^{(\vec A)
\, j}_{bc}(\vec y,x^o)=\int d^3x\, {\cal P}^{(\vec A)\, ij}_{bc}(\vec x,\vec y;
x^o)=0\cr
&{}\cr
\int d^3y\, {\cal P}^{(\vec A)\, ij}_{ab}(\vec x,\vec y;x^o)&{\cal P}^{(\vec A)
\, jk}_{bc}(\vec y,\vec z;x^o)={\cal P}^{(\vec A)\, ik}_{ac}(\vec x,\vec z;x^o)
.\cr}
\form
$$

\noindent This is the form of the projector of Refs.[21a-i] in these functional
spaces.

{}From Eqs.(6-26) one obtains

$$\eqalign{
-{\hat D}^{(\vec A)\, h}_{uv}(\vec z,x^o)&\lbrace \zeta^{(\vec A)}_{ab}(\vec x,
\vec y;x^o),\pi^h_v(\vec z,x^o)\rbrace =\cr
&=\delta^3(\vec x-\vec z)c_{uac}\zeta^{(\vec A)}_{cb}(\vec z,\vec y;x^o)-\zeta
^{(\vec A)}_{ac}(\vec x,\vec z;x^o)c_{ucb}\delta^3(\vec z-\vec y),\cr}
\form
$$

\noindent so that , discarding homogeneous solutions, one gets

$$\eqalign{
&\lbrace \zeta^{(\vec A)}_{ab}(\vec x,\vec y;x^o),\pi^h_v(\vec z,x^o)\rbrace =
\cr
&=\int d^3w\, \zeta^{(\vec A)\, h}_{vs}(\vec z,\vec w;x^o)[\delta^3(\vec x-\vec
w)c_{sac}\zeta^{(\vec A)}_{cb}(\vec w,\vec y;x^o)-\zeta^{(\vec A)}_{ac}(\vec x,
\vec w;x^o)c_{scb}\delta^3(\vec w-\vec y)]=\cr
&=\zeta^{(\vec A)\, h}_{vs}(\vec z,\vec x;x^o)c_{sac}\zeta^{(\vec A)}_{cb}(\vec
x,\vec y;x^o)-\zeta^{(\vec A)\, h}_{vs}(\vec z,\vec y;x^o)\zeta^{(\vec A)}_{ac}
(\vec x,\vec y;x^o)c_{scb}.\cr}
\form
$$

\vfill\eject

\bigskip\noindent
{\bf 7. Coordinatization of the Principal Bundle and Decomposition of the
Gauge Potential.}
\newcount \nfor

\def \form {\global \advance \nfor by 1 \eqno (7.\the\nfor)}
\bigskip

In Section 2 we found that the vector fields on $T^{*}{\cal C}_{fir}^{3;p,
s-1,\delta +1}$

$$
X_a(\vec x,x^o)=-\lbrace .,\Gamma_a(\vec x,x^o)\rbrace =\lbrace .,{\hat {\vec
D}}^{(\vec A)}_{ab}(\vec x,x^o)\cdot {\vec \pi}_b(\vec x,x^o)\rbrace
\form
$$

\noindent satisfy the algebra (see after Eqs.(2-32))

$$
[X_a(\vec x,x^o),X_b(\vec x,x^o)]=\delta^3(\vec x-\vec y)c_{abc}X_c(\vec x,x^o)
\form
$$

\noindent Being the infinitesimal generators of the proper gauge
transformations
, they are the vertical fundamental vector fields tangent to the gauge orbits
(see Appendix B) on the presymplectic submanifold and trivial principal
bundle $\bar \gamma ={\bar \gamma}_R\times {\bar {\cal G}}^{o(P)}_{\infty}$,
where $\bar \gamma$ is the constraint submanifold $\Gamma_a(\vec x,x^o)\approx
0$ in $T^{*}{\cal C}_{fir}^{3;p,s-1,\delta +1}={\cal C}_{fir}^{3;p,s-1,\delta
+1}\times {\cal E}_{ir}^{3;p,s-1,\delta +1}$ (the full phase space is $T^{*}
{\cal C}_{fir}^{4;p,s-1,\delta +1}$ with $A^o_a(\vec x,x^o)$ and $\pi^o_a(\vec
x,x^o)\approx 0$ taken into account), ${\bar \gamma}_R$ is the reduced phase
space of Dirac's observables, ${\bar {\cal G}}_{\infty}^{o(P)}$ is the
suitable restriction of ${\bar {\cal G}}_{\infty}^{3;p,s,\delta}$ to have
${\bar {\cal G}}_{\infty}^{o(P)}={\bar {\cal G}}_{\infty}^o$ (the winding
number, the non-Abelian charges and the center $Z_G$ components, ${\bar {\cal
G}}/{\bar {\cal G}}_{\infty}^{o(P)}=G\times Z$ are not considered at this
stage). Since the gauge group ${\bar {\cal G}}_{\infty}^{o(P)}$, restricted to
a gauge orbit through ${\vec A}_a(\vec x,x^o)\in {\cal C}_{fir}^{3;p,s-1,
\delta +1}$, which is a gauge potential of a connection on $P^{t}=(R^3\times
\lbrace x^o\rbrace )\times G$, is isomorphic to the fiber $G_{(\vec x,x^o)}
\sim G$ over $\vec x$ at $x^o$, then, remembering Eqs.(A-2)
of Appendix A on the group
manifold of G, one can try to introduce for it the following parametrization
in a tubolar neighbourhood of the identity cross section of $P^{t}$ where
canonical coordinates of 1st kind can be safely introduced

$$
X_a(\vec x,x^o)=B_{ba}(\eta (\vec x,x^o)){ {\tilde \delta}\over {\delta \eta_b
(\vec x,x^o)} }
\form
$$

\noindent Here: i) $\eta_a(\vec x,x^o)$ are coordinates on the group manifold
of ${\bar {\cal G}}_{\infty}^{o(P)}$ at fixed $\vec x$ and $x^o$, i.e. they
are coordinates on the fiber $G_{(\vec x,x^o)}\sim G$ of $P^{t}$, which can
be considered as the group manifold of ${\bar {\cal G}}_{\infty}^{o(P)}$;
ii) the
matrix B is the solution of Eqs.(A-3) (its inverse $A=B^{-1}$ has the form of
Eqs.(A-5) if $\eta_a(\vec x,x^o)$ are canonical coordinates of 1st kind for
a neighbourhood of the identity of $G_{(\vec x,x^o)}\sim G$); iii) the
functional derivative $\tilde \delta /\delta \eta_b(\vec x,x^o)$ will be
defined
later on as a directional functional derivative along the lines defining
generalized canonical coordinates of 1st kind on $P^{t}$ (or equivalently
pointwise on ${\bar {\cal G}}^{o(P)}_{\infty}$). To check the validity of
Eqs.(7-3) let us evaluate the commutator (7-2)

$$\eqalign{
[X_a(\vec x,x^o),&X_b(\vec y,x^o)]=\cr
&=B_{ua}(\eta (\vec x,x^o)){ {{\tilde \delta}B_{vb}(\eta (\vec y,x^o))}\over
{\delta \eta_u(\vec x,x^o)} }{ {\tilde \delta}\over {\delta \eta_v(\vec
y,x^o)}}
-\cr
&-B_{ub}(\eta (\vec y,x^o)){ {{\tilde \delta}B_{va}(\eta (\vec x,x^o))}\over
{\delta \eta_u(\vec y,x^o)} }{ {\tilde \delta}\over {\delta \eta_v(\vec
x,x^o)}}
=\cr
&=\delta^3(\vec x-\vec y)[B_{ua}(\eta (\vec x,x^o)){ {\partial B_{vb}(\eta )}
\over {\partial \eta_u)} }{|}_{\eta =\eta (\vec x,x^o)}-\cr
&-B_{ub}(\eta (\vec x,x^o)
){ {\partial B_{va}(\eta )}\over {\partial \eta_u} }{|}_{\eta =\eta (\vec x,x^o
)}]{ {\tilde \delta}\over {\delta \eta_v(\vec x,x^o)} }=\cr
&=\delta^3(\vec x-\vec y)B_{vc}(\eta (\vec x,x^o))c_{cab}{ {\tilde \delta}\over
{\delta \eta_v(\vec x,x^o)} }=\delta^3(\vec x-\vec y)c_{abc}X_c(\vec x,x^o)\cr}
\form
$$

\noindent where we used the following generalization of Eqs.(A-3)

$$
B_{ua}(\eta (\vec x,x^o)){ {\partial B_{vb}(\eta )}
\over {\partial \eta_u)} }{|}_{\eta =\eta (\vec x,x^o)}-B_{ub}(\eta (\vec
x,x^o)
){ {\partial B_{va}(\eta )}\over {\partial \eta_u} }{|}_{\eta =\eta (\vec x,x^o
)}=B_{vc}(\eta (\vec x,x^o))c_{cab}
\form
$$

\noindent which can be interpreted as the generalized Maurer-Cartan equations
on the infinite dimensional group manifold $P^t$
of the gauge group ${\bar {\cal G}}
_{\infty}^{o(P)}$, holding pointwise on each fiber of $P^{t}$ over $\vec x$
at $x^o$ in a suitable tubolar neighbourhood of the identity cross section.

What is needed at this point is a suitable coordinatization of $P^{t}=(R^3
\times \lbrace x^o\rbrace )\times G$ in a tubolar neighbourhood of the
identity cross section, by means of a family, labelled by parameters
$\lambda_a$, of global cross sections $\sigma_{\lbrace \lambda_a\rbrace}
(\vec x,x^o)$ (the time dependence is induced by the gauge potentials),
horizontal
with respect to a given connection ${\cal A}$ with connection 1-form $\omega
^{\cal A}$ on $P^{t}$ (see Appendix B); in this way we would also get a
non-redundant coordinatization of the gauge potentials
${}^{\sigma_{\lbrace \lambda_a
\rbrace}}\, {\vec A}_a=\sigma^{*}_{\lbrace \lambda_a\rbrace}\omega^{\cal A}$,
at least in a neighbourhood of ${}^{\sigma_I}{\vec
A}_a=\sigma^{*}_I\omega^{\cal
A}$ ($\sigma_I$ is the identity cross section) of the corresponding gauge orbit
in ${\cal C}_{fir}^{3;p,s-1,\delta +1}$.

An insight on how to build such a coordinatization of $P^{t}$ is given by the
Green function ${\vec \zeta}^{(\vec A)}_{ab}(\vec x,\vec y;x^o)$ of the
covariant divergence used to solve Gauss' laws. Since it privileges the
straight-lines joining pairs of points in the simply connected base manifold
$R^3$ (these lines are privileged being the geodesics of the flat Riemannian
manifold $R^3$), we can proceed as follows. Let us choose as reference point
an origin in the affine space
$P^t=R^3\times \lbrace x^o\rbrace$, $(\vec x=\vec 0,
x^o)$, so to get a vector space $E^3$; let us join the origin to every other
point with straight-line segments or geodesics (the vector of $E^3$ applied in
the origin). Given a connection ${\cal A}$ on $P^{t}$, let us consider the
${\cal A}$-horizontal lifts of all segments through each point of the
reference fiber $G_o\sim G$ over the origin. Since our connections are fully
irreducible, so that all holonomy bundles coincide with $P^{t}$, and since G
is simply connected, we obtain a family of ${\cal A}$-horizontal global cross
sections $\sigma_b$, parametrized by the elements $b\in G_o\sim G$; they form
a foliation of the manifold $P^{t}$ adapted to the given connection, whose
leaves $\sigma_b$ can be parametrized by means of the coordinates in an
analytic atlas of the analytic manifold $G_o\sim G$; in this way with each
connection 1-form $\omega^{\cal A}$ is associated a privileged set of gauge
potentials ${}^{\sigma_b}{\vec A}_a=\sigma^{*}_b\omega^{\cal A}$. If one
changes
the reference point (the origin) in $R^3$, in general the foliation changes
even keeping the connection ${\cal A}$ fixed due to its holonomy group $\Phi
^{\cal A}$ (this can be seen by considering the geodesic triangles whose
vertices are the old origin, the new one and an other arbitrary point)
and so also the set of privileged gauge potentials changes (they are connected
with the previously privileged gauge potentials by homotopically trivial gauge
transformations); however, the identity cross section belongs to all these
foliations.

Let us put canonical coordinates of 1st kind (see Appendix A) in a
neighbourhood
of the identity of the reference fiber $G_o=G_{(\vec 0,x^o)}\sim G$ and let us
denote them as $\eta_a(\vec 0)$ [$\, (\vec 0, x^o, \eta_a(\vec 0)\, )$ are the
coordinates of the corresponding points in $P^{t}$]. To define coordinates
in a neighbourhood of the identity of another fiber $G_{(\vec x,x^o)}\sim G$
over the point $(\vec x,x^o)$, we use the ${\cal A}$-horizontal lift $\gamma
_{\vec 0,\vec x}$ of the vector $\vec x\in E^3$ through the point $(\vec 0,
x^o,\eta_a(\vec 0))$ in $P^{t}$ and the parallel transport along $\gamma
_{\vec 0,\vec x}$, realized by means of the right action of G on $P^{t}$.

Let $p=(\vec 0,x^o,\eta_a(\vec 0))$ be a point of the fiber $G_o\sim G$ over
$\vec 0$ at time $x^o$ and let $\sigma_{\lbrace \eta (\vec 0)\rbrace}$ be the
global cross section through p, ${\cal A}$-horizontal with respect to the
given connection ${\cal A}$; let ${}^{\sigma_{\lbrace \eta (\vec 0)
\rbrace}}\, {\vec A}_a=\sigma^{*}_{\lbrace \eta (\vec 0)\rbrace}\omega^{\cal
A}$
be the associated gauge potential and let q be the point of the fiber
$G_{(\vec x,x^o)}\sim G$ over $\vec x$ at $x^o$ belonging to the previous
global cross section. We shall assign to q the coordinates on $P^{t}$
$\, (\vec x,x^o,\eta_a(\vec x,x^o))$ with $\eta_a(\vec x,x^o)$ defined by
parallel transport in the following way by means of the Wu-Yang nonintegrable
phase of the second line of Eqs.(6-21) evaluated along the segment $\vec x$ and
depending on the gauge potential ${}^{\sigma_{\lbrace \eta (\vec 0)
\rbrace}}\, {\vec A}_a$:

$$
\eta_a(\vec x,x^o)=\eta_b(\vec 0)\zeta_{ba}^{( {\sigma_{\lbrace \lambda_a
\rbrace}}\, {\vec A} )}(\vec x,\vec 0;x^o).
\form
$$

\noindent For a fiber $G_{(d\vec x,x^o)}\sim G$ infinitesimally near $G_o$ one
has ($\partial^i=\partial /\partial x^i$)

$$\eqalign{
\eta_a(d\vec x,x^o)&\approx \eta_a(\vec 0)+\partial^i\eta_a(\vec 0,x^o)dx^i
\approx \cr
&\approx \eta_b(\vec 0)[\delta_{ba}+{({\hat T}^u)}_{ba}{}^{\sigma_{\lbrace
\eta (\vec 0)\rbrace}}\, A^i_u(\vec o,x^o)dx^i].\cr}
\form
$$

One sees that the identity cross section $\sigma_I$ is ${\cal A}$-horizontal
with respect to every connection ${\cal A}$, because one has

$$\eta_a(\vec 0)=0\quad\quad \Rightarrow \quad\quad \eta_a(\vec x,x^o)=0.
\form
$$

\noindent Moreover $\sigma_I$ remains ${\cal A}$-horizontal also varying the
reference point (the origin).

The defining path $\gamma_{\eta (\vec 0)}(s)$ of the canonical coordinates of
1st kind on $G_o$ (see Appendix A) is in this way extended to a path $\gamma
_{\eta (\vec x,x^o)}(s)=\hat \gamma (\vec x,x^o,s)$ on every fiber $G_{(\vec x,
x^o)}$, and $\hat \gamma (\vec x,x^o,s)$, as a function of $\vec x$ and $s$,
describes a surface connecting the identity cross section for s=0 to the cross
section $\sigma_{\eta (\vec 0)}$ at s=1.

If $\vec \rho =\vec x/|\vec x|$ is the unit vector in the $\vec x$ direction,
one also has

$$
\vec \rho \cdot \vec \partial \eta_a(\vec x,x^o)=-c_{abc}{}^{\sigma_{\lbrace
\eta (\vec 0)\rbrace}}\, {\vec A}_b(\vec x,x^o)\cdot \vec \rho \eta_c(\vec x,
x^o),
\form
$$

\noindent and this implies

$$\eqalign{
&\vec \rho \cdot \vec \partial \eta_a(\vec x,x^o)\,
{\rightarrow}_{\eta \, \rightarrow \, 0}\,\, 0\quad\quad for\, every\, \vec
\rho \cr
&\Rightarrow \quad \vec \partial \eta (\vec x,x^o)\,\,{\rightarrow}_{\eta \,
\rightarrow \, 0}\,\, 0.\cr}
\form
$$

The key point of this construction is that these coordinates are such that a
vertical infinitesimal increment $d\eta_a{|}_{\eta =\eta (\vec x,x^o)}$ of
them along the path $\gamma_{\eta (\vec x,x^o)}(s)$ in the fiber $G_{(\vec x,
x^o)}$ is numerically equal to the horizontal infinitesimal increment $\vec
\partial \eta_a(\vec x,x^o)\cdot d\vec x$ in going from $\vec x$ to $\vec x+
d\vec x$ in the base manifold (see Eqs.(B-5) with $\sigma (\vec x,x^o)=\lbrace
\vec x,x^o;\eta_a(\vec x,x^o)\rbrace$):

$$
d\eta_a{|}_{\eta =\eta (\vec x,x^o)}=\sigma^{*}(\vec x,x^o)d\eta_a
\equiv d\eta_a(\vec x,x^o)=\vec \partial
\eta_a(\vec x,x^o)\cdot d\vec x.
\form
$$

\noindent This construction holds in a suitable neighbourhood of the identity
of
$G_o\sim G$ such that the tubolar neighbourhood of the identity cross section
resulting from this parallel transport intersects each fiber in a neighbourhood
of the identity, which is in the range in which the exponential map is a
diffeomorphism with a neighbourhood of the origin of the Lie algebra $g$. Then
the construction can be extended to the whole $P^{t}$ by using the analytic
atlas of $G_o$, built by right translations of the chart around the identity
with canonical coordinates of 1st kind.

One can now solve the multitemporal equations, i.e. the equations determining
the infinitesimal gauge transformations, of a generic gauge potential ${\vec
A}_a(\vec x,x^o)$ (see Eqs.(2-29)):

$$\eqalign{
X_b(\vec y,x^o)&A^i_a(\vec x,x^o)=-\lbrace A^i_a(\vec x,x^o),\Gamma_b(\vec y,
x^o)\rbrace =B_{cb}(\eta (\vec y,x^o)){ {\tilde \delta A^i_a(\vec x,x^o)}\over
{\delta \eta_c(\vec y,x^o)} }=\cr
&={\hat D}^{(\vec A)i}_{ab}(\vec x,x^o)\delta^3(\vec x-\vec y)=[\delta_{ab}
\partial^i_x-c_{abc}A^i_c(\vec x,x^o)]\delta^3(\vec x-\vec y),\cr}
\form
$$

\noindent where $\tilde \delta /\delta \eta_c(\vec y,x^o)$ is the directional
functional derivative along the path $\gamma_{\eta (\vec x,x^o)}(s)$.

The solution of Eqs.(7-12) is

$$\eqalign{
A^i_a(\vec x,x^o)&={\hat A}^i_a(\vec x,x^o;\eta (\vec x,x^o);\vec \partial
\eta (\vec x,x^o))=\cr
&=A_{ab}(\eta (\vec x,x^o))\, \partial^i\eta_b(\vec x,x^o)+{\hat A}^i_{a,T}
(\vec x,x^o;\eta (\vec x,x^o)),\cr}
\form
$$

\noindent with ${\hat {\vec A}}_{a,T}(\vec x,x^o;\eta (\vec x,x^o))$ satisfying
(see for instance [85]) the equations

$$
{ {\partial {\hat {\vec A}}_{a,T}(\vec x,x^o;\eta )}\over {\partial \eta_b} }
{|}_{\eta =\eta (\vec x,x^o)}=-c_{adc}A_{db}(\eta (\vec x,x^o))\, {\hat {\vec
A}}_{c,T}(\vec x,x^o;\eta (\vec x,x^o));
\form
$$

\noindent as shown in Appendix B, this is the condition of
${\cal A}$-horizontability of the curvature 2-form $\Omega$ of the given
connection 1-form $\omega^{\cal A}$ on $P^{t}$. Indeed by using Eqs.(A-2),
(A-3) and (7-14) one has

$$\eqalign{
B_{cb}(\eta (\vec y,x^o))&{ {\tilde \delta}\over {\delta \eta_c(\vec y,x^o)} }
\lbrack A_{ad}(\eta (\vec x,x^o))\partial^i_x\eta_d(\vec x,x^o)+{\hat
A}^i_{a,T}
(\vec x,x^o;\eta (\vec x,x^o))\rbrack=\cr
&=\delta_{ab}\partial^i_x\delta^3(\vec x-\vec y)+\cr
&+\lbrack { {\partial A_{ad}
(\eta )}\over {\partial \eta_c} }{|}_{\eta =\eta (\vec x,x^o)}-{ {\partial
A_{ac
}(\eta )}\over {\partial \eta_d} }{|}_{\eta =\eta (\vec x,x^o)}\rbrack \cr
&B_{cb}
(\eta (\vec x,x^o))\partial^i\eta_d(\vec x,x^o)\delta^3(\vec x-\vec y)-\cr
&-c_{adv}A_{dc}(\eta (\vec x,x^o)){\hat A}^i_{v,T}(\vec x,x^o;\eta (\vec
x,x^o))
B_{cb}(\eta (\vec y,x^o)\delta^3(\vec x-\vec y)=\cr
&=\delta_{ab}\partial^i_x\delta^3(\vec x-\vec y)-\cr
&-c_{abv}\lbrack A_{vd}(\eta
(\vec x,x^o))\partial^i\eta_d(\vec x,x^o)+{\hat A}^i_{v,T}
(\vec x,x^o;\eta (\vec x,x^o))\rbrack \delta^3(\vec x-\vec y)=\cr
&=[\delta_{ab}\partial^i_x-c_{abc}A^i_c(\vec x,x^o)]\delta^3(\vec x-\vec
y).\cr}
\form
$$

Let us define the quantities

$$\eqalign{
\Theta_a(\eta (\vec x,x^o), \vec \partial \eta (\vec x,x^o))&=A_{ab}(\eta (\vec
x,x^o))\, \vec \partial \eta_b(\vec x,x^o)\cdot d\vec x=\cr
&={\vec \Theta}_a(\eta (\vec x,x^o),\vec \partial \eta (\vec x,x^o))\cdot
d\vec x=\cr
&=\lbrack A_{ab}(\eta )d\eta_b\rbrack {|}_{\eta =\eta (\vec x,x^o)}=
\theta_a(\eta ){|}_{\eta =\eta (\vec x,x^o)},\cr}
\form
$$

\noindent where Eq.(7-11) has been used in the last line; $\theta_a$ are the
Maurer-Cartan 1-forms of G (see Eq.(A-2)); $\Theta_a$ are the generalized
Maurer-Cartan 1-forms on the gauge group ${\cal G}^{3;o(P)}_{\infty}$ (suitable
restriction of ${\cal G}^{3;p,s,\delta}_{\infty}$ to have ${\cal G}^{3;o}
_{\infty}={\cal G}_{\infty}^{3;o(P)}$ as said in Section 6);
the restriction of $\Theta_at^a$ to
the fiber $G_o$ coincides with the canonical 1-form $\omega_G$ of Eq.(A-6).
Therefore, the first term in Eq.(7-13) represents the gauge potential of a pure
gauge (flat) background connection, and coincides with the BRST ghost in the
interpretation of Ref.[86a,b,c]. Following Ref.[86$c_2$], let us write the
exterior derivative on $P^t=R^3\times G$ as $d_P=d+d_G$ (where $d=dx^i\partial
^i$ is the exterior derivative in $R^3$ and $d_G=d\eta_a\partial^a$ the one in
G); one has $d^2_P=d^2=d^2_G=0$ and $dd_G+d_Gd=0$ and one can make the
identification $d_G=s$, $s^2=0$, where s is the BRST operator (s is also called
the fiber or vertical derivative $d_V$ on $P^t$); now one has $A^U(\vec x,x^o)=
U^{-1}(\vec x,x^o)A(\vec x,x^o)U(\vec x,x^o)+U^{-1}(\vec x,x^o)dU(\vec x,x^o)$
($\, A=t^a{\vec A}_a\cdot d\vec x$) with $U:R^3\rightarrow G$, so that the
gauge transformation $U(\vec x,x^o)$ can be thought of as a function $U=U(\vec
x,x^o;\eta_a)$ with $\eta_a$ coordinates of a point, determined by
a well defined functional form of U, on the group
manifold G (by varying $\eta_a$ one varies the gauge transformation); the BRST
ghosts can be written as the $g$-valued 1-forms $c_a$ components of $c=c_at^a=
U^{-1}2sU=U^{-1}2d_GU$; one has $sc=2(sU^{-1})\wedge sU=-2(U^{-1}sU)\wedge (U
^{-1}sU)=-{1\over 2}c\wedge c=-{1\over 2}[t^a,t^b]c_a\wedge c_b=-{1\over 2}
c_{abc}t^cc_a\wedge c_b$, so that $sc_c=-{1\over 2}c_{cab}c_a\wedge c_b$ (like
the Maurer-Cartan equations for $\theta_a$ in G, Eqs.(A-1)); moreover, since
$sA(\vec x,x^o)=d_GA(\vec x,x^o)=0$, one has $2s(-A^U)=-2s(U^{-1}AU+U^{-1}dU)=
dc+c\wedge A^U+A^U\wedge c=t^adc_a-A^U_c\wedge c_b[t^c,t^b]=t^a[dc_a-c_{abc}
A^U_c\wedge c_b]=t^aD^{(A^U)}_{ab}\wedge c_b$ so that $\delta_sA_a=D^{(A^U)}
_{ab}\wedge c_b$; from Eqs.(7-16) one has the identification $\Theta_at^a=
U^{-1}sU={1\over 2}c$.

Therefore, if $d_V$ is the fiber or vertical derivative on $P^t$, whose
restriction to a fiber G coincides with $d_G$, from Eqs.(7-11) and the last
line of Eqs.(7-16) one has the abstract form of the Maurer-Cartan equations
on ${\cal G}_{\infty}^{3;o(P)}$:

$$
d_V\Theta_a=-{1\over 2}c_{abc}\Theta_b\wedge \Theta_c,
\form
$$

\noindent which reproduce Eqs.(B-8) consistently.

Let us remark that in Ref.[87a] it was proposed to identify the vertical
part $\omega_G$ of the connection 1-form $\omega^{\cal A}$ on $P^t$, see
Eq.(B-1), with the BRST ghosts $c=c_at^a$; this proposal was refused in Refs.
[86a,d] due to its lack of dependence on the coordinates of the base
manifold $R^3$. In Ref.[87b] it was proposed to introduce this dependence in
the approach of Ref.[87a] by means of the pullback of $\omega^{\cal A}$ by
using a cross section to get a gauge potential ${}^{\sigma}A=\sigma^{*}\omega
^{\cal A}$; this is just what has been done here: form Eqs.(B-5) and (7-16)
one has

$$
t^a\Theta_a(\eta (\vec x,x^o),\vec \partial \eta (\vec x,x^o))=t^a\theta_a
(\eta ){|}_{\eta =\eta (\vec x,x^o)}=\sigma^{*}\omega_G.
\form
$$

Finally let us note that also in Refs.[88a,b], in the study of BRST cohomology
of classical mechanical systems with constraints, the Hamiltonian BRST operator
has been identified with the vertical derivative ${\bar d}_V$ along the gauge
orbits inside the constraint submanifold (the gauge orbits are the vertical
fibers in the reduction to the reduced phase space of Dirac's observables);
this vertical derivative is also identified with the Cartan-Chevalley-Eilenberg
coboundary operator for the vertical Lie algebra cohomology (see Ref.[88c]);
${\bar d}_V$ is connected with ${\bar {\cal G}}_{\infty}^{3;o(P)}$ in the same
way that $d_V=s=d_G$ is connected with ${\cal G}_{\infty}^{3;o(P)}$.

Eqs.(7-10), (7-13) and (7-16) imply that the pure gauge connection 1-form
$\omega^{{\cal A}_o}$ gives rise to the zero flat gauge potential when
evaluated on the identity cross section $\sigma_I$,

$$\eqalign{
{}^{\sigma_I}A^{(o)}(\vec x,x^o)&=\sigma_I^{*}(\vec x,x^o)\omega^{{\cal A}_o}=
{\hat T}^a\Theta_a(\eta (\vec x,x^o),\vec \partial\eta (\vec x,x^o)){|}_{\eta
=0}=\cr
&={\hat T}^aA_{ab}(\eta (\vec x,x^o))\vec \partial \eta_a(\vec x,x^o)
\cdot d\vec x{|}_{\eta =0}=0.\cr}
\form
$$

\noindent Instead if $\sigma (\vec x,x^o)=\sigma_I(\vec x,x^o)U(\vec x,x^o)=U
(\vec x,x^o)$ is another global cross section obtained from $\sigma_I$ by
means of a gauge transformation $U\in {\cal G}^{3;o(P)}_{\infty}$, one has for
the associated flat gauge potential

$$
{}^{\sigma}A^{(o)}(\vec x,x^o)=\sigma^{*}(\vec x,x^o)\omega^{{\cal A}_o}=U^{-1}
(\vec x,x^o)dU(\vec x,x^o).
\form
$$

\noindent Now ${\cal G}^3/{\cal G}^{3;o(P)}_{\infty}=G\times Z$; Eqs.(3-18)
imply that there are no improper gauge transformations in between ${\cal G}^3
_{\infty}$ and ${\cal G}^{3;o}_{\infty}={\cal G}^{3;o(P)}_{\infty}$
(if a suitable restriction of ${\cal G}^{3;p,s,\delta}$ exists) with ${\cal
G}^3_{\infty}/{\cal G}^{3;o(P)}_{\infty}=Z$; instead G, with $Z_G\subset G$,
generates  only global rigid gauge transformations. This implies that in the
sector with zero winding number, the only gauge transformations connecting
$\sigma_I$ and the global cross sections $\sigma_{I_m}(\vec x,x^o)=I_m\in Z_G$
are global rigid gauge transformations in $G={\cal G}^3/{\cal G}^3_{\infty}$,
so
that also the flat gauge potentials generated by the cross sections $\sigma
_{I_m}$, $I_m\in Z_G$, vanish

$$
{}^{\sigma_{I_m}}A^{(o)}(\vec x,x^o)=\sigma_{I_m}^{*}\omega^{{\cal A}_o}=
I_m^{*}\omega
^{{\cal A}_o}=U^{-1}dU=0,\quad\quad U\in G={\cal G}^3/{\cal G}^3_{\infty}.
\form
$$

\noindent The flat gauge potentials generated by cross sections $\sigma_n$
obtained from $\sigma_I$ by means of gauge transformations with winding number
n, $\sigma_n(\vec x,x^o)=\sigma_I(\vec x,x^o)U_n(\vec x,x^o)=U_n(\vec x,x^o)$,
are

$$
{}^{\sigma_n}A^{(o)}(\vec x,x^o)=\sigma_n^{*}\omega^{{\cal A}_o}=U^{-1}_n(\vec
x
,x^o)dU_n(\vec x,x^o);
\form
$$

\noindent   there is no canonical choice of $\sigma_n$, but ${\cal G}^{3;o}
_{\infty}={\cal G}^{3;o(P)}_{\infty}$ implies that there must exist a global
rigid gauge transformation ${\hat U}_n\in Z={\cal G}^3_{\infty}/{\cal G}^{3;o
(P)}_{\infty}$ such that, if ${\hat \sigma}_n(\vec x,x^o)=\sigma_I(\vec x,x^o)
{\hat U}_n={\hat U}_n$, then

$$
{}^{{\hat \sigma}_n}A^{(o)}(\vec x,x^o)={\hat \sigma}_n^{*}\omega^{{\cal A}_o}
={\hat U}_n^{-1}d{\hat U}_n=0.
\form
$$

Due to Eqs.(A-3) and (7-14) the gauge potential ${\hat {\vec A}}_{a,T}
(\vec x,x^o;\eta (\vec x,x^o))$ is the source of the field strength ${\vec
B}_a$
of the YM theory on $P^{t}$:

$$\eqalign{
F^{ij}_a(\vec x,x^o)&=-\epsilon^{ijk}B^k_a(\vec x,x^o)=\cr
&=\partial^iA^j_a(\vec x,x^o)-\partial^jA^i_a(\vec x,x^o)+c_{abc}A^i_b(\vec x,
x^o)A^j_c(\vec x,x^o)=\cr
&=\lbrack \partial^i{|}_{\eta}{\hat A}^j_{a,T}(\vec x,x^o;\eta )-\partial^j
{|}_{\eta}{\hat A}^i_{a,T}(\vec x,x^o;\eta )+\cr
&+c_{abc}{\hat A}^i_{b,T}
(\vec x,x^o;\eta ){\hat A}^j_{c,T}(\vec x,x^o;\eta )\rbrack {|}_{\eta =\eta
(\vec x,x^o)}+\cr
&+\lbrack ({ {\partial A_{ae}(\eta )}\over {\partial \eta_d} }-{ {\partial
A_{ad}(\eta )}\over {\partial \eta_e} }){|}_{\eta =\eta (\vec x,x^o)}+c_{auv}
A_{ud}(\eta (\vec x,x^o)A_{ve}(\eta (\vec x,x^o))\rbrack \cr
&{}{}\partial^i\eta_d(\vec x
,x^o)\partial^j\eta_e(\vec x,x^o)+\cr
&+\lbrack { {\partial {\hat A}^j_{a,T}(\vec x,x^o;\eta )}\over {\partial
\eta_d}
}{|}_{\eta =\eta (\vec x,x^o)}+c_{abc}A_{bd}(\eta (\vec x,x^o)){\hat A}^j_{c,T}
(\vec x,x^o;\eta (\vec x,x^o))\rbrack \cr
&{}{}\partial^i\eta_d(\vec x,x^o)-\cr
&-\lbrack { {\partial {\hat A}^i_{a,T}(\vec x,x^o;\eta )}\over {\partial
\eta_d}
}{|}_{\eta =\eta (\vec x,x^o)}+c_{abc}A_{bd}(\eta (\vec x,x^o)){\hat A}^i_{c,T}
(\vec x,x^o;\eta (\vec x,x^o))\rbrack \cr
&{}{}\partial^j\eta_d(\vec x,x^o)=\cr
&=\lbrack \partial^i{|}_{\eta}{\hat A}^j_{a,T}(\vec x,x^o;\eta )-\partial^j
{|}_{\eta}{\hat A}^i_{a,T}(\vec x,x^o;\eta )+\cr
&+c_{abc}{\hat A}^i_{b,T}
(\vec x,x^o;\eta ){\hat A}^j_{c,T}(\vec x,x^o;\eta )\rbrack {|}_{\eta =\eta
(\vec x,x^o)}+\cr}
\form
$$

\noindent Let us remark that with a Coulomb gauge decomposition $A^i_a=-{
{\partial^i}\over {\triangle} }\vec \partial \cdot {\vec A}_a+(\delta^{ij}+
{ {\partial^i\partial^j}\over {\triangle} })A^j_a$, like in the Abelian case,
this property is not true ( only in the Abelian case $\partial^i[-{ {\vec
\partial}\over {\triangle} }\cdot \vec A]$ is a pure gauge connection).

To find the transversality properties of the gauge potential ${\vec A}_{a,T}
(\vec x,x^o)={\hat {\vec A}}_{a,T}(\vec x,x^o;$\break
$\eta (\vec x,x^o))$ we need a
non-Abelian analogue of the Hodge decomposition in the Abelian case, where the
analogous transverse gauge potential was a coexact 1-form by using the
codifferential $\delta =
{*}d{*}$ (valid on 1-forms), $\delta^2=0$, see after Eqs.(5-8). See Ref.[82]
for the validity of tensor decompositions like the Hodge one in weighted
Sobolev spaces (they were just introduced to put under control these
decompositions in the case of noncompact manifolds).

Since we have chosen canonical coordinates of 1st kind on the reference fiber
$G_o=G_{(\vec 0,x^o)}$, on each fiber $G_{(\vec x,x^o)}$ we have identified a
definitional privileged path $\gamma_{\eta (\vec x,x^o)}(s)=\hat \gamma (\vec
x,
x^o,s)$. As shown in Appendix A, on $G_o$ it is possible to define a preferred
primitive $\omega^{\gamma_{\eta}}_a(\eta (s)=\eta (\vec 0,s))$, given in Eq.
(A-7), of the nonintegrable Maurer-Cartan 1-forms $\theta_a$, such that the
exterior derivative $d_G$ on the group manifold G restricted to $\gamma_{\eta}=
\gamma_{\eta (\vec 0)}(s)$, $d_{\gamma_{\eta}}$, satisfies $d_{\gamma_{\eta}}
\omega^{\gamma_{\eta}}_a(\eta (s))=\theta_a(\eta (s))$, Eq.(A-8), and $d^2
_{\gamma_{\eta}}=0$, Eqs.(A-9), as a consequence of the Maurer-Cartan equations
(A-1), (A-3) restricted to $\gamma_{\eta}(s)$.

We shall extend this definition to $P^{t}$,
which is the group manifold of ${\cal G}
_{\infty}^{3;o(P)}$ due to the definition of its pointwise multiplication, by
using the vertical derivative (or BRST operator) on $P^{t}$ of Eqs.(7-18); we
shall denote $d_{\hat \gamma}$ the restriction of $d_V$ to $\hat \gamma (\vec
x,
x^o,s)$, i.e. on each fiber $d_{\hat \gamma}$ is the directional derivative
along the path $\gamma _{\eta (\vec x,x^o)}(s)=\hat \gamma (\vec x,x^o,s)$.
Then from Eqs.(7-16) one obtains [$\eta (\vec x,x^o,s)$ are the coordinates of
the points of the path $\gamma_{\eta (\vec x,x^o)}(s)$]:

$$\eqalign{
\Omega^{\hat \gamma}_a(\eta (\vec x,x^o,s))&=\int^{\gamma_{\eta (\vec
x,x^o,s)}}
_{(\hat \gamma )\, I}\Theta_a{|}_{\gamma_{\eta (\vec x,x^o)}}=\cr
&=\int ^{\eta (\vec x,x^o,s)}_{(\hat \gamma )\, 0}\, A_{ab}(\bar \eta (\vec x,
x^o,s)){\cal D}{\bar \eta}_b(\vec x,x^o,s),\cr}
\form
$$

\noindent with the line path integral, evaluated on the surface $\hat \gamma
(\vec x,x^o,s)$ in $P^{t}$ joining the identity cross section $\sigma_I$ and
$\sigma_{\eta(\vec x,x^o,s)}$, denoting a line integral along $\gamma_{\eta
(\vec x,x^o)}(s)$ on each fiber. Then by definition one has

$$
d_{\hat \gamma}\, \Omega^{\hat \gamma}_a(\eta (\vec x,x^o,s))=\Theta_a(\eta
(\vec x,x^o,s), \vec \partial \eta (\vec x,x^o,s)).
\form
$$

In the tubolar neighbourhood of the identity cross section $\sigma_I$ one has
the following realization of $d_{\hat \gamma}$ on functions of $\vec x,\,
\eta_a(\vec x,x^o),\, \vec \partial \eta_a(\vec x,x^o)$ like the functions we
are considering [the coefficient of $dx^i$ is a suitable extension to the first
derivatives of the total derivative with respect to $x^i$ of functions of $\vec
x$ and $\eta_a(\vec x,x^o)$]

$$\eqalign{
d_{\hat \gamma}&=dx^i\, \lbrace { {\partial}\over {\partial x^i} }{|}_{\eta}+
\partial^i\eta_a(\vec x,x^o){ {\partial}\over {\partial \eta_u} }{|}_{\eta =
\eta (\vec x,x^o)}+\cr
&+{1\over 2}c_{abc}[A_{bm}(\eta (\vec x,x^o))\partial^i\eta_m(\vec x,x^o)]
[A_{cn}(\eta (\vec x,x^o)]B_{ua}(\eta (\vec x,x^o)){ {\partial}\over {\partial
\, \partial^j\eta_u} }{|}_{\eta =\eta (\vec x,x^o)}\rbrace =\cr
&=dx^i\, \lbrace { d\over {dx^i}}+\cr
&+{1\over 2}c_{abc}\Theta^i_b(\eta (\vec x,x^o),\vec \partial \eta (\vec
x,x^o))
\Theta^j_c(\eta (\vec x,x^o),\vec \partial \eta (\vec x,x^o))B_{ua}(\eta (\vec
x,x^o)){ {\partial}\over {\partial
\, \partial^j\eta_u} }{|}_{\eta =\eta (\vec x,x^o)}\rbrace ,\cr}
\form
$$

\noindent where $d/dx^i$ is the total derivative with respect to $x^i$ on
functions only of $\vec x$ and $\eta_a(\vec x,x^o)$. $d_{\hat \gamma}$
satisfies

$$
d^2_{\hat \gamma}=0
\form
$$

\noindent due to the generalized Maurer-Cartan equations (B-8) or (7-17) for
${\cal G}_{\infty}^{3;o(P)}$.

By using Eq.(B-9) the generalized canonical 1-form on ${\cal G}^{3;o(P)}
_{\infty}$ in the adjoint representation of G is

$$\eqalign{
\Theta (\eta (\vec x,x^o),\vec \partial \eta (\vec x,x^o))&=\Theta_a(\eta (\vec
x,x^o),\vec \partial \eta (\vec x,x^o)){\hat T}^a=\cr
&=H_b(\eta (\vec x,x^o))\vec \partial \eta_b(\vec x,x^o)\cdot d\vec x=H_b(\eta
(\vec x,x^o))d\eta_b{|}_{\eta =\eta (\vec x,x^o)},\cr}
\form
$$

\noindent with the generalized Maurer-Cartan equations assuming the form of the
zero curvature equations (B-10).

Analogously one defines

$$
\Omega^{\hat \gamma}(\eta (\vec x,x^o,s))=\Omega^{\hat \gamma}_a(\eta (\vec x,
x^o)){\hat T}^a=\int^{\eta (\vec x,x^o,s)}_{(\hat \gamma )\, 0}\, H_b(\bar \eta
(\vec x,x^o,s)){\cal D}{\bar \eta}_b(\vec x,x^o,s).
\form
$$

Due to $d^2_{\hat \gamma}=0$, also the codifferential $\delta_{\hat \gamma}=
{*}d_{\hat \gamma}{*}$ (on 1-forms) satisfies $\delta^2_{\hat \gamma}=0$
(here ${*}$ is the Hodge star operator on $R^3$, since ${\vec A}_{a,T}(\vec x,
x^o)\cdot d\vec x$ is an 1-form on $R^3$). From Eqs.(7-13) one has

$$\eqalign{
{\vec A}_a(\vec x,x^o)\cdot d\vec x&=\Theta_a(\eta (\vec x,x^o),\vec \partial
\eta (\vec x,x^o))+{\hat {\vec A}}_{a,T}(\vec x,x^o;\eta (\vec x,x^o))\cdot
d\vec x=\cr
&=d_{\hat \gamma}\, \Omega^{\hat \gamma}_a(\eta (\vec x,x^o))+
{\hat {\vec A}}_{a,T}(\vec x,x^o;\eta (\vec x,x^o))\cdot
d\vec x=\cr
&=[{d\over {d\vec x}}\Omega^{\hat \gamma}_a(\eta (\vec x,x^o))+
{\hat {\vec A}}_{a,T}(\vec x,x^o;\eta (\vec x,x^o))]\cdot
d\vec x;\cr}
\form
$$

\noindent in this equation one has $d_{\hat \gamma}=d\vec x\cdot {d\over
{d\vec x}}$, see after Eq.(7-27).

Assuming the validity of the Hodge decomposition theorem for these forms
(Ref.[82] should be used to check the consistency of this assumption)
on the simply connected manifold $R^3$ with
exterior derivative $d_{\hat \gamma}$ and codifferential $\delta_{\hat \gamma}$
induced by the BRST global operator $s=d_V$ on $P^{t}=(R^3\times \lbrace x^o
\rbrace )\times G$ and with the local expression (7-27) in our
coordinatization,
one has

$$
\delta_{\hat \gamma}[{\hat {\vec A}}_{a,T}(\vec x,x^o;\eta (\vec x,x^o))\cdot
d\vec x]=0,
\form
$$

\noindent i.e. these 1-forms are $\delta_{\hat \gamma}$-coexact. This implies
the transversality condition

$$
\vec \partial \cdot {\hat {\vec A}}_{a,\perp}
(\vec x,x^o)=(\vec \partial {|}_{\eta}
+\vec \partial \eta_u(\vec x,x^o) { {\partial}\over {\partial \eta_u}
}{|}_{\eta
=\eta (\vec x,x^o)})\cdot {\hat {\vec A}}_{a,T}(\vec x,x^o;\eta (\vec x,x^o))
=0,
\form
$$

\noindent namely by using Eqs.(7-14) and (7-13)

$$\eqalign{
0&=\vec \partial {|}_{\eta}\cdot {\hat {\vec A}}_{a,T}
-\vec \eta_u\cdot c_{abc}A_{bu}{\hat {\vec A}}_{c,T}
=\vec \partial {|}_{\eta}\cdot {\hat {\vec A}}_{a,T}-c_{abc}{\vec
\Theta}_b\cdot
{\hat {\vec A}}_{a,T}=\cr
&=\vec \partial {|}_{\eta}\cdot {\hat {\vec A}}_{a,T}-c_{abc}({\vec \Theta}_b+
{\hat {\vec A}}_{b,T})\cdot {\hat {\vec A}}_{c,T}= (\delta_{ac}\vec \partial
{|}_{\eta}+c_{acb}{\vec A}_b)\cdot {\hat {\vec A}}_{c,T}\cr}
\form
$$

\noindent which is a kind of covariant divergenceless condition. Let us remark
that in Eq.(7-33) we have used the notation ${\hat {\vec A}}_{a,\perp}$, like
in the Abelian case, since with this functional form this quantity is
divergenceless.

Since

$$
{\hat {\vec D}}^{(\vec A)}_{ab}(\vec x,x^o)\cdot {\hat {\vec A}}_{a,\perp}
(\vec x,x^o)=c_{acb}{\vec \Theta}_c(\eta (\vec x,x^o),\vec \partial \eta (\vec
x,x^o))\cdot {\hat {\vec A}}_{b,\perp}(\vec x,x^o)
\form
$$

\noindent one can also write a decomposition like Eqs.(6-48) for ${\hat {\vec
A}}_{a,\perp}$, so that

$$\eqalign{
{\vec A}_a(\vec x,x^o)\cdot d\vec x&=d_{\hat \gamma}\Omega^{\hat \gamma}_a(\eta
(\vec x,x^o))+{\hat {\vec A}}_{a,D\perp}(\vec x,x^o)\cdot d\vec x-\cr
&-d\vec x\cdot \int d^3y\, {\vec \zeta}^{(\vec A)}_{ac}(\vec x,\vec y;x^o)
c_{cuv}{\vec \Theta}_u(\eta (\vec x,x6o),\vec \partial \eta (\vec x,x^o))\cdot
{\hat {\vec A}}_{v,\perp}(\vec y,x^o)\cr}
\form
$$

\noindent with ${\hat {\vec D}}^{(\vec A)}_{ab}(\vec x,x^o)\cdot {\hat {\vec
A}}_{b,D\perp}(\vec x,x^o)\equiv 0$.

We see from Eqs.(7-3) and (7-13) that the natural gauge variables associated
with the proper infinitesimal gauge transformations generated by Gauss' law
1st class constraints are the $\eta_a(\vec x,x^o)$'s and one obtains

$$
\lbrace \eta_a(\vec x,x^o),\Gamma_b(\vec y,x^o)\rbrace =-X_b(\vec y,x^o)\eta_a
(\vec x,x^o)=-B_{ab}(\eta (\vec x,x^o))\delta^3(\vec x-\vec y);
\form
$$

\noindent then Eqs.(7-30) for s=1 and Eqs.(7-29) imply

$$\lbrace \Omega^{\hat \gamma}_a(\eta (\vec x,x^o)),\Gamma_b(\vec y,x^o)\rbrace
=-\delta_{ab}\delta^3(\vec x-\vec y).
\form
$$

\noindent Moreover, from Eqs.(2-29) and (7-13) one gets

$$
\lbrace \eta_a(\vec x,x^o),\eta_b(\vec y,x^o)\rbrace =0.
\form
$$

Eqs.(7-37) give a local expression valid in a tubolar neighbourhood of the
identity cross section. The fact that in ${\cal C}_{fir}^{3;p,s-1,\delta +1}$
there is no Gribov ambiguity (so that operators of the kind of the Faddeev-
Popov ones are invertible), is reflected in the invertibility of the matrix
B, which can only acquire coordinate singularities outside the tubolar
neighbourhood, where, however, one needs different coordinate charts;
therefore, the determinant of the Poisson brackets of the constraints (7-37)
(our form of the Faddeev-Popov determinant) is nonsingular.

Now, since the $\Gamma_a$'s satisfy the algebra (2-32)

$$
\lbrace \Gamma_a(\vec x,x^o),\Gamma_b(\vec y,x^o)\rbrace =c_{abc}\Gamma_c(\vec
x,x^o)\delta^3(\vec x-\vec y),
\form
$$

\noindent one needs to Abelianize them and, in a tubolar neighbourhood of the
identity cross section, this can be done by using the inverse A of the
nonsingular matrix B of Eqs.(7-37) as it can been shown by using the results
of Ref.[89]:

$$
{\tilde \Gamma}_a (\vec x,x^o)=\Gamma_b(\vec x,x^o)\, A_{ba}(\eta (\vec
x,x^o)),
\form
$$

$$
\lbrace {\tilde \Gamma}_a(\vec x,x^o),{\tilde \Gamma}_b(\vec y,x^o)\rbrace =0.
\form
$$

\noindent Then, Eqs.(7-29) imply

$$
\lbrace \eta_a(\vec x,x^o),{\tilde \Gamma}_b(\vec y,x^o)\rbrace =-\delta_{ab}
\delta^3(\vec x-\vec y).
\form
$$

Eqs.(7-43) also give the equations for the determination of the $\eta_a(\vec x,
x^o)$'s in terms of the gauge potentials ${\vec A}_a(\vec x,x^o)$:

$$\eqalign{
\lbrace \eta_a(\vec x,x^o),{\tilde \Gamma}_b(\vec y,x^o)\rbrace &=\lbrace
\eta_a(\vec x,x^o),\Gamma_c(\vec y,x^o)\rbrace A_{cb}(\eta (\vec y,x^o))=\cr
&=-A_{cb}(\eta (\vec y,x^o)){\hat D}^{(\vec A)i}_{cd}(\vec y,x^o)\lbrace \eta_a
(\vec x,x^o),\pi^i_d(\vec y,x^o)\rbrace =\cr
&=A_{cb}(\eta (\vec y,x^o)){\hat D}^{(\vec A)i}_{cd}(\vec y,x^o){ {\delta
\eta_a(\vec x,x^o)}\over {\delta A^i_d(\vec y,x^o)} }=\cr
&=-\delta_{ab}\delta^3(\vec x-\vec y)\cr}
\form
$$

$$
{\hat D}^{(\vec A)i}_{cd}(\vec y,x^o){ {\delta
\eta_a(\vec x,x^o)}\over {\delta A^i_d(\vec y,x^o)} }=-B_{ac}(\eta (\vec
x,x^o))
\delta^3(\vec x-\vec y)
\form
$$

\noindent  where Eqs.(2-28) have been used. Then Eqs.(6-20)-(6-22) and the
assumption that there is no homogeneous covariantly transverse solution
which a priori could be added, one gets

$$\eqalign{
{ {\delta \eta_a(\vec x,x^o)}\over {\delta A^i_b(\vec y,x^o)}}&=-\lbrace
\eta_a(\vec x,x^o),\pi^i_b(\vec y,x^o)\rbrace =\cr
&=B_{ac}(\eta (\vec
x,x^o))\zeta^{(\vec A)i}_{bc}(\vec y,\vec x;x^o)=
=-B_{ac}(\eta (\vec x,x^o))\zeta^{(\vec A)i}_{cb}(\vec x,\vec y;x^o)\cr}
\form
$$

\noindent where we used

$$\eqalign{
{\vec \zeta}^{(\vec A)}_{ab}(\vec x,\vec y;x^o)&={ {\vec x-\vec y}\over {4\pi
{|\vec x-\vec y|}^3} }{(P\, e^{\int_{\vec y}^{\vec x}d\vec z\cdot {\vec A}_u
(\vec z,x^o){\hat T}^u})}_{ab}=\cr
&=-{\vec \zeta}^{(-\vec A)}_{ab}(\vec y,\vec x;x^o)=-{\vec \zeta}^{(\vec A)}
_{ba}(\vec y,\vec x;x^o)\cr}
\form
$$

\noindent because ${\hat T}^u$ is antihermitian.

Eqs.(7-46) have the correct Abelian limit $\delta \eta_a(\vec x,x^o)/\delta
A^i_b(\vec y,x^o)=-\delta_{ab}\, c^i(\vec x-\vec y)$, see Eqs.(5-13);
therefore,
it was correct to discard homogeneous covariantly divergenceless solutions
of Eqs.(7-45).

Eqs.(7-46) are not integrable and again one has to integrate them by using the
privileged paths $\hat \gamma (\vec x,x^o,s)$ between $\sigma_I$ and $\sigma
_{\eta (\vec x,x^o)}$. Using Eq.(7-25) for s=1, Eqs.(7-46) can be rewritten
in the form

$${ {\tilde \delta}\over {\delta {}^{\sigma_{\eta}}A^i_b(\vec y,x^o)} }\Omega
^{\hat \gamma}_a(\eta (\vec x,x^o))=A_{ac}(\eta (\vec x,x^o)){ {\delta \eta_c
(\vec x,x^o)}\over {\delta {}^{\sigma_{\eta}}A^i_b(\vec y,x^o)} }=-
\zeta_{ab}^{({}^{\sigma_{\eta}})\, i}(\vec x,\vec y;x^o),
\form
$$

\noindent where $\tilde \delta /\delta {}^{\sigma_{\eta}}A^i_b(\vec y,x^o)$ is
interpreted as the functional derivative with respect to the family of gauge
potentials ${}^{\sigma_{\eta}}{\vec A}(\vec x,x^o)=\sigma^{*}_{\eta (\vec x,
x^o)}\omega^{\cal A}$ defined by the family of cross sections $\sigma_{\eta
(\vec x,x^o,s)}$ in between $\sigma_I$ for s=0 and $\sigma_{\eta (\vec x,x^o)}$
for s=1 associated with the paths $\hat \gamma (\vec x,x^0,s)=\gamma_{\eta
(\vec x,x^o)}(s)$ [for each s, $\sigma_{\eta (\vec x,x^o,s)}$ intersects all
the paths $\gamma_{\eta (\vec x,x^o)}(s)$ for that value of s as $\vec x$
varies in $R^3$].

The formal solution of Eqs.(7-48) is

$$\eqalign{
\Omega^{\hat \gamma}&(\eta (\vec x,x^o))=\Omega^{\hat \gamma}_a(\eta (\vec x,
x^o)){\hat T}^a=\cr
&=\int_{(\hat \gamma )\, 0}^{\eta (\vec x,x^o)}H_b(\bar \eta (\vec x,x^o))
{\cal D}{\bar \eta}_b(\vec x,x^o)=\cr
&=-\int d^3z\, \int_{(\hat \gamma )\, \vec c(\vec x-\vec z)\cdot {}^{\sigma_I}
\vec A(\vec z,x^o)}^{\vec c(\vec x-\vec z)\cdot {}^{\sigma_{\eta}}\vec A(\vec
z,x^o)}\, {\hat T}^u\zeta_{ub}^{({}^{\sigma_{\eta (s)}}{\bar {\vec A}})}
(\vec x,\vec z;x^o){\cal D}[\vec c(\vec x-\vec z)\cdot {}^{\sigma_{\eta (s)}}
{\bar {\vec A}}_b(\vec z,x^o)],\cr}
\form
$$

\noindent where the line path integral is on the family of quantities $\vec c
(\vec x-\vec z)\cdot {}^{\sigma_{\eta (s)}}{\vec A}_b(\vec z,x^o)$ at fixed
$\vec x$ connecting, along $\hat \gamma (\vec x,x^o,s)$, $\vec c
(\vec x-\vec z)\cdot {}^{\sigma_I}{\vec A}_b(\vec z,x^o)$ at s=0 and $\vec c
(\vec x-\vec z)\cdot {}^{\sigma_{\eta }}{\vec A}_b(\vec z,x^o)$ at s=1.
The integrand $\zeta_{ub}^{({}^{\sigma_{\eta}}\vec A)}(\vec x,\vec z;x^o)$,
having the path along the segment $\vec x-\vec y$, see Eqs.(6-21), is a
functional of $\vec c
(\vec x-\vec z)\cdot {}^{\sigma_{\eta (s)}}{\vec A}_b(\vec z,x^o)$.

To get $\eta_a(\vec x,x^o)$ from $\Omega^{\hat \gamma}_a(\eta (\vec x,x^o))$,
essentially one has to invert Eqs.(A-7) on the group manifold of G with the
matrix A given by Eq.(A-5), i.e. to invert ${\tilde \omega}_a(s)\equiv \omega
_a^{\gamma_{\eta}}(\eta (s))=\int^{\eta (s)}_{(\gamma_{\eta})\, 0}\,
A_{ab}(\bar
\eta )d{\bar \eta}_b$ to $\eta_a(s)={\tilde \eta}_a(\tilde \omega (s))$. By
using the lowest order of Eq.(6-23) in Eq.(7-42) (for $\sigma_{\eta}$ very
near to $\sigma_I$) one sees that to lowest order $\eta_a(\vec x,x^o)=-{1\over
{\triangle}}\vec \partial \cdot {\vec A}_a(\vec x,x^o)+\cdots$, namely the
Coulomb gauge fixing is an approximation of the natural gauge fixing $\eta_a
(\vec x,x^o)\approx 0$, better and better when $\sigma_{\eta}\rightarrow \sigma
_I$ and ${}^{\sigma_{\eta}}\vec A\rightarrow {}^{\sigma_I}\vec A$.

{}From Eqs.(7-13) and (7-46) we have

$$\eqalign{
\lbrace \eta_a(\vec x,x^o),A^i_b(\vec y,x^o)\rbrace &=0\cr
\lbrace \eta_a(\vec x,x^o),\pi^i_b(\vec y,x^o)\rbrace &=B_{ac}(\eta (\vec x,x^o
))\zeta^{(\vec A)\, i}_{cb}(\vec x,\vec y;x^o).\cr}
\form
$$

\vfill\eject

\bigskip\noindent
{\bf {8. The Longitudinal Electric Field, Dirac's Observables and the
Inclusion of Fermions.}}
\newcount \nfor

\def \form {\global \advance \nfor by 1 \eqno (8.\the\nfor)}
\bigskip

We have identified the canonical pairs of conjugate gauge variables as
$A_{a0}$,
$\pi_{ao}$, $\eta_a$, ${\tilde \Gamma}_a$ with the Poisson algebra

$$\eqalign{
\lbrace A_{ao}(\vec x,x^o),\pi_{ao}(\vec y,x^o)\rbrace &=\delta_{ab}\delta^3
(\vec x-\vec y)\cr
&\lbrace \eta_a(\vec x,x^o),{\tilde \Gamma}_b(\vec y,x^o)\rbrace =-\delta_{ab}
\delta^3(\vec x-\vec y)\cr}
\form
$$

The original variables ${\vec A}_a$, ${\vec \pi}_a$ are decomposed as

$$\eqalign{
{\vec A}_a(\vec x,x^o)&={\vec \Theta}_a(\eta (\vec x,x^o),\vec \partial \eta
(\vec x,x^o))+{\hat {\vec A}}_{a,T}(\vec x,x^o;\eta (\vec x,x^o))=\cr
&={\vec \Theta}_a(\eta (\vec x,x^o),\vec \partial \eta (\vec x,x^o))+{\hat
{\vec A}}_{a,\perp}(\vec x,x^o),\cr
&=-{ {\vec \partial}\over {\triangle} }\vec \partial \cdot {\vec A}_a(\vec x,
x^o)+{\vec A}_{a,\perp}(\vec x,x^o),\cr
{\vec \pi}_a(\vec x,x^o)&={\vec \pi}_{a,D\perp}(\vec x,x^o)+\int d^3y\, {\vec
\zeta}^{(\vec A)}_{ab}(\vec x,\vec y;x^o)\Gamma_b(\vec y,x^o)=\cr
&=-{ {\vec \partial}\over {\triangle} }\vec \partial \cdot {\vec \pi}_a(\vec x,
x^o)+{\vec \pi}_{a,\perp}(\vec x,x^o),\cr}
\form
$$

\noindent with ${\vec \pi}^i_{a,\perp}(\vec x,x^o)=(\delta^{ij}+{ {\partial^i
\partial^j}\over {\triangle} })\pi^j_a(\vec x,x^o)=P^{ij}_{\perp}(\vec x)\pi^j
_a(\vec x,x^o)$ and ${\vec A}^i_{a,\perp}(\vec x,x^o)=P^{ij}_{\perp}$\break
$(\vec x)A^j_a(\vec x,x^o)$. In the last line of each formula one has
evidentiated the usual transverse parts.

Eqs.(8-2) and (7-16) imply

$$
\vec \partial \cdot {\vec A}_a(\vec x,x^o)=\vec \partial \cdot {\vec \Theta}_a
(\eta (\vec x,x^o),\vec \partial \eta (\vec x,x^o))=\vec \partial \cdot [A_{ab}
(\eta (\vec x,x^o))\vec \partial \eta_b(\vec x,x^o)].
\form
$$

\noindent Then one gets the following relation between ${\vec A}_{a,\perp}$
and ${\hat {\vec A}}_{a,\perp}$

$$
A^i_{a,\perp}(\vec x,x^o)={\hat A}^i_{a,\perp}(\vec x,x^o)+P^{ij}_{\perp}
(\vec x)\Theta^j_a(\eta (\vec x,x^o),\vec \partial \eta (\vec x,x^o)).
\form
$$

To get the relation between ${\vec \pi}_{a,\perp}$ and ${\vec \pi}_{a,D\perp}$
we note that from the definition of $\Gamma_a$ one has

$$\eqalign{
\vec \partial \cdot {\vec \pi}_a(\vec x,x^o)&=-c_{acb}{\vec A}_c(\vec x,x^o)
\cdot {\vec \pi}_b(\vec x,x^o)-\Gamma_a(\vec x,x^o)=\cr
&=-c_{acb}{\vec A}_c(\vec x,x^o)\cdot [{\vec \pi}_{b,\perp}(\vec x,x^o)- {
{\vec \partial}\over {\triangle} }\vec \partial \cdot {\vec \pi}_b(\vec x,x^o)]
-\Gamma_a(\vec x,x^o),\cr}
\form
$$

\noindent so that using Eqs.(6-31) it follows

$$\eqalign{
Z^{(\vec A)}_{ab}(\vec x,x^o)\vec \partial \cdot {\vec \pi}_b(\vec x,x^o)&=
(\delta_{ab}+c_{abc}{\vec A}_c(\vec x,x^o)\cdot { {\vec \partial}\over
{\triangle} })\vec \partial \cdot {\vec \pi}_b(\vec x,x^o)=\cr
&=-[c_{acb}{\vec A}_c(\vec x,x^o)\cdot {\vec \pi}_{b,\perp}(\vec x,x^o)+
\Gamma_a(\vec x,x^o)].\cr}
\form
$$

\noindent Its solution is

$$\eqalign{
\vec \partial \cdot {\vec \pi}_a(\vec x,x^o)&=-\int d^3y\, G^{(\vec A)}_{Z,
ab}(\vec x,\vec y;x^o)[c_{buv}{\vec A}_u(\vec y,x^o)\cdot {\vec \pi}_{v,\perp}
(\vec y,x^o)+\Gamma_b(\vec y,x^o)]=\cr
&=\int d^3y\, {\vec \partial}_x\cdot {\vec \zeta}^{(\vec A)}_{ab}(\vec x,\vec
y;x^o)[c_{buv}{\vec A}_u(\vec y,x^o)\cdot {\vec \pi}_{v,\perp}
(\vec y,x^o)+\Gamma_b(\vec y,x^o)],\cr}
\form
$$

\noindent where Eqs.(6-32), (6-34) have been used. From Eqs.(6-2) we get

$$\eqalign{
\pi^i_{a,D\perp}(\vec x,x^o)&=\pi^i_{a,\perp}(\vec x,x^o)-\cr
&-{ {\partial_x^i}\over
{\triangle_x} }\int d^3y\, {\vec \partial}_x\cdot {\vec \zeta}^{(\vec A)}
_{ab}(\vec x,\vec y;x^o)[c_{buv}{\vec A}_u(\vec y,x^o)\cdot {\vec
\pi}_{v,\perp}
(\vec y,x^o)+\Gamma_b(\vec y,x^o)]-\cr
&-\int d^3y\, \zeta^{(\vec A)\, i}_{ab}(\vec x,\vec y;x^o)\Gamma_b(\vec y,x^o)
=\cr
&=\int d^3y\, [\delta^{ij}\delta_{av}\delta^3(\vec x-\vec y)-{ {\partial^i_x}
\over {\triangle_x} }{\vec \partial}_x\cdot {\vec \zeta}^{(\vec A)}_{ab}(\vec
x,\vec y;x^o)c_{buv}A^j_u(\vec y,x^o)]\pi^j_{v,\perp}(\vec y,x^o)-\cr
&-P^{ij}_{\perp}(\vec x)\int d^3y\, \zeta^{(\vec A)\, j}_{ab}(\vec x,\vec y;
x^o)\Gamma_b(\vec y,x^o)=\cr
&=\int d^3y\, [\delta^{ij}\delta_{av}\delta^3(\vec x-\vec y)-{ {\partial^i_x}
\over {\triangle_x} }{\vec \partial}_x\cdot {\vec \zeta}^{(\vec A)}_{ab}(\vec
x,\vec y;x^o)c_{buv}A^j_u(\vec y,x^o)]\pi^j_{v,\perp}(\vec y,x^o),\cr}
\form
$$

\noindent due to Eqs.(6-33). Since

$$\eqalign{
\vec \partial \cdot {\vec \pi}_{a,D\perp}(\vec x,x^o)&=-c_{acb}{\vec A}_c(\vec
x
,x^o)\cdot {\vec \pi}_{b,D\perp}(\vec x,x^o)=\cr
&=\int d^3y\, {\vec \partial}_x \cdot {\vec \zeta}^{(\vec A)}_{ab}(\vec x,\vec
y;x^o)c_{buv}{\vec A}_u(\vec y,x^o)\cdot {\vec \pi}_{v,\perp}(\vec y,x^o).\cr}
\form
$$

\noindent one has the inverse relation

$$\eqalign{
\pi^i_{a,\perp}(\vec x,x^o)&=\pi^i_{a,D\perp}(\vec x,x^o)-{ {\partial^i}\over
{\triangle} }[c_{acb}{\vec A}_c(\vec x,x^o)\cdot {\vec \pi}_{b,D\perp}(\vec x,
x^o)]=\cr
&=P^{ij}_{\perp}(\vec x)\pi^j_{a,D\perp}(\vec x,x^o).\cr}
\form
$$

Therefore, using Eqs.(8-2), (8-7),(8-10), we can write

$$
{\vec \pi}_a(\vec x,x^o)={\vec \pi}_{a,\perp}(\vec x,x^o)-{ {{\vec \partial}_x}
\over {\triangle_x} }\int d^3y\, {\vec \partial}_x\cdot
{\vec \zeta}^{(\vec A)}_{ab}(\vec x,\vec y;
x^o) \cdot [c_{buv}{\vec A}_u(\vec y,x^o)\cdot {\vec \pi}_{v,\perp}(\vec
y,x^o)+
\Gamma_b(\vec y,x^o)].
\form
$$

No one of the quantities ${\hat {\vec A}}_{a,\perp}$, ${\vec \pi}_{a,D\perp}$,
${\vec A}_{a,\perp}$, ${\vec \pi}_{a,\perp}$ is gauge invariant. From Eqs.(7-
15) and (6-28) we have with the notations of Eq.(6-45)

$$\eqalign{
\lbrace {\hat A}^i_{a,\perp}(\vec x,x^o),\Gamma_b(\vec y,x^o)\rbrace &=-c_{acb}
{\hat A}^i_{c.\perp}(\vec x,x^o)\delta^3(\vec x-\vec y)={({\hat A}_{\perp}(\vec
x,x^o))}_{ab}\delta^3(\vec x-\vec y),\cr
\lbrace \pi^i_{a,D\perp}(\vec x,x^o),\Gamma_b(\vec y,x^o)\rbrace
&=-c_{acb}\pi^i
_{c,D\perp}(\vec x,x^o)\delta^3(\vec x-\vec y)={(\pi^i_{D\perp}(\vec x,x^o))}
_{ab}\delta^3(\vec x-\vec y),\cr}
\form
$$

\noindent while for the standard transverse quantities we get

$$\eqalign{
\lbrace A^i_{a,\perp}(\vec x,x^o),\Gamma_b(\vec y,x^o)\rbrace &=-c_{acb}P^{ij}
_{\perp}(\vec x)[A^j_c(\vec x,x^o)\delta^3(\vec x-\vec y)]=\cr
&=P^{ij}_{\perp}(\vec
x)[{(A^j(\vec x,x^o))}_{ab}\delta^3(\vec x-\vec y)],\cr
\lbrace \pi^i_{a\perp}(\vec x,x^o),\Gamma_b(\vec y,x^o)\rbrace &=-c_{acb}P^{ij}
_{\perp}(\vec x)[\pi^j_c(\vec x,x^o)\delta^3(\vec x-\vec y)]=\cr
&=P^{ij}_{\perp}
(\vec  x)[{(\pi^j(\vec x,x^o))}_{ab}\delta^3(\vec x-\vec y)],\cr}
\form
$$

{}From Eqs.(7-50), (7-37) and (8-2) one has

$$\eqalign{
&\lbrace \eta_a(\vec x,x^o),\pi^i_b(\vec y,x^o)\rbrace =B_{ac}(\eta (\vec
x,x^o)
)\zeta^{(\vec A)\, i}_{cb}(\vec x,\vec y;x^o)=\cr
&=\lbrace \eta_a(\vec x,x^o),\pi^i_{b,D\perp}(\vec y,x^o)\rbrace +\int d^3y\,
\zeta^{(\vec A)\, i}_{bc}(\vec y,\vec z;x^o)[-B_{ac}(\eta (\vec x,x^o))\delta^3
(\vec x-\vec z)]=\cr
&=\lbrace \eta_a(\vec x,x^o),\pi^i_{b,D\perp}(\vec y,x^o)\rbrace +B_{ac}(\eta
(\vec x,x^o))\zeta^{(\vec A)\, i}_{cb}(\vec x,\vec y;x^o),\cr
&{}\cr
&\lbrace \eta_a(\vec x,x^o),\pi^i_{b,\perp}(\vec y,x^o)\rbrace =B_{ac}(\eta
(\vec x,x^o))P^{ij}_{\perp}(\vec y)\zeta^{(\vec A)\, j}_{cb}(\vec x,\vec y;x^o)
=\cr
&=-B_{ac}(\eta (\vec x,x^o))P^{ij}_{\perp}(\vec y)\zeta^{(\vec A)\, j}_{cb}
(\vec y,\vec x;x^o)=0\cr}
\form
$$

\noindent due to Eqs. (6-33), so that

$$\eqalign{
&\lbrace \eta_a(\vec x,x^o),{\hat A}^i_{b,\perp}(\vec y,x^o)\rbrace =0\cr
&\lbrace \eta_a(\vec x,x^o),\pi^i_{b,D\perp}(\vec y,x^o)\rbrace =0\cr
&\lbrace \eta_a(\vec x,x^o),\pi^i_{b,\perp}(\vec y,x^o)\rbrace =0.\cr}
\form
$$

\noindent Therefore one has

$$\eqalign{
&\lbrace {\hat A}^i_{a,\perp}(\vec x,x^o),\pi^j_{b,D\perp}(\vec y,x^o)\rbrace =
P^{ik}_{\perp}(\vec x)\lbrace {\hat A}^k_{a,\perp}(\vec x,x^o),\pi^j_{b,D\perp}
(\vec y,x^o)\rbrace =\cr
&=P^{ik}_{\perp}(\vec x) \lbrace A^k_a(\vec x,x^o),\pi^j_{b,D\perp}(\vec y,x^o)
\rbrace =\cr
&=P^{ik}_{\perp}(\vec x) \lbrace A^k_a(\vec x,x^o),\pi^j_b(\vec y,x^o)\rbrace
-P^{ik}_{\perp}(\vec x) \int d^3z\, \zeta^{\vec
A)\, j}_{bc}(\vec y,\vec z;x^o)\lbrace A^k_a(\vec x,x^o),\Gamma_c(\vec x,x^o)
=\cr
&=-P^{ij}_{\perp}(\vec x)\delta_{ab}\delta^3(\vec x-\vec y)+c_{auc}P^{ik}
_{\perp}(\vec x)A^k_u(\vec x,x^o)\zeta^{(\vec A)\, j}_{bc}(\vec y,\vec x;x^o)
\cr}
\form
$$

\noindent and Eq.(7-47) imply (this equation is consistent with both types of
transversality of ${\hat {\vec A}}_{a,\perp}$ and ${\vec \pi}_{a,D\perp}$)

$$\eqalign{
\lbrace {\hat A}^i_{a,\perp}(\vec x,x^o)&,\pi^j_{b,D\perp}(\vec y,x^o)\rbrace
=\cr
&-P^{ij}_{\perp}(\vec x)\delta_{ab}\delta^3(\vec x-\vec y)-c_{acd}P^{ik}
_{\perp}(\vec x)[A^k_c(\vec x,x^o)\zeta^{(\vec A)\, j}_{db}(\vec x,\vec
y;x^o)],
\cr}
\form
$$

\noindent while Eqs.(8-10) imply

$$
\lbrace {\hat A}^i_{a,\perp}(\vec x,x^o), \pi^j_{b,\perp}(\vec y,x^o)\rbrace =
P^{jk}_{\perp}(\vec y)\lbrace {\hat A}^i_{a,\perp}(\vec x,x^o), \pi^k_{b,
D\perp}(\vec y,x^o)\rbrace =-P^{ij}_{\perp}(\vec x)\delta_{ab}\delta^3(\vec x-
\vec y),
\form
$$

\noindent to be compared with

$$
\lbrace A^i_{a,\perp}(\vec x,x^o),\pi^j_{b,\perp}(\vec y,x^o)\rbrace =-\delta
_{ab}P^{ij}_{\perp}(\vec x)\delta^3(\vec x-\vec y).
\form
$$

Therefore the natural pairs ${\hat {\vec A}}_{a,\perp}$, ${\vec \pi}_{a,D\perp}
$ arising from the solutions of the constraints and of the multitemporal
equations are
independent from ${\tilde \Gamma}_b$ due to Eqs.(8-15), have tensorial
covariance properties due to Eqs.(8-12), but are not canonical due to
Eqs.(8-17)
and do not have the same kind of transversality; instead the pairs ${\vec A}
_{a,\perp}$, ${\vec \pi}_{a,\perp}$, with the same kind of transversality, are
canonical in the sense of Eqs.(8-19), do not transform as tensors due to Eqs.
(8-13). The strategy will be to extract natural
Dirac's observables from ${\hat {\vec A}}_{a,\perp}$, ${\vec \pi}_{a,D\perp}$
and then to use the observable part of pairs ${\hat {\vec A}}_{a,\perp}$,
${\vec \pi}_{a,\perp}$ in terms of observables.

If we multiply Eqs.(8-12) by $A_{bc}(\eta (\vec y,x^o))$ and we use Eq.(B-9),
we obtain

$$\eqalign{
\lbrace {\hat A}^i_{a,\perp}(\vec x,x^o),{\tilde \Gamma}_b(\vec y,x^o)\rbrace
&=-{ {\tilde \delta {\hat A}^i_{a,\perp}(\vec x,x^o)}\over {\delta \eta_b(\vec
y,x^o)} }=\cr
&=-{(H_b(\eta (\vec x,x^o)))}_{ac}{\hat A}^i_{c,\perp}(\vec x,x^o)\delta^3(\vec
x-\vec y),\cr
\lbrace \pi^i_{a,D\perp}(\vec x,x^o),{\tilde \Gamma}_b(\vec y,x^o)\rbrace &=-
{ {\tilde \delta \pi^i_{a,\perp}(\vec x,x^o)}\over {\delta \eta_b(\vec
y,x^o)} }=\cr
&=-{(H_b(\eta (\vec x,x^o)))}_{ac}\pi^i_{c,\perp}(\vec x,x^o)\delta^3(\vec
x-\vec y).\cr}
\form
$$

These equations are a functional multitemporal generalization of the matrix
equation ${d\over {dt}}U(t,t_o)=hU(t,t_o)$, $U(t_o,t_o)=1$, whose solution
required the introduction of the concept of time-ordering [99];
similar equations
gave origin to path-ordering. In this case we need to introduce a path-ordering
between the configuration with $\eta_a(\vec x,x^o)=0$ (the identity cross
section $\sigma_I$) and a generic one $\eta_a(\vec y,x^o)$ (the cross section
$\sigma_{\eta (\vec y,x^o)}$), inside a suitable tubolar neighbourhood of
$\sigma_I$. The zero curvature condition of Eqs.(B-10) assures that the result
is independent from the path connecting these two configurations (even if in
the evaluation of the solution the path $\hat \gamma (\vec y,x^o,s)$ is
dictated by the directional functional derivatives in Eqs.(8-20)), so that we
will choose the preferred path $\hat \gamma (\vec y,x^o,s)$, $0\leq s\leq 1$.
Therefore, we get the following solution for the dependence of ${\hat {\vec A}}
_{a,\perp}$, ${\vec \pi}_{a,D\perp}$ on the gauge variables $\eta_a$

$$\eqalign{
{\hat {\vec A}}_{a,\perp}(\vec x,x^o)={\hat {\vec A}}_{a,T}(\vec x,x^o;\eta
(\vec x,x^o))=&F_{ab}[\eta (\vec x,x^o)]{\check {\vec A}}_{b,\perp}(\vec
x,x^o),
\cr
{\vec \pi}_{a,D\perp}(\vec x,x^o)={\hat {\vec \pi}}_{a,DT}(\vec x,x^o;\eta
(\vec x,x^o))&=F_{ab}[\eta (\vec x,x^o)]{\check {\vec \pi}}_{b,D\perp}(\vec x,
x^o),\cr}
\form
$$

\noindent where the Cauchy data are ${\hat {\vec A}}_{a,\perp}$, ${\vec \pi}
_{a,D\perp}$ evaluated by using the identity cross section $\sigma_I$
starting from a connection 1-form

$$\eqalign{
{\check {\vec A}}_{a,\perp}(\vec x,x^o)&={\hat {\vec A}}_{a,T}(\vec x,x^o;0),
\quad\quad \lbrace {\check {\vec A}}_{a,\perp}(\vec x,x^o),{\tilde \Gamma}_b
(\vec y,x^o)\rbrace =0,\cr
{\check {\vec \pi}}_{a,D\perp}(\vec x,x^o)&={\hat {\vec \pi}}_{a,DT}(\vec
x,x^o;
0),\quad\quad \lbrace {\check {\vec \pi}}_{a,D\perp}(\vec x,x^o),{\tilde
\Gamma}
_b(\vec y,x^o)\rbrace =0.\cr}
\form
$$

These are natural Dirac's observables of YM theory, in the sense of the
symplectic decoupling looked for with the Shanmugadhasan canonical
transformation, since they have zero
Poisson brackets with ${\tilde \Gamma}_b$ being Cauchy data of Eqs.(8-20)
and also with $\eta_a$ due to Eqs.(8-15). The matrix $F[\eta (\vec x,x^o)]$ is
($(l)$ denotes an arbitrary path)

$$\eqalign{
F_{ab}[\eta (\vec x,x^o)]&= {(P\, e^{\int^{\eta_a(\vec x,x^o)}_{(l)\,
0}H_u(\bar
\eta ){\cal D}{\bar \eta}_u})}_{ab}=\cr
&={(P\, e^{\int^{\eta_a(\vec x,x^o)}_{(\hat \gamma )\, 0}H_u(\bar \eta ){\cal
D}{\bar \eta}_u})}_{ab}={(P\, e^{\Omega^{\hat \gamma}(\eta (\vec
x,x^o))})}_{ab}
\cr}
\form
$$

In the limit $\eta \rightarrow 0$, due to Eqs.(7-10) and (7-19), the
transversality conditions (6-22) and (7-33) become

$$\eqalign{
&(\vec \partial \delta_{ab}+c_{acb}{\vec A}_c(\vec x,x^o))\cdot {\vec \pi}_{b,
D\perp}(\vec x,x^o)\equiv 0 \, \rightarrow \cr
&(\vec \partial \delta_{ab}+c_{acb}{\check {\vec A}}_{c,\perp}(\vec x,x^o))
\cdot {\check {\vec \pi}}_{b,D\perp}(\vec x,x^o)={\hat {\vec D}}^{({\check
{\vec A}})}_{ab}(\vec x,x^o)\cdot {\check {\vec \pi}}_{b,D\perp}(\vec x,x^o)
\equiv 0\cr
&{}\cr
&\vec \partial \cdot {\check {\vec A}}_{a,\perp}(\vec x,x^o)=0.\cr}
\form
$$

Therefore natural Dirac's observables of pure YM theory are ${\check {\vec A}}
_{a,\perp}$ and ${\check {\vec \pi}}_{a,D\perp}$, i.e. quantities evaluated on
the identity cross section $\sigma_I$. This means that to separate the gauge
degrees of freedom and the 1st class constraints from Dirac's observables in
the sense of the Shanmugadhasan canonical transformation we have to evaluate
the gauge potentials and the electric fields (the momenta) of the YM theory on
$P^{t}=(R^3\times \lbrace x^o\rbrace )\times G$, and therefore the gauge
potentials and field strengths of YM theory on $P^4=M^4\times G$, on the
respective identity cross section starting from the respective connection
1-forms and curvature 2-forms. If "$\, {\check {}}\, $" denotes such
evaluation,
from Eqs.(8-2) and (8-11) one has for the corresponding gauge potential and
electric field

$$\eqalign{
{\check {\vec A}}_a(\vec x,x^o)&={\check {\vec A}}_{a,\perp}(\vec x,x^o),\cr
{\check {\vec \pi}}_a(\vec x,x^o)&={\check {\vec \pi}}_{a,\perp}-\cr
&-{ {{\vec \partial}_x}\over {\triangle_x} }\int d^3y\, {\vec \partial}_x\cdot
{\vec \zeta}^{({\check {\vec A}})}_{ab}(\vec x,\vec y;x^o)[c_{buv}{\check
{\vec A}}_{u,\perp}(\vec y,x^o)\cdot {\check {\vec \pi}}_{v,\perp}(\vec y,x^o)
+{\check \Gamma}_b(\vec x,x^o)],\cr
&{}\cr
{\check \pi}^i_{a,\perp}(\vec x,x^o)&=P^{ij}_{\perp}(\vec x){\check \pi}^j
_{a,D\perp}(\vec x,x^o),\cr}
\form
$$

\noindent with

$$\eqalign{
\lbrace {\check A}^i_{a,\perp}(\vec x,x^o),{\check \pi}^j_{b,\perp}(\vec x,
x^o)\rbrace &=-P^{ij}_{\perp}(\vec x)\delta_{ab}\delta^3(\vec x-\vec y)\cr
\lbrace {\check A}^i_{a,\perp}(\vec x,x^o),{\tilde \Gamma}_b(\vec y,x^o)
\rbrace &=\lbrace {\check \pi}^i_{a,\perp}(\vec x,x^o),{\tilde \Gamma}_b(\vec
y,x^o)\rbrace =0.\cr}
\form
$$

\noindent These are final Dirac's observables of pure YM theory; the gauge
degrees of freedom are $A_{ao}$, $\pi^o_a$, $\eta_a$, ${\tilde \Gamma}_a$.

In presence of fermions, Gauss' law 1st class constraints $\Gamma^T_a$ are
given in Eqs.(4-21)

$$\Gamma^T_a(\vec x,x^o)=\Gamma_a(\vec x,x^o)+i\psi^{\dagger}_{\alpha u}(\vec
x,x^o){(T^a)}_{uv}\psi_{\alpha v}(\vec x,x^o)\approx 0,
\form
$$

\noindent and the discussion of the pure YM part does not change. By using the
Dirac brackets (4-29), one obtains the following transformation rules of the
fermionic fields (${\tilde \Gamma}^T_a=\Gamma_b^TA_{ba}(\eta )$)

$$\eqalign{
\lbrace \psi_{\alpha a}(\vec x,x^o),\Gamma^T_b(\vec y,x^o)\rbrace &= {(T^b)}
_{ac}\psi_{\alpha c}(\vec x,x^o)\delta^3(\vec x-\vec y)\cr
\lbrace \psi_{\alpha a}(\vec x,x^o),{\tilde \Gamma}^T_b(\vec x,x^o)\rbrace &=
-{ {\tilde \delta \psi_{\alpha a}(\vec x,x^o)}\over {\delta \eta_b(\vec
x,x^o)}}
=\cr
&={(T^cA_{cb}(\eta (\vec x,x^o)))}_{ad}\psi_{\alpha d}(\vec x,x^o)\delta^3(\vec
x-\vec y)=\cr
&={(H_b(\eta (\vec x,x^o)))}_{ad}\psi_{\alpha d}(\vec x,x^o)\delta^3
(\vec x-\vec y)\cr}
\form
$$

\noindent so that the solution of the functional multitemporal equations,
taking into account the zero curvature conditions of Eqs.(B-10) and Eqs.(8-23),
is

$$\eqalign{
\psi_{\alpha a}(\vec x,x^o)&={\hat \psi}_{\alpha a}(\vec x,x^o;\eta (\vec x,
x^o))
={(P\, e^{-\Omega^{\hat \gamma (\rho)}(\eta (\vec x,x^o))})}_{ab}{\check \psi}
_{\alpha b}(\vec x,x^o)=\cr
&={(P\, e^{-\Omega^{\hat \gamma}_c(\eta (\vec x,x^o))T^c})}_{ab}{\check \psi}
_{\alpha b}(\vec x,x^o)=F^{(\rho )}_{ab}[\eta (\vec x,x^o)]{\check \psi}
_{\alpha b}(\vec x,x^o)\cr
&{}\cr
\psi^{\dagger}_{\alpha a}(\vec x,x^o)&={\check \psi}^{\dagger}_{\alpha b}(\vec
x,x^o){(F^{(\rho )\, -1})}_{ba}[\eta (\vec x,x^o)]={\check \psi}^{\dagger}
_{\alpha b}(\vec x,x^o)F^{(\rho )}_{ba}[-\eta (\vec x,x^o)],\cr}
\form
$$

\noindent with

$$
\eqalign{
&{\check \psi}_{\alpha a}(\vec x,x^o)={\hat \psi}_{\alpha a}(\vec
x,x^o;0),\quad
\quad \lbrace {\check \psi}_{\alpha a}(\vec x,x^o),{\tilde \Gamma}^T_b(\vec y,
x^o)\rbrace =0\cr
&\lbrace {\check \psi}_{a\alpha}(\vec x,x^o),{\check \psi}^{\dagger}_{b\beta}
(\vec y,x^o)\rbrace =-i\delta_{ab}\delta_{\alpha\beta}\delta^3(\vec x-\vec y).
\cr}
\form
$$

\noindent Again Dirac's (Grassmann-valued) observables are the Cauchy data of
the multitemporal equations and are associated with the identity cross section
of the associated bundle E of Dirac fields.

To extract the gauge invariant part of the interaction Hamiltonian (4-31) one
needs to use the Baker-Haussdorff formula [90]

$$
e^{-x}\, y\, e^x=\sum_{n=0}^{\infty}{1\over {n!}}{}^{\hat {}}[yx^n]
\form
$$

\noindent
valid for x,y belonging to a Lie algebra $g$ and with the notation ${}^{\hat
{}}[xy]=[x,y]$, ${}^{\hat {}}[xyz]=[{}^{\hat {}}[xy],z]=[[x,y],z]$ and so on.
In this way one obtains ($T^a$ are in the representation $\rho$, while ${\hat
T}^a$ are in the adjoint representation)

$$\eqalign{
e^{\Omega^{\hat \gamma}_bT^b}\, T^a\, e^{-\Omega^{\hat \gamma}_cT^c}&=
\sum_{n=0}^{\infty}{ {{(-)}^n}\over {n!} }\, \alpha_{b_1}\cdots \alpha_{b_n}
{}^{\hat {}}[T^aT^{b_1}\cdots T^{b_n}]=\cr
&=T^c\, \sum_{n=0}^{\infty}{ {{(-)}^n}\over {n!} }\, c_{cd_1b_1}\alpha_{b_1}
\cdots c_{d_{n-1}ab_n}\alpha_{b_n}=\cr
&T^c\, \sum_{n=0}^{\infty}{ {{(-)}^n}\over
{n!} }\, {(\alpha_{b_1}T^{b_1})}_{cd_1}\cdots {(\alpha_{b_n}T^{b_n})}_{d_{n-1}
a}=\cr
&=T^c\, {(e^{-\Omega^{\hat \gamma}_b{\hat T}^b})}_{ca}.\cr}
\form
$$

Therefore taking into account Eqs.(7-31) (with $d/d\vec x\rightarrow \vec
\partial$), (8-23) and (8-30) we have

$$\eqalign{
i\psi^{\dagger}_u(\vec x,x^o)\vec \alpha \cdot &{[\vec \partial +{\vec A}_a
(\vec x,x^o)T^a]}_{uv}\psi_v(\vec x,x^o)=\cr
&{}\cr
&=i\psi^{\dagger}_u(\vec x,x^o)\vec \alpha \cdot {[\vec \partial +({\vec
\Theta}_a(\eta (\vec x,x^o),\vec \partial \eta (\vec x,x^o))+{\hat {\vec A}}
_{a,\perp}(\vec x,x^o))T^a]}_{uv}\psi_v(\vec x,x^o)=\cr
&=i\psi^{\dagger}_u(\vec x,x^o)\vec \alpha \cdot [\vec \partial +F^{(\rho )}
_{ur}[-\eta (\vec x,x^o)]\vec \partial \Omega^{\hat \gamma}_a(\eta (\vec x,
x^o)){(T^a)}_{rs}F^{(\rho )}_{sv}[\eta (\vec x,x^o)]+\cr
&+ ({\vec \Theta}_a(\eta (\vec x,x^o),\vec \partial \eta (\vec x,x^o))+{\hat
{\vec A}}_{a,\perp}(\vec x,x^o))\cr
&F^{(\rho )}_{ur}[-\eta (\vec x,x^o)]{(T^a)}_{rs}
F^{(\rho )}_{sv}[\eta (\vec x,x^o)] ]{}_{uv}{\check \psi}_v(\vec x,x^o)=\cr
&=i{\check \psi}^{\dagger}_a(\vec x,x^o)\vec \alpha \cdot [\vec \partial +
{\check {\vec A}}_{c,\perp}(\vec x,x^o)T^c+\cr
&+2T^c{(P\, e^{\Omega^{\hat \gamma}_u(\eta (\vec x,x^o){\hat T}^u})}_{cd}{\vec
\Theta}_d(\eta (\vec x,x^o),\vec \partial
\eta (\vec x,x^o))]{}_{ab}{\check \psi}_b(\vec x,x^o)\cr
&{\rightarrow}_{\eta \, \rightarrow \, 0}\,\, i{\check \psi}^{\dagger}_a(\vec
x,
x^o)\vec \alpha \cdot {(\vec \partial +{\check {\vec A}}_c(\vec x,x^o)T^c)}
_{ab}{\check \psi}_b(\vec x,x^o).\cr}
\form
$$

Let us conclude this Section with the remark that the relevance of the global
identity cross section of the trivial principal bundle $P^t$ in the
construction
of global Dirac's observables seems to indicate that global Dirac's observables
probably do not exist for theories based on non-trivial principal bundles,
which, therefore, cannot be re-expressed globally only in terms of
measurable quantities as it will be done in next Section for the YM theory
with fermions.

\bigskip
\vfill\eject

\bigskip\noindent
{\bf {9. Physical Lagrangian and Hamiltonian.}}
\newcount \nfor

\def \form {\global \advance \nfor by 1 \eqno (9.\the\nfor)}
\bigskip

Let us now consider the action $S_T$ with the Lagrangian density ${\cal L}_T
(\vec x,x^o)$ of Eqs.(4-3) and the Dirac Hamiltonian $H_{DT}$ of Eqs.(4-31)
after the elimination of the second class constraints (4-18) with the Dirac
brackets (4-29). Since ${\cal L}_T$ is gauge invariant, we can evaluate it by
using the gauge potentials ${\check A}_{a\mu}$ generated by the identity cross
section $\sigma_I$ from a connection 1-form; let ${\check {\cal L}}_T(\vec x,
x^o)$ be the resulting Lagrangian density and let ${\check H}_{cT}$ be the
associated canonical Hamiltonian.

By using Eqs.(2-16) and (4-3) and omitting the surface term in Eqs.(4-3),
the original Lagrangian density is

$$\eqalign{
{\cal L}_T(\vec x,x^o)
&={1/{2g^2}}\sum_a \lbrace ( {\dot {\vec A}}_a(\vec x,x^o)
-{\hat {\vec D}}^{(\vec A)}_{ab}(\vec x,x^o)A^o_a(\vec x,x^o)){}^2-{\vec B}_a
^2(\vec x,x^o)\rbrace +\cr
&+i\bar \psi (\vec x,x^o)\gamma^{\mu}(\partial_{\mu}+A_{a\mu}(\vec x,x^o)T^a)
\psi (\vec x,x^o)-m\bar \psi (\vec x,x^o)\psi (\vec x,x^o),\cr}
\form
$$

\noindent while, from Eqs.(2-17), (4-9) and (4-10), the equations of motion
are

$$\eqalign{
L^{oT}_a(\vec x,x^o)&={\hat {\vec D}}^{(\vec A)}_{ab}(\vec x,x^o)\cdot {\dot
{\vec A}}_b(\vec x,x^o)-({\hat {\vec D}}^{(\vec A)}_{ac}\cdot {\hat {\vec D}}
^{(\vec A)}_{cb})(\vec x,x^o)A^o_b(\vec x,x^o)+\cr
&+ig^2\psi^{\dagger}(\vec x,x^o)
T^a\psi (\vec x,x^o){\buildrel \circ \over =}0,\cr
&{}\cr
{\vec L}^T_a(\vec x,x^o)&=(\delta_{ab}\partial^o+c_{aub}A^o_u(\vec x,x^o))
(-{\hat {\vec D}}^{(\vec A)}_{bc}(\vec x,x^o)A^o_c(\vec x,x^o)+{\dot {\vec
A}}_b(\vec x,x^o))-\cr
&-{\hat {\vec D}}^{(\vec A)}_{ab}(\vec x,x^o)\times [(
{\hat {\vec D}}^{(\vec A)}_{bc}(\vec x,x^o)-{1\over 2}c_{bdc}{\vec A}_d(\vec
x,x^o))\times {\vec A}_c(\vec x,x^o)]+\cr
&+i\psi (\vec x,x^o)\vec \alpha T^a\psi (\vec x,x^o){\buildrel \circ \over
=}0,\cr
&{}\cr
L^T_{\psi}(\vec x,x^o)&=
\bar \psi (\vec x,x^o)[i({\buildrel \leftarrow \over {\partial_{\mu}} }-
A_{a\mu}(\vec x,x^o)T^a)\gamma^{\mu}+m]{\buildrel \circ \over =}0,\cr
L^T_{\bar \psi}(\vec x,x^o)&=
[i(\partial_{\mu}+A_{a\mu}(\vec x,x^o)T^a)\gamma^{\mu}-m]\psi (\vec x,x^o)
{\buildrel \circ \over =}0.\cr}
\form
$$

Using Eqs.(7-13), (7-31) and (7-14) we get ($\dot \eta =\partial^o \eta$)

$$\eqalign{
\partial^oA^i_a(\vec x,x^o)&= \partial^i{\dot \Omega}^{\hat \gamma}_a(\eta
(\vec
x,x^o))+\partial^o{|}_{\eta}{\hat A}^i_{a,T}(\vec x,x^o;\eta (\vec x,x^o))+\cr
&+{\dot \eta}_b{ {\partial {\hat A}^i_{a,T}(\vec x,x^o;\eta (\vec x,x^o))}\over
{\partial \eta_b} }{|}_{\eta =\eta (\vec x,x^o)}=\cr
&=\partial^i{\dot \Omega}^{\hat \gamma}_a(\eta (\vec x,x^o))+[\partial^o{|}
_{\eta}\delta_{ac}-c_{adc}A_{db}(\eta (\vec x,x^o)) {\dot \eta}_b]{\hat A}^i
_{a,T}(\vec x,x^o;\eta (\vec x,x^o)).\cr}
\form
$$

\noindent Since from Eqs.(7-25) one obtains ($\partial^o{|}_{\eta}\Omega^{\hat
\gamma}_a$ is the derivative with respect to the time dependence of the
integrand in Eq.(7-30))

$$
{\dot \Omega}^{\hat \gamma}_a(\eta (\vec x,x^o))=A_{ab}(\eta (\vec x,x^o))
{\dot \eta}_b(\vec x,x^o)+\partial^o{|}_{\eta}\Omega^{\hat \gamma}_a(\eta (\vec
x,x^o)),
\form
$$

\noindent and since the time independency of the identity cross section
$\sigma_I$ implies

$$
{\dot \eta}_a(\vec x,x^o){|}_{\eta =0}={(\partial^o)}^n\eta (\vec x,x^o){|}
_{\eta =0}=0,\quad \quad \Rightarrow \,\,\, {\dot \Omega}^{\hat \gamma}_a(\eta
(\vec x,x^o)){|}_{\eta =0}=0,
\form
$$

\noindent then these results, together with $\partial^o{|}_{\eta}\Omega^{\hat
\gamma}_a(\eta (\vec x,x^o)){|}_{\eta =0}=0$, imply the following expression
and limit of ${\dot A}^i_a$:

$$\eqalign{
{\dot A}^i_a(\vec x,x^o)&= \partial^o{|}_{\eta}{\hat A}^i_{a,T}(\vec x,x^o;
\eta (\vec x,x^o))+(\delta_{ab}\partial^i+c_{acb}{\hat A}^i_{a,T}(\vec x,x^o;
\eta (\vec x,x^o))){\dot \Omega}^{\hat \gamma}_b(\eta (\vec x,x^o))=\cr
&=\partial^o{|}_{\eta}{\hat A}^i_{a,T}(\vec x,x^o;\eta (\vec x,x^o))+{\hat D}
^{({\hat A}_T)\, i}_{ab}(\vec x,x^o){\dot \Omega}^{\hat \gamma}_b(\eta (\vec
x,x^o))\, {\rightarrow}_{\eta \rightarrow 0}\cr
&{\rightarrow}_{\eta \rightarrow 0}\,\, \partial^o{\check A}^i_{a,\perp}(\vec
x,x^o).\cr}
\form
$$

Since, due to Eqs.(6-35), the equation $L^{oT}_a(\vec x,x^o){\buildrel \circ
\over =}0$ can be written as ($J^{\mu}_a(\vec x,x^o)=i\bar \psi (\vec x,x^o)
\gamma^{\mu}T^a\psi (\vec x,x^o)$ from Eqs.(4-10))

$$\eqalign{
\triangle^{(\vec A)}_{ab}(\vec x,x^o)A^o_b(\vec x,x^o)&=({\hat {\vec D}}_{ac}
^{(\vec A)}\cdot {\hat {\vec D}}^{(\vec A)}_{cb})(\vec x,x^o)A^o_b(\vec x,x^o)
=\cr
&={\hat {\vec D}}^{(\vec A)}_{ab}(\vec x,x^o)\cdot {\dot {\vec A}}_b(\vec
x,x^o)
+g^2J^o_a(\vec x,x^o)-L^{oT}_a(\vec x,x^o),\cr}
\form
$$

\noindent then its solution is

$$
A^o_a(\vec x,x^o)=\int d^3y\, G^{(\vec A)}_{\triangle ,ab}(\vec x,\vec y;x^o)
[{\hat {\vec D}}^{(\vec A)}_{bc}(\vec y,x^o)\cdot {\dot {\vec A}}_c(\vec y,x^o)
+g^2J^o_a(\vec y,x^o)-L^{oT}_a(\vec y,x^o)],
\form
$$

\noindent or else (also here as in the previous formula we discard homogeneous
solutions)

$$
{\dot {\vec A}}_a(\vec x,x^o)=-\int d^3y\, {\vec \zeta}^{(\vec A)}_{ab}(\vec x,
\vec y;x^o)[\triangle^{(\vec A)}_{bc}(\vec y,x^o)A^o_c(\vec y,x^o)-g^2J^{o}_b
(\vec y,x^o)+L^{oT}_b(\vec y,x^o)],
\form
$$

\noindent We obtain

$$\eqalign{
&\triangle^{(\vec \Theta +{\hat {\vec A}}_T)}_{ab}(\vec x,x^o)A^o_b(\vec
x,x^o)=
\cr
&={\hat {\vec D}}_{ab}^{(\vec \Theta +{\hat {\vec A}}_T)}(\vec x,x^o)\cdot [
\partial^o{|}_{\eta}{\hat {\vec A}}_{b,T}(\vec x,x^o;\eta (\vec x,x^o))+
{\hat {\vec D}}_{bc}^{(\vec \Theta +{\hat {\vec A}}_T)}(\vec x,x^o){\dot
\Omega}
^{\hat \gamma}_c(\eta (\vec x,x^o))-\cr
&-c_{buc}{\vec \Theta}_u(\eta (\vec x,x^o),\vec \partial \eta (\vec x,x^o))
{\dot \Omega}^{\hat \gamma}_c(\eta (\vec x,x^o))]+g^2J^o_a(\vec x,x^o)-L^{oT}_a
(\vec x,x^o),\cr}
\form
$$

\noindent so that

$$\eqalign{
A^o_a(\vec x,x^o)&={\dot \Omega}^{\hat \gamma}_a(\vec x,x^o)+\int d^3y\,
G^{(\vec \Theta +{\hat {\vec A}}_T)}_{\triangle ,ab}(\vec x,\vec y;x^o)\cr
&[{\hat {\vec D}}_{ab}^{(\vec \Theta +{\hat {\vec A}}_T)}(\vec y,x^o)\cdot
(\partial^o{|}_{\eta}{\hat {\vec A}}_{b,T}(\vec y,x^o;\eta (\vec y,x^o))-
c_{buc}{\vec \Theta}_u(\eta (\vec y,x^o),\vec \partial \eta (\vec y,x^o))\cr
&{\dot \Omega}^{\hat \gamma}_c(\eta (\vec y,x^o)))
+g^2J^o_a(\vec y,x^o)-L^{oT}_a(\vec y,x^o)]\cr}
\form
$$

\noindent and finally

$$\eqalign{
A^o_a{|}_{\eta =0}&(\vec x,x^o)=\cr
&=\int d^3y\, G^{({\check {\vec A}}_{\perp})}
_{\triangle ,ab}(\vec x,\vec y;x^o)[{\hat {\vec D}}
^{({\check {\vec A}}_{\perp})}_{bc}(\vec y,x^o)\cdot \partial^o{\check {\vec
A}}_{c,\perp}(\vec y,x^o)+\cr
&+g^2J^o_a{|}_{\eta =0}(\vec y,x^o)-L^{oT}_a{|}_{\eta
=0}(\vec y,x^o)]=\cr
&=\int d^3y\, G^{({\check {\vec A}}_{\perp})}
_{\triangle ,ab}(\vec x,\vec y;x^o)[{\hat {\vec D}}
^{({\check {\vec A}}_{\perp})}_{bc}(\vec y,x^o)\cdot \partial^o{\check {\vec
A}}_{c,\perp}(\vec y,x^o)+\cr
&+ig^2{\check \psi}^{\dagger}(\vec y,x^o)T^b\check \psi (\vec y,x^o)-
L^{oT}_b{|}_{\eta =0}(\vec y,x^o)]\cr}
\form
$$

\noindent where the next Eqs.(9-14) have been used.
Let us remark that the use of $L^{oT}_a{\buildrel \circ \over =}0$ is
equivalent to the phase space constraint $\Gamma^T_a\approx 0$.

For the fermionic terms Eqs.(8-29), (8-33) imply

$$\eqalign{
{\dot \psi}_a(\vec x,x^o)&=\partial^o(F^{(\rho )}_{ab}[\eta (\vec x,x^o)]
{\check \psi}_b(\vec x,x^o))=\cr
&=F^{(\rho )}_{ab}[\eta (\vec x,x^o)]\partial^o{\check \psi}_b(\vec x,x^o)-
{({\dot \Omega}^{\hat \gamma}(\eta (\vec x,x^o))F^{(\rho )}
[\eta (\vec x,x^o)])}_{ab}{\check \psi}_b
(\vec x,x^o){\rightarrow}_{\eta \rightarrow 0} \cr
&{\rightarrow}_{\eta \rightarrow 0}\,\, \partial^o{\check \psi}_a(\vec x,x^o)
\cr
&{}\cr
i\psi^{\dagger}_u(\vec x,x^o)&{(\partial^o+A^o_a(\vec x,x^o)T^a)}_{uv}\psi_v
(\vec x,x^o)=\cr
&=i{\check \psi}^{\dagger}_u(\vec x,x^o)[\delta_{uv}\partial^o
-{(F^{(\rho )}[-\eta (\vec x,x^o)]{\dot \Omega}^{\hat \gamma}(\eta (\vec x,
x^o))F^{(\rho )}[\eta (\vec x,x^o))])}_{uv}+\cr
&+{(T^c)}_{uv}{(e^{-\Omega^{\hat
\gamma}(\eta (\vec x,x^o))})}_{ca}A^o_a(\vec x,x^o)]\check \psi (\vec x,x^o)
\rightarrow_{\eta \rightarrow 0\,\, L^{oT}_a{\buildrel \circ \over =}0} \cr
&{\rightarrow}_{\eta \rightarrow 0\,\, L^{oT}_a{\buildrel \circ \over =}0}
\,\,\, i{\check \psi}^{\dagger}_u(\vec x,x^o)(\partial^o+\cr
&+T^a\int d^3y\, G^{({\check
{\vec A}}_{\perp})}_{\triangle ,ab}(\vec x,\vec y;x^o)[{\hat {\vec D}}
^{({\check {\vec A}}_{\perp})}_{bc}(\vec y,x^o)\cdot \partial^o{\check {\vec
A}}_{c,\perp}(\vec y,x^o)+\cr
&+ig^2{\check \psi}^{\dagger}(\vec y,x^o)T^b\check \psi (\vec y,x^o)]){}_{uv}
{\check \psi}_v (\vec x,x^o)\cr}
\form
$$

$$\eqalign{
J^o_a(\vec x,x^o)&=i\psi^{\dagger}(\vec x,x^o)T^a\psi (\vec x,x^o)=\cr
&=i{\check \psi}^{\dagger}_u(\vec x,x^o)F^{(\rho )}_{ur}[-\eta (\vec x,x^o)]
{(T^a)}_{rs}F^{(\rho )}_{sv}[\eta
(\vec x,x^o)]{\check \psi}_v(\vec x,x^o)=\cr
&=i{\check \psi}^{\dagger}_u(\vec x,x^o){(T^c)}_{uv}{(e^{-\Omega^{\hat \gamma}
_b(\eta (\vec x,x^o){\hat T}^b})}_{ca}{\check \psi}_v(\vec x,x^o)
{\rightarrow}_{\eta \rightarrow 0}\,\, i{\check \psi}^{\dagger}(\vec x,x^o)T^a
{\check \psi}(\vec x,x^o)\cr
&{}\cr
{\vec J}_a(\vec x,x^o)&=i\psi^{\dagger}(\vec x,x^o)\vec \alpha T^a\psi (\vec x,
x^o){\rightarrow}_{\eta \rightarrow 0}\,\, i{\check \psi}^{\dagger}(\vec x,x^o)
\vec \alpha T^a\check \psi (\vec x,x^o).\cr}
\form
$$

Using Eqs.(9-1), (9-6), (9-12), (9-13), (8-33), we obtain

$$\eqalign{
{\cal L}_T(\vec x,x^o)&= {1\over {2g^2}}\sum_a\, \lbrace ({\dot {\vec A}}_a
(\vec x,x^o)-{\hat {\vec D}}^{(\vec A)}_{ab}(\vec x,x^o)A^o_b(\vec x,x^o)){}^2
-{\vec B}_a^2(\vec x,x^o)\rbrace +\cr
&+i\bar \psi (\vec x,x^o)\gamma^{\mu}(\partial_{\mu}+A_{a\mu}(\vec x,x^o)T^a)
\psi (\vec x,x^o)-m\bar \psi (\vec x,x^o)\psi (\vec x,x^o)=\cr}
$$

\vfill \eject

$$\eqalign{
&={1\over {2g^2}}\lbrace [\partial^o{|}_{\eta}{\hat {\vec A}}_{a,T}
(\vec x,x^o;\eta (\vec x,x^o)))
+{\hat {\vec D}}^{({\hat {\vec A}}_T)}_{ab}(\vec x,x^o){\dot \Omega}_b^{\hat
\gamma}(\eta (\vec x,x^o))-\cr
&-{\hat {\vec D}}^{(\vec \Theta +{\hat {\vec A}}_T)}
(\vec x,x^o)A^o_b(\vec x,x^o)]{}^2-{\vec B}^2_a(\vec x,x^o)\rbrace +\cr
&+i\bar \psi (\vec x,x^o)\gamma^{\mu}(\partial_{\mu}+A_{a\mu}(\vec x,x^o)T^a)
\psi (\vec x,x^o)-m\bar \psi (\vec x,x^o)\psi (\vec x,x^o)=\cr
&{}\cr
&={1\over {2g^2}}\lbrace [\partial^o{|}_{\eta}{\hat {\vec A}}_{a,T}(\vec x,x^o;
\eta (\vec x,x^o))+c_{aub}{\vec \Theta}_u(\eta (\vec x,x^o),\vec \partial \eta
(\vec x,x^o))A^o_b(\vec x,x^o)-\cr
&-{\hat {\vec D}}_{ab}^{(\vec \Theta +{\hat {\vec
A}}_T)}(\vec x,x^o)(A^o_b(\vec x,x^o)-{\dot \Omega}^{\hat \gamma}_b(\eta (\vec
x,x^o)))]{}^2-{\vec B}^2_a(\vec x,x^o)\rbrace +\cr
&+i\bar \psi (\vec x,x^o)\gamma^{\mu}(\partial_{\mu}+A_{a\mu}(\vec x,x^o)T^a)
\psi (\vec x,x^o)-m\bar \psi (\vec x,x^o)\psi (\vec x,x^o){\rightarrow}
_{\eta \rightarrow 0,\, L^{oT}_a{\buildrel \circ \over =}0} \cr
&{}\cr
&{\rightarrow}_{\eta \rightarrow 0,\, L^{oT}_a{\buildrel \circ \over =}0}
\,\,\,\, {\check {\cal L}}_T^{'}(\vec x,x^o)\cr}
\form
$$

\noindent with

$$\eqalign{
{\check {\cal L}}_T^{'}(\vec x,x^o)
&= {1\over {2g^2}}(\, [\partial^o{\check {\vec A}}_{a,
\perp}(\vec x,x^o)-{\hat {\vec D}}_{ab}^{({\check {\vec A}}_{\perp})}(\vec x,
x^o)\int d^3y\, G^{({\check {\vec A}}_{\perp})}_{\triangle ,bc}(\vec x,\vec y;
x^o)\cr
&({\hat {\vec D}}_{cd}^{({\check {\vec A}}_{\perp})}(\vec y,x^o)\cdot
\partial^o
{\check {\vec A}}_{d,\perp}(\vec y,x^o)+ig^2{\check \psi}^{\dagger}(\vec y,x^o)
T^c\check \psi (\vec y,x^o))]{}^2-\cr
&-{\check {\vec B}}_a^2(\vec x,x^o)\, )
+{\check \psi}^{\dagger}(\vec x,x^o)\lbrace i[\partial^o -\vec \alpha \cdot
(\vec \partial_x+{\check {\vec A}}_{a,\perp}(\vec x,x^o)T^a)]+\cr
&+iT^a\int d^3y\,
G^{({\check {\vec A}}_{\perp})}_{\triangle ,ab}(\vec x,\vec y;x^o)[{\hat {\vec
D}}_{bc}^{({\check {\vec A}}_{\perp})}(\vec y,x^o)\cdot \partial^o{\check {\vec
A}}_{c,\perp}(\vec y,x^o)+\cr
&+ig^2{\check \psi}^{\dagger}(\vec y,x^o)T^b\check \psi
(\vec y,x^o)]-m\beta \rbrace \check \psi (\vec x,x^o)=\cr
&{}\cr
&={1\over {2g^2}}(\, [\int d^3y\, {\cal P}_{ab}^{({\check {\vec A}}_{\perp})\,
ij}(\vec x,\vec y;x^o)\partial^o{\check A}^j_{b,\perp}(\vec y,x^o)-\cr
&-g^2{\hat
{\vec D}}^{({\check {\vec A}}_{\perp})}_{ab}(\vec x,x^o)\int d^3y\, G^{({\check
{\vec A}}_{\perp})}_{\triangle ,bc}(\vec x,\vec y;x^o)i{\check \psi}^{\dagger}
(\vec y,x^o)T^c\check \psi (\vec y,x^o)]{}^2-\cr
&-\sum_a{\check {\vec B}}^2_a(\vec x,x^o)\, )
+{\check \psi}^{\dagger}(\vec x,x^o)\lbrace i[\partial^o -\vec \alpha \cdot
(\vec \partial_x+{\check {\vec A}}_{a,\perp}(\vec x,x^o)T^a)]+\cr
&+iT^a\int d^3y\,
G^{({\check {\vec A}}_{\perp})}_{\triangle ,ab}(\vec x,\vec y;x^o)[{\hat {\vec
D}}_{bc}^{({\check {\vec A}}_{\perp})}(\vec y,x^o)\cdot \partial^o{\check {\vec
A}}_{c,\perp}(\vec y,x^o)+\cr
&+ig^2{\check \psi}^{\dagger}(\vec y,x^o)T^b\check \psi
(\vec y,x^o)]-m\beta \rbrace \check \psi (\vec x,x^o)=\cr}
$$

\vfill\eject

$$\eqalign{
&={1\over {2g^2}}[ \int d^3y_1\, {\cal P}_{ab}^{({\check {\vec A}}_{\perp}
)\, ij}(\vec x,\vec {y_1};x^o)\, \partial^o{\check A}^j_{b,\perp}(\vec {y_1},
x^o)\cr
&\int d^3y_2\, {\cal P}_{ac}^{({\check {\vec A}}_{\perp}
)\, ik}(\vec x,\vec {y_2};x^o)\, \partial^o{\check A}^k_{b,\perp}(\vec {y_2},
x^o)\, -\sum_a\, {\check {\vec B}}^2_a(\vec x,x^o)] +\cr
&+\int d^3y_1\, {\cal P}_{ab}^{({\check {\vec A}}_{\perp})\, ij}(\vec x,\vec
{y_1};x^o)\, \partial^o{\check A}^j_{b,\perp}(\vec {y_1},x^o)\, {\hat D}^{(
{\check {\vec A}}_{\perp})\, i}_{ac}(\vec x,x^o)\cr
&\int d^3y_2\, G^{({\check {\vec
A}}_{\perp})}_{\triangle ,cd}(\vec x,\vec {y_2};x^o)\, i{\check \psi}^{\dagger}
(\vec {y_2},x^o)T^d\check \psi (\vec {y_2},x^o)+\cr
&+{{g^2}\over 2}{\hat {\vec D}}^{({\check {\vec A}}_{\perp})}_{ab}(\vec x,x^o)
\int d^3y_1\, G^{({\check {\vec A}}_{\perp})}_{\triangle ,bc}(\vec x,\vec
{y_1};
x^o)\, i{\check \psi}^{\dagger}(\vec {y_1},x^o)T^c\check \psi (\vec {y_1},x^o)
\cdot \cr
&\cdot {\hat {\vec D}}^{({\check {\vec A}}_{\perp})}_{au}(\vec x,x^o)\int
d^3y_2
\, G^{({\check {\vec A}}_{\perp})}_{\triangle ,uv}(\vec x,\vec {y_2};x^o)\, i
{\check \psi}^{\dagger}(\vec {y_2},x^o)T^v\check \psi (\vec {y_2},x^o)+\cr
&+{\check \psi}^{\dagger}(\vec x,x^o)(\, i[ \partial^o -\vec \alpha \cdot
(\vec \partial_x+{\check {\vec A}}_{a,\perp}(\vec x,x^o)T^a)]+\cr
&+iT^a\int d^3y\,
G^{({\check {\vec A}}_{\perp})}_{\triangle ,ab}(\vec x,\vec y;x^o)[{\hat {\vec
D}}_{bc}^{({\check {\vec A}}_{\perp})}(\vec y,x^o)\cdot \partial^o{\check {\vec
A}}_{c,\perp}(\vec y,x^o)+\cr
&+ig^2{\check \psi}^{\dagger}(\vec y,x^o)T^b\check \psi
(\vec y,x^o)]-m\beta ) \check \psi (\vec x,x^o).\cr}
\form
$$

\noindent Let us note that in the first line of this equation, due to the
transversality of ${\check {\vec A}}_{\perp}$, one could replace ${\hat {\vec
D}}_{cd}^{({\check {\vec A}}_{\perp})}(\vec y,x^o)$ with $-{({\check {\vec A}}
_{\perp}(\vec y,x^o))}_{cd}$, see Eqs.(6-46).

By making integrations by part and using transversality and the second line of
Eqs.(6-36), from Eqs. (9-16) one gets

$$\eqalign{
\int d^3x\,& {\check {\cal L}}_T^{'}(\vec x,x^o)=\cr
&{1\over {2g^2}}\int d^3x\, [ \partial^o{\check A}^i_{a,\perp}(\vec x,x^o)\int
d^3y\, {\cal P}^{({\check {\vec A}}_{\perp})\, ik}_{ac}(\vec x,\vec y;x^o)
\partial^o{\check A}^k_{c,\perp}(\vec y,x^o)-\cr
&-\int d^3y_1\, G^{({\check {\vec A}}_{\perp})}_{\triangle ,bc}(\vec x,\vec
{y_1};x^o){\hat {\vec D}}^{({\check {\vec A}}_{\perp})}_{cd}(\vec {y_1},x^o)
\cdot \partial^0{\check {\vec A}}_{d,\perp}(\vec {y_1},x^o)\lbrace -{\hat {\vec
D}}^{({\check {\vec A}}_{\perp})}_{ba}(\vec x,x^o)\rbrace \cr
&\int d^3y_2\, {\cal P}_{au}^{({\check {\vec A}}_{\perp})\, ik}(\vec x,\vec
{y_2};x^o)\partial^o{\check A}^k_{u,\perp}(\vec {y_2},x^o)-\sum_a{\check {\vec
B}}^2_a(\vec x,x^o)\, ]-\cr
&-\int d^3x\, [{\hat D}^{({\check {\vec A}}_{\perp})\, i}_{ca}(\vec x,x^o)\int
d^3y_1\, {\cal P}^{({\check {\vec A}}_{\perp})\, ij}_{ab}(\vec x,\vec y;x^o)
\partial^o{\check A}^j_{b,\perp}(\vec y,x^o)]\cdot \cr
&\cdot \int d^3y_2\, G^{({\check {\vec A}}_{\perp})}_{\triangle ,cd}(\vec x,
\vec y;x^o)\, [i{\check \psi}^{\dagger}(\vec {y_2},x^o)T^d\check \psi (\vec
{y_2},x^o)]-\cr}
$$

\vfill\eject

$$\eqalign{
&-{ {g^2}\over 2}\int d^3x\, [\triangle^{({\check {\vec A}}_{\perp})}_{ub}
(\vec x,x^o)\int d^3y_1\, G^{({\check {\vec A}}_{\perp})}_{\triangle ,bc}(\vec
x,\vec {y_1};x^o)\, [i{\check \psi}^{\dagger}(\vec {y_1},x^o)T^c\check \psi
(\vec {y_1},x^o)]]\cr
&\int d^3y_2\, G^{({\check {\vec A}}_{\perp})}_{\triangle ,uv}(\vec x,\vec
{y_2}
;x^o)\, [i{\check \psi}^{\dagger}(\vec {y_2},x^o)T^v\check \psi (\vec
{y_2},x^o)
]+\cr
&+\int d^3x\, [i{\check \psi}(\vec x,x^o)T^a\check \psi (\vec x,x^o)]\, \int
d^3y\, [(-{\hat {\vec D}}^{({\check {\vec A}}_{\perp})}_{cb}(\vec y,x^o)
G^{({\check {\vec A}}_{\perp})}_{\triangle ,ab}(\vec x,\vec y;x^o))\cdot \cr
&\cdot \partial^o{\check {\vec A}}_{c,\perp}(\vec y,x^o)+G^{({\check {\vec A}}
_{\perp})}_{\triangle ,ab}(\vec x,\vec y;x^o)\, g^2[i{\check \psi}^{\dagger}
(\vec y,x^o)T^b\check \psi (\vec y,x^o)]]+\cr
&+\int d^3x {\check \psi}^{\dagger}(\vec x,x^o)(i[\partial^o-\vec \alpha \cdot
(\vec \partial +{\check {\vec A}}_{a,\perp}(\vec x,x^o)T^a)]-m\beta)\check
\psi (\vec x,x^o)=\cr
&{}\cr
&={1\over {2g^2}}\int d^3xd^3y \partial^o{\check A}^i_{a,\perp}(\vec x,x^o)
{\cal P}_{ab}^{({\check {\vec A}}_{\perp})\, ij}(\vec x,\vec y;x^o)\partial^o
{\check A}^j_{b,\perp}(\vec y,x^o)-\cr
&-{1\over {2g^2}}\int d^3x\, \sum_a {\check {\vec B}}_a^2(\vec x,x^o)-\cr
&-\int d^3xd^3y\, [i{\check \psi}^{\dagger}(\vec x,x^o)T^a\check \psi (\vec x,
x^o)]{\vec \zeta}^{({\check {\vec A}}_{\perp})}_{ab}(\vec x,\vec y;x^o)\cdot
\partial^o{\check {\vec A}}_{b,\perp}(\vec y,x^o)+\cr
&+{{g^2}\over 2}\int d^3xd^3y\, [i{\check \psi}^{\dagger}(\vec x,x^o)T^a\check
\psi (\vec x,x^o)]\, G^{({\check {\vec A}}_{\perp})}_{\triangle ,ab}(\vec x,
\vec y;x^o)\, [i{\check \psi}^{\dagger}(\vec y,x^o)T^b\check \psi (\vec y,x^o)
]+\cr
&+\int d^3x {\check \psi}^{\dagger}(\vec x,x^o)(i[\partial^o-\vec \alpha \cdot
(\vec \partial +{\check {\vec A}}_{a,\perp}(\vec x,x^o)T^a)]-m\beta)\check
\psi (\vec x,x^o)=\cr
&=\int d^3x\, {\check {\cal L}}_T(\vec x,x^o);\cr
&{}\cr
{\check S}_T&=\int dx^o\int d^3x\, {\check {\cal L}}_T^{'}(\vec x,x^o)=\int
dx^o
\int d^3x\, {\check {\cal L}}_T(\vec x,x^o);\cr}
\form
$$

\noindent this final Lagrangian density ${\check {\cal L}}_T(\vec x,x^o)$
has the correct Abelian limit, Eqs.(5-51), since
$\triangle^{({\check {\vec A}}_{\perp})}\rightarrow -\triangle$, $T^a
\rightarrow -i$ and the third line of the last expression becomes a surface
term.

The physical equations of motion can be obtained from Eqs.(9-2) with the same
procedure, instead of evaluating them as
the Euler-Lagrange equations associated with
Eqs.(9-16). Using Eqs.(9-3), (9-6), (9-8), (9-12), (9-14), one obtains
(one uses the simplified notation ${\vec K}_a[\vec A(\vec x,x^o)]=
{\hat {\vec D}}^{(\vec A)}_{ab}(\vec x,x^o)\times [(
{\hat {\vec D}}^{(\vec A)}_{bc}(\vec x,x^o)-{1\over 2}c_{bdc}{\vec A}_d(\vec
x,x^o))\times {\vec A}_c(\vec x,x^o)]$):

$$\eqalign{
{\vec L}^T_a(\vec x,x^o)&=(\delta_{ab}\partial^o+c_{aub}A^o_u(\vec x,x^o))[
\partial^o{|}_{\eta}{\hat {\vec A}}_{b,T}(\vec x,x^o;\eta (\vec x,x^o))+\cr
&+{\hat {\vec D}}_{bc}^{({\hat {\vec A}}_T)}(\vec x,x^o){\dot \Omega}^{\hat
\gamma}_c(\eta (\vec x,x^o))-{\hat {\vec D}}_{bc}^{(\vec \Theta +
{\hat {\vec A}}_T)}(\vec x,x^o)A^o_c(\vec x,x^o)]-\cr
&-{\vec K}_a[\vec A(\vec x,x^o)]+g^2{\vec J}_a(\vec x,x^o)=\cr
&=(\delta_{ab}\partial^o{|}_{\eta}+\delta_{ab}{\dot \eta}_w{ {\partial}\over
{\partial \eta_w} }{|}_{\eta =\eta (\vec x,x^o)}+c_{aub}A^o_u(\vec x,x^o))[
\partial^o{|}_{\eta}{\hat {\vec A}}_{b,T}(\vec x,x^o;\eta (\vec x,x^o))-\cr
&-c_{bvc}{\vec \Theta}_v(\eta (\vec x,x^o);\vec \partial \eta (\vec x,x^o))
{\dot \Omega}^{\hat \gamma}_c(\eta (\vec x,x^o))-\cr
&-{\hat {\vec D}}_{bc}^{(\vec \Theta +{\hat {\vec A}}_T)}(\vec x,x^o)(A^o_c
(\vec x,x^o)-{\dot \Omega}^{\hat \gamma}_c(\eta (\vec x,x^o)))]-{\vec K}_a
[\vec A(\vec x,x^o)]+g^2{\vec J}_a(\vec x,x^o)=\cr
&=(\partial^o{|}_{\eta}){}^2{\hat {\vec A}}_{a,T}(\vec x,x^o;\eta (\vec
x,x^o))+
\cr
&+c_{aub}(A^o_u(\vec x,x^o)-{\dot \Omega}^{\hat \gamma}_u(\eta (\vec x,x^o)))
\partial^o{|}_{\eta}{\hat {\vec A}}_{b,T}(\vec x,x^o;\eta (\vec x,x^o))-\cr
&-c_{aub}(\partial^o{|}_{\eta}A_{uw}(\eta (\vec x,x^o))){\dot \eta}_w(\vec x,
x^o)-\cr
&-c_{bvc}(\delta_{ab}{\dot \eta}_w(\vec x,x^o){ {\partial}\over {\partial
\eta_w} }{|}_{\eta =\eta (\vec x,x^o)}+\cr
&+c_{aub}A^o_u(\vec x,x^o))({\vec \Theta}
_v(\eta (\vec x,x^o);\vec \partial \eta (\vec x,x^o)){\dot \Omega}^{\hat
\gamma}
_c(\eta (\vec x,x^o)))-\cr
&-c_{auc}(\partial^o{|}_{\eta}{\hat {\vec A}}_{u,T}(\vec x,
x^o;\eta (\vec x,x^o)))(A^o_c(\vec x,x^o)-{\dot \Omega}^{\hat \gamma}_c(\eta
(\vec x,x^o)))-\cr
&-{\hat {\vec D}}_{ab}^{(\vec \Theta +{\hat {\vec A}}_T)}(\vec x,
x^o)\partial^o{|}_{\eta}(A^o_b(\vec x,x^o)-{\dot \Omega}^{\hat \gamma}_b(\eta
(\vec x,x^o)))-\cr
&-c_{aub}A^o_u(\vec x,x^o){\hat {\vec D}}_{bc}^{(\vec \Theta +{\hat {\vec A}}
_T)}(\vec x,x^o)(A^o_c(\vec x,x^o)-{\dot \Omega}^{\hat \gamma}(\eta (\vec x,
x^o)))-\cr
&-c_{auc}{\dot \eta}_w(\vec x,x^o)({ {\partial {\vec \Theta}_u(\eta (\vec x,x^o
);\vec \partial \eta (\vec x,x^o))}\over {\partial \eta_w} }{|}_{\eta =\eta
(\vec x,x^o)}-\cr
&-c_{udr}A_{dw}(\eta (\vec x,x^o)){\hat {\vec A}}_{r,T}(\vec x,x^o;
\eta (\vec x,x^o)))(A^o_c(\vec x,x^o)-{\dot \Omega}^{\hat \gamma}_c(\eta (\vec
x,x^o)))-\cr
&-{\dot \eta}_w(\vec x,x^o){\hat {\vec D}}_{ac}^{(\vec \Theta +{\hat {\vec A}}
_T)}(\vec x,x^o)
{ {\partial}\over {\partial \eta_w} }{|}_{\eta =\eta (\vec x,x^o)}(A^o_c
(\vec x,x^o)-{\dot \Omega}^{\hat \gamma}_c(\eta (\vec x,x^o)))-\cr
&-{\vec K}_a[\vec A(\vec x,x^o)]+g^2{\vec J}_a(\vec x,x^o){\rightarrow}_{\eta
\rightarrow 0,\,\, L^{oT}_a{\buildrel \circ \over =}0}\cr
&{\rightarrow}_{\eta \rightarrow 0,\,\, L^{oT}_a{\buildrel \circ \over =}0}
{\check {\vec L}}^T_a(\vec x,x^o)\cr}
\form
$$

\noindent with

$$\eqalign{
{\check {\vec L}}^T_a(\vec x,x^o)&={(\partial^o)}^2{\check {\vec A}}_{a,\perp}
(\vec x,x^o)+\cr
&-{\hat {\vec D}}^{({\check {\vec A}}_{\perp})}_{ab}(\vec x,x^o)\times [(
{\hat {\vec D}}^{({\check {\vec A}}_{\perp})}_{bc}(\vec x,x^o)-{1\over 2}
c_{bdc}{\check {\vec A}}_{d\perp}(\vec x,x^o))\times {\check {\vec A}}_{c,
\perp}(\vec x,x^o)]+\cr
&+ig^2{\check \psi}^{\dagger}(\vec x,x^o)\vec \alpha T^a\check \psi (\vec
x,x^o)
\cr
&-\int d^3z\, (\delta^3(\vec x-\vec z)[\partial^o{\hat {\vec D}}_{au}^{({\check
{\vec A}}_{\perp})}(\vec x,x^o)+c_{anu}\partial^o{\check {\vec A}}_{b,\perp}
(\vec x,x^o)]+\cr
&c_{amu}{\vec \zeta}^{({\check {\vec A}}_{\perp})}_{mn}(\vec x,\vec z;x^o)[
{\hat {\vec D}}_{nr}^{({\check {\vec A}}_{\perp})}(\vec z,x^o)\cdot \partial^o
{\check {\vec A}}_{r,\perp}(\vec z,x^o))+\cr
&+ig^2{\check \psi}^{\dagger}(\vec z,x^o)T^n\check \psi (\vec z,x^o)])\cdot \cr
&\int d^3y\, G_{\triangle ,ub}^{({\check {\vec A}}_{\perp})}(\vec x,\vec y;x^o)
[{\hat {\vec D}}_{bc}^{({\check {\vec A}}_{\perp})}(\vec y,x^o)\cdot \partial^o
{\check {\vec A}}_{c,\perp}(\vec y,x^o))+\cr
&+ig^2{\check \psi}^{\dagger}(\vec y,x^o)T^b\check \psi (\vec y,x^o)]=\cr
&={\bar \sqcup}{\check {\vec A}}_{a,\perp}(\vec x,x^o)+\cdots
{\buildrel \circ \over =}0.\cr}
\form
$$

\noindent where Eqs.(6-36) have been used. Analogously one obtains

$$\eqalign{
\beta {\check L}^T_{\psi}(\vec x,x^o)&=[i(\partial^o-\vec \alpha \cdot (\vec
\partial +{\check {\vec A}}_{a,\perp}(\vec x,x^o)T^a)+\cr
&+iT^a\int d^3y\, G^{({\check {\vec A}}_{\perp})}_{\triangle ,ab}(\vec x,\vec
y;
x^o)[{\hat {\vec D}}_{bc}^{({\check {\vec A}}_{\perp})}(\vec y,x^o)\cdot
\partial^o {\check {\vec A}}_{c,\perp}(\vec y,x^o))+\cr
&+ig^2{\check \psi}^{\dagger}(\vec y,x^o)T^b\check \psi (\vec y,x^o)]-m\beta ]
\check \psi (\vec x,x^o){\buildrel \circ \over =}0\cr}
\form
$$

These equations of motion and the corresponding Hamilton equations will be
studied elsewhere, since in this paper we are primarily interested in the
problem of Dirac's observables.

Using Eqs.(9-17), we can now evaluate the canonical momenta

$$\eqalign{
{\check \pi}^i_{a,\perp}(\vec x,x^o)&={ {\delta {\check S}_T}\over {\delta
\partial^o{\check A}^i_{a,\perp}(\vec x,x^o)} }=\cr
&{1\over {g^2} }P^{ij}_{\perp}(\vec x)\int d^3y\, {\cal P}_{ab}^{({\check
{\vec A}}_{\perp})\, jk}(\vec x,\vec y;x^o)\partial^o{\check A}^k_{b,\perp}
(\vec y,x^o)+\cr
&+P^{ij}_{\perp}(\vec x)\int d^3y\, \zeta^{({\check {\vec A}}_{\perp})\, j}
_{ab}(\vec x,\vec y;x^o)\, i{\check \psi}^{\dagger}(\vec y,x^o)T^b\check \psi
(\vec y,x^o),\cr
&{}\cr
&\vec \partial \cdot {\hat {\vec \pi}}_{a,\perp}(\vec x,x^o)\equiv 0,\cr
&{}\cr
&\lbrace {\check A}^i_{a,\perp}(\vec x,x^o),{\check \pi}^j_{b,\perp}(\vec y,
x^o)\rbrace =-\delta_{ab}P^{ij}_{\perp}(\vec x)\delta^3(\vec x-\vec y).\cr}
\form
$$

To invert this equation in $\partial^o{\check {\vec A}}_{a,\perp}$, one applies
the following operator

$$\eqalign{
P^{mn}_{\perp}&(\vec z)\int d^3x\, [\delta^{ni}\delta_{ca}\delta^3(\vec z-\vec
x)+{\hat D}^{({\check {\vec A}}_{\perp})\, n}_{cu}(\vec z,x^o)G^{({\check
{\vec A}}_{\perp})}_{\triangle ,uv}(\vec z,\vec x;x^o){\hat D}^{({\check {\vec
A}}_{\perp})\, i}_{va}(\vec x,x^o)]P^{ij}_{\perp}(\vec x)\cr
&\int d^3y\, [\delta^{jk}\delta_{ab}\delta^3(\vec x-\vec y)-{\hat D}^{({\check
{\vec A}}_{\perp})\, j}_{ar}(\vec x,x^o)G^{({\check {\vec A}}_{\perp})}
_{\triangle ,rs}(\vec x,\vec y;x^o){\hat D}^{({\check {\vec A}}_{\perp})\, k}
_{sb}(\vec y,x^o)]\cr
&\cdot \partial^o{\check A}^k_{b,\perp}(\vec y,x^o)=\cr
&=\partial^o{\check A}^m_{c.\perp}(\vec z,x^o)-\int d^3xd^3y\, P^{mn}_{\perp}
(\vec z){\hat D}^{({\check {\vec A}}_{\perp})\, n}_{cu}(\vec z,x^o)G_{\triangle
,uv}^{({\check {\vec A}}_{\perp})}(\vec z,\vec x;x^o)\cr
&[{\hat D}^{({\check {\vec A}}_{\perp})\, i}_{va}(\vec x,x^o)P^{ij}_{\perp}
(\vec x){\hat D}^{({\check {\vec A}}_{\perp})\, j}_{ar}(\vec x,x^o)]\,
G^{({\check {\vec A}}_{\perp})}_{\triangle ,rs}(\vec x,\vec y;x^o){\hat D}_{sb}
^{({\check {\vec A}}_{\perp})\, k}(\vec y,x^o) \partial^o{\check A}^k_{b,
\perp}(\vec y,x^o)=\cr
&=\partial^o{\check A}^m_{c,\perp}(\vec z,x^o),\cr}
\form
$$

\noindent because Eqs.(6-35), (6-29) and (6-38) imply

$$\eqalign{
{\hat D}^{({\check {\vec A}}_{\perp})\, i}_{va}(\vec x,x^o)&(\delta^{ij}+
{ {\partial^i\partial^j}\over {\triangle} }){\hat D}^{({\check {\vec A}}
_{\perp})\, j}_{ar}(\vec x,x^o)=\cr
&=\triangle^{({\check {\vec A}}_{\perp})}_{vr}(\vec x,x^o)+K_{va}({\check
{\vec A}}_{\perp})(\vec x,x^o){1\over {\triangle} }K_{ar}({\check {\vec A}}
_{\perp})(\vec x,x^o)=0.\cr}
\form
$$

\noindent Therefore, one gets

$$\eqalign{
\partial^0{\check A}^i_{a,\perp}&(\vec x,x^o)=\cr
&=g^2P^{ij}_{\perp}(\vec x)\int d^3y\, [\delta^{jk}\delta_{ab}\delta^3(\vec x-
\vec y)+{\hat D}^{({\check {\vec A}}_{\perp})\, j}_{au}(\vec x,x^o)G^{({\check
{\vec A}}_{\perp})}_{\triangle ,uv}(\vec x,\vec y;x^o){\hat D}^{({\check {\vec
A}}_{\perp})\, k}_{vb}(\vec y,x^o)]\cr
&\cdot P^{kn}_{\perp}(\vec y)\,\,
[{\check \pi}^n_{b,\perp}(\vec y,x^o)-i\int d^3z\,
\zeta^{({\check {\vec A}}_{\perp})\, n}_{bc}(\vec y,\vec z;x^o){\check \psi}
^{\dagger}(\vec z,x^o)T^c\check \psi (\vec z,x^o)],\cr}
\form
$$

\noindent so that from Eqs.(9-19) we get the following canonical Hamiltonian

$$\eqalign{
{\check H}_{cT}&=\int d^3x\, [{\check {\vec \pi}}_{a,\perp}(\vec x,x^o)\cdot
\partial^o{\check {\vec A}}_{a,\perp}(\vec x,x^o)+\check \psi (\vec x,x^o)
\pi (\vec x,x^o)-\cr
&-{\check {\bar \psi}}(\vec x,x^o){\check {\bar \pi}}(\vec x,x^o)-{\check
{\cal L}}_T(\vec x,x^o)]=\cr
&={{g^2}\over 2}\int d^3xd^3y\, [{\check \pi}^i_{a,\perp}(\vec x,x^o)-P^{ij}
_{\perp}(\vec x)\int d^3v\, \zeta^{({\check {\vec A}}_{\perp})\, j}_{ar}(\vec
x,\vec v;x^o)\, i{\check \psi}^{\dagger}(\vec v,x^o)T^r\check \psi (\vec v,x^o)
]\cr
&[\delta^{ik}\delta_{ab}\delta^3(\vec x-\vec y)+{\hat D}^{({\check {\vec A}}
_{\perp})\, i}_{au}(\vec x,x^o)G^{({\check {\vec A}}_{\perp})}_{\triangle ,uv}
(\vec x,\vec y;x^o){\hat D}^{({\check {\vec A}}_{\perp})\, k}_{vb}(\vec y,x^o)
]\cr
&[{\check \pi}^k_{b,\perp}(\vec y,x^o)-P^{kn}_{\perp}(\vec y)\int d^3w \zeta
^{({\check {\vec A}}_{\perp})\, n}_{bc}(\vec y,\vec w;x^o)\, i{\check \psi}
^{\dagger}(\vec w,x^o)T^c\check \psi (\vec w,x^o)]+\cr
&+{1\over {2g^2}}\int d^3x\, \sum_a{\check {\vec B}}_a^2(\vec x,x^o)+\cr
&+\int d^3x\, {\check \psi}^{\dagger}(\vec x,x^o)[i\vec \alpha
\cdot (\vec \partial +{\check {\vec A}}_{a,\perp}(\vec x,x^o)T^a)+m\beta ]
\check \psi (\vec x,x^o)-\cr
&-{{g^2}\over 2}\int d^3xd^3y\, [i{\check \psi}^{\dagger}(\vec x,x^o)T^a\check
\psi (\vec x,x^o)]G^{({\check {\vec A}}_{\perp})}_{\triangle ,ab}(\vec x,\vec
y;x^o)\, [i{\check \psi}^{\dagger}(\vec y,x^o)T^b\check \psi (\vec y,x^o)]=\cr
&={{g^2}\over 2}\int d^3xd^3y\, {\check \pi}^i_{a,\perp}(\vec
x,x^o)[\delta^{ik}
\delta_{ab}\delta^3(\vec x-\vec y)+{\hat D}^{({\check {\vec A}}_{\perp})\, i}
_{au}(\vec x,x^o)G^{({\check {\vec A}}_{\perp})}_{\triangle ,uv}(\vec x,\vec y;
x^o)\cr
&{\hat D}^{({\check {\vec A}}_{\perp})\, k}_{vb}(\vec y,x^o)]{\check \pi}^k_{b,
\perp}(\vec y,x^o)+{1\over {2g^2}}\int d^3x\, \sum_a
{\check {\vec B}}_a^2(\vec x,x^o)-\cr
&-g^2\int d^3xd^3y\, P^{ij}_{\perp}(\vec x)\int d^3v\, \zeta^{({\check {\vec
A}}_{\perp})\, j}_{ar}(\vec x,\vec v;x^o)[i{\check \psi}^{\dagger}(\vec v,x^o)
T^r\check \psi (\vec v,x^o)]\cr
&\cdot [\delta^{ik}\delta_{ab}\delta^3(\vec x-\vec y)+{\hat D}^{({\check {\vec
A}}_{\perp})\, i}_{au}(\vec x,x^o)G^{({\check {\vec A}}_{\perp})}_{\triangle ,
uv}(\vec x,\vec y;x^o){\hat D}^{({\check {\vec A}}_{\perp})\, k}_{vb}(\vec y,
x^o)]{\check \pi}^k_{b,\perp}(\vec y,x^o)+\cr}
$$

$$
\eqalign{
&+{{g^2}\over 2}\int d^3xd^3y\, P^{ij}_{\perp}(\vec x)\int d^3v\, \zeta_{ur}
^{({\check {\vec A}}_{\perp})\, j}(\vec x,\vec v;x^o)[i{\check \psi}^{\dagger}
(\vec v,x^o)T^r\check \psi (\vec v,x^o)]\cr
&[\delta^{ik}\delta_{ab}\delta^3(\vec x-\vec y)+{\hat D}^{({\check {\vec A}}
_{\perp})\, i}_{au}(\vec x,x^o)G^{({\check {\vec A}}_{\perp})}_{\triangle ,uv}
(\vec x,\vec y;x^o){\hat D}^{({\check {\vec A}}_{\perp})\, k}_{vb}(\vec y,x^o)]
\cr
&P^{kn}_{\perp}(\vec y)\int d^3w\, \zeta_{bc}^{({\check {\vec A}}_{\perp})\, n}
(\vec y,\vec w;x^o)[i{\check \psi}^{\dagger}(\vec w,x^o)T^c\check \psi (\vec w,
x^o)]-\cr
&-{{g^2}\over 2}\int d^3xd^3y\, [i{\check \psi}^{\dagger}(\vec x,x^o)T^a\check
\psi (\vec x,x^o)]G^{({\check {\vec A}}_{\perp})}_{\triangle ,ab}(\vec x,\vec
y;
x^o)[i{\check \psi}^{\dagger}(\vec y,x^o)T^b\check \psi (\vec y,x^o)]+\cr
&+\int d^3x\, {\check \psi}^{\dagger}(\vec x,x^o)[i\vec \alpha
\cdot (\vec \partial +{\check {\vec A}}_{a,\perp}(\vec x,x^o)T^a)+m\beta ]
\check \psi (\vec x,x^o)=\cr
&{}\cr
&={1\over 2}\int d^3x \sum_a[g^2{\check {\vec \pi}}^2_{a,\perp}(\vec x,x^o)+
{1\over {g^2}}{\check {\vec B}}^2_a(\vec x,x^o)]-\cr
&-{{g^2}\over 2}\int d^3xd^3y\, [{({\check {\vec A}}_{\perp}(\vec x,x^o))}_{ua}
\cdot {\check {\vec \pi}}_{a,\perp}(\vec x,x^o)]G^{({\check {\vec A}}_{\perp})}
_{\triangle ,uv}(\vec x,\vec y;x^o)[{({\check {\vec A}}_{\perp}(\vec y,x^o))}
_{vb}\cdot {\check {\vec \pi}}_{b,\perp}(\vec y,x^o)]-\cr
&-g^2\int d^3xd^3y\, [{({\check {\vec A}}_{\perp}(\vec x,x^o))}_{ua}
\cdot {\check {\vec \pi}}_{a,\perp}(\vec x,x^o)]G^{({\check {\vec A}}_{\perp})}
_{\triangle ,uv}(\vec x,\vec y;x^o)[i{\check \psi}^{\dagger}(\vec y,x^o)T^v
\check \psi (\vec y,x^o)]-\cr
&-{{g^2}\over 2}\int d^3xd^3y\, [i{\check \psi}^{\dagger}(\vec x,x^o)T^u\check
\psi (\vec x,x^o)]G_{\triangle ,uv}^{({\check {\vec A}}_{\perp})}(\vec x,\vec
y;
x^o)[i{\check \psi}^{\dagger}(\vec y,x^o)T^v\check \psi (\vec y,x^o)]+\cr
&+\int d^3x\, {\check \psi}^{\dagger}(\vec x,x^o)[i\vec \alpha
\cdot (\vec \partial +{\check {\vec A}}_{a,\perp}(\vec x,x^o)T^a)+m\beta ]
\check \psi (\vec x,x^o)=\cr
&{}\cr
&={1\over 2}\int d^3x \sum_a[g^2{\check {\vec \pi}}^2_{a,\perp}(\vec x,x^o)+
{1\over {g^2}}{\check {\vec B}}^2_a(\vec x,x^o)]-\cr
&-{{g^2}\over 2}\int d^3xd^3y\, [{({\check {\vec A}}_{\perp}(\vec x,x^o))}_{au}
\cdot {\check {\vec \pi}}_{u,\perp}(\vec x,x^o)+i{\check \psi}^{\dagger}(\vec
x,
x^o)T^a\check \psi (\vec x,x^o)]G^{({\check {\vec A}}_{\perp})}_{\triangle ,ab}
(\vec x,\vec y;x^o)\cr
&[{({\check {\vec A}}_{\perp}(\vec y,x^o))}_{bv}\cdot {\check {\vec \pi}}_{v,
\perp}(\vec y,x^o)+i{\check \psi}^{\dagger}(\vec y,x^o)T^b\check \psi (\vec y,
x^o)]+\cr
&+\int d^3x\, {\check \psi}^{\dagger}(\vec x,x^o)[i\vec \alpha
\cdot (\vec \partial +{\check {\vec A}}_{a,\perp}(\vec x,x^o)T^a)+m\beta ]
\check \psi (\vec x,x^o),\cr}
\form
$$

\noindent where Eqs.(6-36) and (9-23) have been used to get ${\hat D}_{vb}
^{({\check {\vec A}}_{\perp})\, k}(\vec y,x^o)P^{kn}_{\perp}(\vec y)\zeta
^{({\check {\vec A}}_{\perp})\, n}_{bc}(\vec y,\vec w;x^o)=0$ and similar
results
after integrations by parts; also the notation of Eqs.(6-46) has been used.
The first expression of  this canonical Hamiltonian shows the correct Abelian
limit of Eqs.(5-39), (5-40), because in this limit the extra terms in the first
lines disappear for transversality reasons, $P^{ij}_{\perp}(\vec x)
\zeta_{ar}^{({\check {\vec A}}_{\perp})\, j}(\vec x,\vec v;x^o)\rightarrow
0$.

By introducing the following notation for the non-Abelian charge density of
Eqs.(4-40)

$$\eqalign{
Q^T_a&=g\int d^3x\, \rho_a^{(\vec A,\vec \pi )}(\vec x,x^o)\cr
\rho_a^{(\vec A,\vec \pi )}(\vec x,x^o)&={(\vec A(\vec x,x^o))}_{ab}\cdot {\vec
\pi}_b(\vec x,x^o)+i\psi^{\dagger}(\vec x,x^o)T^a\psi (\vec x,x^o),\cr
&{}\cr
&\lbrace \rho_a^{(\vec A,\vec \pi )}(\vec x,x^o),\rho_b^{(\vec A,\vec \pi )}
(\vec y,x^o)\rbrace =c_{abc}\rho_c^{(\vec A,\vec \pi )}(\vec x,x^o)\delta^3
(\vec x-\vec y)\cr
&\lbrace \rho_a^{({\check {\vec A}}_{\perp},{\check {\vec \pi}}_{\perp})}(\vec
x,x^o),\rho_b^{({\check {\vec A}}_{\perp},{\check {\vec \pi}}_{\perp})}
(\vec y,x^o)\rbrace =c_{abc}\rho_c^{({\check {\vec A}}_{\perp},{\check {\vec
\pi}}_{\perp})}(\vec x,x^o)\delta^3(\vec x-\vec y)\cr}
\form
$$

\noindent one arrives at the final form of the canonical Hamiltonian

$$\eqalign{
{\check H}_{cT}&={1\over 2}\int d^3x\, \sum_a [g^2{\check {\vec \pi}}^2_{a,
\perp}(\vec x,x^o)+{1\over {g^2}}{\check {\vec B}}^2_a(\vec x,x^o)]-\cr
&-{{g^2}\over 2}\int d^3xd^3y\, \rho_a^{({\check {\vec A}}_{\perp},{\check
{\vec \pi}}_{\perp})}(\vec x,x^o)G^{({\check {\vec A}}_{\perp})}_{\triangle ,
ab}(\vec x,\vec y;x^o)\rho_b^{({\check {\vec A}}_{\perp},{\check
{\vec \pi}}_{\perp})}(\vec y,x^o)+\cr
&+\int d^3x\, {\check \psi}^{\dagger}(\vec x,x^o)[i\vec \alpha \cdot (\vec
\partial +{\check {\vec A}}_{a,\perp}(\vec x,x^o)T^a)+m\beta ]\check \psi
(\vec x,x^o).\cr}
\form
$$

Let us now consider $H_{DT}=H_{cT}+\int d^3x\, [\lambda_{ao}(\vec x,x^o)
\pi^o_a(\vec x,x^o)-A_{ao}(\vec x,x^o)\Gamma^T_a(\vec x,x^o)]$. From Eqs.(8-11)
, (8-27) and with the notation of Eq.(6-46), (9-26), we have

$$\eqalign{
\sum_a\, {\vec \pi}^2_a(\vec x,x^o)&=\sum_a\, \lbrace {\vec \pi}_{a,\perp}(\vec
x,x^o)+{ {{\vec \partial}_x}\over {\triangle_x} }\int d^3y\, {\vec \partial}_x
\cdot {\vec \zeta}^{(\vec A)}_{ab}(\vec x,\vec y;x^o) [\rho_b(\vec y,x^o)-
\Gamma^T_b(\vec y,x^o)]\rbrace {}^2=\cr
&=\sum_a\, \lbrace {\vec \pi}_{a,\perp}(\vec x,x^o)+{ {{\vec \partial}_x}\over
{\triangle_x} }\int d^3y\, {\vec \partial}_x\cdot {\vec \zeta}^{(\vec A)}_{ab}
(\vec x,\vec y;x^o) \rho_b(\vec y,x^o) \rbrace {}^2-\cr}
$$

$$
\eqalign{
&-\lbrace {\vec \pi}_{a,\perp}(\vec x,x^o)+{ {{\vec \partial}_x}\over
{\triangle_x} }\int d^3y\, {\vec \partial}_x\cdot {\vec \zeta}^{(\vec A)}_{ab}
(\vec x,\vec y;x^o) \rho_b(\vec y,x^o) \rbrace \cdot \cr
&\cdot { {{\vec \partial}_x}\over
{\triangle_x} }\int d^3z\, {\vec \partial}_x\cdot {\vec \zeta}^{(\vec A)}_{ac}
(\vec x,\vec z;x^o)\Gamma^T_c(\vec z,x^o)+\cr
&+\sum_a \lbrack { {{\vec \partial}_x}\over
{\triangle_x} }\int d^3y\, {\vec \partial}_x\cdot {\vec \zeta}^{(\vec A)}_{ab}
(\vec x,\vec y;x^o)\Gamma^t_b(\vec y,x^o)\rbrack {}^2=\cr
&=\sum_a {\vec \pi}^2_{a,\perp}(\vec x,x^o)+2{\vec \pi}_{a,\perp}(\vec x,x^o)
\cdot { {{\vec \partial}_x}\over
{\triangle_x} }\int d^3y\, {\vec \partial}_x\cdot {\vec \zeta}^{(\vec A)}_{ab}
(\vec x,\vec y;x^o)\rho_b(\vec y,x^o)+\cr
&+\sum_a \lbrack { {{\vec \partial}_x}\over
{\triangle_x} }\int d^3y\, {\vec \partial}_x\cdot {\vec \zeta}^{(\vec A)}_{ab}
(\vec x,\vec y;x^o)\rho_b(\vec y,x^o)\rbrack {}^2-\cr
&-\lbrack {\vec \pi}_{a,\perp}(\vec x,x^o)+{ {{\vec \partial}_x}\over
{\triangle_x} }\int d^3y\, {\vec \partial}_x\cdot {\vec \zeta}^{(\vec A)}_{ab}
(\vec x,\vec y;x^o)\rho_b(\vec y,x^o)\rbrack \cdot \cr
&\cdot { {{\vec \partial}_x}\over
{\triangle_x} }\int d^3z\, {\vec \partial}_x\cdot {\vec \zeta}^{(\vec A)}_{ac}
(\vec x,\vec z;x^o)\Gamma^T_c(\vec z,x^o)+\cr
&+\sum_a \lbrack { {{\vec \partial}_x}\over
{\triangle_x} }\int d^3y\, {\vec \partial}_x\cdot {\vec \zeta}^{(\vec A)}_{ab}
(\vec x,\vec y;x^o)\Gamma^T_b(\vec y,x^o)\rbrack {}^2=\cr}
\form
$$

After discarding surface terms from some integrations by parts (for instance of
the kind $\int d^3x\, [(\vec \partial /\triangle )F(\vec x)]\cdot [(\vec
\partial /\triangle )G(\vec x)]=\int d^3x\, F(\vec x)(1/\triangle )G(\vec x)$),
one gets (here $\rho_a=\rho_a^{(\vec A,\vec \pi )}$)

$$\eqalign{
H_{DT}&=\int d^3x\, \lbrace {{g^2}\over 2}\sum_a{\vec \pi}^2_{a,\perp}(\vec x,
x^o)+{1\over {2g^2}}\sum_a{\vec B}_a^2(\vec x,x^o)+\cr
&+{{g^2}\over 2}\int d^3y_1d^3y_2\, [{\vec \partial}_x\cdot {\vec \zeta}^{(\vec
A)}_{ab}(\vec x,\vec {y_1};x^o)\rho_b(\vec {y_1},x^o)]{1\over {\triangle_x} }
[{\vec \partial}_x\cdot {\vec \zeta}^{(\vec A)}_{ac}(\vec x,\vec {y_2};x^o)
\rho_c(\vec {y_2},x^o)]+\cr
&+i\psi^{\dagger}(\vec x,x^o)\vec \alpha \cdot (\vec \partial +{\vec A}_a(\vec
x,x^o)T^a)\psi (\vec x,x^o)+m\psi^{\dagger}(\vec x,x^o)\beta
\psi (\vec x,x^o)-\cr
&-{{g^2}\over 2}\int d^3y_1d^3y_2\, [{\vec \partial}_x\cdot {\vec \zeta}^{(\vec
A)}_{ab}(\vec x,\vec {y_1};x^o)\rho_b(\vec {y_1},x^o)]{1\over {\triangle_x} }
[{\vec \partial}_x\cdot {\vec \zeta}^{(\vec A)}_{ac}(\vec x,\vec {y_2};x^o)
\Gamma^T_c(\vec {y_2},x^o)]+\cr
&+{{g^2}\over 2}\int d^3y_1d^3y_2\, [{\vec \partial}_x\cdot {\vec \zeta}^{(\vec
A)}_{ab}(\vec x,\vec {y_1};x^o)\Gamma^T_b(\vec {y_1},x^o)]{1\over {\triangle_x}
}[{\vec \partial}_x\cdot {\vec \zeta}^{(\vec A)}_{ac}(\vec x,\vec {y_2};x^o)
\Gamma^T_c(\vec {y_2},x^o)]-\cr
&-A_{ao}(\vec x,x^o)\Gamma^T_a(\vec x,x^o)+\lambda_{ao}(\vec x,x^o)\pi^o_a(\vec
x,x^o)\rbrace .\cr}
\form
$$

If we evaluate $H_{DT}$ on the identity cross section $\sigma_I$, from Eqs.(9-
27) and (9-29) we get

$$\eqalign{
{\check H}_{DT}&={\check H}_{cT}
+\int d^3x\, (\lambda_{ao}(\vec x,x^o){\check \pi}^o_a(\vec x,x^o)-\cr
&-\int d^3y\, [{\check A}_{ao}(\vec x,x^o)\delta^3(\vec x-\vec y)+\cr
&+{{g^2}\over 2}\int
d^3z ({ 1\over {\triangle_x} }[{\vec \partial}_x\cdot {\vec \zeta}_{bc}^{(
{\check {\vec A}}_{\perp})}(\vec x,\vec z;x^o)\rho_c^{({\check {\vec A}}
_{\perp},{\check {\vec \pi}}_{\perp})}(\vec z,x^o)]){\vec \partial}_x\cdot
{\vec \zeta}_{ba}^{({\check {\vec A}}_{\perp})}(\vec x,\vec y;x^o)]
{\check \Gamma}^T_a(\vec y,x^o))+\cr
&+{{g^2}\over 2}\int d^3y_1d^3y_2\, [{\vec \partial}_x\cdot {\vec \zeta}
^{({\check {\vec A}}_{\perp})}_{ab}(\vec x,\vec {y_1};x^o){\check \Gamma}^T_b
(\vec {y_1},x^o)]\cr
&{1\over {\triangle_x}
}[{\vec \partial}_x\cdot {\vec \zeta}^{({\check {\vec A}}_{\perp})}_{ac}(\vec x
,\vec {y_2};x^o){\check \Gamma}^T_c(\vec {y_2},x^o)]=\cr
&{}\cr
&={\check H}_{cT}+\int d^3x\, (\lambda_{ao}(\vec x,x^o){\check \pi}^o_a(\vec x,
x^o)-\cr
&-[{\check A}_{ao}(\vec x,x^o)-{{g^2}\over 2}\int d^3y\, G^{({\check {\vec A}}
_{\perp})}_{\triangle ,ab}(\vec x,\vec y;x^o)\rho_b^{({\check {\vec A}}_{\perp}
,{\check {\vec \pi}}_{\perp})}(\vec y,x^o)]{\check \Gamma}^T_a(\vec x,x^o))-\cr
&-{{g^2}\over 2}\int d^3xd^3y\, {\check \Gamma}^T_a(\vec x,x^o)G^{({\check
{\vec A}}_{\perp})}_{\triangle ,ab}(\vec x,\vec y;x^o){\check \Gamma}^T_b(\vec
y,x^o),\cr}
\form
$$

\noindent where Eqs.(6-39), (6-40) and an integration by parts have been used.
This equation has to be compared with Eqs.(5-39) of the Abelian case.

If, along the lines of the Abelian case, Eqs.(5-41), we introduce the
generalized Coulomb gauge-fixing $\eta_a(\vec x,x^o)\approx 0$, then from
Eqs.(8-15), (9-29), (7-41), (7-43), we obtain

$$\eqalign{
\partial^o\eta_a&(\vec x,x^o)=\lbrace \eta_a(\vec x,x^o),H_{DT}\rbrace
\approx \cr
&\approx (A_{bo}(\vec x,x^o)+{{g^2}\over 2}\int d^3zd^3y\, [{\vec \partial}_z
\cdot {\vec \zeta}^{(\vec A)}_{uv}(\vec z,\vec y;x^o)\rho_v^{(\vec A,\vec \pi
)}(\vec y,x^o)]\cr
&{1\over {\triangle_z}}[{\vec \partial}_z\cdot {\vec \zeta}^{(\vec
A)}_{ub}(\vec z,\vec x;x^o)])B_{ba}(\eta (\vec x,x^o)){|}_{\eta \approx 0}
\approx \cr
&\approx {\check A}_{ao}(\vec x,x^o)(\vec x,x^o)-{{g^2}\over 2}\int d^3y\,
G^{({\check {\vec A}}_{\perp})}_{\triangle ,ab}(\vec x,\vec y;x^o)\rho_b
^{({\check {\vec A}}_{\perp},{\check {\vec \pi}}_{\perp})}(\vec y,x^o)\approx
0;\cr}
\form
$$

\noindent this turns out to be the natural generalized temporal gauge. Its
time constancy determines the Dirac multipliers $\lambda_{ao}$:

$$\eqalign{
\partial^o&[{\check A}_{ao}(\vec x,x^o)(\vec x,x^o)-{{g^2}\over 2}\int d^3y\,
G^{({\check {\vec A}}_{\perp})}_{\triangle ,ab}(\vec x,\vec y;x^o)\rho_b
^{({\check {\vec A}}_{\perp},{\check {\vec \pi}}_{\perp})}(\vec y,x^o)]=\cr
&=\lbrace {\check A}_{ao}(\vec x,x^o)(\vec x,x^o)-{{g^2}\over 2}\int d^3y\,
G^{({\check {\vec A}}_{\perp})}_{\triangle ,ab}(\vec x,\vec y;x^o)\rho_b
^{({\check {\vec A}}_{\perp},{\check {\vec \pi}}_{\perp})}(\vec y,x^o),H_{DT}
\rbrace \approx \cr
&\approx \lambda_{ao}(\vec x,x^o)-\lbrace {{g^2}\over 2}\int d^3y\,
G^{({\check {\vec A}}_{\perp})}_{\triangle ,ab}(\vec x,\vec y;x^o)\rho_b
^{({\check {\vec A}}_{\perp},{\check {\vec \pi}}_{\perp})}(\vec y,x^o),
H_{dt}\rbrace {|}_{\eta \approx 0}\approx 0.\cr}
\form
$$

We shall study elsewhere the non-trivial problem of finding the analogue of
the Abelian canonical transformation (5-42)-(5-43), of getting a separation
of ${\check H}_{DT}$ in physical and pure gauge parts like in Eqs.(5-45),
and, subsequently of analyzing similar separations for the generators of the
Poincar\'e group like at the end of Section 5.

As shown by Eqs.(9-31) the standard temporal gauge $A_{ao}(\vec x,x^o)=0$ is
not allowed, also in absence of fermions. Therefore, for G=SU(2), our physical
Hamiltonian cannot coincide either with the one of Ref.[91a] evaluated moreover
in the Coulomb gauge, or with the one of Ref.[91b] (notwithstanding the
technique for finding gauge invariants used in this paper has similarities
with our approach), or with the results of Ref.[91c] obtained starting from a
polar decomposition of ${\vec A}_a$, or of Refs.[47]; see also Ref.[91d] for a
different canonical approach, but always in the temporal gauge.

For the non-Abelian charges Eqs.(4-40), (8-25), (8-27), (9-14) imply

$$\eqalign{
&-{1\over {g^2}}\int d^3x\, \vec \partial \cdot {\vec E}_a(\vec x,x^o)
=-\int d^3x\,
\vec \partial \cdot {\vec \pi}_a(\vec x,x^o){\buildrel \circ \over =}\cr
{\buildrel \circ \over =}Q^T_a&=
{1\over g}\int d^3x\, G^o_{1a}(\vec x,x^o)=\cr
&={1\over {g^2}}c_{abc}\int d^3x\, F^{ok}_b(\vec x,x^o)A^k_c(\vec x,x^o)+i\int
d^3x\, \psi^{\dagger}(\vec x,x^o)T^a\psi (\vec x,x^o)=\cr
&=\int d^3x\, \rho_a^{(\vec A,\vec \pi )}(\vec x,x^o)=\cr
&=\int d^3x\, [{(\vec A
(\vec x,x^o))}_{ab}\cdot {\vec \pi}_b(\vec x,x^o)+i\psi^{\dagger}(\vec x,x^o)
T^a\psi (\vec x,x^o)]\rightarrow_{\eta \rightarrow 0,L^{oT}_a{\buildrel \circ
\over =}0} \cr
&{\rightarrow}_{\eta \rightarrow 0,L^{oT}_a{\buildrel \circ \over =}0}\,\,\,
{\check Q}^T_a\cr}
\form
$$

\noindent with

$$\eqalign{
{\check Q}^T_a&=\int d^3x\, [{({\check {\vec A}}_{\perp}(\vec x,x^o))}_{ab}
\cdot ({\check {\vec \pi}}_{b,\perp}(\vec x,x^o)+{ {{\vec \partial}_x}\over
{\triangle_x}}\int d^3y\, {\vec \partial}_x\cdot {\vec \zeta}^{({\check {\vec
A}}_{\perp})}_{bc}(\vec x,\vec y;x^o)\cr
&[{({\check {\vec A}}_{\perp}(\vec y,x^o))}_{cd}\cdot {\check {\vec \pi}}_{d,
\perp}(\vec y,x^o)+i{\check \psi}^{\dagger}(\vec y,x^o)T^c\check \psi (\vec y,
x^o)])+i{\check \psi}^{\dagger}(\vec x,x^o)T^a\check \psi (\vec x,x^o)]=\cr
&=\int d^3x\, [\rho_a^{({\check {\vec A}}_{\perp},{\check {\vec \pi}}_{\perp})}
(\vec x,x^o)+\cr
&+{({\check {\vec A}}_{\perp}(\vec x,x^o))}_{ab}\cdot
{ {{\vec \partial}_x}\over
{\triangle_x}}\int d^3y\, {\vec \partial}_x\cdot {\vec \zeta}^{({\check {\vec
A}}_{\perp})}_{bc}(\vec x,\vec y;x^o)\rho_c^{({\check {\vec A}}_{\perp},{\check
 {\vec \pi}}_{\perp})}(\vec y,x^o)]=\cr
&=\int d^3x\, \rho_a^{({\check {\vec A}}_{\perp},{\check {\vec \pi}}_{\perp})}
(\vec x,x^o)=\cr
&=\int d^3x\, [{({\check {\vec A}}_{\perp}(\vec x,x^o))}_{ab}\cdot {\check
{\vec \pi}}_{\perp}(\vec x,x^o)+i{\check \psi}^{\dagger}(\vec x,x^o)T^a\check
\psi (\vec x,x^o)].\cr}
\form
$$

\noindent after an integration by parts. As expected [35], Eqs.(9-26) imply

$$
\eqalign{
&\lbrace {\check Q}^T_a,{\check Q}^T_b\rbrace =c_{abc}\,
{\check Q}^T_c\cr
&\lbrace {\check A}^i_{a,\perp}(\vec x,x^o),{\check Q}^T_b\rbrace =c_{abc}
{\check A}^i_{c,\perp}(\vec x,x^o)\cr
&\lbrace {\check \pi}^i_{a,\perp}(\vec x,x^o),{\check Q}^T_b\rbrace =c_{abc}
{\check \pi}^i_{c,\perp}(\vec x,x^o)\cr
&\lbrace {\check \psi}_{a\alpha}(\vec x,x^o),{\check Q}^T_b\rbrace =[T^b{\check
\psi}_{\alpha}(\vec x,x^o)]{}_a\cr
&\lbrace {\check \psi}^{\dagger}_{a\alpha}(\vec x,x^o),{\check Q}^T_b\rbrace
=-[{\check \psi}^{\dagger}_{\alpha}(\vec x,x^o)T^b]{}_a.\cr}
\form
$$

Since, from Eqs.(4-38)-(4-40), the non-Abelian charges $Q^T_a$ are constants
of the motion, the same is true of the ${\check Q}^T_a$. The equation
$\lbrace {\check Q}^T_a,{\check H}_{cT}\rbrace =0$ can be used to deduce the
transformation properties of the Green function $G^{({\check {\vec A}}_{\perp}}
_{\triangle ,ab}(\vec x,\vec y;x^o)$: since from Eqs.(6-35) we have
$\lbrace \triangle^{({\check {\vec A}}_{\perp}}_{au}(\vec x,x^o)
G^{({\check {\vec A}}_{\perp}}_{\triangle ,uc}(\vec x,\vec y;x^o),{\check
Q}^T_b\rbrace =0$, with the help of Eqs.(6-36) we get

$$
\eqalign{
&\lbrace G^{({\check {\vec A}}_{\perp})}_{\triangle ,ac}(\vec x,\vec y;x^o),
{\check Q}^T_b\rbrace =\cr
&=-\int d^3z\, G^{({\check {\vec A})}_{\perp}}_{\triangle ,au}(\vec x,\vec z;
x^o)\, \lbrace {\hat {\vec D}}^{({\check {\vec A}}_{\perp})}_{uv}(\vec z,
x^o)\cdot {\hat {\vec D}}^{({\check {\vec A}}_{\perp})}_{vs}(\vec z,x^o),
{\check Q}^T_b\rbrace \,  G^{({\check {\vec A}}_{\perp})}_{\triangle ,sc}
(\vec z,\vec y;x^o)=\cr
&=\int d^3z\, [{\vec \zeta}_{av}^{({\check {\vec A}}_{\perp})}(\vec x,\vec z;
x^o)G^{({\check {\vec A}}_{\perp})}_{\triangle ,uc}(\vec z,\vec y;x^o)+
G^{({\check {\vec A}}_{\perp})}_{\triangle ,av}(\vec x,\vec z;x^o){\vec \zeta}
^{({\check {\vec A}}_{\perp})}_{uc}(\vec z,\vec y;x^o)]\cdot \cr
&\cdot c_{vud}c_{dbr}{\check {\vec A}}_{r,\perp}(\vec z,x^o)=
c_{abu}G^{({\check {\vec A}}_{\perp})}_{\triangle ,uc}(\vec x,\vec y;x^o)-
G^{({\check {\vec A}}_{\perp})}_{\triangle ,au}(\vec x,\vec y;x^o)c_{ubc}.\cr}
\form
$$

To find the classical superselection sectors labelled by $\sum_a{\check Q}
^{T\, 2}_a$, as said in Section 3, one has to select the subset of Dirac's
observables which has vanishing Poisson brackets with ${\check Q}^T_a$;
Eqs.(9-35) then automatically require that only even functions of
Grassmann-valued Dirac's observables belong to each superselection sector.

For the topological charge Eqs.(8-25), (8-27), (9-14) imply

$$\eqalign{
Q_T&={1\over 4}\int d^3x\,  F_{a\mu\nu}(\vec x,x^o){*}F_a^{\mu\nu}(\vec
x,x^o)=-\int d^3x\, {\vec E}_a(\vec x,x^o)\cdot {\vec B}_a(\vec x,x^o)=\cr
&=-g^2\int d^3x\, {\vec \pi}_a(\vec x,x^o)\cdot {\vec B}_a(\vec x,x^o)
{\rightarrow}_{\eta \rightarrow 0,L^{oT}_a{\buildrel \circ \over =}0} \cr
&{\rightarrow}_{\eta \rightarrow 0,L^{oT}_a{\buildrel \circ \over =}0}
\,\,\,{\check Q}_T\cr}
\form
$$

\noindent with

$$\eqalign{
{\check Q}_T&=-g^2\int d^3x\, [{\check {\vec \pi}}_{a,\perp}(\vec x,x^o)+\cr
&+{ {{\vec \partial}_x}\over {\triangle_x}}\int d^3y\, {\vec \partial}_x\cdot
{\vec \zeta}^{({\check {\vec A}}_{\perp})}_{ab}(\vec x,\vec y;x^o)
\rho_b^{({\check {\vec A}}_{\perp},{\check {\vec \pi}}_{\perp})}(\vec y,x^o)]
\cdot {\check {\vec B}}_a(\vec x,x^o)=\cr
&=-g^2\int d^3x\, {\check {\vec B}}_a(\vec x,x^o)\cdot [{\check {\vec \pi}}
_{a,\perp}(\vec x,x^o)+\cr
&+{({\check {\vec A}}_{\perp}(\vec x,x^o))}_{ab}\cdot
{1\over {\triangle_x}}\int d^3y\, {\vec \partial}_x\cdot {\vec \zeta}^{({\check
 {\vec A}}_{\perp})}_{bc}(\vec x,\vec y;x^o)
\rho_c^{({\check {\vec A}}_{\perp},{\check {\vec \pi}}_{\perp})}(\vec y,x^o)]
,\cr}
\form
$$

\noindent where, after an integration by parts, the Bianchi identity in the
limit $\eta \rightarrow 0$, i.e.\break ${\hat {\vec D}}_{ab}^{({\check
 {\vec A}}_{\perp})}(\vec x,x^o)\cdot {\check {\vec B}}_a(\vec x,x^o)\equiv
0$, has been used.

By using Eq.(6-44) and the rescaling $A={\sl g}\tilde A$ (for sake of
simplicity we go on to use the notation A also after the rescaling),
the action (9-17) and
the Hamiltonian (9-27) can be rewritten in the following forms, which show
explicitly the absence of singularities in ${\sl g}$:

$$
\eqalign{
{\check S}_T&=\int dx^o\, \int d^3x\, {\check {\cal L}}_T(\vec x,x^o)=\cr
&=\int dx^o\, \lbrace {1\over 2}\int d^3y_1d^3y_2\, {\dot {\check A}}^i
_{\perp a}({\vec {y_1}})\cr
&\lbrack \delta^{ij}\delta_{ab}\delta^3({\vec {y_1}}-{\vec {y_2}})+
g^2{({\check A}^i_{\perp}({\vec {y_1}}))}_{au}c({\vec {y_1}}-
{\vec {y_2}}){({\check A}^j_{\perp}({\vec {y_2}}))}_{ub}-\cr
&-2g^3\int d^3z\, c({\vec {y_1}}-\vec z)[\partial^h_zc(\vec z-{\vec {y_2}})]
{({\check A}^i_{\perp}({\vec {y_1}}))}_{ac}{({\check A}_{\perp}^h(\vec z))}
_{cu} \cdot \cr
&\cdot{(Pexp[g\int^{\vec z}_{{\vec {y_2}}}d\vec w\cdot{\check {\vec A}}_{\perp}
(\vec w)])}_{ud}{({\check A}^j_{\perp}({\vec {y_2}}))}_{db}+\cr
&+g^4\int d^3z_1d^3z_2\, c({\vec {y_1}}-{\vec {z_1}})[\partial^h_{z_1}c({\vec
{z_1}}-{\vec {z_2}})][\partial^k_{z_2}c({\vec {z_2}}-{\vec {y_2}})]
{({\check A}_{\perp}^i({\vec {y_1}}))}_{ac}{({\check A}_{\perp}^h({\vec
{z_1}}))
}_{cu}\cdot \cr
&\cdot {(Pexp[g\int^{{\vec {z_1}}}_{{\vec {z_2}}}d\vec w\cdot {\vec {\check A}}
_{\perp}(\vec w)])}_{uv}{({\check A}^k_{\perp}({\vec {z_2}}))}_{vr}{(Pexp
[g\int^{{\vec {z_2}}}_{{\vec {y_2}}}d\vec w\cdot {\vec {\check A}}_{\perp}
({\vec {z_2}})])}_{rd}{({\check A}^j_{\perp}({\vec {y_2}}))}_{db}\rbrack \cdot
\cr
&\cdot {\dot {\check A}}^j_{\perp b}({\vec {y_2}})-\cr
&-{1\over 2}\int d^3x\, \lbrack (\partial^i{\check A}^j_{\perp a}(\vec x))
\partial^i{\check A}^j_{\perp a}(\vec x)+2gc_{abc}(\partial^i{\check A}^j
_{\perp a}(\vec x)){\check A}^i_{\perp b}(\vec x){\check A}^j_{\perp c}(\vec x)
+\cr
&+{1\over 2}g^2c_{abc}c_{auv}{\check A}^i_{\perp b}(\vec x){\check A}^j_{\perp
c}(\vec x){\check A}^i_{\perp u}(\vec x){\check A}^j_{\perp v}(\vec x)\rbrack
+\cr
&+\int d^3x\, {\check \psi}^{\dagger}_a(\vec x){\lbrack
i(\partial^o\, I-\vec
\alpha \cdot (\vec \partial\, I+g{\vec {\check A}}_{\perp u}(\vec x)T^u))-
m\beta\rbrack }_{ab}{\check \psi}_b(\vec x)-\cr
&-{1\over 2}\int d^3y_1d^3y_2\, (i{\check \psi}^{\dagger}({\vec {y_1}})T^a
{\check \psi}({\vec {y_1}}))\lbrack \delta_{ab}c({\vec {y_1}}-{\vec {y_2}})-\cr
&-2g\int d^3z\, c({\vec {y_1}}-\vec z)[\partial^h_zc(\vec z-{\vec {y_2}})]
{({\check A}^h_{\perp}(\vec z))}_{au}{(Pexp[g\int^{\vec z}_{{\vec {y_2}}}
d\vec w\cdot {\vec {\check A}}_{\perp}(\vec w)])}_{ub}+\cr
&+g^2\int d^3z_1d^3z_2\, c({\vec {y_1}}-{\vec {z_1}})[\partial^h_{z_1}c(
{\vec {z_1}}-{\vec {z_2}})][\partial^k_{z_2}c({\vec {z_2}}-{\vec {y_2}})]
{(Pexp[g\int^{{\vec {z_1}}}_{{\vec {z_2}}}d\vec w\cdot {\vec {\check
A}}_{\perp}
(\vec w)])}_{uv}\cdot \cr
&\cdot {({\check A}^k_{\perp}({\vec {z_2}}))}_{vd}
{(Pexp[g\int^{{\vec {z_2}}}_{{\vec {y_2}}}d\vec w\cdot {\vec {\check
A}}_{\perp}(\vec w)])}_{db}\rbrack \lbrack {\check \psi}^{\dagger}({\vec
{y_2}})
T^b{\check \psi}({\vec {y_2}})+2{({\vec {\check A}}_{\perp}({\vec {y_2}}))}
_{bc}\cdot {\dot {\vec {\check A}}}_{\perp c}({\vec {y_2}})\rbrack \rbrace \cr}
\form
$$

\vfill \eject

$$
\eqalign{
{\check H}_{cT}&={1\over 2}\int d^3y_1d^3y_2\, {\check \pi}^i_{\perp a}
({\vec {y_1}})\cdot \cr
&\cdot \lbrack \delta^{ij}\delta_{ab}\delta^3({\vec {y_1}}-{\vec {y_2}})-g^2
{({\check A}^i_{\perp}({\vec {y_1}}))}_{au}c({\vec {y_1}}-{\vec
{y_2}}){({\check
A}^j_{\perp}({\vec {y_2}}))}_{ub}+\cr
&+2g^3\int d^3z\, c({\vec {y_1}}-\vec z)[\partial^h_zc(\vec z-{\vec {y_2}})]
{({\check A}^i_{\perp}({\vec {y_1}}))}_{ac}{({\check A}^h_{\perp}(\vec z))}
_{cu}\cdot \cr
&\cdot {(Pexp[g\int^{\vec z}_{{\vec {y_2}}}d\vec w\cdot {\vec {\check A}}
_{\perp}(\vec w)])}_{ud}{({\check A}^j_{\perp}({\vec {y_2}})}_{db}-\cr
&-g^4\int d^3z_1d^3z_2\, c({\vec {y_1}}-{\vec {z_1}})[\partial^h_{z_1}c({\vec
{z_1}}-{\vec {z_2}})][\partial^k_{z_2}c({\vec {z_2}}-{\vec {y_2}})]{({\check
A}^i_{\perp}({\vec {y_1}}))}_{ac}{({\check A}_{\perp}^h({\vec {z_1}}))}_{cu}
\cdot \cr
&\cdot {(Pexp[g\int^{{\vec {z_1}}}_{{\vec {z_2}}}d\vec w\cdot {\vec {\check
A}}_{\perp}(\vec w)])}_{uv}{({\check A}^k_{\perp}({\vec {z_2}}))}_{vr}{(Pexp
[g\int^{{\vec {z_2}}}_{{\vec {y_2}}}d\vec w\cdot {\vec {\check A}}_{\perp}
(\vec w)])}_{rd}{({\check A}^j_{\perp}({\vec {y_2}}))}_{db}\rbrack \cdot \cr
&\cdot {\check \pi}^j_{\perp b}({\vec {y_2}})+\cr
&+{1\over 2}\int d^3x \lbrack (\partial^i{\check A}^j_{\perp a}(\vec
x))\partial
^i{\check A}^j_{\perp a}(\vec x)+2gc_{abc}(\partial^i{\check A}^j_{\perp a}
(\vec x)){\check A}^i_{\perp b}(\vec x){\check A}^j_{\perp c}(\vec x)+\cr
&+{1\over 2}g^2c_{abc}c_{auv}{\check A}^i_{\perp b}(\vec x){\check A}^j
_{\perp c}(\vec x){\check A}^i_{\perp u}(\vec x){\check A}^j_{\perp v}(\vec x)
\rbrack +\cr
&+\int d^3x\, {\check \psi}^{\dagger}_a(\vec x){\lbrack i\vec \alpha \cdot
(\vec \partial I+g{\vec {\check A}}_{\perp u}(\vec x)T^u)+m\beta \rbrack}_{ab}
{\check \psi}_b(\vec x)+\cr
&+{1\over 2}g^2\int d^3y_1d^3y_2\, \lbrack 2{\vec {\check \pi}}_{\perp c}({\vec
{y_1}})\cdot {({\vec {\check A}}_{\perp}({\vec {y_1}}))}_{ca}+i{\check \psi}
^{\dagger}({\vec {y_1}})T^a{\check \psi}({\vec {y_1}})\rbrack
\lbrack \delta_{ab}c({\vec {y_1}}-{\vec {y_2}})-\cr
&-2g\int d^3z\, c({\vec {y_1}}-\vec z)[\partial^h_zc(\vec z-{\vec {y_2}})]
{({\check A}^h_{\perp}(\vec z))}_{au}{(Pexp[g\int^{\vec z}_{{\vec {y_2}}}
d\vec w\cdot {\vec {\check A}}_{\perp}(\vec w)])}_{ub}+\cr
&+g^2\int d^3z_1d^3z_2\, c({\vec {y_1}}-{\vec {z_1}})[\partial^h_{z_1}c({\vec
{z_1}}-{\vec {z_2}})][\partial^k_{z_2}c({\vec {z_2}}-{\vec {y_2}})]{({\check
A}^h_{\perp}({\vec {z_1}}))}_{au}\cdot \cr
&\cdot {(Pexp[g\int^{{\vec {z_1}}}_{{\vec {z_2}}}
d\vec w\cdot {\vec {\check A}}_{\perp}(\vec w)])}_{uv}
{({\check A}^k_{\perp}({\vec {z_2}}))}_{vr}\cr
&{(Pexp[g\int^{{\vec {z_2}}}
_{{\vec {y_2}}}d\vec w\cdot {\vec {\check A}}_{\perp}(\vec w)])}_{rb}\rbrack
(i{\check \psi}^{\dagger}({\vec {y_2}})T^b{\check \psi}({\vec {y_2}}))\cr}
\form
$$

{}From the discussion of Sections 2, 3, 7, and from Eqs.(7-19)-(7-23), one can
define a different set of Dirac's observables
${\check {\vec A}}^{(m)}_{a,\perp ,n}(\vec x,
x^o)$ [${\check {\vec A}}_{a,\perp}={\check {\vec A}}^{(0)}_{a,\perp ,0}$] for
each value $n\in Z$ of the winding number and for each index $m=0,1,..,dimZ_G
-1$, labelling the elements of the center $Z_G$ of G. Instead for the fermion
fields we have only the dependence on the winding number, ${\hat \psi}_{a
\alpha , (n)}(\vec x,x^o)$. In fact, instead that
around the identity cross section $\sigma_I$, one can repeat the whole
construction around each one of the cross sections defined in Eqs.(7-19)-(7-
23) in accord with the approach of Refs.[61].
Therefore the final action in terms of these configuration coordinates
is

$$
{\check S}_D=\sum_{n=-\infty}^{+\infty}\, \sum_{m=0}^{dimZ_G-1}\, ({\check S}
^{(n)}_{T(m)}+\theta {\check Q}^{(n)}_{T(m)}),
\form
$$

\noindent with ${\check S}_T$ and ${\check Q}_T$ given by Eqs.(9-39) and
(9-38) respectively.

\bigskip
\vfill\eject

\bigskip\noindent
{\bf {10. Covariantization and a Possible Ultraviolet Cutoff.}}
\newcount \nfor

\def \form {\global \advance \nfor by 1 \eqno (10.\the\nfor)}
\bigskip

Till now the whole construction of Dirac's observables, both in the Abelian
and non-Abelian case, has been manifestly Lorentz non-covariant, since a 3+1
splitting has been privileged. To recover some kind of covariance, i.e. Wigner
covariance, like for some relativistic two-body system [14] and for the Nambu
string [16], one has to follow a strategy which was essentially initiated by
Dirac himself [1a]. We have chosen boundary conditions such that the ten
functionals of the field configuration defining the generators of the
Poincar\'e
group, assumed globally realized (so that the global momentum map reformulation
of the first Noether theorem [69] can be applied; let us remark that in Ref.
[69d] there is a momentum map reformulation of the second Noether theorem
with only first class constraints and a different approach to the
covariantization of the phase space of classical field theory), are finite:
$P^{\mu} < \infty$, $J^{\mu\nu} < \infty$. Therefore, we have selected field
configurations for the coupled system of gauge potentials and matter fields,
which admit well defined values of the Poincar\'e Casimirs $P^2$ and $W^2=-P^2
{\hat {\vec S}}^2$ (${\hat {\vec S}}$ is the rest-frame Thomas spin [14-16]).
The first implication is that the constraint submanifold in phase space has a
first stratification structure, being the disjoint union of sets identified
by the various kinds of Poincar\'e orbits admitted by the system. The stratum
of the $P^2$- or mass-stratification for $P^2 > 0$ will be further stratified
due to the two kinds of orbits of the Thomas spin (${\hat {\vec S}}^2=0$ or
$\not= 0$): this is the spin-stratification. Let us remark that a priori
there is also the stratification induced by all possible types of gauge
symmetries, when the chosen functional space of connections allows reducible
connections; with our choice this stratification is absent. The mass- and
spin-stratifications could also be introduced at the level of the configuration
space in the Lagrangian approach, but they are more easily described in the
Hamiltonian approach.

Let us remark that the requirement $P^{\mu} < \infty$, $J^{\mu\nu} < \infty$,
presupposes the possibility of a regularization of the classical self-energy
problems (see for instance Ref.[32] for the Abelian case with charged scalar
particles) connected with the classical charge radius $\sim e^2$, when matter
particles are replaced by matter fermionic fields; at the pseudoclassical
level, in which the charges of the particles are treated by means of Grassmann
variables (pseudoclassical remnant of quantized charges), the pseudoclassical
charge radius, and then the self-energy, vanishes since $e^2_{PS}=0$ [44] (
$e_{PS}$ is the pseudoclassical charge which becomes $e$ after quantization).
It is plausible that with fermionic Grassmann-valued fields this kind of
regularization still works at the pseudoclassical level; however, this point
needs further investigation and it has not yet been studied which kind of
regularization would correspond to a consistent quantization of these
pseudoclassically regularized models. While in this Section, disregarding this
problem, we will find a classical basis for a ultraviolet cutoff, the
classical basis for the infrared problem is still practically unexplored: there
is an interconnection among the existence of the exceptional Poincar\'e
orbit $P^{\mu}=0$ (see the surface terms in Ref.[40]), the accumulation of
the time-like and space-like orbits toward it and the behaviour under
Lorentz boosts of the Coulomb clouds (either before or after the
decoupling of the gauge degrees of freedom).

The first step of the strategy is to study the classical spectrum of the
system to see which kinds of Poincar\'e orbits are allowed; if space-like
orbits $P^2 < 0$ are allowed, they have to be discarded already at the
classical level to avoid tachionic effects.

The second step is the treatment of the time-like ($P^2 > 0$, $P^o > 0$ or
$< 0$), light-like ($P^2=0$, $P^0 > 0$ or $< 0$) and $P^{\mu}=0$ orbits. For
the time-like orbits one has to reformulate the classical theory of the system
on space-like hypersurfaces, as it will be done later on; actually in this
case the instant form of dynamics of Dirac [92] is the natural one. For
light-like orbits one has to add $P^2\approx 0$, as a functional of the fields,
to the original constraints and to study anew the new constraint submanifold
(it defines the stratum of the original constraint submanifold containing
light-like orbits); the theory has to be reformulated on light-cones, because
now the front form of dynamics of Dirac [92] is the natural one. Also for the
orbit $P^{\mu}=0$ one has to add $P^{\mu}\approx 0$ to the original
constraints,
but a clear understanding of the geometrical and dynamical aspects of this
case is still lacking. In what follows we shall restrict ourselves to the
time-like case $P^2 > 0$, since the light-like one can be formulated along
similar lines and this will be done elsewhere; also we shall limit ourselves
to the Abelian case, whose reformulation on space-like hypersurfaces when
scalar relativistic particles instead of fermionic fields
are present, was done in Ref.
[93]; the introduction of fermionic Grassmann-valued fields and/or of
non-Abelian gauge potentials is straightforward and will be done elsewhere.
Then, being in the realm of special relativity, the gauge freedom in the
foliation of Minkowski space-time with space-like hypersurfaces will be reduced
to the natural (in the sense of intrinsic) family of hyperplanes perpendicular
to the total momentum $P^{\mu}$ ($P^2 > 0$) of the system: this can be called
the natural Wigner foliation for the case $P^2 > 0$.

The third step is to find a canonical transformation from the original
variables
to a new basis containing $P^{\mu}$, a conjugated center-of-mass coordinate
$X^{\mu}$ (functional of the field configuration) and a set of field relative
variables, like for the Nambu string [15-16]. Then a second canonical
transformation [14,16] will boost (at rest in the case $P^2 > 0$) the relative
variables with the standard Wigner boost pertaining to the chosen kind of
Poincar\'e orbit ($P^2 > 0$ or $P^2=0$): this new canonical basis will contain
$P^{\mu}$, a new Lorentz non-covariant center-of-mass coordinate ${\tilde X}
^{\mu}$ and relative field variables with Wigner covariance. In the case
$P^2 > 0$ some of the relative variables (those corresponding to relative-time-
like variables and to their conjugated variables) will be Lorentz scalars and
the remaining relative variables will be spin-1 Wigner three-vectors. The
center-of-mass variables ${\tilde X}^{\mu}$, $P^{\mu}$, with another canonical
transformation [14], will be replaced with the total mass, the conjugated
rest-frame time  and six Dirac's observables describing the independent
Euclidean
Cauchy data for the center-of-mass

$$\eqalign{
\eta &\sqrt {P^2},\quad \quad P\cdot \tilde X/\eta \sqrt {P^2},\quad\quad
\eta =sign\, P^o,\cr
\vec z&=\eta \sqrt {P^2}({\tilde x}^i-{ {P^i}\over {P^o}}{\tilde x}^o)=\cr
&=\eta \sqrt {P^2}\lbrack x^i-{{P^i}\over {P^o}}x^o+{{S^{io}}\over {\eta
\sqrt {P^2}}}+(S^{ik}-{{P^i}\over {P^o}}S^{ok}){ {P_k}\over {\eta
\sqrt {P^2}(\eta \sqrt {P^2}+P^o)} }\rbrack \cr
\vec k&=\vec P/\eta \sqrt {P^2},\cr}
\form
$$

\noindent where $S^{\mu\nu}$ is the spin part of $J^{\mu\nu}$ after the center-
of-mass decomposition. $\vec z$ is not a three-vector under Lorentz boosts and
${\vec z}^{'}=\vec z/\eta \sqrt {P^2}$ is the classical analogue of the
Newton-Wigner operator in presence of spin (it is canonical,
$\lbrace z^i,z^j\rbrace =0$, but not covariant). Let us note that when some
mass parameter is available one could redefine $\vec z$ and $\vec k$ so to
have the standard dimensions, which have been altered by the canonical
transformation due to the necessity of commuting with the rest frame time.

Let us consider a foliation of Minkowski space-time with a family of space-like
hypersurfaces $\Sigma (\tau )$, whose points are given coordinates $z^{\mu}
(\tau ,\vec \sigma )=z^{\mu} (\sigma^A)$, $A=\tau ,r$, $r=1,2,3$; let $\partial
_A=(\partial_{\tau}=\partial /\partial \tau $; $\partial_r=\partial /\partial
\sigma^r)$ denote the derivatives with respect to the parameters $\tau ,\vec
\sigma$, whose space has Lorentz signature; let $z^{\mu}_A(\sigma^B)=\partial_A
z^{\mu}(\sigma^B)$ be the associated vierbeins ($\partial_Bz^{\mu}_A-\partial_A
z^{\mu}_B=0$ are the Gauss-Weingarten equations for the embedding of $\Sigma
(\tau )$ in the flat Minkowski space-time) and $g_{AB}(\sigma
)=z^{\mu}_A(\sigma
)\eta_{\mu\nu}z^{\nu}_B(\sigma )$ the metric in the parametric space; since
$\Sigma (\tau )$ is space-like, one has $g_{\tau \tau}(\sigma ) > 0$; let
$\gamma_{rs}(\sigma )=-g_{rs}(\sigma )$ denote the Euclidean 3-metric
and $\gamma^{rs}(\sigma )$ its inverse; let $g=-det(g_{AB})={(det(z^{\mu}_A))
}^2$, $\gamma =det(\gamma_{rs})$; let $g^{AB}$ be the inverse of the metric
$g_{AB}$ ($g^{\tau \tau}=\gamma /g$, $g^{\tau r}=\gamma g_{\tau u}\gamma^{ur}/
g$; $g^{rs}=-\gamma^{rs}+g_{\tau u}g_{\tau v}\gamma^{ur}\gamma^{vs}/g$). If
$l^{\mu}(\sigma )$ is the normal unit vector to $\Sigma (\tau )$ in $(\tau ;
\vec \sigma )$ [$l^2(\sigma )=1$], then $l^{\mu}(\sigma )=\epsilon^{\mu\alpha
\beta\gamma}z_{1\alpha}(\sigma )z_{2\beta}(\sigma )z_{3\gamma}(\sigma )/
\sqrt {\gamma (\sigma )}$ and $\eta^{\mu\nu}=z^{\mu}_A(\sigma )g^{AB}(\sigma )
z^{\nu}_B(\sigma )=l^{\mu}(\sigma )l^{\nu}(\sigma )-\gamma^{rs}(\sigma )z^{\mu}
_r(\sigma )z^{\nu}_s(\sigma )$ and $z^{\mu}_{\tau}(\sigma )=\sqrt { {{g(\sigma
)}\over {\gamma (\sigma )}} }l^{\mu}(\sigma )-g_{\tau r}(\sigma )\gamma^{rs}
(\sigma )z^{\mu}_s(\sigma )$; the volume element is $d^4z=z^{\mu}_{\tau}(\sigma
)d\tau d^2\Sigma_{\mu}=d\tau z^{\mu}_{\tau}(\sigma )l_{\mu}(\sigma )\sqrt
{\gamma (\sigma )}d^3\sigma =\sqrt {g(\sigma )}d\tau d^3\sigma$; let $z^A_{\mu}
(\sigma )$ be the inverse vierbeins satisfying $g^{AB}(\sigma )=z^A_{\mu}
(\sigma )\eta^{\mu\nu}z^B_{\nu}(\sigma )$.

The electromagnetic gauge potential $A_{\mu}(z(\sigma ))=z^A_{\mu}(\sigma )
A_A(\sigma )$ has the field strength $F_{\mu\nu}(z(\sigma ))=z^A_{\mu}(\sigma )
z^B_{\nu}(\sigma )F_{AB}(\sigma )$; its inverse expressions $A_A(\sigma )=
z^{\mu}_A(\sigma )A_{\mu}(z(\sigma ))$ are the new configuration space
variables together with $z^{\mu}(\sigma )$, and one has
$F_{AB}(\sigma )=z_A^{\mu}(\sigma )z^{\nu}_B(\sigma )$\break
$F_{\mu\nu}(z(\sigma ))$. The variables $A_A(\sigma )$ are Lorentz scalars.
The action is

$$
S=\int d\tau L(\tau)=
-{1\over 4}\int d\tau d^3\sigma \, \sqrt {g(\sigma )}g^{AC}(\sigma )g^{BD}
(\sigma )F_{AB}(\sigma )F_{CD}(\sigma )
\form
$$

\noindent and the canonical momenta are

$$\eqalign{
\rho_{\mu}(\tau ,\vec \sigma )&={ {\partial L}\over {\partial z^{\mu}_{\tau}
(\tau ,\vec \sigma )} }=\cr
&={ {\sqrt {g(\tau ,\vec \sigma )}}\over 4}[(g^{\tau \tau}z_{\tau \mu}+
g^{\tau r}z_{r\mu})(\tau ,\vec \sigma )g^{AC}(\tau ,\vec \sigma )
g^{BD}(\tau ,\vec \sigma )F_{AB}(\tau ,\vec \sigma )F_{CD}(\tau ,\vec \sigma )
-\cr
&-2[z_{\tau \mu}(\tau ,\vec \sigma )(g^{A\tau}g^{\tau C}g^{BD}+g^{AC}
g^{B\tau}g^{\tau D})(\tau ,\vec \sigma )+\cr
&+z_{r\mu}(\tau ,\vec \sigma )
(g^{Ar}g^{\tau C}+g^{A\tau}g^{rC})(\tau ,\vec \sigma )g^{BD}
(\tau ,\vec \sigma )]f_{AB}(\tau ,\vec \sigma )F_{CD}(\tau ,\vec \sigma )],\cr
\pi^{\tau}(\tau ,\vec \sigma )&={ {\partial L}\over {\partial \partial_{\tau}
A_{\tau}(\tau ,\vec \sigma )} }=0,\cr
\pi^r(\tau ,\vec \sigma )&={ {\partial L}\over {\partial \partial_{\tau}A_r
(\tau ,\vec \sigma )} }=-{ {\gamma (\tau ,\vec \sigma )}\over
{\sqrt {g(\tau ,\vec \sigma )}} }\gamma^{rs}(\tau ,\vec \sigma )(F_{\tau s}+
g_{\tau v}\gamma^{vu}F_{us})(\tau ,\vec \sigma )=\cr
&={ {\gamma (\tau ,\vec \sigma )}\over
{\sqrt {g(\tau ,\vec \sigma )}} }\gamma^{rs}(\tau ,\vec \sigma )(E_s(\tau ,\vec
\sigma )-g_{\tau v}(\tau ,\vec \sigma )\gamma^{vu}(\tau ,\vec \sigma )
\epsilon_{ust} B_t(\tau ,\vec \sigma )).\cr}
\form
$$

The canonical Hamiltonian is

$$
H_c=-\int d^3\sigma \, A^{\tau}(\tau ,\vec \sigma )\Gamma (\tau ,\vec \sigma )
\form
$$

\noindent with

$$
\Gamma(\tau ,\vec \sigma )=-\partial^r\pi^r(\tau ,\vec \sigma ),
\form
$$

\noindent while the basic Poisson brackets are

$$\eqalign{
\lbrace z^{\mu}(\tau ,\vec \sigma ),\rho_{\nu}(\tau ,\vec \sigma^{'} )\rbrace
&=-\eta^{\mu}_{\nu}\delta^3(\vec \sigma -\vec \sigma^{'}),\cr
\lbrace A_A(\tau ,\vec \sigma ),\pi^B(\tau ,\vec \sigma^{'} )\rbrace &=
-\eta^B_A\delta^3(\vec \sigma -\vec \sigma^{'}).\cr}
\form
$$

The ten conserved Poincar\'e generators are

$$\eqalign{
P^{\mu}&=\int d^3\sigma \, \rho^{\mu}(\tau ,\vec \sigma ),\cr
J^{\mu\nu}&=\int d^3\sigma \, (z^{\mu}(\tau ,\vec \sigma )\rho^{\nu}
(\tau ,\vec \sigma )-z^{\nu}(\tau ,\vec \sigma )\rho^{\mu}(\tau ,\vec \sigma
)).
\cr}
\form
$$

The Lorentz scalar constraint $\pi^{\tau}(\tau ,\vec \sigma )\approx 0$ is
generated by the gauge invariance of S; its time constancy will produce the
only secondary constraint $\Gamma (\tau ,\vec \sigma )\approx 0$ as usual.
The invariance of S under arbitrary $(\tau ,\vec \sigma )$-reparametrizations
gives rise to the constraints

$$
H_{\mu}(\tau ,\vec \sigma )=\rho_{\mu}(\tau ,\vec \sigma )-T_{\tau \tau}(\tau ,
\vec \sigma )l_{\mu}(\tau ,\vec \sigma )+T_{\tau r}(\tau ,\vec \sigma )
z^r_{\mu}(\tau ,\vec \sigma )\approx 0,
\form
$$

\noindent where

$$\eqalign{
T_{\tau \tau}(\tau ,\vec \sigma )
&=-{1\over 2}({1\over {\sqrt {\gamma}} }\pi^rg_{rs}\pi^s-{ {\sqrt
{\gamma}}\over 2}\gamma^{rs}\gamma^{uv}F_{ru}F_{sv})(\tau ,\vec \sigma ),\cr
T_{\tau r}(\tau ,\vec \sigma )&=F_{rs}(\tau ,\vec \sigma )\pi^s(\tau ,\vec
\sigma ),\cr}
\form
$$

\noindent are the energy density and the Poynting vector respectively. The
constraints $H_{\mu}\approx 0$ describe the arbitrariness of the foliation:
physical results do not depend on its choice.

The six constraints $H_{\mu}(\tau ,\vec \sigma )\approx 0$, $\pi^{\tau}
(\tau ,\vec \sigma )\approx 0$, $\Gamma (\tau ,\vec \sigma )\approx 0$ are
first class with the only non vanishing Poisson brackets

$$
\lbrace H_{\mu}(\tau ,\vec \sigma ),H_{\nu}(\tau ,\vec \sigma^{'} )\rbrace =
z^r_{\mu}(\tau ,\vec \sigma )F_{rs}(\tau ,\vec \sigma )z^s_{\nu}(\tau ,\vec
\sigma )\Gamma (\tau ,\vec \sigma )\delta^3(\vec \sigma -\vec \sigma^{'}).
\form
$$

The Dirac Hamiltonian is

$$
H_D=\int d^3\sigma \, [\lambda^{\mu}(\tau ,\vec \sigma )H_{\mu}(\tau ,\vec
\sigma )+\lambda_{\tau}(\tau ,\vec \sigma )\pi^{\tau}(\tau ,\vec \sigma )-
A^{\tau}(\tau ,\vec \sigma )\Gamma (\tau ,\vec \sigma )].
\form
$$

\noindent Let us remark that the simplicity of Eqs.(10-10) is due to the use
of Cartesian coordinates [94]: if we had used the constraints $H_l(\tau ,\vec
\sigma )=l^{\mu}(\tau ,\vec \sigma )H_{\mu}(\tau ,\vec \sigma )$, $H_r
(\tau ,\vec \sigma )=z^{\mu}_r(\tau ,\vec \sigma )H_{\mu}(\tau ,\vec \sigma )$
(i.e. nonholonomic coordinates), so that their associated Dirac multipliers
$\lambda_l(\tau ,\vec \sigma )$, $\lambda_r(\tau ,\vec \sigma )$ would have
been
the lapse and shift functions of general relativity, one would have obtained
the universal algebra of Ref.[1a].

Using Eqs.(5-8) and (5-10) in this coordinatization, i.e. $A^r(\tau ,\vec
\sigma )={ {\partial}\over {\partial \sigma^r} }\eta (\tau ,\vec \sigma )+
A^r_{\perp}(\tau ,\vec \sigma )$, $\eta (\tau ,\vec \sigma )=-{1\over
{\triangle_{\sigma}} }{ {\partial}\over {\partial \vec \sigma} }\cdot \vec
A(\tau ,\vec \sigma )$, $\pi^r(\tau ,\vec \sigma )=\pi^r_{\perp}(\tau ,\vec
\sigma )+{1\over {\triangle_{\sigma}} }{ {\partial}\over {\partial \sigma^r}}
\Gamma (\tau ,\vec \sigma )$, one gets

$$\eqalign{
H_{\mu}(\tau ,\vec \sigma )&=H_{\perp \mu}(\tau ,\vec \sigma )+{ {g^{rs}(\tau ,
\vec \sigma )}\over {\sqrt {\gamma (\tau ,\vec \sigma )}} }
[\pi^r_{\perp}{1\over {\triangle
_{\sigma}} }\partial^s\Gamma +{1\over 2}({1\over {\triangle_{\sigma}} }\partial
^r\Gamma )({1\over {\triangle_{\sigma}} }\partial^s\Gamma )](\tau ,\vec \sigma
)
l_{\mu}(\tau ,\vec \sigma )+\cr
&+F_{rs}(\tau ,\vec \sigma )({1\over {\triangle_{\sigma}} }\partial^s\Gamma
(\tau ,\vec \sigma )\, )z^r_{\mu}(\tau ,\vec \sigma ),\cr
&{}\cr
H_{\perp \mu}(\tau ,\vec \sigma )&=\rho_{\mu}(\tau ,\vec \sigma )-T_{\perp
\tau \tau}(\tau ,\vec \sigma )l_{\mu}(\tau ,\vec \sigma )+T_{\perp \tau r}
(\tau ,\vec \sigma )z^r_{\mu}(\tau ,\vec \sigma ),\cr}
\form
$$

\noindent with $T_{\perp \tau \tau}$, $T_{\perp \tau r}$ obtained from $T_{\tau
\tau}$, $T_{\tau r}$ with the replacement $\pi^r\rightarrow \pi^r_{\perp}$.
Then

$$\eqalign{
H_D&=H_{\perp D}+\int d^3\sigma \,[\lambda_{\tau}(\tau ,\vec \sigma )\pi^{\tau}
(\tau ,\vec \sigma )-A^{\tau}(\tau ,\vec \sigma )\Gamma (\tau ,\vec \sigma
)+\cr
&+\lambda^{\mu}(\tau ,\vec \sigma )({ {g_{rs}(\tau ,\vec \sigma )}\over {\sqrt
{\gamma (\tau ,\vec \sigma )}} }
[\pi^r_{\perp}{1\over {\triangle_{\sigma}}}\partial^s\Gamma +
{1\over 2}({1\over {\triangle_{\sigma}}}\partial^r\Gamma )({1\over {\triangle
_{\sigma}}}\partial^s\Gamma)](\tau ,\vec \sigma )l_{\mu}(\tau ,\vec \sigma
)+\cr
&+F_{rs}(\tau ,\vec \sigma )({1\over {\triangle_{\sigma}}}\partial^s\Gamma )
(\tau ,\vec \sigma )z^r_{\mu}(\tau ,\vec \sigma )\, )]\cr}
\form
$$

\noindent and, after the decoupling of the gauge degrees of freedom, the
physical Hamiltonian is

$$
H_{\perp D}=\int d^3\sigma \, \lambda^{\mu}(\tau ,\vec \sigma )H_{\perp \mu}
(\tau ,\vec \sigma ).
\form
$$

Since we are working in the framework of special relativity, we can restrict
ourselves to space-like hyperplanes by imposing the gauge-fixing constraints

$$
z^{\mu}(\tau ,\vec \sigma )-x^{\mu}(\tau )-b^{\mu}_r(\tau )\sigma^r\approx
0
\form
$$

\noindent Here $b^{\mu}_r=z^{\mu}_r(\tau ,\vec \sigma )=z^{\mu}_r(\tau)$ are
three orthonormal space-like 4-vectors containig six independent degrees of
freedom due to $g_{rs}(\tau ,\vec \sigma )=-\gamma_{rs}(\tau ,\vec \sigma )
\approx b^{\mu}_r(\tau )\eta_{\mu\nu}b^{\nu}_s(\tau )=-\delta_{rs}$. With
$x^{\mu}(\tau )$ we have only 10 degrees of freedom left, describing the family
of space-like hyperplanes and its parametrization. Now one has

$$\eqalign{
l^{\mu}(\tau ,\vec \sigma )&=l_{\mu}(\tau )=\epsilon^{\mu\alpha\beta\gamma}
b_{1\alpha}(\tau )b_{2\beta}(\tau )b_{3\gamma}(\tau ),\cr
\eta^{\mu\nu}&=l^{\mu}(\tau )l^{\nu}(\tau )-b^{\mu}_r(\tau )b^{\nu}_r(\tau )
\cr}
\form
$$

By putting $b^{\mu}_o(\tau )=l^{\mu}(\tau )$, we may build a tetrad $b^{\mu}_A
(\tau )$, A=0,1,2,3, with $\eta^{\mu\nu}-b^{\mu}_A(\tau )\eta^{AB}b^{\nu}_B
(\tau )=0$; as said, there are only six independent degrees of freedom $\phi
_{\lambda}(\tau )$, $\lambda =1,..,6$, in the $b^{\mu}_A$. Their six conjugated
momenta $T_{\lambda}(\tau )$, $\lbrace \phi_{\lambda}(\tau ),T_{\lambda^{'}}
(\tau )\rbrace =\delta_{\lambda \lambda^{'}}$, have to be extracted from
$\int d^3\sigma \, \sigma^v\rho^{\mu}(\tau ,\vec \sigma )$, see later on.
In Refs. [95,2b] (see also Appendix A of Ref.[96]), Hanson and Regge found
Dirac brackets for quantities $b^{\mu}_A(\tau )$, $S^{\mu\nu}$, compatible
with the constraints $\eta^{\mu\nu}-b^{\mu}_A(\tau )\eta^{AB}b_B^{\nu}(\tau )
\approx 0$; in this way they avoided to work with the variables
$\phi_{\lambda}$
, $T_{\lambda}$.

If $L({\dot x}^{\mu},b^{\mu}_A,{\dot b}^{\mu}_A,..)$ is a Lagrangian
depending on a center-of-mass variable $x^{\mu}(\tau )$, on a tetrad $b^{\mu}
_A(\tau )$ sitting on $x^{\mu}(\tau )$ and other relative variables, and we add
the holonomic constraint $\eta^{\mu\nu}-b^{\mu}_A(\tau )\eta^{AB}b_B^{\nu}
(\tau )\approx 0$ (so that $b^{\mu}_A(\tau )$ becomes an orthonormal frame at
$x^{\mu}(\tau )$), then we could replace L with $\tilde L({\dot x}^{\mu},
\phi_{\lambda},{\dot \phi}_{\lambda},..)$ and we would have $T_{\lambda}=
\partial \tilde L/\partial {\dot \phi}_{\lambda}$; for functions $f(\phi
_{\lambda},T_{\lambda})$, $g(\phi_{\lambda},T_{\lambda})$ one would have the
Poisson brackets $\lbrace f,g\rbrace ={ {\partial f}\over {\partial \phi
_{\lambda}} }{ {\partial g}\over {\partial T_{\lambda}} }-{ {\partial f}\over
{\partial T_{\lambda}} }{ {\partial g}\over {\partial \phi_{\lambda}} }$.

Let us introduce a generalized angular velocity

$$\eqalign{
\sigma^{\mu\nu}(\tau )&=-\sigma^{\nu\mu}(\tau )=
\eta^{AB}b_A^{\mu}(\tau ){\dot b}^{\nu}_B(\tau )=l^{\mu}(\tau ){\dot l}^{\nu}
(\tau )-b^{\mu}_r(\tau ){\dot b}^{\nu}_r(\tau )\equiv \cr
&\equiv a^{\mu\nu}{}_{\lambda}(\phi(\tau ) ){\dot \phi}_{\lambda}(\tau ),\quad
\quad a^{\mu\nu}{}_{\lambda}=-a^{\nu\mu}{}_{\lambda}\cr}
\form
$$

\noindent The inverse relation $\eta_{\mu\nu}b^{\mu}_A(\tau )b^{\nu}_B(\tau )-
\eta_{AB}=0$ gives

$$
{\dot b}^{\mu}_A(\tau )=\eta_{\nu\rho}b^{\rho}_A(\tau )\sigma^{\nu\mu}(\tau )
=\eta_{\nu\rho}b^{\rho}_A(\tau )a^{\nu\mu}{}_{\lambda}(\phi (\tau ) ){\dot
\phi}
_{\lambda}(\tau ).
\form
$$

\noindent If we introduce the angular change $\delta \theta^{\mu\nu}(\tau )=
-\delta \theta^{\nu\mu}(\tau )=\eta^{AB}b^{\mu}_A(\tau )\delta b^{\nu}_B(\tau
)$, we get

$$
\delta \sigma^{\mu\nu}(\tau )={d\over {d\tau }}\delta \theta^{\mu\nu}(\tau )-
\delta \theta^{\mu\rho}(\tau )\sigma_{\rho}{}^{\nu}(\tau )+\delta \theta^{\rho
\nu}(\tau )\sigma^{\mu}{}_{\rho}.
\form
$$

\noindent If, moreover, we require that $\delta \theta^{\mu\nu}\equiv
a^{\mu\nu}
{}_{\lambda}(\phi )\delta \phi_{\lambda}$, so that $\delta b^{\mu}_A=\eta_{\nu
\rho}b_A^{\rho}\delta \theta^{\nu\mu}=$\break
$\eta_{\nu\rho}b^{\rho}_Aa^{\nu\mu}
{}_{\lambda}(\phi )\delta \phi_{\lambda}$, Eqs.(10-19) imply

$$
{ {\partial a^{\mu\nu}{}_{\lambda}}\over {\partial \phi_{{\lambda}^{'}}
} }-{ {\partial a^{\mu\nu}{}_{{\lambda}^{'}}}\over {\partial \phi_{\lambda}} }
+a^{\mu\rho}{}_{{\lambda}^{'}}\eta_{\rho\sigma}a^{\sigma\nu}{}_{\lambda}-
a^{\mu\rho}{}_{\lambda}\eta_{\rho \sigma}a^{\sigma\nu}{}_{{\lambda}^{'}}=0
\form
$$

If we assume that an inverse $b_{\mu\nu ,\lambda}(\phi )=-b_{\nu\mu ,\lambda}
(\phi )$ of $a^{\mu\nu}{}_{\lambda}(\phi )$ exists with the properties

$$\eqalign{
&a^{\mu\nu}{}_{\lambda}b_{\mu\nu ,\lambda^{'}}=2\delta_{\lambda \lambda^{'}}\cr
&a^{\mu\nu}{}_{\lambda}b_{\alpha\beta ,\lambda}=\eta^{\mu}_{\alpha}\eta^{\nu}
_{\beta}-\eta^{\mu}_{\beta}\eta^{\nu}_{\alpha},\cr}
\form
$$

\noindent then Eqs.(10-20) saturated with $b^{\alpha\beta}{}_{'\lambda}$,
$b^{\gamma \delta}{}_{,\lambda^{'}}$, $b_{\mu \nu ,\lambda^{"}}$ plus the
$\tau$ derivative of the first of Eqs.(10-21) imply

$$
b^{\alpha\beta}{}_{,\lambda}{ {\partial b^{\gamma\delta}{}_{,\lambda^{'}}}
\over {\partial \phi_{\lambda}} }-b^{\gamma\delta}{}_{,\lambda} { {\partial
b^{\alpha\beta}{}_{,\lambda^{'}}}\over {\partial \phi_{\lambda}} }=
C^{\alpha\beta\gamma\delta}_{\mu\nu}b^{\mu\nu}{}_{,\lambda^{'}}
\form
$$

\noindent where

$$
C^{\alpha\beta\gamma\delta}_{\mu\nu}=\eta^{\nu}_{\gamma}\eta^{\alpha}_{\delta}
\eta^{\mu\beta}+\eta^{\mu}_{\gamma}\eta^{\beta}_{\delta}\eta^{\nu\alpha}-
\eta^{\nu}_{\gamma}\eta^{\beta}_{\delta}\eta^{\mu\alpha}-\eta^{\mu}_{\gamma}
\eta^{\alpha}_{\delta}\eta^{\nu\beta}
\form
$$

\noindent are the structure constants of the Lorentz algebra. Therefore, Eqs.
(10-20), (10-22) are the Maurer-Cartan equations of the Lorentz group.

One has $\delta \phi_{\lambda}={1\over 2}\eta^{AB}b^{\mu}_A\delta b^{\nu}_B
b_{\mu\nu,\lambda}(\phi )$. Now $\tilde L$ can depend on ${\dot
\phi}_{\lambda}$
only through $\sigma^{\mu\nu}$. If we define $S^{\mu\nu}=-\partial \tilde L/
\partial \sigma_{\mu\nu}$, then we have

$$T_{\lambda}={ {\partial \tilde L}\over {\partial {\dot \phi}_{\lambda}} }=
{1\over 2}{ {\partial \sigma^{\mu\nu}}\over {\partial {\dot \phi}_{\lambda}} }
{ {\partial \tilde L}\over {\partial \sigma^{\mu\nu}} }=-{1\over 2}a^{\mu\nu}
{}_{\lambda}(\phi)S_{\mu\nu}\quad \Rightarrow \,\, S^{\mu\nu}=-b^{\mu\nu}
{}_{\lambda}(\phi )T_{\lambda}
\form
$$

\noindent In general, $S^{\mu\nu}$ is the spin part of the total angular
momentum $J^{\mu\nu}=x^{\mu}P^{\nu}-x^{\nu}P^{\mu}+S^{\mu\nu}$ of the
system. Since, if $F(b,S)=f(\phi ,T)$, we have

$${ {\partial f}\over {\partial \phi_{\lambda}} }={ {\partial F}\over
{\partial b^{\mu}_A} }{ {\partial b^{\mu}_A}\over {\partial \phi_{\lambda}} }
+{1\over 2}{ {\partial F}\over {\partial S^{\mu\nu}} }{ {\partial S^{\mu\nu}}
\over {\partial \phi_{\lambda}} },\quad\quad { {\partial f}\over {\partial
T_{\lambda}} }={1\over 2}{ {\partial F}\over {\partial S^{\mu\nu}} }{ {\partial
S^{\mu\nu}}\over {\partial T_{\lambda}} },$$

\noindent then we get

$$
\lbrace f(\phi ,T),g(\phi ,T)\rbrace =-S^{\mu\nu}\eta^{\rho\sigma}{ {\partial
f}
\over {\partial S^{\mu\rho}} }{ {\partial g}\over {\partial S^{\nu\sigma}} }
-b^{\mu}_A\eta^{\alpha\beta}({ {\partial f}\over {\partial b^{\alpha}_A} }
{ {\partial g}\over {\partial S^{\mu\beta}} }-{ {\partial f}\over {\partial
S^{\mu\beta}} }{ {\partial g}\over {\partial b^{\alpha}_A} }).
\form
$$

\noindent These brackets

$$\eqalign{
&\lbrace b^{\mu}_A(\tau ),b^{\nu}_B(\tau )\rbrace =0\cr
&\lbrace S^{\mu\nu}(\tau ),b^{\rho}_A(\tau )\rbrace =\eta^{\rho\nu}b^{\mu}_A
(\tau )-\eta^{\rho\mu}b^{\nu}_A(\tau )\cr
&\lbrace S^{\mu\nu}(\tau ),S^{\alpha\beta}(\tau )\rbrace =C^{\mu\nu\alpha\beta}
_{\gamma\delta}S^{\gamma\delta}(\tau )\cr}
\form
$$

\noindent are Dirac brackets for the ten constraints $\eta^{\mu\nu}-b^{\mu}_A
\eta^{AB}b^{\nu}_B=0$ (the other ten constraints to form twenty second class
constraints are hidden in the relation $S^{\mu\nu}=-b^{\mu\nu}{}_{,\lambda}
T_{\lambda}$), since

$$
\lbrace b^{\rho}_C,\eta^{\mu\nu}-b^{\mu}_A\eta^{AB}b^{\nu}_B\rbrace =\lbrace
S^{\gamma\delta},\eta^{\mu\nu}-b^{\mu}_A\eta^{AB}b^{\nu}_B\rbrace =0.
\form
$$

\noindent After this digression, one sees that the time constancy of the
gauge-fixing constraints (10-15) gives

$$\eqalign{
z^{\mu}_{\tau}(\tau ,\vec \sigma )&-{\dot x}^{\mu}(\tau )-{\dot b}^{\mu}_r
(\tau )\sigma^r=\cr
&=\lbrace z^{\mu}(\tau ,\vec \sigma ),H_{\perp D}\rbrace -{\dot x}^{\mu}(\tau )
-{\dot b}^{\mu}_r(\tau )\sigma^r=\cr
&=-\lambda^{\mu}(\tau ,\vec \sigma )-{\dot x}^{\mu}(\tau )-{\dot b}^{\mu}_r
(\tau )\sigma^r\approx 0\cr}
\form
$$

\noindent which implies

$$\eqalign{
\lambda^{\mu}(\tau ,\vec \sigma )&=-{\dot x}^{\mu}(\tau )-{\dot b}^{\mu}_r(\tau
)\sigma^r={\tilde \lambda}^{\mu}(\tau )+{\tilde \lambda}^{\mu\nu}(\tau
)b_{r\nu}
(\tau )\sigma^r\cr
&{\tilde \lambda}^{\mu}(\tau )=-{\dot x}^{\mu}(\tau )\cr
&{\tilde \lambda}^{\mu\nu}(\tau )={\dot b}^{\mu}_r(\tau )b^{\nu}_r(\tau )=
{1\over 2}({\dot b}^{\mu}_r(\tau )b^{\nu}_r(\tau )-{\dot b}^{\nu}(\tau )b^{\mu}
_r(\tau ))=-{\tilde \lambda}^{\nu\mu}(\tau ),\cr}
\form
$$

\noindent and

$$\eqalign{
{\hat T}_{\perp \tau \tau}(\tau ,\vec \sigma )&={1\over 2}({\vec \pi}^2_{\perp}
+{\vec B}^2)(\tau ,\vec \sigma ),\cr
{\hat T}_{\perp \tau r}(\tau ,\vec \sigma )&=F_{rs}(\tau ,\vec \sigma )\pi^r
_{\perp}(\tau ,\vec \sigma )=\epsilon_{rst}\pi^s_{\perp}(\tau ,\vec \sigma )
B_t(\tau ,\vec \sigma ),\cr}
\form
$$

After the elimination of all degrees of freedom of the hypersurface except ten
with the Dirac brackets of the resulting second class constraints, the Dirac
Hamiltonian becomes

$$\eqalign{
H_{\perp D}&=\int d^3\sigma \,\lambda^{\mu}(\tau ,\vec \sigma )[\rho_{\mu}
(\tau ,\vec \sigma )-T_{\perp \tau \tau}(\tau ,\vec \sigma )l_{\mu}(\tau ,\vec
\sigma )+T_{\perp \tau r}(\tau ,\vec \sigma )z^r_{\mu}(\tau ,\vec \sigma )]
\equiv \cr
&\equiv -{\dot x}^{\mu}(\tau )[\int d^3\sigma \, \rho_{\mu}(\tau ,\vec \sigma
)-
l_{\mu}(\tau )\int d^3\sigma \, {\hat T}_{\perp \tau \tau}(\tau ,\vec \sigma )
-b_{r\mu}(\tau )\int d^3\sigma \, {\hat T}_{\perp \tau r}(\tau ,\vec \sigma )]
-\cr
&-{\dot b}^{\mu}_r(\tau )[\int d^3\sigma \, \sigma^v\rho_{\mu}(\tau ,\vec
\sigma )-l_{\mu}(\tau )\int d^3\sigma \, \sigma^v{\hat T}_{\perp \tau \tau}
(\tau ,\vec \sigma )-\cr
&-b_{s\mu}(\tau )\int d^3\sigma \, \sigma^v {\hat T}_{\perp
\tau s}(\tau ,\vec \sigma )]=\cr
&={\tilde \lambda}^{\mu}(\tau )\chi_{\mu}(\tau )+{\tilde \lambda}^{\mu\nu}(\tau
)\chi_{\mu\nu}(\tau ),\cr}
\form
$$

\noindent with

$$\eqalign{
\chi_{\mu}(\tau )&=P_{\mu}-l_{\mu}(\tau ){1\over 2}\int d^3\sigma \, ({\vec
\pi}^2_{\perp}+{\vec B}^2)(\tau ,\vec \sigma )-b_{r\mu}(\tau )\int d^3\sigma
\, {({\vec \pi}_{\perp}\times \vec B)}_r(\tau ,\vec \sigma ),\cr
\chi_{\mu\nu}(\tau )&=S_{\mu\nu}-{1\over 2}(b_{v\mu}l_{\nu}-b_{v\nu}l_{\mu})
(\tau )\int d^3\sigma \, \sigma^v{1\over 2}({\vec \pi}^2_{\perp}+{\vec B}^2)
(\tau ,\vec \sigma )-\cr
&-{1\over 2}(b_{v\mu}b_{s\nu}-b_{v\nu}b_{s\mu})(\tau )
\int d^3\sigma \, \sigma^v{({\vec \pi}_{\perp}\times \vec B)}_s(\tau ,\vec
\sigma )\cr}
\form
$$

\noindent where

$$
S_{\mu\nu}={1\over 2}\int d^3\sigma \, \sigma^v(b_{v\mu}(\tau )\rho_{\nu}
(\tau ,\vec \sigma )-b_{v\nu}(\tau )\rho_{\mu}(\tau ,\vec \sigma )).
\form
$$

Therefore we are left with the ten independent degrees of freedom contained in
the variables $x^{\mu}(\tau )$, $b^{\mu}_r(\tau )$ for the hyperplanes, with
conjugate variables $P^{\mu}$ and $S^{\mu\nu}$. To get the Wigner foliation,
i.e. the one with the hyperplanes orthogonal to the total momentum [$l^{\mu}
(\sigma )\sim P^{\mu}$], we add the following six independent
gauge-fixing constraints

$$
b^{\mu}_r(\tau )-L^{\mu}{}_r(P,{\buildrel \circ \over P})\approx 0\quad\quad
\Rightarrow \,\,\, l^{\mu}(\tau )\approx { {P^{\mu}}\over {\eta \sqrt {P^2}} }
\form
$$

\noindent where [14]

$$
L^{\mu}{}_{\nu}(P,{\buildrel \circ \over P})=\epsilon^{\mu}{}_{\nu}(P/\eta
\sqrt {P^2})=\eta^{\mu}_{\nu}+2{ {P^{\mu}{\buildrel \circ \over P}_{\nu}}\over
{P^2} }-{ {(P^{\mu}+{\buildrel \circ \over P}^{\mu})(P_{\nu}+
{\buildrel \circ \over P}_{\nu})}\over {(P+{\buildrel \circ \over P},
{\buildrel \circ \over P})} }
\form
$$

\noindent is the standard Wigner boost for $P^2 > 0$, $P^{\mu}=L^{\mu}{}_{\nu}
(P,{\buildrel \circ \over P}){\buildrel \circ \over P}^{\nu}$, ${\buildrel
\circ \over P}^{\mu}=(\eta \sqrt {P^2};\vec 0)$. Now the indices "r" become
spin-1 Wigner indices and transform under the Wigner rotations induced by
Lorentz transformations [14]. Therefore, Dirac's observables ${\vec A}_{\perp}
$, ${\vec \pi}_{\perp}$ have Wigner covariance already at this level: the
gauge-
fixings (10-34) is equivalent to boost at rest the absolute Dirac's observables
before finding a decomposition in center-of-mass and relative variables. Here
we see at work the globality associated with the implementation of the
Poincar\'e group (remember the problems of Ref.[12b]).

The time constancy of this new gauge-fixing gives

$$\eqalign{
{d\over {d\tau} }&(b^{\mu}_r(\tau )-L^{\mu}{}_r(P,{\buildrel \circ \over P}))=
\lbrace b^{\mu}_r(\tau )-L^{\mu}{}_r(P,{\buildrel \circ \over P}),
H_{\perp D}\rbrace =\cr
&={\tilde \lambda}^{\alpha\beta}\lbrace b^{\mu}_r(\tau ),S_{\alpha\beta}\rbrace
={\tilde \lambda}^{\alpha\beta}(\tau )(\eta^{\mu}_{\alpha}b_{r\beta}(\tau )-
\eta^{\mu}_{\beta}b_{r\alpha}(\tau ))=2{\tilde \lambda}^{\mu\beta}(\tau )
b_{r\beta}(\tau )\approx 0;\cr}
\form
$$

\noindent so that ${\tilde \lambda}^{\alpha\beta}(\tau )=0$ and one has

$$\eqalign{
H_{\perp D}&\equiv {\tilde \lambda}(\tau ){\hat \chi}_{\mu}(\tau ),\cr
&{}\cr
{\hat \chi}_{\mu}(\tau )&={ {P_{\mu}}\over {\eta \sqrt {P^2}} }[\eta \sqrt
{P^2}-{1\over 2}\int d^3\sigma \, ({\vec \pi}^2_{\perp}+{\vec B}^2)(\tau ,
\vec \sigma )]+\cr
&+\epsilon_{r\mu}(P/\eta \sqrt {P^2})\int d^3\sigma \, {({\vec \pi}
_{\perp}\times \vec B)}_r(\tau ,\vec \sigma )\approx 0\cr}
\form
$$

\noindent and the only left degrees of freedom of the hyperplane are $x^{\mu}
(\tau )$, $P^{\mu}$; if we replace them with the variables of Eqs.(10-1), the
four first class constraints become

$$\eqalign{
\chi_{\tau}&=\eta \sqrt {P^2}-V^{\tau}\approx 0,\quad \quad V^{\tau}={1\over 2}
\int d^3\sigma \, ({\vec \pi}^2_{\perp}+{\vec B}^2)(\tau ,\vec \sigma )\cr
&\vec V=\int d^3\sigma \, {\vec \pi}_{\perp}(\tau ,\vec \sigma )\times {\vec B}
(\tau ,\vec \sigma )\approx 0,\quad \quad V^{\mu}=(V^{\tau};\vec V)\approx
P^{\mu},\cr}
\form
$$

\noindent where $\vec V\approx 0$ expresses the vanishing of the total
3-momentum of the field configuration, when described using the Wigner
foliation
. $\chi_{\tau}$ is the mass-spectrum constraint and plays the role of
Hamiltonian
. To eliminate completely the degrees of freedom of the hyperplane, one should
add four new gauge-fixing constraints, whose natural forms are

$$\eqalign{
&{ {P\cdot x}\over {\eta \sqrt {P^2}} }-Z^{\tau}\approx 0\cr
&\vec z-\vec Z\approx 0.\cr}
\form
$$

\noindent Here $(Z^{\tau};\vec Z)$ should be those functionals of the field
configuration which play the role of the non-covariant $(P\cdot x/\eta \sqrt
{P^2};\vec z)$ associated with the hyperplane. Therefore, we have indirectly
shown that there must exist a center-of-mass decomposition also for classical
gauge field theory and that it is the lacking ingredient to get the final,
still manifestly Wigner covariant, either Lagrangian or Hamiltonian
description of its Dirac's observables on which the subsequent discussion is
based.

Even if the canonical transformation from the basis $A_{\mu}(\vec x,x^o)$,
$\pi_{\mu}(\vec x,x^o)$ to a center-of-mass basis $X^{\mu}$, $P^{\mu}$,
$a^{\mu}
(\vec x,x^o)$, $q^{\mu}(\vec x,x^o)$ has still to be found, we got a
unification of the description of particles, strings and classical gauge fields
in special relativity. All extended special relativistic systems with first
class constraints show a breaking of Lorentz covariance in the description of
the center-of-mass, which is universally described by the variables of Eqs.
(10-1); this is due to the fact that usually the mass spectrum of the system is
given by a first class constraint solved in $\eta \sqrt {P^2}$ and this
variable
is in the basis (10-1); therefore, for each branch n of the mass spectrum,
$\eta \sqrt {P^2}-g_n(...)\approx 0$, one could pass from the basis (10-1) to a
Shanmugadhasan basis in which $\eta \sqrt {P^2}$ is replaced by $\eta \sqrt
{P^2}-g_n(..)$.

The problems of the relativistic center-of-mass is very complex and many
options appear in the literature. Following Ref.[97] for the case $P^2 >0$,
$P^o > 0$, there are three definitions of center-of-mass which appear relevant
for our discussion (the three definitions coincide in the rest frame): \hfill
\break
i) the canonical noncovariant position $\vec z$ of Eqs. (10-1) (actually
${\vec z}^{'}=\vec z/
\sqrt {P^2}$), also called center of spin (see also Refs.[98,99]), which is a
classical analogue of the generalized Foldy-Wouthuysen mean position operator
[100], i.e. the generalization of the Newton-Wigner position operator [101]
in presence of spin;\hfill\break
ii) the M\"oller center-of-mass $\vec R$ [102], which, for a system of point
particles, corresponds to the standard non-relativistic definition with the
Newtonian masses replaced by dynamical masses (it is a kind of center of energy
): it is neither covariant nor canonical ($\lbrace R^i,R^j\rbrace \not= 0$);
\hfill\break
iii) the Fokker center of inertia $\vec Q$ [103,98], which is defined as
${\vec Q}^{(o)}={\vec R}^{(o)}={\vec z}^{(o)}$ in the rest frame and then
Lorentz transformed to an arbitrary frame: it is covariant ($Q^{\mu}$ is a
four-vector) but not canonical ($\lbrace Q^i,Q^j\rbrace \not= 0$). It can be
shown [97] that ${\dot {\vec z}}^{'}={\dot {\vec R}}={\dot {\vec Q}}=c^2\vec P/
P^o$ (we have reintroduced the factors of c).

Therefore, $\vec Q$ is the only space vector which is associated with an
invariantly defined world-line; instead, the space-vectors ${\vec z}^{'}$ and
$\vec R$ define world-lines whose objective space-time location depends on the
observer, i.e. on the chosen reference frame. Since $\vec Q=\vec R+{ {\vec S
\times \vec P}\over {\sqrt {P^2}P^o} }={\vec z}^{'}+{ {\vec S\times \vec P}
\over {\sqrt {P^2}(\sqrt {P^2}c^2+P^o)} }$
(so that ${\vec z}^{'}=\vec R+{ {\vec S\times
\vec P}\over {P^0(\sqrt {P^2}c^2+P^o)} }={ {P^o\vec R+\sqrt {P^2}c^2\vec Q}
\over {\sqrt {P^2}c^2+P^o} }$) [97], it can be shown that, in a given reference
frame, the world-lines associated with $\vec R$ (and also those associated with
${\vec z}^{'}$)
by the whole set of observers (i.e. by changing reference frame with
all possible Lorentz transformations) fill a "world-tube" having the $\vec Q$
world-line as a central axis and a radius (it measures a distance orthogonal to
${\dot {\vec Q}}=c^2\vec P/P^o$; here the spin is the Thomas spin, i.e. the
classical relative angular momentum in the rest frame)

$$
\rho ={ {|{\hat {\vec S}}|}\over {\sqrt {P^2}c} }={ {\sqrt {-W^2}}\over
{cP^2}};
\form
$$

\noindent every time an extended relativistic system belongs to an irreducible
Poincar\'e representation its radius is determined by the associated Poincar\'e
Casimirs. This value is exactly the same which one naively obtains for the
minimal radius of a relativistic matter bulk of mass $\sqrt {P^2}$ and
intrinsic
angular momentum S, if the linear velocity of the peripheral points has not to
exceed the velocity of light; as shown firstly by M\"oller [102], if an
extended
relativistic body has the classical energy density everywhere positive
definite,
then its spatial extension must be wider than the diameter of the world-tube;
therefore, we have some kind of non-locality as a spin effect.
Let us remark that the case $\vec S=0$ (like those with $P^2=0$ or $P^{\mu}=0$)
has to be analyzed separately by adding $\vec S\approx 0$ to the original
constraints.

The following remarks are relevant at this point: i) in all relativistic
extended systems with first class constraints (a their subset is always used
to eliminate the temporal degrees of freedom, either relative times or
temporal components of gauge potentials), all final Dirac's
observables describing
absolute positions will inherit the non-covariance of the canonical center-of-
mass $\vec z$: only Dirac's observables describing relative degrees of freedom
will have well defined Wigner covariance (if we use variables from the point
of view of the Wigner foliation; the absolute, Wigner covariant, Dirac's
observables ${\vec A}_{\perp}$, ${\vec \pi}_{\perp}$ are not natural);
ii) even if $\vec z$ is a Dirac's
observable, it cannot be classically measurable if we adopt a covariant
definition of measurability, i.e. no measure can depend on the reference
frame in a non-covariant way; this implies that we cannot localize the three
degrees of freedom of the center-of-mass $\vec z$ inside the world-tube (it
would be interesting to have a classical relativistic theory of measurement
taking into account these non-local spin effects and the correct counting of
the
independent Cauchy data, i.e. Dirac's observables, of the system; see also
later on); iii) the classical non-testability of the interior of the world-tube
would give an answer to the standard criticism of the classical picture based
on the quantum effect of pair production: actually this effect
happens inside the
world-tube and both pair production and the radius $\rho$ are effects of the
Lorentz signature of Minkowski space-time, i.e. of the existence of the light
cone; iv) the M\"oller argument on the energy density being definite positive
only with extensions larger than $\rho$, seems to imply that the world-tube is
a remnant in flat Minkowski space-time of the energy conditions of general
relativity.

Therefore, if we have a field configuration in an irreducible Poincar\'e
representation with $P^2 > 0$ and $W^2=-P^2{\hat {\vec S}}^2\not= 0$, and we
try
to quantize its center-of-mass variable ${\vec z}^{'}=\vec z/\sqrt {P^2}$,
$\vec P=\sqrt {P^2}\vec k$, a natural generalization of Heisemberg
indetermination relations taking into account the veto for a relativistic
localization inside the world-tube is ($|{\hat {\vec S}}|\rightarrow \hbar  s$
with s the maximal eigenvalue)

$$
\triangle z^{{'}\, i}\geq { {\hbar }\over {\triangle P^i} }\geq { {\hbar
s}\over
{\sqrt {P^2}c} }\quad \quad \Rightarrow \,\,\, \triangle P\leq {1\over s}
\sqrt {P^2}c.
\form
$$

This implies non-localizability inside the Compton wave length of the field
configuration and a possible definition of a ultraviolet cutoff $\Lambda =
{1\over s}\sqrt {P^2}c$ of the type looked for by Dirac and Yukawa, as a spin
effect induced by the Lorentz signature. The sectors $P^2 > 0$, ${\hat {\vec
S}}=0$ and $P^2=0$ have to be treated separately, as already said; often only
a finite number of degrees of freedom is present in these sectors.

At the quantum level a related open problem is whether ${\vec z}^{'}$, and
every
absolute Dirac's observable position operator, can be assumed to be
self-adjoint like the four-momentum and the angular momentum (no problem seems
to arise about relative position operators). The previous discussion seems to
suggest that to avoid non-covariant statements one should allow only wave
packets
constant inside the world-tube (all the pseudo-world-lines associated with all
possible reference frames would be put on the same level) and this would imply
non-self-adjointness of ${\vec z}^{'}$.
A further problem comes from Hegerfeldt's
theorems on the quantization of the Newton-Wigner position, see Ref.[104] and
the references quoted therein. There are two basic options:
i) either ${\vec z}^{'}$
is a self-adjoint operator (i.e. one has good localization properties) and then
one finds a violation of Einstein causality in the sense that
$\alpha )$ compact support-, $\beta )$ Gaussian-, $\gamma )$ exponential- wave
packets will spread at later times with velocity higher than c (only at the
level of wave packets with power tails the possibility appears of avoiding the
violation);
ii) or ${\vec z}^{'}$ is not a self-adjoint operator, so that Einstein
causality is preserved due to bad localization properties (power tails): this
second option is preferred by Hegerfeldt, who says that these problems should
be experimentally verifiable. Again one faces the lack of a genuine
either classical or quantum
relativistic theory of position measurements (which after the contraction
$c\rightarrow \infty$ would produce the nonrelativistic theory); after all in
high energy experiments the indetermination in the positions of the detected
particles on the space-like surface of the detector allows only the
determination
of world-tubes which can be traced backward towards the interaction region.

All these problems form a big unsolved puzzle. Do we have really to consider
only
wave packets constant inside the world-tube and with suitable power tails? If
yes, the position operators would behave differently from the momentum, energy
and angular momentum operators, because most of the mathematical wave
packets in the Hilbert space would be physically forbidden (there are
similarities with the situation of Glauber states which are concentrated on the
boundary of the space of coherent states). Whichever is the physical solution
to these problems (it has also been suggested that non-commutative geometry and
maybe the Manin quantum plane have to be introduced to treat these problems),
it is
connected with the definition of the ultraviolet cutoff and with the
foundations of the path integral approach, which heavily uses non-relativistic
position operator concepts in its definition.

\bigskip
\vfill\eject

\bigskip\noindent
{\bf {11. Conclusions.}}
\newcount \nfor

\def \form {\global \advance \nfor by 1 \eqno (11.\the\nfor)}
\bigskip

We have found global
Dirac's observables of classical YM theory with Grassmann-
valued fermions and formulated a non-Lorentz-covariant physical action and
Hamiltonian, which do not present singularities in ${\sl g}$
and whose associated
equations of motion could be treated perturbatively. The basic observables
are transverse due to the existence of the BRST operator and satisfy Poisson
brackets with a transversality projector like in the Abelian case.
The main feature of both the action and the Hamiltonian is
their non-locality and non-polynomiality; they contain the explicit
realization of the Mitter-Viallet-Babelon abstract metric [21b-i] in a suitable
functional space in which the Gribov ambiguity is absent.
One important technical point would be to find the right class of functional
spaces (a restriction of the quoted weigthed Sobolev spaces) so that all
needed boundary conditions on the gauge potentials and the gauge
transformations (like the absence of improper non-rigid ones, i.e.
${\bar {\cal G}}^o_{\infty}={\bar {\cal G}}_{\infty}^{o(P)}$) are satisfied.
For instance the existence of global Dirac's observables, at least at our
naive level, seems to indicate that ${\bar {\cal G}}^o_{\infty}={\bar {\cal G}}
_{\infty}^{o(P)}$ behaves like a closed local (analytic) Lie subgroup of a
certain Hilbert-Lie group ${\bar {\cal G}}$ determined by the chosen
functional space. This choice will also determine properties of the Green
function (6-21), like for instance which kind of prescription is needed for
having Eqs.(6-23) satisfying the Poisson brackets (6-26) and which
asymptotic behaviour has to be expected for ${\vec \zeta}^{(A)}_{ab}(\vec x,
\vec y;x^o)$. Also, once the correct functional space has been identified, the
problems of the mutual relation of the Gribov ambiguity and of the
stability subgroups ${\cal G}^{\cal A}$ and ${\cal G}^{\Omega}$ have to be
re-analized in detail in more general spaces, also to try to identify the
physical meaning, if any, of connections admitting these stability subgroups.

Since the Green function (6-23) privileges the geodesics of flat space, in the
dynamics with Dirac's observables geodesic triangles should be important,
confirming the suggestion of Ref.[105] of approximating the loop space
connected with Wilson loops with such triangles. In connection with these
problems, it would be interesting to be able to riformulate the dynamics with
Dirac's observables on the lattice.

In any case, we  now have a formulation of the (pseudo)classical basis of QED
and QCD with fermions.  Dirac's observables associated with the non-Abelian
charges, ${\check Q}^T_a$ [non-vanishing for $\epsilon \not= 0$ in Eqs.(2-40)
ii)] have been found: they satisfy the correct algebra and rotate global
Dirac's observables with color indices. A classical implementation of the
concept of superselection rule requires the selection of the subspace of the
space of all functions of Dirac's observables consisting of those functions
which have zero Poisson bracket with all ${\check Q}^T_a$: this subspace is
labelled by the values of the Casimirs, like $\sum_a({\check Q}^T_a){}^2$,
and contains only color-scalar functions of Dirac's observables, in
particular only even functions of the Grassmann-valued "unobservable" fields
${\check \psi}_{a\alpha}(\vec x,x^o)$. Probably this superselection subspace
does not admit a symplectic structure, so that there is not a complete
canonical basis of the superselected observables in contrast to the
situation with the original observables. Therefore it is not clear how to
quantize the classical superselection sector: maybe one is forced to quantize
Dirac's observables and then to impose superselection rules at the quantum
level. Since no gauge degree of freedom is left, one could start the Haag-
Kastler program of "local observables" at this stage, or even better after
having covariantized the description to take into account the non-localities
induced by the implementation of the Poincar\'e group.

Also the topological charge has been expressed in terms
of Dirac's observables and a discussion of the role of the winding number
and of the center of the gauge group has been given. The procedure for
finding Dirac's observables requires a trivial principal bundle over a
simply connected base manifold with a semisimple, compact, connected, simply
connected structure group: does it imply that global Dirac's observables
do not exist when one has a non-trivial principal bundle? For a non simply
connected base manifold it is suggested that harmonic one-forms could be
the classical basis in the Hamiltonian formalism of the Bohm-Aharonov phase.

At the classical level a formulation of the problem of color confinement
could be that, notwithstanding one starts with ${\check Q}^T_a\not= 0$, one
requires that the only relevant dynamical sector must be defined by the
conditions ${\check Q}^T_a=0$, for instance added to the Lagrangian with
Lagrange multipliers. Otherwise, one has to study the bound-state equations
at the quantum level to see whether ${\check Q}^T_a=0$ is emerging
dynamically for the bound states (maybe in the form of charge screening
like in Ref.[59]).

Work is in
progress to extend these results to the electroweak sector of the standard
model, in which some of Gauss' laws have to be treated differently due to
the Higgs mechanism: they have to be solved with respect to Higgs momenta and
not with respect to longitudinal gauge potentials. It would be interesting to
have  the solution for Dirac's observables in the electroweak sector
to compare flavour charges of bound states to color ones.

A still open problem both for the Nambu string and for YM theory is to find a
further canonical transformation to a canonical basis adapted to the Thomas
spin (and therefore to the Poincar\'e Casimir $W^2$) to extract the
independent transverse degrees of freedom.
Also the center-of-mass decomposition of classical field theory, whose
existence has been shown in Section 10 studying the covariantization
problem, has to be found. Moreover, the classical background of the
infrared singularities and its connection with the exceptional Poincar\'e
orbit $P^{\mu}=0$ has to be explored.

However the main problem is how to quantize and regularize a non-local and
non-polynomial field theory. Our approach to covariantization suggests an
ultraviolet cutoff in the spirit of Dirac and Yukawa ideas, but not yet how
to utilize it. In the QED case, the problem is connected with the one of
finding a regularization of the Coulomb gauge (see Ref.[45a] for some comments
on this problem) both in the canonical and in the path integral
quantizations.

In any case there is another unsolved preliminary problem: quantum field theory
does not have a particle interpretation; this is forced on it by means of
the asymptotic Fock spaces, which are build from tensor products of free
one-particle states, each one belonging to a Hilbert space with an
independent scalar product; but in this way we allow the possibility that
one "in" particle be in the absolute future of another "in" particle: the one-
particle Hilbert spaces are completely unrelated and this is the basic origin
of the spurious solutions of the Bethe-Salpeter equations describing
excitations
in the relative energy of the two particles. Instead in the description of
relativistic two-body systems and of the Nambu string, always there are enough
first class constraints to eliminate all the existing relative energies,
so that all relative times become gauge variables in the sense of general
relativity: they describe the freedom of the observer to describe n-particle
systems with arbitrary relative-time delays in the Cauchy data, when no
pair of particles is relatively time-like.
Therefore, it seems that, if we wish to have all the tools consistent with
special relativity and with the absence of physical degrees of freedom
connected with relative energies, one has to define Fock spaces on space-like
hypersurfaces (for instance on the space-like
hyperplanes orthogonal to $P^{\mu}$
of the Wigner foliation when $P^2 > 0$) with suitable scalar products,
extensions of the non-local ones of Ref.[14] for two scalar relativistic
particles: this would be a concrete realization of the Tomonaga-Schwinger
formulation [106] of field theory on space-like hypersurfaces with the
description of the asymptotic states relativistically consistent with
the Cauchy data formulation of Dirac's observables of a relativistic theory
with first class constraints. Then one should reformulate asymptotic conditions
and reduction formulas in this non-local setting and try to develop
perturbative expansions in which relative energies do not propagate and
the previous ultraviolet cutoff could be consistently used.

All relevant physical systems from Newtonian particles and Newtonian gravity
till all formulations of general relativity are described by singular
Lagrangians and by a Hamiltonian formulation with first class constraints;
this fact singles out the presymplectic approach as a unifying mathematical
tool and emphasizes the role of Dirac's observables as the independent Cauchy
data of every theory. The limitation of the manifestly covariant Lagrangian
approach is that it does not have a natural method for finding a complete
basis of gauge invariant observables; since the search of such a basis of
Dirac's observables is based on the Shanmugadhasan canonical transformation
in the Hamiltonian approach, to define the analoguous procedure at the
Lagrangian level one needs to pull-back canonical transformations with the
inverse Legendre transformation and then to study Lie-B\"acklund
transformations (i.e. velocity dependent Lagrangian transformations) in a
reformulation of the singular Lagrangian theory in the infinite jet
bundle; but this is a largely unexplored area also for regular Lagrangians.
The quantization of presymplectic theories reformulated
in terms of Dirac's observables is far for being trivial due to the
non-local and non-polynomial character of the resulting theories, as a
consequence of the Lorentz signature and of the global implementation of the
Poincar\'e group. In particular one has to invent some
kind of regularization independent from power counting but connected in some
way with the standard methods for the renormalizable theories as QED and YM;
the natural starting point should be the regularization of the Coulomb gauge
in electrodynamics with fermions, with its non-local four fermion
self-interaction and with the associated modified temporal gauge.
All physical theories are now presented in a form similar to that of
general relativity, for which the standard regularizations do not work, so
that any new result would be extremely important. Finally, the techniques
developed in this paper should be useful for finding the solution of the
constraints of tetrad gravity.

\bigskip
I would like to thank Dr.M.Mintchev, who let me know the paper of Dirac
in Ref.[17], which inspired this work, and Profs. G.Dell'Antonio and
C.Reina for fruitful discussions.
\vfill\eject\noindent

{{\bf APPENDIX A.: Some Results on Lie Group Manifolds.}}
\newcount \nfor

\def \form {\global \advance \nfor by 1 \eqno (A.\the\nfor)}
\bigskip

Let us consider [49]
the analytic atlas of the compact, connected, simply connected,
semisimple Lie group G, with compact semisimple Lie algebra $g$, and a
coordinate chart with canonical coordinates of 1st kind for a neighbourhood
$N_I$ of the identity $I\in G$. Since the exponential map exp is a
diffeomorphism of a neighbourhood of $0\in g$ onto a neighbourhood of $I\in G$
containing $N_I$, these coordinates are defined by considering the
one-parameter
subgroup $\gamma_{\eta}(s)=exp(s{\eta}_at^a)$ of G, whose tangent vector at I
is
$\eta =\eta_at^a\in g$; then the element $\gamma_{\eta}(1)=exp(\eta_at^a)\in G$
reached at s=1 is given coordinates $\lbrace \eta_a\rbrace$.

Given a basis of left invariant vector fields $Y_a$, $a=1,..,dim\, g$, and the
dual basis of left invariant (or Maurer-Cartan) 1-forms $\theta_a$ on G
($i_{Y_a}\theta_b=\delta_{ab}$), one has

$$
\eqalign{
[Y_a,Y_b]&=c_{abc}Y_c,\quad \quad Y_a{|}_I=t^a\in g\cr
d\theta_a&=-{1\over 2}c_{abc}\theta_b\wedge \theta_c,\quad \quad \theta_a{|}_I=
t_a\in g^{*}\cr}
\form
$$

\noindent where the second set of equations is called the Maurer-Cartan
structure equations and where $g^{*}$ is the dual of the Lie algebra $g$
(one has $TG\sim g$ and $T^{*}G\sim g^{*}$). In $N_I$ one has

$$
\eqalign{
Y_a&=B_{ba}(\eta ){ {\partial} \over {\partial \eta_b} },\quad \quad
A(\eta )=B^{-1}(\eta )\cr
\theta_a&=A_{ab}(\eta )d\eta_b,\quad \quad A(0)=B(0)=1\cr}
\form
$$

\noindent and Eqs.(A-1) become

$$
\eqalign{
{ {\partial A_{ac}(\eta )}\over {\partial \eta_b} }-{ {\partial A_{ab}(\eta )}
\over {\partial \eta_c} }&=-c_{auv}A_{ub}(\eta )A_{vc}(\eta )\cr
Y_bB_{ac}(\eta )-Y_cB_{ab}(\eta )&
=B_{ub}{ {\partial B_{ac}(\eta )}\over {\partial\eta_u} }-B_{uc}(\eta )
{ {\partial B_{ab}(\eta )}\over {\partial \eta_u} }=B_{au}(\eta )c_{ubc}\cr}
\form
$$

Since $\eta_a$ are canonical coordinates of first kind one has (this can be
taken as a definition of these coordinates)

$$A_{ab}(\eta )\eta_b=\eta_a
\form
$$

\noindent so that

$$A(\eta )={ {e^{T\eta}-1}\over {T\eta} },\quad with\, {(T\eta )}_{ab}={({\hat
T}^c)}_{ab}\eta_c=c_{abc}\eta_c.
\form
$$

The canonical 1-form on G is defined as

$$\omega_G=\theta_at^a\quad\quad  [=A_{ab}(\eta )d\eta_bt^a\,\, in\,  N_I];
\form
$$

\noindent one can also write $\omega_G=a^{-1}(\eta )d_Ga(\eta )$
for some $a(\eta )\in G$, where $d_G$ is the exterior derivative on G (in $N_I$
one has $d_G=d\eta_a\partial /\partial \eta_a=d\eta_a\partial_a$). The Maurer-
Cartan structure equations can be rewritten in the form $d_G\omega_G+{1\over 2}
[\omega_G,\omega_G]=d_G\omega_G+{1\over 2}[t^a,t^b]\theta_a\wedge \theta_b=0$.
These equations say that the 1-forms $\theta_a$ are not integrable on G (their
integral along a line joining two elements of G depends on the line); however,
in the neighbourhood $N_I$ there is the preferred line $\gamma_{\eta}(s)$
defining the canonical coordinates of 1st kind and then one can define

$$
\eqalign{
\omega^{\gamma_{\eta}}_a(\eta (s))&=\int_{(\gamma_{\eta})\,
I}^{\gamma_{\eta}(s)
}\theta_a{|}_{\gamma_{\eta}}=\cr
&=\int_{(\gamma_{\eta})\, 0}^{\eta (s)}A_{ab}(\bar \eta )d{\bar \eta}_b=\int
_{(\gamma_{\eta})\, 0}^sA_{ab}(\eta (s)){ {d\eta_b(s)}\over {ds} }ds\cr}
\form
$$

\noindent if $\eta_a(s)$ are the coordinates of the points of
$\gamma_{\eta}(s)$
(with $\eta_a(1)=\eta_a$). If $d_{\gamma_{\eta}}=$\break
$ds { {d\eta_a(s)}\over {ds}}
{ {\partial}\over {\partial \eta_a}}{|}{\eta =\eta (s)}$ is the directional
derivative along $\gamma_{\eta }$ (the restriction of $d_G$ to $\gamma_{\eta}
(s)$), one has

$$d_{\gamma_{\eta}}\omega_a^{\gamma_{\eta}}(\eta (s))=\theta_a(\eta (s)),
\form$$

\noindent and the restriction of the Maurer-Cartan equations to $\gamma_{\eta}
(s)$ becomes

$$d_{\gamma_{\eta}}\theta_a(\eta (s))=0\quad \Rightarrow \quad d^2_{\gamma
_{\eta}}=0.
\form
$$

The analytic atlas ${\cal V}$ for the group manifold G is built starting from
the neighbourhood $N_I$ of I with canonical coordinates of 1st kind by left
multiplication by elements of G: ${\cal V}=\cup_{a\in G}\lbrace a\cdot N_I
\rbrace$.

\vfill\eject

\bigskip\noindent
{{\bf APPENDIX B. Connection 1-forms on Principal G-bundles.}}
\newcount \nfor

\def \form {\global \advance \nfor by 1 \eqno (B.\the\nfor)}
\bigskip

The trivial principal G-bundle $P^4=M^4\times G$ ($P^3=R^3\times G$) will be
coordinatized by using for the group manifold G an analytic atlas built from a
chart $N_I$ around $I\in G$ with canonical coordinates of 1st kind as in
Appendix A; if p=(x,a) [$p\in P^{4\, or\, 3}$, $x\in M^4\, or\, R^3$, $a\in G$]
is a point in a tubolar neighbourhood ${\cal N}_I$ of the identity cross
section $\sigma_I$ such that its restriction to each fiber G coincides with
$N_I$, one has $p=(x^{\mu},\eta_a)$.

The $g$-valued connection 1-form $\omega^{\cal A}$ [24a,b,85]
on $P^4=M^4\times G$
[$P^3=R^3\times G$], defining a connection ${\cal A}$ (i.e. the distribution
of spaces $H^{\cal A}_p$ of ${\cal A}$-horizontal vectors in each point
$p\in P^4$ or $p\in P^3$), is given in this tubolar neighbourhood of $\sigma_I$
by

$$
\eqalign{
\omega^{\cal A}&=\omega^{\cal A}_at^a=(\theta_a-{\cal A}_a)t^a=\omega_G-{\cal
A}_{a\mu}(x,\eta )dx^{\mu}t^a\quad on\, P^4\cr
&[=\omega_G+{\vec {\cal A}}_a(\vec x,x^o,\eta )\cdot d\vec xt^a\quad on\,
P^3]\cr}
\form
$$

\noindent where $\omega_G=\theta_at^a$ is the canonical 1-form on G defined in
Appendix A. With each left invariant vector field $Y_a$ on G such that $Y_a
{|}_I=t^a\in g$ (see Eqs.(A-1)), one can associate a vertical vector field $X
_{(t^a)}$ on $P^4$ [$P^3$] (i.e. along the fibers; its projection on the base
manifold $M^4$ [$R^3$] vanishes), called a fundamental vector field, such that
its restriction to every fiber $G_{(x)}$ over x satisfies $X_{(t^a)}{|}_{G
_{(x)}}=Y_a=B_{ba}(\eta )\partial /\partial \eta_b$. Then one has $\omega^{\cal
A}(X_{(t^a)})=i_{X_{(t^a)}}\omega^{\cal A}=t^a$ and, from the definition of
connection 1-form, $\omega^{\cal A}_{|p}(X_{|p}^{H^{\cal A}})=0$, $p\in P^4$
[$p\in P^3$], for every ${\cal A}$-horizontal vector at p. Another
representation of $\omega^{\cal A}$ on $P^4$ [$P^3$] is $\omega^{\cal A}=a^{-1}
(\eta )d_Pa(\eta )-a^{-1}(\eta ){\tilde {\cal A}}_{a\mu}(x)dx^{\mu}t^aa(\eta )$
[$\omega^{\cal A}=a^{-1}(\eta )d_Pa(\eta )+a^{-1}(\eta ){\vec {\tilde {\cal
A}}}_a(\vec x,x^o)\cdot d\vec xt^aa(\eta )$] with $a(\eta )\in G$ and $d_{P^4}=
d+d_G=dx^{\mu}\partial /\partial x^{\mu}+d\eta_a\partial /\partial \eta_a=dx
^{\mu}\partial_{\mu}+d\eta_a\partial_a$ [$d_{P^3}=d\vec x\cdot \vec \partial +
d\eta_a\partial_a$] the exterior derivative on $P^4$ [$P^3$].

The $g$-valued curvature 2-form $\Omega^{\cal A}$ associated with $\omega^{\cal
A}$ is (${\cal D}^{\cal A}$ is the exterior covariant differentiation: ${\cal
D}^{\cal A}\omega^{\cal A}_{|p}(X_{|p},Y_{|p})=d_P\omega^{\cal A}_{|p}(X_{|p}
^{H^{\cal A}},Y_{|p}^{H^{\cal A}})$)

$$
\eqalign{
\Omega^{\cal A}&=\Omega^{\cal A}_{a\mu\nu}dx^{\mu}\wedge dx^{\nu}t^a={\cal D}
^{\cal A}\omega^{\cal A}=d_{P^4}\omega^{\cal A}+{1\over 2}[\omega^{\cal A},
\omega^{\cal A}]=\cr
&=\lbrace -{1\over 2}\lbrack \partial_{\mu}{\cal A}_{a\mu}(x,\eta )-\partial
_{\nu}{\cal A}_{a\mu}(x,\eta )+c_{abc}{\cal A}_{b\mu}(x,\eta ){\cal
A}_{c\nu}(x,
\eta )\rbrack dx^{\mu}\wedge dx^{\nu}-\cr
&-\lbrack \partial_b{\cal A}_{a\mu}(x,\eta )+c_{adc}A_{db}(\eta ){\cal
A}_{c\mu}
(x,\eta )\rbrack d\eta_b\wedge dx^{\mu}\rbrace \, t^a\cr}
\form
$$

\noindent and an analogous expression for $P^3$. The ${\cal A}$-horizontability
of the curvature 2-form implies the following property of ${\cal A}_{a\mu}(x,
\eta )$:

$${ {\partial}\over {\partial \eta_b} }{\cal A}_{a\mu}(x,\eta )=-c_{adc}A_{db}
(\eta ){\cal A}_{c\mu}(x,\eta ),
\form
$$

\noindent so that the only non-vanishing components of $\Omega^{\cal A}$ are

$$\Omega^{\cal A}_{a\mu\nu}(x,\eta )=-{1\over 2}\lbrack \partial_{\mu}{\cal A}
_{a\nu}(x,\eta )-\partial_{\nu}{\cal A}_{a\mu}(x,\eta )+c_{abc}{\cal A}_{b\mu}
(x,\eta ){\cal A}_{c\nu}(x,\eta )\rbrack .
\form
$$

Given a global cross section $\sigma :M^4\rightarrow P^4$, whose
coordinatization is $\sigma (x)=$\break
$\lbrace x^{\mu},\eta_a(x)\rbrace$, $x\in M^4$,
one has (analogous formulas hold for $P^3$):

$$\eqalign{
A(x)&=A_{a\mu}(x)dx^{\mu}t^a=\sigma^{*}\omega^{\cal A}=\lbrack \theta_a(\eta )
{|}_{\eta =\eta (x)}-{\cal A}_{a\mu}(x,\eta (x))dx^{\mu}\rbrack \, t^a=\cr
&=\lbrack A_{ab}(\eta (x)){ {\partial \eta_b(x)}\over {\partial x^{\mu}} }-
{\cal A}_{a\mu}(x,\eta (x))\rbrack dx^{\mu}t^a\cr}
\form
$$

$$
\eqalign{
F(x)&=F_{a\mu\nu}(x)dx^{\mu}\wedge dx^{\nu}\, t^a=2\sigma^{*}\Omega^{\cal A}
=\cr
&=\lbrack \partial_{\mu}A_{a\nu}(x)-\partial_{\nu}A_{a\mu}(x)+c_{abc}A_{b\mu}
(x)A_{c\nu}(x)\rbrack dx^{\mu}\wedge dx^{\nu}\, t^a\cr}
\form
$$

\noindent due to Eqs.(B-6) and to the fact that $t^a\theta_a(\eta ){|}_{\eta =
\eta (x)}$ is a flat connection with zero curvature; in this check one uses
Eqs. (B-3) and (A-3) which are valid on each fiber $G_{(x)}$ (see for instance
[85]):

$${ {\partial}\over {\partial \eta_b} }{\cal A}_{a\mu}(x,\eta ){|}_{\eta =\eta
(x)}=-c_{adc}A_{db}(\eta (x)){\cal A}_{c\mu}(x,\eta (x)),
\form
$$

$$
{ {\partial A_{ac}(\eta )}\over {\partial \eta_b} }{|}_{\eta =\eta (x)}
-{ {\partial A_{ab}(\eta )}\over {\partial \eta_c(x)} }{|}_{\eta =\eta (x)}
=-c_{auv}A_{ub}(\eta (x))A_{vc}(\eta (x))
\form
$$

\noindent The analogue of the second line of Eqs.(A-3) is given in Eqs.(7-5).

This last equations can be considered as generalized Maurer-Cartan equations
for the principal G-bundle, taking into account all the fibers simultaneously.
If we multiply them by ${\hat T}^C$ (the generators of $g$ in the adjoint
representation) and we define  a generalized canonical 1-form in this
representation ${\tilde \omega}_G(x)={\hat T}^a\theta_a(x)$ whose form in
${\cal N}_I$ is ${\tilde \omega}_G(x)=H_a(\eta (x))d\eta_a(x)$ with

$$H_a(\eta (x))={\hat T}^bA_{ba}(\eta (x)),
\form
$$

\noindent then the generalized Maurer-Cartan equations can be written as a
zero curvature condition

$${ {\partial H_a(\eta )}\over {\partial \eta_b} }{|}_{\eta =\eta (x)}
-{ {\partial H_b(\eta )}\over {\partial \eta_a)} }{|}_{\eta =\eta (x)}
+[H_a(\eta (x)),H_b(\eta (x))]=0.
\form
$$

The Bianchi identities ${\hat D}^{(A)}{*}F\equiv 0$ follow from ${\cal D}^{\cal
A}\Omega^{\cal A}=d_P\Omega^{\cal A}+[\omega^{\cal A},\omega^{\cal A}]\equiv
0$.

In the application to YM theory Eqs.(B-7), (B-8) are evaluated in the adjoint
representation ($t^a\rightarrow {\hat T}^a$).

\vfill\eject

\bigskip
\newcount \nref

\def \ref {\global \advance \nref by 1
\ifnum \nref<10 \item {$~(\the\nref)~$}
\else \item {$(\the\nref)~$}\fi}

\noindent
\centerline{{\bf REFERENCES}}
\bigskip\noindent

\ref a) P.A.M.Dirac, Can.J.Math. $\underline 2$ (1950),129 ; "Lectures on
     Quantum Mechanics", Belfer Graduate School of Science, Monographs Series,
     Yeshiva University, New York, N.Y., 1964.\hfill\break
     b) J.L.Anderson and P.G.Bergmann, Phys.Rev. $\underline {83}$ (1951),
1018.
     \hfill\break
     P.G.Bergmann and J.Goldberg, Phys.Rev. $\underline {98}$ (1955), 531.
\ref a) E.C.G.Sudarshan and N.Mukunda, "Classical Mechanics: a Modern
     Perspective", Wiley, New York, N.Y., 1974.\hfill\break
     b) A.J.Hanson, T.Regge and C.Teitelboim, "Constrained Hamiltonian
Systems",
     in Contributi del Centro Linceo Interdisciplinare di Scienze Matematiche,
     Fisiche e loro Applicazioni, n.22, Accademia Nazionale dei Lincei,
     Roma, 1975.\hfill\break
     c) K.Sundermeyer, "Constraint Dynamics with Applications to Yang-Mills
     Theory, General Relativity, Classical Spin, Dual String Model", Lecture
     Notes in Physics Vol.169, Springer,Berlin, 1982.\hfill\break
     d) A.Ashtekar, "New Perspectives in Canonical Gravity", Bibliopolis,
     Napoli, 1986; "Lectures on Non-Perturbative Canonical Gravity",
     World Scientific, Singapore,\break 1991.\hfill\break
     e) G.Longhi and L.Lusanna (Eds.), "Constraint's Theory and Relativistic
     Dynamics", Proc.Firenze Workshop 1986, World Scientific, Singapore,
     1987.\hfill\break
     f) D.M.Gitman and I.V.Tyutin, "Quantization of Fields with Constraints",
     Springer, Berlin, 1990.\hfill\break
     g) J.Govaerts, "Hamiltonian Quantization and Constrained Dynamics",
     Leuwen University Press, Leuwen, 1991.\hfill\break
     h) M.Henneaux and C.Teitelboim, "Quantization of Gauge Systems", Princeton
     Univ. Press, Princeton, 1992.
\ref G.Longhi, L.Lusanna and J.M.Pons, J.Math.Phys. $\underline {30}$ (1989),
     1893.\hfill\break
     R.De Pietri, L.Lusanna and M.Pauri, "Generalized Newtonian Gravities
     as 'Gauge' Theories of the Extended Galilei Group", Parma Univ.
     preprint 1994.
\ref A.Lichnerowicz, C.R.Acad.Sci.Paris, Ser. A, $\underline {280}$ (1975),
     523.\hfill\break
     W.Tulczyiew, Symposia Math. $\underline {14}$ (1974), 247.\hfill\break
     J.\`Sniatycki, Ann.Inst.H.Poincar\`e $\underline {20}$ (1984), 365.
     \hfill\break
     M.J.Gotay, J.M.Nester and G.Hinds, J.Math.Phys. $\underline {19}$
     (1978), 2388.\hfill \break
     M.J.Gotay and J.M.Nester, Ann.Inst.Henri Poincar$\grave e$ $\underline
     {A30}$ (1979), 129 and $\underline {A32}$ (1980), 1.\hfill \break
     M.J.Gotay and J.$\grave S$niatycki, Commun.Math.Phys. $\underline
     {82}$ (1981), 377.\hfill \break
     M.J.Gotay, Proc.Am.Math.Soc. $\underline {84}$ (1982), 111; J.Math.
     Phys. $\underline {27}$ (1986), 2051.\hfill \break
     G.Marmo, N.Mukunda and J.Samuel, Riv.Nuovo Cimento $\underline 6$
     (1983), 1.\hfill\break
     M.J.Bergvelt and E.A.De Kerf, Physica $\underline {139A}$ (1986), 101 and
     125.
\ref a) L.Lusanna, Nuovo Cimento $\underline {B52}$ (1979), 141.\hfill\break
     b) L.Lusanna, Phys.Rep. $\underline {185}$ (1990), 1.\hfill\break
     c) L.Lusanna, Riv. Nuovo cimento $\underline {14}$ (1991), 1.\hfill\break
     d) L.Lusanna, J.Math.Phys. $\underline {31}$ (1990), 2126; "Multitemporal
     Relativistic Particle Mechanics: a Gauge Theory without Gauge-Fixings",
     in Proc. IV Marcel Grossmann Meeting on General Relativity (R.Ruffini
Ed.),
     Elsevier, Amsterdam, 1986.\hfill\break
     e) L.Lusanna, J.Math.Phys. $\underline {31}$ (1990), 428.\hfill\break
     f) M.Chaichian, D.Louis Martinez and L.Lusanna, "Dirac's Constrained
     Systems: The Classification of Second-Class Constraints", Helsinki
     Univ. preprint HU-TFT-93-5 (1993), to appear in Annals of Physics.
\ref I.A.Batalin and G.A.Vilkovisky, Nucl.Phys.$\underline {B234}$ (1984),
     106.
\ref S.Shanmugadhasan, J.Math.Phys. $\underline {14}$ (1973), 677.
\ref L.Lusanna, "The Shanmugadhasan Canonical Transformation, Function
     Groups and the Extended Second Noether Theorem", Int.J.Mod.
     Phys. $\underline {A8}$ (1993), 4193.
\ref L.Lusanna, "Classical Observables of Gauge Theories from the
     Multitemporal Approach", talk given at the Conference 'Mathematical
     Aspects of Classical Field Theory', Seattle 1991, in Contemporary
     Mathematics $\underline {132}$ (1992), 531.
\ref L.Lusanna, "Dirac's Observables: from Particles to Strings and Fields",
     talk given at the International Symposium on 'Extended Objects and
     Bound States', Karuizawa 1992, (O.Hara, S.Ishida and S.Naka Eds.),
     World Scientific, Singapore, 1993.
\ref M.Henneaux, Phys.Rep. $\underline {126}$ (1985), 1.\hfill\break
     M.Henneaux and C.Teitelboim, Commun.Math.Phys. $\underline {115}$
     (1988), 213.
\ref a) R.Haag and D.Kastler, J.Math.Phys. $\underline 5$ (1964), 848.\hfill
     \break
     S.Doplicher, in "Ideas and Methods in Quantum and Statistical Physics",
     Vol.2, (S.Albeverio, J.E.Fenstad, H.Holden and T.Lindstrom Eds.),
Cambridge
     Univ.Press, Cambridge, 1990.\hfill\break
     b) R.Haag, "Local Quantum Physics: Fields, Particles, Algebras", Springer,
     Berlin, 1992.
\ref F.Strocchi, "Gauss' Law in Local Quantum Field Theory",in 'Field Theory,
     Quantization and Statistical Physics', (E.Tirapegui Ed.), Reidel,
     Dordrecht, 1981.\hfill\break
     G.Morchio and F.Strocchi, in "Fundamental Problems of Gauge Field Theory",
     (G.\break
     Velo and A.S.Wightman Eds.), NATO ASI 141B, Plenum, New York, 1986.
\ref G.Longhi and L.Lusanna, Phys.Rev. $\underline {D34}$ (1986), 3707.
\ref F.Colomo, G. Longhi and L.Lusanna, Int.J.Mod.Phys. $\underline {A5}$
     (1990), 3347; Mod.Phys.\break Lett. $\underline {A5}$ (1990), 17.
\ref F.Colomo and L.Lusanna, Int.J.Mod.Phys. $\underline {A7}$ (1992),
     1705 and 4107.
\ref P.A.M.Dirac, Can.J.Phys. $\underline {33}$ (1955), 650.
\ref I.Goldberg, Phys.Rev. $\underline {112}$ (1958), 1361; $\underline
     {139B}$ (1965), 1665.\hfill\break
     I.Goldberg and E.Marx, Nuovo Cimento $\underline {LVII B}$ (1968), 485.
     \hfill\break
     E.Marx, Int.J.Theor.Phys. $\underline 3$ (1970), 467 and $\underline
     6$ (1972), 307.\hfill\break
     N.S.Han and V.N.Pervushin, Fortschr.Physik $\underline {37}$ (1989), 611.
     \hfill\break
     L.V.Prokhorov and S.V.Shabanov, Int.J.Mod.Phys. $\underline {A7}$ (1992),
     7815.
\ref S.Mandelstam, Ann.Phys.(N.Y.) $\underline {19}$ (1962), 1.
\ref V.Moncrief, J.Math.Phys. $\underline {20}$ (1979), 579.
\ref a) I.M.Singer, a) Commun.Math.Phys. $\underline {60}$ (1978), 7; b) Phys.
     Scripta $\underline {24}$ (1981), 817.\hfill\break
     b) P.K.Mitter, in "Recent Developments in Gauge Theories",
     NATO School 1979, (G.'t'Hooft, C.Itzykson, A.Jaffe, H.Lehmann, P.K.Mitter,
     I.M.Singer and R.Stora Eds.),
     Plenum, New York,1980.\hfill\break
     c) P.K.Mitter and C.M.Viallet, Commun.Math.Phys. $\underline {79}$ (1981),
     457.\hfill\break
     d) M.Daniel and C.M.Viallet, Phys.Lett. $\underline {76B}$ (1978), 458.
     \hfill\break
     e) O.Babelon and C.M.Viallet, Phys.Lett. $\underline {85B}$ (1979), 246
and
     $\underline {103B}$ (1981), 45;
     Commun.Math.Phys. $\underline {81}$ (1981), 51.\hfill\break
     f) O.Babelon, F.A.Schaposnik and C.M.Viallet, Phys.Lett. $\underline
     {177B}$ (1986), 385.\hfill\break
     g) S.N.Vergeles, Lett.Math.Phys. $\underline 7$ (1983), 399.\hfill\break
     h) C.M.Viallet, in XXII Karpacz Winter School of Theoretical Physics,
     1986.\hfill\break
     i) M.Grabiak, B.M\"uller and W.Greiner, Ann.Phys.(N.Y.) $\underline
     {172}$ (1986), 213.\hfill\break
     l) M.S.Narasimhan and T.R.Ramadas, Commun.Math.Phys. $\underline {67}$
     (1979), 21.
\ref a) C.N.Yang and R.L.Mills, Phys.Rev.$\underline {96}$ (1954), 191.\hfill
     \break
     b) C.N.Yang, Phys.Rev.Lett. $\underline {33}$ (1974), 445.\hfill\break
     T.T.Wu and C.N.Yang, Phys.Rev. $\underline {D12}$ (1975), 3845.
\ref a) R.Jackiw, "Topological Investigations of Quantized Gauge Theories",
     in 'Current Algebra and Anomalies", (S.B.Treiman, R.Jackiw, B.Zumino
     and E.Witten Eds.), World Scientific, Singapore, 1985; it is an updated
     version of the contribution in "Relativity, Groups and Topology II",
     Les Houches 1983, (B.S.DeWitt and R.Stora Eds.), North-Holland,
     Amsterdam, 1984.\hfill\break
     b) R.Jackiw, Rev.Mod.Phys. $\underline {56}$ (1980), 661.\hfill\break
     c) E.S.Abers and B.W.Lee, Phys.Rep.$\underline {9C}$ (1973), 1.
     \hfill\break
     d) N.Christ and T.D.Lee, Phys.Rev. $\underline {D22}$ (1980), 939.\hfill
     \break
     e) G.Mack, Fortschritt der Physik $\underline {29}$ (1981), 135.\hfill
     \break
     f) C.Itzykson and J.B.Zuber, "Quantum Field Theory", McGraw-Hill,
     Singapore, 1987.\hfill\break
     g) J.Leite Lopez, "Gauge Field Theories: An Introduction", Pergamon,
     Oxford, 1981.\hfill\break
     h) K.Huang, "Quarks, Leptons and Gauge Fields", World Scientific,
     Singapore, 1982.\hfill\break
     i) L.O'Raifeartaigh, "Group Structure of Gauge Theories", Cambridge
     Univ.Press, Cambridge, 1986.
\ref a) S.Kobayashi and K.Nomizu, "Foundations of Differential Geometry",
     Interscience, New York, 1963.\hfill\break
     b) M.Daniel and C.M.Viallet, Rev.Mod.Phys. $\underline {52}$ (1980), 175.
     \hfill\break
     c) T.Eguchi, P.B.Gilkey and A.J.Hanson, Phys.Report $\underline {66}$
     (1980), 214.\hfill\break
     d) D.D.Bleecker, "Gauge Theory and Variational Principles",
Addison-Wesley,
     Reading, MA, 1981.\hfill\break
     e) B.Doubrovine, S.Novikov and A.Fomenko, "G\'eom\'etrie contemporaine",
     \'Editions \break MIR, Moscou, traduction francaise 1982.\hfill\break
     f) A.Trautman, 'Differential Geometry for Physicists', Bibliopolis,
     Napoli, 1984.\hfill\break
     g) P.Cotta Ramusino and C.Reina, J.Geom.Phys. $\underline 1$ (1984), 121.
     \hfill\break
     h) B.Boos and D.D.Bleecker, "Topology and Analysis", Springer, Berlin,
     1985.\hfill\break
     i) R.Schmidt, "Infinite Dimensional Hamiltonian Systems", Bibliopolis,
     Napoli, 1987.\hfill\break
     l) R.Abraham, J.E.Marsden and T.Ratiu, "Manifolds, Tensor Analysis and
     Applications", Springer, Berlin, 1988.\hfill\break
     m) E.Binz, J.\'Sniatycki and H.Fischer, "Geometry of Classical Fields",
     North-Holland, Amsterdam, 1988.\hfill\break
     n) K.B.Marathe and G.Martucci, J.Geom.Phys. $\underline 6$ (1989),
1.\hfill
     \break
     o) K.B.Marathe and G.Martucci, "The Mathematical Foundations of Gauge
     Theories", North-Holland, Amsterdam, 1992.\hfill\break
     q) "Encyclopedic Dictionary of Mathematics", (S.Iyanaga and Y.Kawada
Eds.),
     MIT Press, Cambridge, Massachusetts, 1977.
\ref a) S.J.Avis and C.J.Isham, in Recent Developments in Gravitation,
     Carg\`ese 1978, (M.Levy and S.Deser Eds.), Plenum Press, New York N.Y.,
     1979.\hfill\break
     b) C.J.Isham and G.Kunstatter, Phys.Lett. $\underline {102B}$ (1981), 417
     and J.Math.Phys. $\underline {23}$ (1982), 1668.\hfill\break
     c) M.Asorey, J.Math.Phys. $\underline {22}$ (1981), 179.\hfill\break
     M.Asorey and L.J.Boya, Int.J.Theor.Phys. $\underline {18}$ (1979), 295.
     \hfill\break
     D.H.Mayer and K.S.Viswanathan, Commun.Math.Phys. $\underline {67}$ (1979),
     199.
\ref J.\'Sniatycki, Rep.Math.Phys. $\underline {25}$ (1988), 291.\hfill\break
     E.Binz and J.\'Sniatycki, Class.Quantum Gravity $\underline 3$ (1986),
     1191.
\ref L.F.Abbot and S.Deser, Phys.Lett. $\underline {116B}$ (1982), 259.
\ref A.Heil, A.Kersch, N.A.Papadopoulos and B.Reifenh\"auser, Ann.Phys.(N.Y.)
     $\underline{217}$ \break (1992), 173.
\ref a) P.G.Bergmann and E.J.Flaherty jr., J.Math.Phys. $\underline {19}$
     (1978), 212.\hfill\break
     b) P.Forg\'acs and N.S.Manton, Commun.Math.Phys. $\underline {72}$ (1980),
     15.\hfill\break
     c) R.Jackiw, Phys.Rev.Lett. $\underline {41}$ (1978), 1635.\hfill\break
     d) R.Jackiw and N.S.Manton, Ann.Phys.(N.Y.) $\underline {127}$ (1980),
257.
     \hfill\break
     e) M.Molelekoa, J.Math.Phys. $\underline {26}$ (1985), 192.
     and $\underline {27}$ (1986), 746.
\ref a) S.Sciuto, Phys.Reports $\underline {49}$ (1979), 181.\hfill\break
     b) M.Ademollo, E.Napolitano and S.Sciuto,Nucl.Phys.$\underline {B134}$
     (1978), 477.
\ref A.A.Belavin, A.M.Polyakov, A.S.Schwartz and Yu.S.Tyupkin, Phys.Lett.
     $\underline {59B}$ (1975), 85.
\ref C.Teitelboim, D.Villarroel and C.van Weert, Rivista Nuovo Cimento
     $\underline 3$ (1980), 1.
\ref J.Tafel and A.Trautman, J.Math.Phys. $\underline {24}$ (1983), 1087.
\ref R.Sugano, Y.Kagraoka and T.Kimura, Int.J.Mod.Phys. $\underline 7$ (1992),
     61.
\ref R.Benguria, P.Cordero and C.Teitelboim, Nucl.Phys. $\underline {B122}$
     (1977), 61.
\ref P.J.Steinhart, Ann.Phys. (N.Y.) $\underline {128}$ (1980), 425.
\ref a) S.Schlieder, Nuovo Cimento $\underline {63A}$ (1981), 137.\hfill\break
     b) B.D.Bramson, Proc.R.Soc.London $\underline {A341}$ (1975), 463.
\ref T.Regge and C.Teitelboim, Ann.Phys.(N.Y.) $\underline {88}$ (1974), 286;
     Phys.Lett. $\underline {53B}$ (1974), 101.
\ref J.L.Gervais, B.Sakita and S.Wadia, Phys.Lett. $\underline {63B}$ (1976),
     55.\hfill\break
     S.R.Wadia, Phys.Rev.$\underline {D15}$ (1977), 3615.
\ref J.L.Gervais and D.Zwanziger, Phys.Lett. $\underline {94B}$ (1980), 389.
\ref a) A.Ashtekar, in Proc. of the Oregon Conference on 'Mass and Asymptotic
     Behaviour of Space-time', (F.Flaherty and J.IsenbergEds.), Springer,
     Berlin, 1984.\hfill\break
     b) A.Ashtekar and R.O.Hansen, J.Math.Phys. $\underline {19}$ (1978), 1542.
     \hfill\break
     A.Ashtekar and A.Magnon, J.Math.Phys. $\underline {25}$ (1984), 2682.
     \hfill\break
     A.Ashtekar, in 'General Relativity and Gravitation', (A.Held Ed.), Plenum,
     New York, 1980.
\ref J.\'Sniatycki and G.Schwarz, "An Invariance Argument for Confinement",
     Calgary Univ. preprint n.754, 1993.\hfill\break
     J.\'Sniatycki and G.Schwarz, "The Existence and Uniqueness of
     Solutions of Yang-Mills Equations with Bag Boundary Conditions" and
     "Yang-Mills and Dirac Fields in a Bag", Calgary Univ. preprints
     n. 746 and 753, 1993.
\ref T.P.Cheng and L.F.Li, "Gauge Theory of Elementary Particle Physics",
     Oxford Univ.Press, New York,1984.
\ref R.Casalbuoni, Nuovo Cim. $\underline {33A}$ (1976), 115 and 389.\hfill
     \break
     F.A.Berezin and M.S.Marinov, Ann.Phys. (N.Y.) $\underline {104}$ (1977),
     336.\hfill\break
     A.Barducci, R.Casalbuoni and L.Lusanna, Nuovo Cim. $\underline {35A}$
     (1976), 377; Nuovo Cim.Lett. $\underline {19}$ (1977), 581; Nucl.Phys.
     $\underline {B124}$ (1977), 93; Nucl.Phys. $\underline {B180}$ [FS2]
     (1981), 141.
\ref a) F.Strocchi and A.S.Wightman, J.Math.Phys. $\underline {15}$ (1974),
     2198.\hfill\break
     b) F.Strocchi, Phys.Rev. $\underline {D17}$ (1978), 2010; Phys.Lett.
     $\underline {62B}$ (1976), 60.\hfill\break
     c) D.Bucholz, Phys.Lett. $\underline {174B}$ (1986), 331.\hfill\break
     d) J.A.Swieca, Phys.Rev. $\underline {D13}$ (1976), 312.\hfill\break
     F.Strocchi, Commun.Math.Phys. $\underline {56}$ (1977), 57.\hfill\break
     e) G.Morchio and F.Strocchi, Commun.Math.Phys. $\underline {99}$ (1985),
     153; J.Math.Phys. $\underline {28}$ (1987), 622; in "Selected Topics
     in QFT and Mathematical Physics", (J.Niederle and J.Fischer Eds.), World
     Scientific, Singapore, 1990.
\ref a) G.Mack, Fortschritt der Physik $\underline {29}$ (1981), 135.\hfill
     \break
     b) G.Mack, Phys.Lett. $\underline {78B}$ (1978), 263.\hfill\break
\ref a) J.Goldstone and R.Jackiw, Phys.Letters $\underline {74B}$ (1978), 81.
     \hfill\break
     V.Baluni and B.Grossman, Phys.Lett. $\underline {78B}$ (1978), 226.
     \hfill\break
     b) A.G.Izergin, V.E.Korepin, M.A.Semenov-Tyan-Shanskii and L.D.Faddeev,
     Theor.\break Math.Phys. $\underline {38}$ (1979), 1.
\ref a) A.P.Balachandran, "Gauge Symmetries, Topology and Quantization",
     Lectures at the Summer Course on "Low Dimensional Quantum Field Theories
     for Condensed Matter Physicists", ICTP, Trieste, 1992.\hfill\break
     b) A.P.Balachandran, G.Marmo, B.S.Skagerstam and A.Stern, "Classical
     Topology and Quantum States", World Scientific, Singapore, 1991.
\ref a) V.S.Varadarajan, "Lie Groups, Lie Algebras and Their Representations",
     Springer, Berlin, 1984.\hfill\break
     b) S.Helgason, "Differential Geometry and Symmetric Spaces", Academic
     Press, New York, 1962.
\ref P.A.Horv\'athy and J.H.Rawnsley, Phys.Rev. $\underline {D32}$ (1985), 968;
     J.Math.Phys. $\underline {27}$ (1986), 982.
\ref A.M.Polyakov, Phys.Lett. $\underline {59B}$ (1975), 82.\hfill\break
     G.'t Hooft, Phys.Rev.Lett. $\underline {37}$ (1976), 8.\hfill\break
     R.Jackiw and C.Rebbi, Phys.Rev.Lett. $\underline {37}$ (1976), 172.\hfill
     \break
     C.G.Callen, R.F.Dashen and D.J.Gross, Phys.Lett. $\underline {63B}$
     (1976), 34.
\ref C.J.Isham, Phys.Lett. $\underline {106B}$ (1981), 188.
\ref S.Coleman, in "The Whys of Subnuclear Physics", (A.Zichichi Ed.), Plenum
     Press, New York, 1977.
\ref R.J.Crewther, Acta Phys. Austriaca, Suppl.$\underline {19}$ (1978), 47.
\ref R.Bott and L.W.Tu, "Differential Forms in Algebraic Topology", Springer,
     Berlin, 1982.
\ref R.Jackiw, I.Muzinich and C.Rebbi, Phys.Rev.
     $\underline {D17}$ (1978), 1576.
\ref a) J.L.Gervais and B.Sakita, Phys.Rev. $\underline {D16}$ (1977), 3507.
     \hfill\break
     b) K.M.Bitar and S.J.Chang, Phys.Rev. $\underline {D17}$ (1978), 486.
\ref C.Cronstr\"om and J.Mickelsson, J.Math.Phys. $\underline {24}$ (1983),
     2528.
\ref a) J.Polonyi, in "Frontiers in Nonperturbative Field Theory", (Z.
     Horvath, L.Palla and A.Patk\'ov Eds.), World Scientific, Singapore, 1989.
     \hfill\break
     b) J.Polonyi, in "Quark Gluon Plasma", (R.C.Hwa Ed.) World Scientific,
     Singapore, 1990.\hfill\break
     c) K.Johnson, L.Lellouch and J.Polonyi, Nucl.Phys. $\underline {B367}$
     (1991), 675.
\ref a) G.'t Hooft, Nucl.Phys. $\underline {B153}$ (1979), 141;Phys. Scripta
     $\underline {24}$ (1981), 890.\hfill\break
     b) G.'t Hooft, Nucl.Phys. $\underline {B138}$ (1978), 1.\hfill\break
     c) G.'t Hooft, Nucl.Phys. $\underline {B190}$ (1981), 455; Phys.Scripta
     $\underline {25}$ (1982), 133.\hfill\break
     $d_1$) P.van Baal, in "Probabilistic Methods in Quantum Field Theory and
     Quantum Gravity", (P.H.Damgaard, H.H\"uffel and A.Rosenblum Eds.), Plenum
     Press, New York, 1990.\hfill\break
     $d_2$) P.van Baal and J.Koller, Ann.Phys. (N.Y.) $\underline {174}$
(1987),
     299.\hfill\break
     $d_3$) P.van Baal, Commun.Math.Phys. $\underline {85}$ (1982), 529.
\ref G.t'Hooft, Nucl.Phys. $\underline {B79}$ (1974), 276.\hfill\break
     A.Polyakov, Jept Lett. $\underline {20}$ (1974), 194.
\ref E.Witten, Phys.Lett. $\underline {86B}$ (1979), 283.
\ref J.de Siebenthal, Comment.Math.Helv. $\underline {31}$ (1956), 41.
\ref A.P.Balachandran, G.Marmo, B.S.Skagerstam and A.Stern, "Gauge Symmetries
     and Fibre Bundles", Lect.Note Phys. n.188, Springer, Berlin, 1983.
\ref a) M.A.Solov'ev, JEPR Lett. $\underline {44}$ (1986), 366; Theor. Math.
     Phys.$\underline {78}$ (1989), 117.\hfill\break
     b) T.P.Killingback, Commun.Math.Phys. $\underline {100}$ (1985), 267.
     Phys.Lett. $\underline {138B}$ (1984), 87.
\ref a) J.Milnor, in "Relativity, Groups and Topology II", Les Houches 1983,
     (B.S.DeWitt \break
     and R.Stora Eds.), Elsevier, Amsterdam, 1984.\hfill\break
     b) A.Trautman, in "General Relativity and Gravitation", vol.I, (A.Held,
     Ed.), Plenum, New York, 1980.\hfill\break
     c) H.Omori, Proc.Sympos.Pure Math. XV (1970), 11. "Infinite Dimensional
     Lie Transformation Groups", Lecture Notes Math. n.427, Springer, Berlin,
     1974.\hfill\break
     d) R.Schmid, "Diffeomorphism Groups and Physical Systems", Lectures at the
     Int.\break
     Seminar on Diffeomorphism Groups and Physical Systems, Clausthal
     1985, A.\break Sommerfeld Institute for Mathematical Physics.\hfill\break
     e) P.Cotta Ramusino and C.Reina, J.Geom.Phys. $\underline 1$ (1984), 121.
     \hfill\break
     f) G.Valli, J.Geom.Phys. $\underline 4$ (1987), 335. (geodesics in gauge
     groups).\hfill\break
     g) A.Pressley and G.Segal, "Loop Groups", Clarendon, Oxford, 1986.
\ref A.E.Fischer, Commun.Math.Phys.$\underline {113}$ (1987), 231.
\ref a) J.M.Arms, J.E.Marsden and V.Moncrief, Commun.Math.Phys. $\underline
     {78}$ (1981), 455.\hfill\break
     b) J.M.Arms, $\alpha$) J.Math.Phys. $\underline {20}$ (1979), 443.
     $\beta$) Math.Proc.Camb.Phil.Soc. $\underline {90}$ (1981), 361.
     $\gamma$) Acta Phys.Pol. $\underline
     {B17}$ (1986), 499 (review paper).\hfill\break
     c) V.Moncrief, in "Differential Geometric Methods in Mathematical
     Physics", P.L.\break
     Garcia, A.Perez-Rend\'on and J.M.Souriau Eds.), Lecture
     Notes Math. n.835,  p.276, Springer, Berlin, 1980 (review paper); J.Math.
     Phys.$\underline {20}$ (1979), 579.
\ref a) J.M.Souriau, "Structure des syst$\grave e$mes dynamiques", Dunod,
Paris,
     1970.\hfill \break
     b) B.Konstant, "Quantization and Unitary Representations", Lecture Notes
     in Math., Vol.170, Springer, Berlin, 1970.\hfill \break
     c) J.E.Marsden and A.Weinstein, Rep.Math.Phys. $\underline 5$ (1974), 121.
     \hfill\break
     d) M.J.Gotay, J.Isenberg, J.E.Marsden, R.Montgomery, J.$\grave S$niatycki
     and P.B.Yasskin, Momentum Maps and Classical Relativistic Fields: the
     Lagrangian and Hamiltonian Structure of Classical Field Theories with
     Constraints, Berkeley Math.Dept. 1991.
\ref a) F.A.Doria, Commun.Math.Phys. $\underline {79}$ (1981), 435.\hfill\break
     b) F.A.Doria, J.Math.Phys. $\underline {22}$ (1981), 2943.\hfill\break
     c) F.A.Doria and J.Abrah$\tilde a$o, J.Math.Phys. $\underline {19}$
     (1978), 1650.\hfill\break
     d) A.Karlhede, Class.Quantum Grav. $\underline 3$ (1986), L27;
$\underline
     7$ (1990), 449.
\ref H.G.Loos, J.Math.Phys. $\underline 8$ (1967), 2114.
\ref a) W.Kondracki and J.S.Rogulski, "On the Stratification of the Orbit
     Space for the Action of Automorphisms on Connections", Warszawa: preprint
     PAN (1983); "On the Notion of Stratification", Warszawa: preprint PAN
     (1983); "On Conjugacy Classes of Closed Subgroups", Warszawa: preprint
     PAN (1983).\hfill\break
     b) W.Kondracki and P.Sadowski, J.Geom.Phys. $\underline 3$ (1986), 421.
     \hfill\break
     c) V.Berzi and M.Reni, Int.J.Theor.Phys. $\underline {26}$ (1987), 151.
     \hfill\break
     d) M.C.Abbati, R.Cirelli and A.Mania`, J.Geom.Phys. $\underline 6$ (1989),
     537.
\ref V.N.Gribov, Nucl.Phys. $\underline {B139}$ (1978), 1. Materials for the
     XII Winter School of the Leningrad Nuclear Research Institute (1977).
\ref a) A.Heil, A.Kersch, N.A.Papadopoulos, B.Reifenh\"auser, F.Scheck and
     H.Vogel,\break J.Geom.Phys. $\underline 6$ (1989), 237.\hfill\break
     b) A.Heil, A.Kersch, B.Reifenh\"auser and H.Vogel, Eur.J.Phys. $\underline
     9$ (1988), 200.\hfill\break
     c) A.Heil, N.A.Papadopoulos, B.Reifenh\"auser and F.Scheck, Nucl.Phys.
     $\underline {B293}$ (1987), 445.\hfill\break
     d) A.Heil, A.Kersch, N.A.Papadopoulos, B.Reifenh\"auser and F.Scheck,
     J.Geom.Phys. $\underline 7$ (1990), 489.\hfill\break
     e) A.Heil, A.Kersch, N.A.Papadopoulos, B.Reifenh\"auser and F.Scheck,
     Ann.Phys.\break
     (N.Y.) $\underline {200}$ (1990), 206.\hfill\break
\ref M.Asorey and P.K.Mitter, Commun.Math.Phys. $\underline {80}$ (1981),
     43; Phys.Lett. $\underline {153B}$ (1985), 147; "On Geometry, Topology and
     $\theta$-sectors in a regularized Quantum Yang-Mills Theory", Cern
     preprint TH.3424, 1982 (unpublished); Ann.Inst.H.Poincar\'e $\underline
     {45}$ (1986), 61.\hfill\break
     M.Asorey and F.Falceto, Phys.Lett. $\underline {206B}$ (1988), 485; Ann.
     Phys. (N.Y.) $\underline {196}$ (1989), 209; Nucl.Phys. $\underline
     {B327}$ (1989), 427.\hfill\break
     L.D.Faddeev and S.L.Shatasvili, Theor. Math. Phys. $\underline {60}$
     (1984), 770; Phys.Lett. $\underline {167B}$ (1986), 225.\hfill\break
     R.Jackiw, Phys.Rev.Lett. $\underline {54}$ (1985), 159; in "Current
     Algebra and Anomalies" (S.B.\break
     Treiman, R.Jackiw, B.Zumino and E.Witten
     Eds.), World Scientific, Singapore, 1985.\hfill\break
     J.Mickelsson, "Current Algebra and Groups", Plenum, New York, 1989; in
     "Topological and Geometrical Methods in Field Theory", (J.Mickelsson
     and O.Pekonen Eds.), World Scientific, Singapore, 1992.\hfill\break
     "Symposium on Anomalies, Geometry and Topology" (W.A.Bardeen and A.R.White
     Eds.), World Scientific, Singapore, 1985.
\ref A.Chodos and V.Moncrief, J.Math.Phys. $\underline {21}$ (1980), 364.
\ref a) J.Schwinger, Phys.Rev. $\underline {125}$ (1962), 1043, $\underline
     {127}$ (1962), 324, $\underline {130}$ (1963), 402.\hfill\break
     b)R.D.Peccei, Phys.Rev. $\underline {D17}$ (1978), 1097.\hfill\break
     c) C.Benders, T.Eguchi and H.Pagels, Phys.Rev.  $\underline {D17}$
     (1978), 1086.\hfill\break
     d)S.Wadia and T.Yoneda, Phys.Lett. $\underline {66B}$ (1977), 341.\hfill
     \break
     e) L.F.Abbott and T.Eguchi, Phys.Lett. $\underline {72B}$ (1977), 215.
     \hfill\break
     f) A.Jevicki and N.Papanicolaou, Phys.Lett. $\underline {78B}$
     (1978), 438.
\ref a) D.Zwanziger, Phys.Lett. $\underline {114B}$ (1982), 337.
     Nucl.Phys. $\underline {B209}$ (1982), 336; $\underline
     {B323}$ (1989), 513; $\underline {B345}$ (1990), 461.
     In "Fundamental Problems of Gauge Field Theory", NATO-ASI series n.B141,
     (G.Velo and A.S.Wightman Eds.), Plenum, New York, 1986.\hfill\break
     b) G.Dell'Antonio and D.Zwanziger,  Nucl.Phys. $\underline {B326}$
     (1989),  333. "All Gauge Orbits
     and Some Gribov Copies Encompassed by the Gribov Horizon", in
     "Probabilistic Methods in Quantum Field Theory and Quantum Gravity",
     (P.H.Damgaard, H.H\"uffel and A.Rosemblum Eds.), Carg\`ese 1989,
     NATO ASI B224, Plenum, New York, 1990; Commun.Math.Phys. $\underline
     {138}$ (1991), 291.\hfill\break
     c) M.A.Semenov-Tyan-Shanskii and V.A.Franke, Zap.Nauch.Sem.Leningrad
     \break Otdeleniya Matematicheskogo Instituta im. V.A.Steklov AN SSSR,
     vol.120,p.159, 1982 (in russian).
\ref P.van Baal, Nucl.Phys. $\underline {B369}$ (1992), 259.
\ref F.S.Henyey, Phys.Rev. $\underline {D20}$ (1979), 1460.
\ref V.Rivasseau, "From Perturbative to Constructive Renormalization",
     Princeton Univ. Press, Princeton, 1991.
\ref M.Cantor, Bull.Am.Math.Soc. $\underline 5$ (1981), 235.
\ref J.York, J.Math.Phys. $\underline {14}$ (1973), 456.\hfill\break
     N.O'Murchadha and J.York, Phys.Rev. $\underline {D10}$ (1974), 428.
\ref a) J.Segal, J.Funct.Anal. $\underline {33}$ (1979), 175.\hfill\break
     b) J.Ginibre and G.Velo, Commun.Math.Phys. $\underline {82}$
     (1981), 1; Ann.Inst.H.Poincar\'e $\underline {XXXVI}$ (1982), 59
     (also scalar fields).\hfill\break
     c) D.Eardley and V.Moncrief, Commun.Math.Phys. $\underline {83}$
     (1982), 171 and 193.\hfill\break
     d) J.Sniatycki, Commun.Math.Phys. $\underline {141}$ (1991), 593.\hfill
     \break
     e) V.Georgiev and P.P.Schirmer, Commun.Math.Phys. $\underline {148}$
     (1992), 425.
\ref A.Karlhede, Class.Quantum Grav. $\underline 3$ (1986), L27.
\ref a)L.Bonora and P.Cotta-Ramusino, Commun.Math.Phys. $\underline {87}$
     (1983), 589.\hfill\break
     $b_1$) R.Stora, in "Progress in Gauge Field Theories", (G.'t Hooft,
     A.Jaffe, H.Lehmann, \break
     P.K.Mitter, I.M.Singer and R.Stora Eds.), NATO ASI B...,
     Plenum, New York, 1984.\hfill\break
     $b_2$) R.Stora, in "Non-Perturbative Methods", (S.Narison Ed.), World
     Scientific, Singapore, 1986.\hfill\break
     $c_1$) B.Zumino, in "Relativity, Groups, Topology II", Les Houches 1983,
     (B.S.de\break
     Witt and R.Stora Eds.), North Holland, Amsterdam, 1984 [reprinted in
     "Current Algebra and Anomalies", (S.B.Treiman, R.Jackiw, B.Zumino and
     E.Witten Eds.),  World Scientific, Singapore, 1985].\hfill\break
     $c_2$) B.Zumino, in "Symposium on Anomalies, Geometry, Topology", (W.A.
     \break Bardeen and A.R.White Eds.), World Scientific, Singapore, 1985.
     \hfill\break
     d) M.Quir\'os, F.J.de Urries, J.Hoyo.., M.L.Maz\'on and E.Rodriguez,
     J.Math.Phys. $\underline {22}$ (1981), 1767.
\ref a)  J.Thierry-Mieg, J.Math.Phys. $\underline {21}$ (1980), 2834; Nuovo
     Cimento $\underline {A56}$ (1980), 396.\hfill\break
     b) R.F.Pickens, J.Phys. $\underline {A19}$ (1986), L219.
\ref a) M.Henneaux and C.Teitelboim, Commun.Math.Phys. $\underline {115}$
     (1988), 213.\hfill\break
     b) J.Fisch, M.Henneaux, J.Stasheff and C.Teitelboim, Commun.Math.Phys.
     $\underline {120}$ (1989), 379.\hfill\break
     c) D.Kastler and R.Stora, J.Geom.Phys. $\underline 2$ (1985), 1.
\ref S.Hwang, Nucl.Phys. $\underline {B351}$ (1991), 425.
\ref a) F.J.Dyson, Phys.Rev. $\underline {75}$ (1949), 486.\hfill\break
     b) W.Magnus, Commun.Pure Appl.Math. $\underline {VII}$ (1954), 649.\hfill
     \break
     c) G.H.Weiss and A.A.Maradudin, J.Math.Phys. $\underline 3$ (1962), 771.
     \hfill\break
     d) R.M.Wilcox, J.Math.Phys. $\underline 8$ (1967), 962.\hfill\break
     e) I.Bialynicki-Birula, B.Mielnik and J.Pleba\'nski, Ann.Phys.(N.Y.)
     $\underline {51}$ (1969), 187;\hfill\break
     B.Mielnik and J.Pleba\'nski, Ann.Inst.H.Poincar\'e $\underline {A12}$
     (1970), 215.\hfill\break
     f) J.D.Dollard and C.N.Friedman, "Product Integration", vol.10 of
     Encyclopedia of mathematics and Its Applications, Addison, Reading, 1979.
     \hfill\break
     g) J.Czyz, "The Baker-Campbell-Hausdorf Formula: Derivation,
     Generalization and Applications, Math.Inst., Polish Academy of Sciences,
     preprint 477 (1990).\hfill\break
     h) R.Gilmore, "Lie Groups, Lie Algebras and Some of Their Applications",
     Wiley, New York, 1974.
\ref a) T.D.Lee, "Particle Physics and Introduction to Field Theory", Harwood,
     New York, 1981.\hfill\break
     b) S.S.Chang, Nucl.Phys. $\underline {B172}$ (1980), 335.\hfill\break
     c) Yu.A.Simonov, Sov.J.Nucl.Phys. $\underline {41}$ (1985), 835.
     \hfill\break
     d) K.Johnson, "The Yang-Mills Ground State", in "QCD - 20 Years Later",
     Aachen, June 1992.
\ref P.A.M.Dirac, Rev.Mod.Phys. $\underline {21}$ (1949), 392.
\ref L.Lusanna, in "Gauge Field Theories", XVIII Karpacz Winter School 1981,
     \break (W. Garczy\'nski Ed.), Harwood, Chur, 1986.
\ref K.Kuchar, Found.Phys. $\underline {16}$ (1986), 193.
\ref A.J.Hanson and T.Regge, Ann.Phys. (N.Y.) $\underline {87}$ (1974), 498.
\ref A.Barducci, L.Lusanna and E.Sorace, Nuovo Cimento $\underline {46B}$
     (1978), 287.
\ref M.Pauri, "Invariant Localization and Mass-Spin Relations in the
     Hamiltonian Formulation of Classical Relativistic Dynamics", Parma Univ.
     preprint IFPR-T-019, 1971 (unpublished); in "Group Theoretical Methods
     in Physics", (K.B.Wolf Ed.), Lecture Notes Phys. n.135, Springer, Berlin,
     1980.
\ref M.H.L.Pryce, Proc.Roy.Soc.(London) $\underline {195A}$ (1948), 6.
\ref G.N.Fleming, Phys.Rev. $\underline {137B}$ (1965), 188; $\underline
{139B}$
     (1965), 963.
\ref L.L.Foldy and S.A.Wouthuysen, Phys.Rev. $\underline {78}$ (1950), 29.
\ref T.D.Newton and E.P.Wigner, Rev.Mod.Phys. $\underline {21}$ (1949), 400.
\ref C.M\"oller, Ann.Inst.H.Poincar\'e $\underline {11}$ (1949), 251; "The
Theory
     of Relativity", Oxford Univ.Press, Oxford, 1957.
\ref A.D.Fokker, "Relativiteitstheorie", p.171, Noordhoff, Groningen, 1929.
\ref G.C.Hegerfeldt, Nucl.Phys. B (Proc.Suppl.) $\underline 6$ (1989), 231.
\ref J.Anandan, in "Conference on Differential Geometric Methods in
     Theoretical Physics", (G.Denardo and H.D.Doebner Eds.), World Scientific,
     Singapore, 1983.
\ref a) S.Tomonaga, Prog.Theor.Phys. $\underline 1$ (1946), 27.\hfill\break
     Z.Koba, T.Tati and S.Tomonaga, Prog.Theor.Phys. $\underline 2$ (1947),
     101 and 198.\hfill\break
     S.Kanesawa and S.Tomonaga, Prog.Theor.Phys. $\underline 3$ (1948), 1 and
     101.\hfill\break
     S.Tomonaga, Phys.Rev. $\underline {74}$, (1948) 224.\hfill\break
     b) J.Schwinger, Phys.Rev. $\underline {73}$ (1948), 416; $\underline {74}$
     (1948), 1439.

\vfill\eject
\bye